\documentclass[authoryear,preprint,12pt]{elsarticle}

\usepackage{amssymb, color}
 \usepackage{hyperref} 
\usepackage{rotating}
\usepackage{txfonts}
\usepackage{textcomp}
\usepackage{epstopdf}
\usepackage{lineno}
\usepackage{natbib}
\usepackage{subfig}
\usepackage{graphicx}
\usepackage{soul}
\newcommand{\quest}[1]{{\color[rgb]{0,0,0} #1}}
 \newcommand{\add}[1]{\quest{#1}}
\newcommand{\change}[2]{\quest{\hphantom{\vphantom{#1}}#2}}
\newcommand{\delete}[1]{\hphantom{\vphantom{#1}}}

\usepackage{lineno}

\journal{Advances in Space Research}

\graphicspath{{./}{figures/}}

\begin{document}

\begin{frontmatter}

%% Title, authors and addresses

%% use the tnoteref command within \title for footnotes;
%% use the tnotetext command for theassociated footnote;
%% use the fnref command within \author or \address for footnotes;
%% use the fntext command for theassociated footnote;
%% use the corref command within \author for corresponding author footnotes;
%% use the cortext command for theassociated footnote;
%% use the ead command for the email address,
%% and the form \ead[url] for the home page:
%% \title{Title\tnoteref{label1}}
%% \tnotetext[label1]{}
%% \author{Name\corref{cor1}\fnref{label2}}
%% \ead{email address}
%% \ead[url]{home page}
%% \fntext[label2]{}
%% \cortext[cor1]{}
%% \affiliation{organization={},
%%             addressline={},
%%             city={},
%%             postcode={},
%%             state={},
%%             country={}}
%% \fntext[label3]{}

% \title{Current Understanding and Predictive Capabilities of the Particle Radiation Environment in Heliosphere}
\title{Particle Radiation Environment in the Heliosphere: Status, limitations and recommendations}

%% use optional labels to link authors explicitly to addresses:
%% \author[label1,label2]{}
%% \affiliation[label1]{organization={},
%%             addressline={},
%%             city={},
%%             postcode={},
%%             state={},
%%             country={}}
%%
%% \affiliation[label2]{organization={},
%%             addressline={},
%%             city={},
%%             postcode={},
%%             state={},
%%             country={}}

\author[inst1,inst2]{Jingnan Guo}
\ead{jnguo@ustc.edu.cn}
%0000-0002-8707-076X 
\affiliation[inst1]{organization={Deep space Exploration Laboratory/School of Earth and Space Sciences, University of Science and Technology of China, Hefei, 230026, China}}
\affiliation[inst2]{organization={CAS Center for Excellence in Comparative Planetology USTC, Hefei, China}}

\author[inst3]{Bingbing Wang}
\ead{bw0121@uah.edu}   
\affiliation[inst3]{organization={Center for Space Plasma and Aeronomic Research (CSPAR), University of Alabama in Huntsville, Huntsville, AL 35899, USA}}

\author[inst4]{Kathryn Whitman}
\ead{kathryn.whitman@nasa.gov}
\affiliation[inst4]{organization={University of Houston, 4800 Calhoun Rd, Houston, TX 77204, USA}}

\author[inst5]{Christina Plainaki}
\ead{christina.plainaki@asi.it}
\affiliation[inst5]{organization={Italian Space Agency, Rome, Italy}}

\author[inst3,inst6]{Lingling Zhao}
\ead{lz0009@uah.edu}
%0000-0002-4299-0490
\affiliation[inst6]{organization={Department of Space Science, University of Alabama in Huntsville, Huntsville, AL 35899, USA}}

\author[inst7,inst8]{Hazel M. Bain}
\ead{hazel.bain@noaa.gov}
%0000-0003-2595-3185
\affiliation[inst7]{organization={Cooperative Institute for Research in Environmental Sciences, University of Colorado Boulder, CO, USA}}
\affiliation[int8]{organization={NOAA, Space Weather Prediction Center, Boulder, CO, USA}}

\author[inst9]{Christina Cohen}
\ead{cohen@srl.caltech.edu}
%0000-0002-0978-8127
\affiliation[inst9]{organization={California Institute of Technology, Pasadena, CA, USA}}

\author[inst10]{Silvia Dalla}
\ead{SDalla@uclan.ac.uk}
%0000-0002-7837-5780
\affiliation[inst10]{organization={Jeremiah Horrocks Institute, University of Central Lancashire, Preston, PR1 2HE, UK}}

\author[inst11]{Mateja Dumbovic}
\ead{mdumbovic@geof.hr}
\affiliation[inst11]{organization={University of Zagreb, Faculty of Geodesy, Hvar Observatory, Zagreb, Croatia}}

\author[inst12,inst13]{Miho Janvier}
% ORCID: 0000-0002-6203-5239
\ead{miho.janvier@universite-paris-saclay.fr}
\affiliation[inst12]{organization={European Space Agency, ESTEC, Noordwijk, The Netherlands}}
\affiliation[inst13]{organization={Université Paris-Saclay, CNRS, Institut d’Astrophysique Spatiale, 91405 Orsay, France}}

\author[inst9]{Insoo Jun}
\ead{insoo.jun@jpl.nasa.gov}
\affiliation[inst9]{organization={California Institute of Technology, Pasadena, CA, USA}}

\author[inst14]{Janet Luhmann}
\ead{jgluhman@ssl.berkeley.edu}
\affiliation[inst14]{organization={Space Sciences Laboratory of the University of California, Berkeley, USA}}

\author[inst15]{Olga E. Malandraki}
\ead{omaland@noa.gr}
\affiliation[inst15]{organization={Institute for Astronomy, Astrophysics, Space Applications and Remote Sensing (IAASARS), National Observatory of Athens, Greece}}

\author[inst16]{M. Leila Mays}
\ead{m.leila.mays@nasa.gov}
\affiliation[inst16]{organization={NASA Goddard Space Flight Center, USA}}

\author[inst17]{Jamie S. Rankin}
\ead{jsrankin@princeton.edu}
\affiliation[inst17]{organization={Department of Astrophysical Sciences, Princeton University, Princeton, NJ 08540, USA}}

\author[inst18]{Linghua Wang}
\ead{wanglhwang@pku.edu.cn}
%0000-0001-7309-4325
\affiliation[inst18]{organization={School of Earth and Space Sciences, Peking University, Beijing, 100871, China}}

\author[inst19]{Yihua Zheng}
\ead{yihua.zheng@nasa.gov}
\affiliation[inst19]{organization={Space Weather Laboratory, NASA Goddard Space Flight Center, Greenbelt, MD, USA}}

\begin{abstract}
Space weather is a multidisciplinary research area connecting \delete{science and} scientists from across heliophysics domains \change{toward}{seeking} a coherent understanding of our space environment that can also serve modern life and society’s needs. 
COSPAR’s ISWAT (International Space Weather Action Teams) ‘clusters’ focus attention on different areas of space weather study while ensuring the coupled system is broadly addressed via regular communications and interactions. 
The ISWAT cluster ``H3: Radiation Environment in the Heliosphere'' (\url{https://www.iswat-cospar.org/h3}) has been working to provide a scientific platform to understand, characterize and predict the energetic particle radiation in the heliosphere with the practical goal of mitigating radiation risks associated with areospace activities, satellite industry and human space explorations. 
In particular, present approaches help us understand the physical phenomena at large, optimizing the output of multi-view-point observations and pushing current models to their limits.

In this paper, we review the scientific aspects of the radiation environment in the heliosphere covering four different radiation types: Solar Energetic Particles (\change{SEP}{SEPs}), Ground Level Enhancement (GLE, \change{an extreme}{a} type of SEP events \add{with energies high enough to trigger the enhancement of ground-level detectors}), Galactic Cosmic Rays (\change{GCR}{GCRs}) and Anomalous Cosmic Rays (\change{ACR}{ACRs}). 
We focus on related advances in the research community in the past 10-20 years and what we still lack in terms of understanding and predictive capabilities. 
Finally we also consider some recommendations related to the improvement of both observational and modeling capabilities in the field of space radiation environment.
\end{abstract}

%%Graphical abstract
%\begin{graphicalabstract}
%\includegraphics{grabs}
%\end{graphicalabstract}

%%Research highlights
%\begin{highlights}
%\item Research highlight 1
%\item Research highlight 2
%\end{highlights}

\begin{keyword}
%% keywords here, in the form: keyword \sep keyword
Space Weather \sep Space Radiation \sep Solar Energetic Particles \sep Galactic Cosmic Rays \sep Ground Level Enhancement \sep Anomalous Cosmic Rays
%% PACS codes here, in the form: \PACS code \sep code
%\PACS 0000 \sep 1111
%% MSC codes here, in the form: \MSC code \sep code
%% or \MSC[2008] code \sep code (2000 is the default)
%\MSC 0000 \sep 1111
\end{keyword}

\end{frontmatter}

%\linenumbers

% \citep{} with reference in parenthesis at the end of the sentence  
% \cite{} for text with in-line reference

\section{Introduction and Motivation}
\label{sec:intro}

\begin{figure}[ht!]
\centering
\includegraphics[trim=5 5 5 5,clip, width=\textwidth]{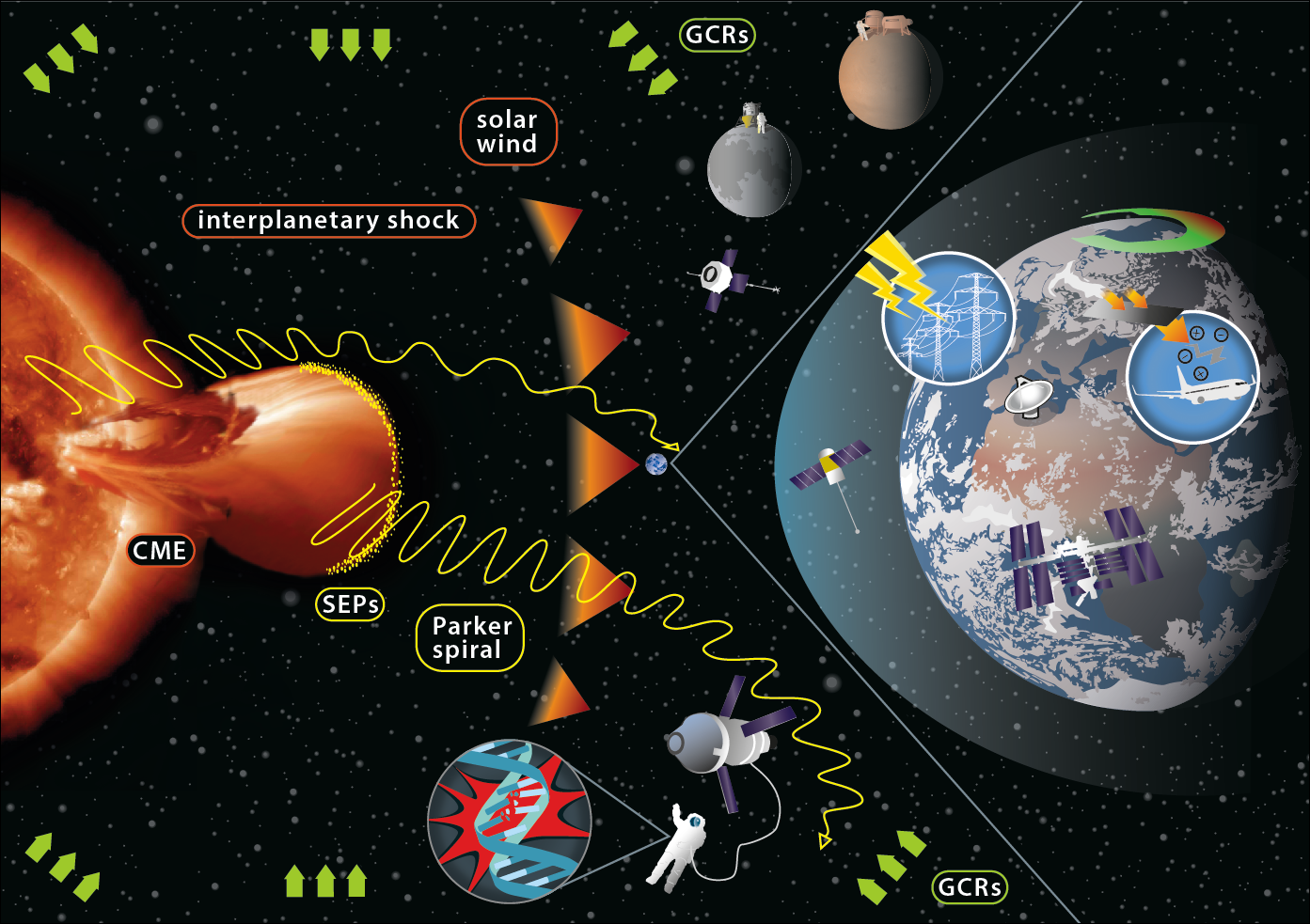}
\caption{A cartoon illustration of Space weather phenomena and impacts. More discussions can be found in the main text.}
\label{fig:SW_intro}
\end{figure}

Space weather studies generally address the conditions on the Sun and in the heliosphere and how they impact the interplanetary and planetary space environment, influence the performance and reliability of space-borne and ground-based technological systems and endanger human life or health, as illustrated in Fig.~\ref{fig:SW_intro} (also see the US National Space Weather Plan for the up-to-date definition of space weather \url{http://www.nswp.gov/}). 

In the past years, space weather has appropriately become a multidisciplinary research area connecting scientists from across all heliophysics domains from solar physics to aeronomy. The international members of this community work together under the umbrella of the COSPAR Panel on Space Weather, the International Space Weather Action Teams (ISWAT, \url{https://iswat-cospar.org/}) which were recently established. 
Each ISWAT Action Team is an international group of interested researchers charged with addressing a specific and focused space weather-related task. Action Teams are in turn organized into various ``clusters'' grouped by domain, phenomena or impact, and work in coordinated efforts to improve the scientific understanding of space weather phenomena and our society’s resilience to the effects of space weather.  
The goal of the ISWAT H3 cluster (\url{https://iswat-cospar.org/h3}) is to understand, characterize and predict the fluxes of the major sources of energetic particle radiation in the heliosphere including Solar Energetic Particles (SEPs), Galactic Cosmic Rays (GCRs) and Anomalous Cosmic Rays (ACRs). 

\subsection{Introduction to the heliospheric radiation environment}
\begin{figure}[ht!]
\centering
\includegraphics[width=1.05\textwidth]{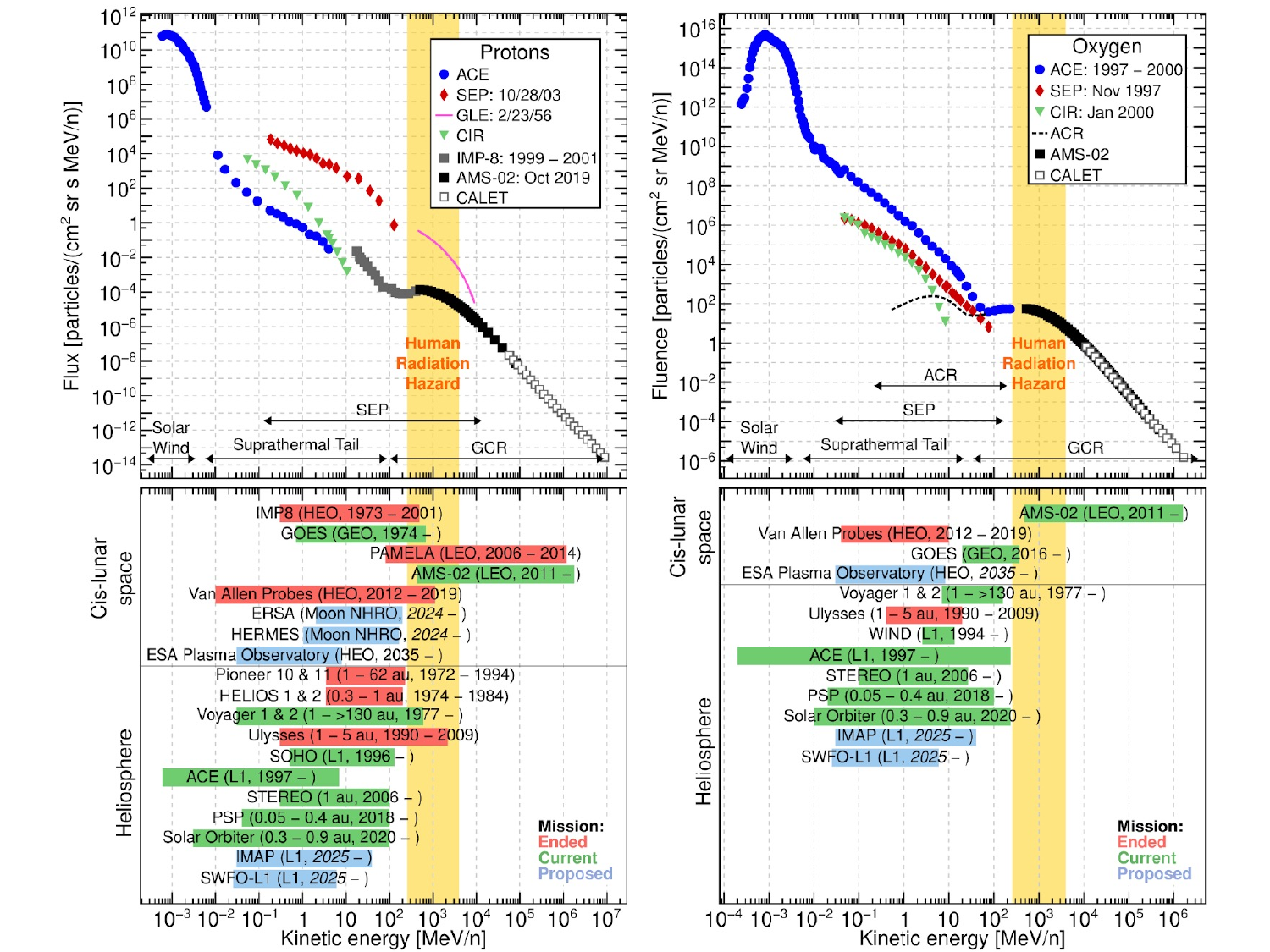}
\caption{Proton ({\it top left}) and oxygen ({\it top right}) fluence spectra  showing the energy ranges of various particle populations in the heliosphere. The coverages of in situ particle detectors ({\it bottom panels}) are also shown. The vertical yellow band highlights the energy range most relevant for human radiation exposure \add{behind shielding in deep space environment}. (Image credit: Claudio Corti \citep{Corti2022}).}
\label{fig:spec}
\end{figure}

SEPs are protons, heavier ions and electrons that are accelerated by solar flares and shocks driven by Coronal Mass Ejections (CMEs) \citep[see][and references therein]{cohen2021SEP}, and have energies from a few keV up to several GeV (Fig. \ref{fig:spec}). In \change{the largest}{some very large} SEP events, the presence of protons of energies up to tens of \change{GeVs}{GeV} can contribute to the Earth's surface particle fluxes and be detected through ground-based neutron monitors, in so-called Ground Level Enhancement (GLE) events. 
SEP events occur sporadically, leading to intense but temporary increases in radiation levels in the heliosphere that can commence with little (few minutes) warning following observed solar activity. Their occurrence rate follows the 11-year solar activity cycle, with more events taking place during solar maximum \add{and also during the declining phase of the the cycle, especially those leading to GLEs \citep[e.g.,][]{shea1993history}}. 
Being electrically charged, SEPs gyrate around and follow the interplanetary magnetic field lines (with the nominal condition described as Parker spirals) such that the largest particle fluxes tend to be related to flares and CME-shocks that had a direct magnetic connection to the observer \citep[e.g.,][]{klein2017acceleration}. However, SEPs can be distributed more widely in the heliosphere indicating that additional transport processes, often modeled as diffusion, take place either near the Sun or in the solar wind \citep[e.g.,][]{desai2016large}. 
The physics of these processes is poorly understood at present, as are the processes accelerating SEPs. In particular, the roles of reconnection in solar flares versus acceleration by CME-driven shocks, which become spatially extended sources in themselves, continues to be under debate. In addition, theories of particle acceleration by shocks may require the presence of a ``seed population'', i.e., suprathermal particles, at the Sun and/or in the interplanetary medium that are difficult to quantify \citep[e.g.,][]{cohen2021SEP}. Thus, the successful prediction of the contribution of SEPs to the radiation environment in the heliosphere is challenging because there is much uncertainty in fundamental processes involved, and little forewarning of when a SEP event will occur. For instance, we don't know when an active region will erupt or what flare size it will create or how big a CME will be or if any of that will accelerate particles and to what energies.  
More detailed discussions concerning the current understanding and limitations as well as forecasting capability of SEPs and GLE events can be found in Sections \ref{sec:SEP} and \ref{sec:GLE}, respectively.

GCRs are particles with high energies (\change{hundreds of MeV to GeV energies}{from tens of MeV up to 10$^8$ GeV with their fluxes peaking around GeV/nuc}\delete{, also shown in Fig. 2}) most likely accelerated by supernova-driven shocks, that enter the heliosphere from the interstellar medium \citep{blasi2013origin}. Once in the heliosphere, they are subject to solar modulation processes that are dominated by the large-scale structure of the \change{interplanetary}{heliospheric} magnetic field and its variation during the solar activity cycle \citep[e.g.,][]{potgieter13}. This gives rise to 11-year \delete{(and 22-year)} solar-cycle \add{(and 22-year solar magnetic cycle) }variations in the GCR intensity which is anti-correlated with solar activity \delete{and related to the configuration of the global heliospheric magnetic field}. \add{GCRs are nearly isotropic in the interplanetary space. However, a small anisotropy has been observed and it may also change following the solar magnetic cycle \citep[e.g.,][]{Modzelewska2019}.} The long-term solar cycle variations in the GCR flux are reasonably well characterized in comparison to SEPs. However, some observations such as the lag of the GCR variation after the solar activity cycle still cannot be fully explained and the detailed quantification of the modulation is still needed. %Extensive work is being carried out to understand the details of the modulation mechanisms. 
Meanwhile, short term depressions of the GCR flux that result from transient disturbances in the solar wind (such as interplanetary shocks and CMEs, or stream interaction regions (SIRs)), are poorly understood and difficult to predict \citep[e.g.,][]{richardson11, richardson18}.  When assessing radiation risk, the risks from GCRs pose a bigger challenge than SEPs for long-term space explorations such as a mission to Mars \citep{cucinotta2013safe, guo2021radiation} as they contribute cumulatively throughout the mission and particles with energy above hundreds of MeV/nuc are very difficult to shield against \add{(see the energy range of GCRs shown in Fig. \ref{fig:spec})}. More detailed discussions on the current understanding and open questions related to GCRs can be found in Section \ref{sec:GCR}.

ACRs are a population of predominantly singly ionized ions that generally include the elements H, He, C, N, O, Ne, and Ar located at the low-energy part of the GCR spectra (Fig. \ref{fig:spec}). They are believed to result from interstellar neutrals that have entered the heliosphere and are ionized by photonionization, electron impact ionization, or charge exchange to become pickup ions \citep{Fisk1974}. These pickup ions are then convected into the outer heliosphere, where they are accelerated at or near the heliospheric termination shock and can reach energies of about 1 to 100 MeV/nuc \citep{McComas2006}. However, the acceleration mechanism of ACRs is still under debate. Since ACRs are transported in the same heliospheric environment as GCRs, they also experience the solar modulation effect \citep{Fu2021}. Presently the differences in the ACR and GCR modulation by the solar activity are not fully understood. They are probably related to the different source origin and rigidity dependence of these two populations.
More detailed discussions on the current understanding and open questions concerning ACRs can be found in Section \ref{sec:ACR}.

\subsection{Space radiation risks}

\begin{figure}[ht!]
\centering
\includegraphics[width=0.95\textwidth]{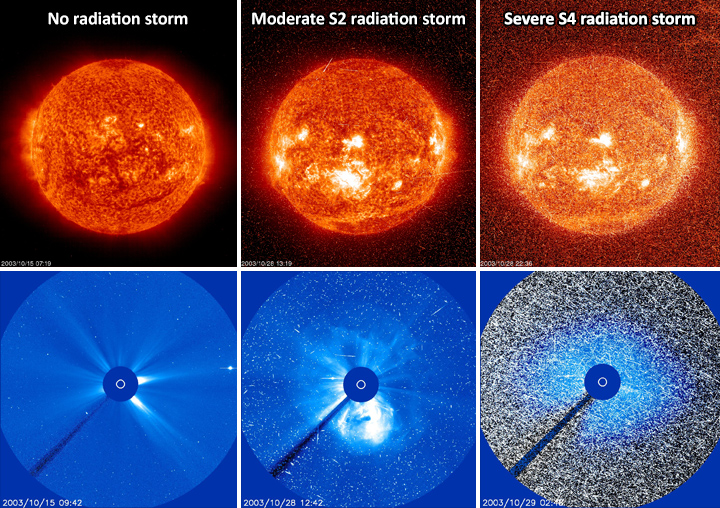}
\caption{\change{Illustration of various levels of radiation storms and signals are triggered by energetic particles hitting the CCD of the imaging telescopes of the SOHO satellites during solar radiations storms.}{Normal (S0, left column), Moderate (S2, middle column), and Severe (S4, right column) radiation storms illustrated using signals triggered by the Solar and Heliospheric Observatory (SOHO) satellites.} Image is taken from \url{https://www.spaceweatherlive.com/en/help/what-is-a-solar-radiation-storm.html}. }
\label{fig:radiation_storm}
\end{figure}

Space radiation is a major concern for the safety of robotic and human exploration both in the near-Earth environment and towards deep space and other planets such as Mars. 
Near Earth, the National Oceanic and Atmospheric Administration (NOAA) uses a five-level system, the solar radiation storm scale i.e. the S-scale\footnote{\url{https://www.swpc.noaa.gov/noaa-scales-explanation}}, to indicate the severity of a solar radiation storm and the anticipated impacts to satellites, \change{HF}{high frequency (HF)} communications, astronauts and flight crew and passengers on high flying aircraft at high latitudes. This scale ranges from S1 to S5, corresponding to minor, moderate, strong, severe and extreme levels of radiation. Fig. \ref{fig:radiation_storm} illustrates various levels of radiation storms and camera signals triggered by energetic particles. The severity is determined by examining the $>$10 MeV proton flux: the threshold to be exceeded is 10, 10$^2$, 10$^3$, 10$^4$, 10$^5$ pfu \add{(1pfu = 1 particle $\cdot$ cm$^{-2}$ $\cdot$ s$^{-1}$ $\cdot$ ster$^{-1}$)} for S1, S2, S34, S4, and S5 storms, respectively, as observed by the GOES spacecraft particle sensors. The severity of these storms is anti-correlated with their occurrence frequency. 

With varying degrees of impacts, space radiation over a wide energy range and particle types can affect modern technological systems in various ways.
For example, sudden enhancement of energetic protons can penetrate into the atmosphere and ionize the D-layer of the ionosphere. This process prevents the \change{High Frequency (HF)}{HF} radio waves from reaching the much higher E, F1 and F2 layers where these radio signals normally refract and bounce back to Earth, thus causing degraded HF radio communications. SEPs may also produce high latitude atmospheric chemistry changes including ozone depletion \citep{Maliniemi2022}. 
Both SEPs and GCRs can affect satellite, spacecraft and their instruments in various ways and cause temporary or permanent damages\citep{stassinopoulos1988space}. 
Single Event Effects (SEE) refer to the deposition of charge in spacecraft circuits which can cause, for example, upset, latch up or burn out. %Internal Charging (IC) and Electrostatic Discharge (ESD) can result in electrical breakdown. Surface Charging (SC) from charged particles collecting on spacecraft surfaces can produce high voltages which leads to arcing and electromagnetic interference. 
Internal Charging (IC) or Internal Electrostatic Discharge (IESD) is a phenomena where energetic particles deposit their charges in materials inside the spacecraft structure, ultimately causing electrical breakdown. 
Another phenomenon called Surface charging (SC) occurs when the incoming \change{electrons are below about 100 keV which are collected}{electrons with energies below about 100 keV accumulate} on spacecraft surfaces and produce surface discharges leading to arcing and electromagnetic interference.
Total Ionizing Dose (TID) effects can lead to long term radiation damage. With the above effects, energetic protons can degrade solar panel efficiency, on board electronic circuitry can malfunction and the protons \change{will}{can} create noise in star-tracking systems.

Space radiation also poses a radiation hazard for human space exploration endeavours. By showing the energy range of SEPs, GCRs and ACRs as measured by current and upcoming in situ particle detectors, Fig. \ref{fig:spec} highlights a current and prospective future gap in energetic particle measurements in the critical energy range concerning human radiation risks \citep{Corti2022}. 
As shown, GCRs in the energy range of 250 MeV/nuc to 4 GeV/nuc \citep{slaba2014gcr, dobynde2021beating} and SEP protons above 100 MeV \citep{mertens2019} are the most significant contributors to radiation dose in humans behind shielding.
For a potential human mission to Mars which generally requires a long mission duration (about 3 years), both modeled results and measurements show that NASA's exposure limits are approached or exceeded \citep{CUCINOTTA20171, guo2021radiation}. Chronic exposure to the GCR radiation environment does not immediately endanger the astronauts' life, but it increases the probability of late-term consequences \citep[e.g.,][]{cucinotta2006, Kennedy2014}, such as development of cancer and cataracts or damage to the central nervous system and/or cardiovascular system and hereditary effects \citep{iancu2018space}. Alternatively, intense SEP events, apart from contributing to the above stochastic \change{effect}{effects}, can be also associated with deterministic effects as they may deliver short-term, intense dosages that cause radiation poisoning or even death.   

Even for aviation activities, radiation  poses potential risks \citep{sihver2015radiation}. During the most energetic SEP events, i.e., those which result in a GLE event at Earth, the radiation environment at aviation altitudes can be significantly enhanced, particularly for flight routes over in the geomagnetic polar regions \citep{copeland08, meier2020radiation, dobynde2023aircraft}. In recent years, the International Civil Aviation Organisation (ICAO) has requested new space weather advisories specifically tailored to the needs of the aviation industry, including a requirement for advisories notifying operators and airlines of SEP events which increase the effective dose rates at commercial flight levels above preset threshold levels \citep[e.g.,][]{bain23noaa}. Therefore, improved scientific understanding, modeling and forecasting of these events will be required for both crews at aviation altitudes and in space \citep[e.g.,][]{hands2022new, bain23improved}.

\subsection{The importance of studying space radiation}

Following the brief discussions in the previous sections, it is therefore important to study energetic particle radiation in the heliosphere driven by the space weather forecasting requirements and also by scientific interest in understanding particle energization and transport processes. We stress the following important aspects for studying the space radiation environment (which will also be discussed in detail in Sections \ref{sec:SEP} - \ref{sec:ACR}): 
\begin{itemize}
    \item To constrain physical mechanisms of SEP acceleration at the Sun 
    \item To understand SEP transport processes in the heliosphere
    \item To understand the large-scale heliospheric environment through which GCRs and ACRs propagate
    \item To forecast and nowcast SEP occurrence, fluence, energy range and ionization effects which may pose radiation risks to aviation and space industries
    \item To understand and predict the variability of GCR radiation (both of its long-term and short-term modulations) for better mitigating long-term radiation risks
\end{itemize}

\subsection{Context and structure of this paper}

In this paper, we review the scientific aspects of the radiation environment in the heliosphere with a focus on advances in the past 10 years concerning our current scientific understanding and predictive capabilities. We also discuss our limitations in knowledge and open questions in the field, and offer considerations related to the planning of future space observations.%, in particular in view of human space exploration.

Sections \ref{sec:SEP} - \ref{sec:ACR} will provide an overview of our current understanding of energetic particles in the heliosphere including SEPs, GLE events, GCRs and ACRs. As GLE events are SEP events with fluxes at relativistic proton energies high enough to generate secondary particles that trigger signals in ground-based particle detectors,  they are discussed separately as different methods are often required to study the interaction of GLE particles with planetary magnetospheres and atmospheres. 

In section \ref{sec:Recommendations}, we will stress our knowledge gaps in observing, modeling, physical understanding, and forecasting capabilities of space particle radiation and we offer suggestions for narrowing these gaps and moving the field forward in the next 5 to 10 years. 

\section{Solar Energetic Particles}\label{sec:SEP}

\begin{figure}[ht!]
\centering
\includegraphics[trim=5 5 5 5,clip, width=\textwidth]{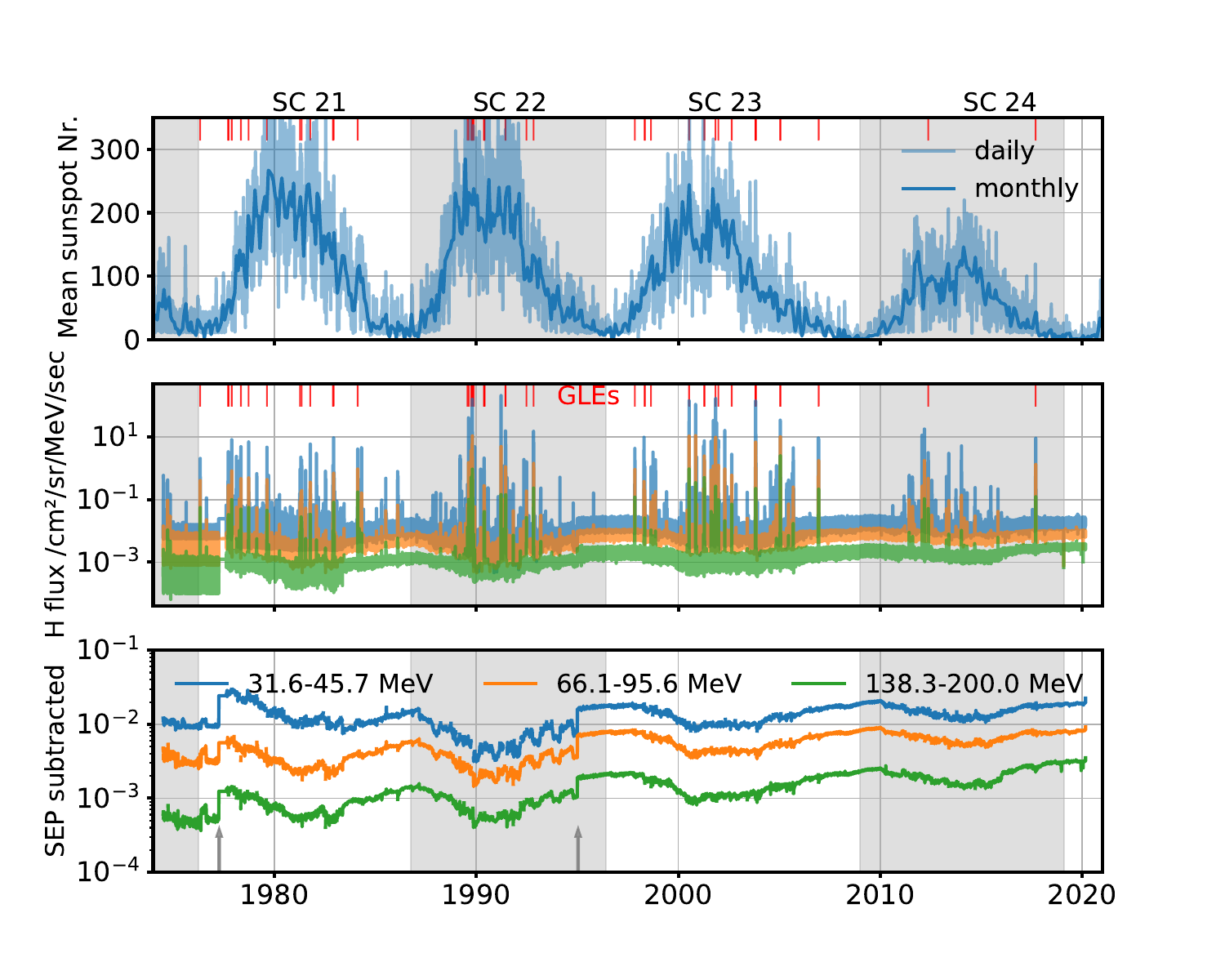}
\caption{Top: The daily and monthly averaged value of sunspot numbers over different solar cycles. Middle: The 5-min resolution proton fluxes binned into three different energies (31.6-45.7 MeV, 66.1-95.6 MeV, 138.3-200 MeV, as plotted in different colors) as calibrated from GOES-series measurements by the SEPEM service for data before 2016 \citep{crosby2015sepem} and by applying the GOES recalibrated effective energies \citep{sandberg2014cross} for data from 2016 to 2020. 
Bottom: Daily-averaged GCR flux with SEPs removed. The gray arrows mark two noticeable \change{discontinuity}{discontinuities} of the background flux resulting from change of GCR effective energy across different instruments (whilst the calibration makes sure that the SEP effective energy is consistent over time). 
To differentiate adjacent solar cycles, even cycles are shaded in grey in all three panels. 
Times of historical GLEs are marked by red vertical lines in the top two panels.
The monthly sunspot data \add{in the top panel} are downloaded from the World Data Center SILSO, Royal Observatory of Belgium, Brussels (\url{http://www.sidc.be/silso/monthlyssnplot}). The SEPEM V2.0 data \add{in the lower panels} are obtained from the ESA's Solar Energetic Particle Environment Modelling (SEPEM) application server (\url{http://sepem.eu}). }
\label{fig:pfl}
\end{figure}

\subsection{Introduction}
\label{sec:sep_intro}
Solar Energetic Particles (SEPs) are transient enhancements (with a duration varying from hours to days) in ions (mainly protons) and electrons in the space environment associated with solar activity. 
SEPs have been observed in the Earth’s space environment since the early years of the space age \cite[e.g.,][]{arnoldy1968energetic}, either via intentional detection with various forms of dosimeters and particle telescopes, or as backgrounds in other forms of data-including images. Improvements in detector technology subsequently led to species identifications (including ion composition), charge states, and anisotropies (directionality). 

The detection of SEPs in the last decades has revealed that SEPs span a very broad energy range that is from the suprathermal energy range (a few 10s of keV/nuc) up to a few GeV/nuc, for ions, or up to a few MeV for electrons. Most SEP protons have energy in between the solar wind plasma energies and the higher energies ($>$100 MeV) where the Anomalous and Galactic Cosmic Ray populations (see sections \ref{sec:GCR} and \ref{sec:ACR}) dominate the heliospheric fluxes. Fig. \ref{fig:spec}, as already introduced in Section \ref{sec:intro}, illustrates the energy range of the SEPs and GCRs, and typical relative fluxes, within the heliospheric particle energy spectrum.

The bottom panels of Fig. \ref{fig:spec} show that SEP fluxes are currently monitored in space by instruments onboard the following spacecraft near Earth\footnote{GOES stands for Geostationary Operational Environmental Satellite\citep{sauer89, Rodriguez14, kress20}; SOHO stands for Solar and Heliospheric Observatory \citep{domingo1995soho}; ACE is for Advanced Composition Explorer \citep{stone1998ace}; WIND refers to the Global Geospace Science Wind satellite \citep{acuna1995wind}; AMS-02 is the Alpha Magnetic Spectrometer \citep{kounine2012ams} that is mounted on the International Space Station (ISS). GSO is geosynchronous orbit; LEO is low earth orbit.}: GOES 1-17 (since 1974, at GSO and also serving for real-time forecasting), SOHO (since 1996, at the Lagrangian point L1), ACE (since 1997, at L1), WIND (since 1994, at L1) and AMS-02 (since 2011, at LEO). 
At LEO altitude, strong magnetic shielding of SEPs by Earth's magnetosphere can modulate SEP fluxes, causing them to be different from those measured at GSO by GOES or at L1.
Fig. \ref{fig:pfl} shows the 5-min resolution proton fluxes measured by the GOES series of spacecraft from 1974 until 2020 over more than four solar cycles. To account for the different instrument responses, GOES data have been calibrated with IMP8 (in service from 1973 to 2001) data and re-binned into different energies \citep{sandberg2014cross, crosby2015sepem}. 
The figure illustrates the episodic nature of the occurrence of the SEP populations (spikes in particle flux in the middle panel) and the solar activity cycle dependence of SEP events (top panel), superposed on a background tied to the slowly varying GCR fluxes (bottom panel, Section \ref{sec:GCR}) anti-correlated with long-term solar activities.

%Although the SEPEM energy channels give fluxes at the stated energies during SEP events, this energy may vary during background times due to the broad channel response of the underlying radiation monitors from GOES (and earlier SMS) and the hardness of the GCR spectra. Whilst this is likely to be a shift upwards in energy it is not easy to quantify without the underlying instrument response functions and it may differ due to differing widths of the underlying channel responses. Variations in the background can be assumed to result from GCR evolution over the solar cycle but additional sources of background such as thermal instrument noise cannot be discounted.

Both flares and the resulting heliospheric disturbances associated with coronal mass ejections (CMEs) are considered sources for accelerating particles \citep{reames2013two, reames2015sources, cliver2016flare}. However, the relative role of flare or CME association with particle energization is still under debate. 
The exact transport effects of SEPs in the heliosphere and how they can modify SEP properties are unclear.
Understanding the physical mechanisms responsible for the energization of SEPs and their transport through the heliosphere is key towards better forecasts of the heliospheric radiation environment. 

\subsection{Recent Progress and Current Understanding} \label{sec:sep_progress}

In the last decades, important progress has been made on the interconnection, within the space weather chain, of the various episodes of a SEP event related to the generation, acceleration, and propagation of solar particles \citep[see, for instance, ][and references therein]{desai2016large, klein2017acceleration, cohen2021SEP}. 
However, both our knowledge on the exact physical mechanisms taking place within the various regions of the solar atmosphere and/or in the interplanetary space, and our capacity to predict relativistic SEP events are still limited \citep{Anastasiadis2019, cliver2022extreme}. 

\subsubsection{SEP acceleration sources}

\begin{figure}[ht!]
\subfloat{\includegraphics[width=0.487\columnwidth]{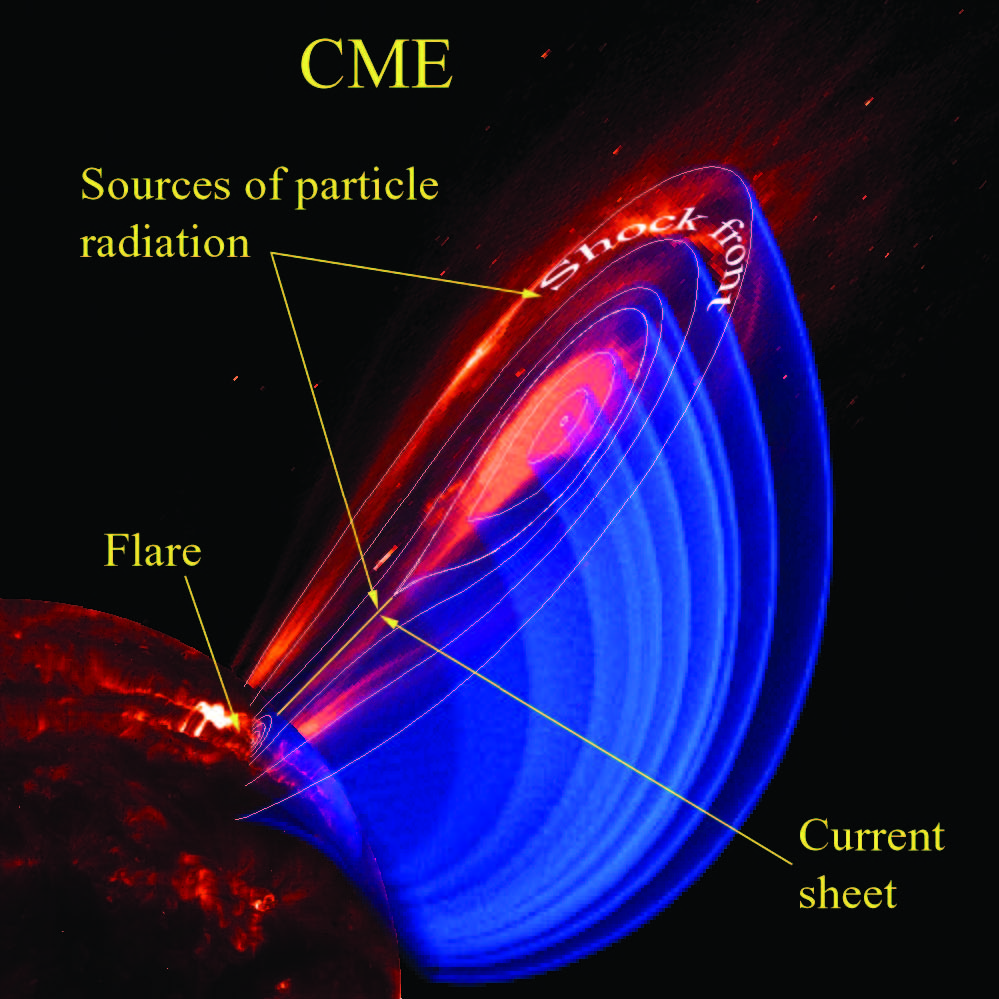}}
\subfloat{\includegraphics[width=0.48\columnwidth]{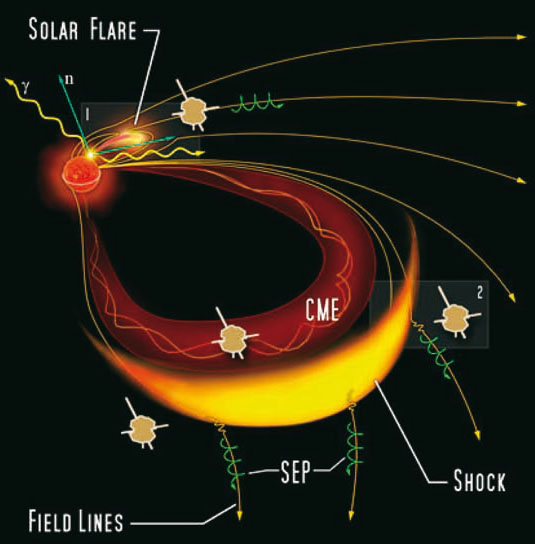}}\\
\caption{Left: Illustration of a generic flare and CME system related to the production of energetic particles close to the Sun (Image from \citet{NRC2006} as adapted from \citet{kohl2006ultraviolet}; Copyright 2006, Springer). %\url{http://sepem.eu} ). 
Right: Illustration of SEPs associated with the interplanetary shock and the SEP propagation from the acceleration sites along interplanetary magnetic field lines \citep[Image from][]{lin2006solar}. } \label{fig:SEP_acc}
\end{figure}

It was widely accepted that flares are often associated with relatively weak, electron-rich, and $^3$He-rich events with impulsive (sudden) onsets and less than a day duration, whilst the CME-driven shocks are more responsible for the spatially (up to circumsolar) and temporally (up to several days) extended enhancements known as gradual events \citep{reames1999particle, reames2013two, reames2015sources, desai2016large}. 
As shown in Fig. \ref{fig:SEP_acc}, the conceptual picture of SEP acceleration includes the 1) ``flare source'', where energization is associated with the flare processes via e.g., wave-particle interactions and magnetic reconnection and 2) the ``shock source'', where theoretical work suggested ambient plasma compression and collisionless shock can energize particles via shock drift acceleration at quasi-perpendicular shocks or diffusive shock acceleration at quasi-parallel shocks. Shock acceleration may take place in the solar corona close to the original eruption site or in the interplanetary space as the CME shock propagates outward. Interested readers can find more details on the acceleration processes in a review article by \citet{klein2017acceleration} and references therein. 

%seed particles
The suprathermal particles in the corona and solar wind \citep[e.g.,][]{mason2012ssr, wang2015} can provide seed populations for the acceleration/generation of SEPs. In fact, the composition of SEPs carries key information on the seed populations and particle acceleration process in these events. Gradual SEPs generally exhibit the large proton-to-electron ratios, as well as the ion compositions and charge states similar to the corona and solar wind. This indicates that their seed populations can directly originate from the corona and solar wind, while their formation processes likely lead to the preferential acceleration of protons \citep{cohen2021SEP}.

Impulsive electron/$^3$He-rich SEPs are typically characterized by small proton-to-electron ratios \citep{wangLY2012}, strong enhancements of $^3$He, significant enhancements of heavy nuclei such as Fe, extreme enhancements of ultra-heavy nuclei up to $>$ 200 amu, and high heavy-ion ionization states \citep{mason2007}. Thus, this more common type of SEPs would favor the preferential acceleration of electrons and heavy ions such as $^3$He, Fe, etc \citep[e.g.,][]{liu2006, mason2007}.

However, the traditional view that impulsive/$^3$He-rich/small SEPs are associated with flares whilst gradual SEPs are related to CMEs has been challenged \citep[e.g.,][]{mason1999particle, cohen1999new, Wiedenbeck2012} and the relative contribution of flare-related mechanisms and CME-driven shocks to particle energization remains an open question \citep{Cane2003, Cane2006role}.
The difficulty in disentangling the role of the two types of processes is due to the fact that a SEP event is often associated with the onsets of both a solar flare and a CME. Both flare and CME-shock acceleration may contribute to the energized SEPs during the same event and their relative role is likely to depend on particle energy \citep{Dier2015}. In addition, the transport conditions of SEPs in interplanetary space further blur the cause/effect associations \citep[e.g.,][]{klein2017acceleration}. 
A number of studies have concluded that at low proton energies (up to $\sim$10s of MeV) energisation at CME-driven shocks plays a key role in particle acceleration in the corona and the interplanetary space \citep[e.g.,][and references therein]{desai2016large}. 
The sources of acceleration at relativistic particle energies, for example during GLEs, might include flare processes, CME shock acceleration or both \citep[e.g.,][]{Mas2009,Asc2012}.
High-sensitivity observations show that solar energetic electron events only have a weak association with solar flares \citep{wangLY2012}. 

%\subsubsection{Solar source observations}
% source observations
One important way to better understand the energization close to the Sun and the injection process into interplanetary space is to use remote-sensing solar images and radio burst observations provided by ground- and spacecraft-based solar telescopes. 
For instance, X-ray and $\gamma$-ray emissions can be produced by accelerated particles during solar flares (via bremsstrahlung emission or nuclear collisions) and their observations provide direct information on particle acceleration at the solar corona \citep{cliver1989solar}. Therefore, they give complementary diagnostics to the escaping SEPs seen by in situ spacecraft. 
% HXR: RHESSI and the Chinese Advanced Space-based Solar Observatory \citep[ASO-S, launched in 2022 October][]{gan2019advanced}. 
Meanwhile, radio emission at different wavelengths provides important information of the source and injection, such as shock propagation can be closely related to type II radio emission and escaping non-thermal electrons are associated with type III bursts. Thus, radio emissions can give direct proof of the release of SEPs into interplanetary space \citep{kouloumvakos2015properties}. 
Moreover, coronal observations of the flare or CME eruption processes and shock dynamics, via e.g., coronagraph images, and their association with the properties of SEPs give important information on the acceleration and injection process of SEPs \citep{kahler1994injection, tylka2005shock, Dier2015}.

\subsubsection{Transport in the heliosphere}
% transport: mechanisms
Once the energized particles reach interplanetary space, they  travel preferentially along the magnetic field and observers with the best magnetic connectivity normally see the earliest onset of SEPs. However, turbulence in the interplanetary magnetic field plays an important role: it produces particle scattering and may cause transport across the average magnetic field so that observers not magnetically connected to the acceleration site also observe the SEPs.  Traditionally, only particle scattering has been taken into account, described as pitch-angle scattering. Recently, however, the possibility that turbulence may produce perpendicular transport of SEPs via magnetic field line meandering has also been studied \citep{Lai2016,Ber2020}. %This process may help to explain why observers not magnetically connected to the acceleration site also observe the SEPs.
Drifts associated with gradient and curvature of the Parker spiral field of interplanetary space \citep{Dal2013,Dal2020, Mar2013} may also contribute to particle transport across the average magnetic field.
The relative roles of different factors for cross-field transport are still not fully understood and may differ for different event scenarios. 

%Heliospheric disturbances 
The heliospheric current sheet (HCS), fast-slow solar wind stream interaction regions (SIRs), the magnetic structures of CMEs and even shocks themselves may also influence the transport of the SEPs in the heliosphere. 
See Section \ref{sec:GCR} and Temmer et al. 2023 this issue, for more descriptions and discussions of heliospheric disturbance structures. 
During large gradual SEPs, gradients of particle fluxes (often observed as a jump in the intensity time profile) have been observed in association with the pass-by of these disturbances \citep[e.g.,][]{guo2018September} and this is likely because perpendicular diffusion of SEPs could be damped at magnetic discontinuities within these structures \citep{strauss2016}.
Alternatively, \citet{Wat2022} showed that transport along the HCS is likely to play an important role in the transport of relativistic protons. 
Consequently, predicting SEPs assuming that they travel primarily along the nominal interplanetary magnetic field lines to reach a certain observer is further challenged in the presence of heliospheric disturbances. 

% transport: Anisotropy observations
Direct observation of the particle scattering in the interplanetary space is impossible, although anisotropy and pitch angle information (which needs to combine the particle direction and the vector magnetic field measurement) can provide considerable insight into the transport physics. 
For example, counter-streaming SEP fluxes may indicate ‘reservoirs’ produced by local magnetic trapping, while the pitch angle distribution indicates the role of scattering during transport to the detection site.
Studies in the recent years using multiple-view observations combined with directionality measurements by the Solar Terrestrial Relations Observatory (STEREO, since 2007) telescopes have made considerable progress in better understanding the transport physics \citep[e.g.,][]{dresing2014statistical, gomez2015circumsolar, strauss2015aspects}.
However, these observations are mostly based at 1 AU solar distance with limited spatial resolution. The observed anisotropy is a combined result of the connectivity to the particle acceleration/injection site and different transport processes accumulated over the long propagation path from the Sun to the observer. 
More measurements of this feature closer to the Sun with advanced detection techniques allowing for better angular resolution such as by Solar Orbiter's Energetic Particle Detector \citep[EPD,][]{rodriguez2020energetic} and the Integrated Science Investigations of the Sun (IS$\odot$IS) instrument suite on Parker Solar Probe \citep[PSP, since 2018,][]{mccomas2016integrated}, can help constrain SEP event sources, inform early transport processes and advance modeling efforts.

% transport: Onset analysis
The in situ SEP onset information is also important for understanding the release and transport processes \citep[e.g.,][]{Rouillard2012}.
Although particle scattering has been widely recognized to play an important role in SEP transport, velocity dispersion analysis (VDA) has been commonly used to diagnose the first arriving SEPs based on either single-view observations or multi-view observations for the same event \citep[e.g.,][]{krucker1999origin, Kollhoff2021}. 
The VDA assumes that these first-arriving particles with different energies are injected at the same time $t_{rel}$ and travel nearly scatter-free along the same path $L$ to the observer (with the arrival time $t_{arr}(E)$) such that the release time and the path length can be derived by fitting $t_{rel} = t_{arr}(E) - L/v(E)$. 
For some events, the VDA predicted release time and propagation length are consistent with the expected physical process indicating that these particles experienced mostly scatter-free condition during the transport while for many other events, the fitted values are not sensible suggesting that cross-field transport should be taken into account and/or there may have been different release times for particles with different energies \citep{laitinen2015correcting}. 

\subsubsection{SEP Spatial distribution}
% spatial distribution
The spatial distribution of SEPs in the heliosphere, in the radial, longitudinal and latitudinal directions, depends on the energy of particles, the transport conditions, the connectivity to the source and the duration of the particle injection. 
Previously, the SEP spatial distribution was examined by statistical studies using observations near-Earth \citep[e.g.,][]{lin1995wind, Cane2003, reames2013two} and also by multi-spacecraft missions such as Helios (from 1974 to 1984) \citep[e.g.,][]{wibberenz2006multi, lario2006radial}. 
For single-event studies, if multiple spacecraft can be in service at various well-separated locations, multi viewpoint studies can provide simultaneous in situ measurements at radial and longitudinal separations and reveal the injection and transport processes of SEPs. 
Studies in the last decade using the STEREO twin spacecraft have gained insight into the longitudinal distribution of SEPs \citep{Rouillard2012, dresing2014statistical,
lario2013longitudinal, lario2016longitudinal, richardson2014, Xie2019}. 
It has been shown that the SEP peak intensity is dependent on the longitudinal separation between the observer and the source region and their relationship can be fitted by e.g., a Gaussian expression with slight east–west asymmetry. However, it does not always make sense to fit three Gaussian parameters based on, in most cases, only three observers (two STEREO viewers and one from Earth). Future multispacecraft studies (using more heliospheric locations) are needed to reveal the longitudinal extent of SEPs. 
Moreover, the relative role of extended source injection and cross-field transport for the longitudinal distribution for SEPs is \change{not fully understood and is often case dependent}{is often case dependent and not fully understood}. 
We have limited observations on the latitudinal distribution of SEPs which has been sampled by the Ulysses spacecraft. Data from this mission showed that SEPs have relatively easy access to high latitudes, although their onsets are significantly delayed \citep{Dal2003a}.

%Transport models predict a lower limit of {R}$^{-1.7\pm 0.1}$ for empirically modeling the radial dependence of SEP peak intensities \citep{he2017propagation}.
The radial dependence of peak intensities are suggested to follow $R^{-\alpha}$ whilst $\alpha$ may range between less than $1$ to larger than $5$ derived by different studies using different datasets (that cover different particle energy and heliospheric distances) whereby different conclusions are obtained. 
For instance, \citet{lario2006radial} suggests that the smaller the mean free path of the particles, the larger the decrease of both peak intensities and fluences with radial distance; the smaller the energy of the particles, the larger the decrease of both peak intensities and fluences with radial distance. 
In contrast, \citet{fu2022first} shows that the higher the energy of protons, the larger decrease of peak intensity with larger radial distances. 

More observations by ongoing Solar Orbiter \citep[SolO, since 2020,][]{muller2020solar}, PSP \citep[since 2018,][]{fox2016psp} and BepiColombo \citep[launched in 2018 delivering the two spacecraft to Mercury][]{milillo2020} as well as future interplanetary missions will be essential to understand the spatial distribution of SEPs.

\subsubsection{SEP energy distribution}
The energy distribution of SEPs is another important and combined outcome of both the acceleration process and transport effects. 
The energy range over which SEPs are observed at 1 AU varies considerably between different events.  
In the largest events, the presence of protons of energies up to a few GeV can be detected through neutron monitors at the Earth’s surface, in so-called GLE events (marked in Fig. \ref{fig:pfl}, see Section \ref{sec:GLE}) whilst most SEPs have energies up to tens of MeV. 
SEP spectra can be described by a single power law distribution, or a double power law (band function) or a power law with an exponential decay at high energies.  
The decay of the flux at high energies was considered as a consequence of a finite lifetime and a finite size of the shock \citep{ellison1985shock}, and the spectral break should be a direct consequence of the acceleration process in the source region \citep{mason2012}. However, some other studies suggest that a double power spectrum can result from a single power-law spectrum at the source which gradually forms as particles propagate out from the Sun \citep{li2015scatter}. 

More spectra observations with different radial distances from the Sun would be helpful to better understand the SEP spectral evolution in relation to the acceleration/transport effects. 

\begin{figure}[ht!]
\subfloat{\includegraphics[width=0.63\columnwidth]{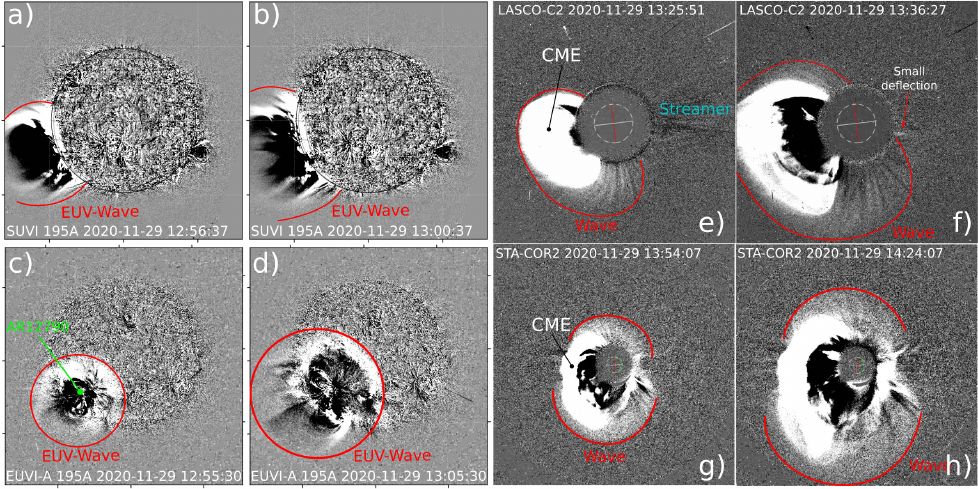}}\\
\subfloat{\includegraphics[width=0.63\columnwidth]{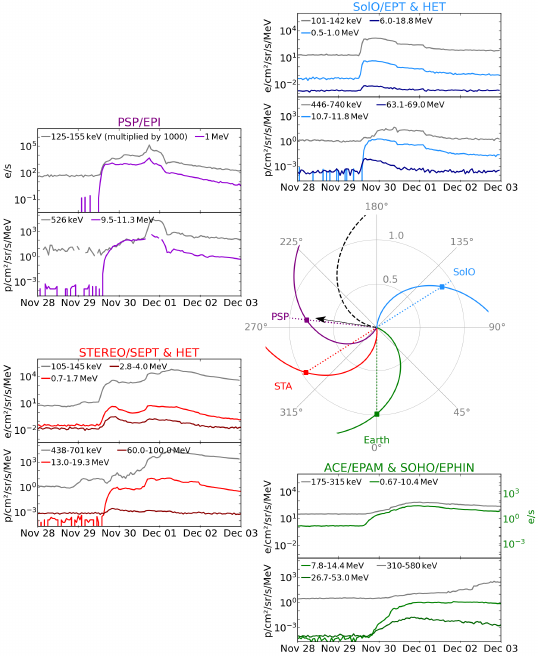}}\\
\caption{The remote, in situ and multi-view observations of a wide-spread SEP event on 2020 November 29 (Credit: \citet{Kollhoff2021}, reproduced with permission from ESO). Top: Remote sensing observations of the EUV wave, the CME, and the white-light shock wave from two different viewpoints: Earth and STEREO-A. Bottom: overview of the energetic particle measurements at PSP, SolO, STEREO-A, and near-Earth spacecraft. } \label{fig:kollhoff}
\end{figure}

\vspace{0.5cm}
%\subsubsection{The overall goal}
In synergy, the remote, in situ and multi-view observations with both energy and direction resolutions could allow tracing sequences of phenomena from the flare or coronal eruption at the Sun, to radio signatures of near-Sun shock formation, and to the energy-resolved SEP intensity time profile observations at different points in space for different directions \citep[e.g.,][and Fig. \ref{fig:kollhoff}]{Kollhoff2021, klein2022relativistic}. 
With the new observations near Earth and from PSP, Solar orbiter as well as various planetary missions, we expect to make important progresses on understanding the exact physics of SEP energization and transport during this new Solar Cycle (SC25). 

\subsection{Nowcasting and forecasting capability}
\label{sec:sep_forecasting_capability}

SEP forecasts can only be made using information that is accessible before or at the start of an SEP event and the related measurements may include solar magnetograms, optical imaging, EUV imaging, soft X-ray measurements, coronagraph imaging of CMEs from single or multiple vantage points, ground and space-based radio observations in the wavelengths pertinent for Type II, III, and IV radio bursts, in situ energetic proton and electron observations, and in situ measurements of solar wind density, temperature, and magnetic field.
In real-time and at the geosynchronous orbit, energetic particle measurements are available from the NOAA's GOES spacecraft \citep{sauer89, Rodriguez14, kress20}. These measurements have been used in operational real-time forecasting for several decades (see Fig. \ref{fig:pfl}) and provide a long term dataset for model validation purposes.
At the time of writing, real-time SEP observations are provided by the Space Environment In-situ Suite (SEISS) on GOES-16 and 17, the first ones of the GOES-R series spacecraft, measuring protons with energies from 1 to $\ge$500 MeV in 14 energy channels. Real-time information on the solar event responsible for particle acceleration is crucial to develop predictive capabilities of the particle radiation environment.
Besides, soft-X-ray data from GOES-16 are also available in real time.
The GOES-16 Solar Ultraviolet Imager (SUVI) instrument monitors the full solar disk in the extreme ultraviolet (EUV) wavelength.
Obtaining real-time CME data is a critical development needed for improvement of SEP forecasting capability \cite{Whitman22}. White light coronagraph data from SOHO/LASCO (currently more than 25 years old and the solo coronagraph near Earth) used to image and fit CME ejecta can take 4-6 hours, sometimes much longer, before being available for a forecaster or model (Temmer et al. 2023, this issue). Planned for launch in 2024, the next GOES mission will include an operational compact coronagraph, followed in 2025 with a similar instrument on NOAA's Space Weather Follow-On Lagrange 1 (SWFO-L1). Coronagraph latency from SWFO-L1 is expected to be 15-30 minutes.

A recent forecast verification study \citep{bain21} from the NOAA Space Weather Prediction Center (SWPC) highlights the center's performance and skill with regard to forecasting SEP events (defined in this study as the period when the GOES $\ge$ 10 MeV integral proton channel exceeds 10 p.f.u., i.e., an S1 storm) during SC 23 and 24. 
In particular the study evaluates forecasters' skill in assigning a probabilistic forecast for an S1 storm in the next 1, 2 and 3 days and, separately, in issuing a shorter term (minutes to hours) deterministic \change{forecast Warning products}{warning} for an imminent event. The forecast products used in this study are retrieved from the NOAA archive and are based on the observations and tools that were available to forecasters in real-time.
The study provides a baseline for our current ability to forecast SEP events and highlights a few key points. SWPC probabilistic forecasts have improved from SC 23 to 24, with True Skill Scores increasing for day 1 (from 0.47 to 0.61), day 2 (from 0.16 to 0.34) and day 3 (from 0.06 to 0.13) forecasts. For the SC 24, SWPC $\ge$10 MeV proton warnings have a probability of detection (POD) of 0.91 and a false alarm ratio (FAR) of 0.24, with a median lead time of 88 min, which is better than many models currently available in the research domain. This was an improvement over SC 23. SWPC also issues warning products for higher energy, $\ge$ 100 MeV protons. In SC 24 the $\ge$ 100 MeV proton warnings had a POD of 0.53 and a FAR 0.38, with a median lead time of 10 min. Results such as these highlight the difficulty of issuing a forecast with significant lead time while trying to limit the number of false alarms. It is also challenging to determine in advance how intense and energetic and event will be, as evidenced by the decreasing performance from $\ge$ 10 MeV to $\ge$ 100 MeV warning products. 

%However, it remains a challenge to forecast a SEP event with significant lead time and persistence (the expectation that the environmental state will persist from one day to the next).

Robust and accurate SEP models are needed to support forecasting endeavors. Recent years have seen the development of sophisticated modelling tools aiming to describe particle acceleration and propagation. A recent study by \citet{Whitman22} reviewed current SEP modeling capabilities. In this paper, the authors summarize 36 different SEP models, ranging from physics-based to empirical to machine learning approaches. For each model, the authors note the models' inputs/outputs, their limitations and caveats as well as the degree of validation that has been carried out. Furthermore, the authors endeavor to assess the degree of readiness for these models, i.e., a measure of the maturity of the field, with regard to moving research models into operations. 

The paper found that the different models made use of a wide variety of observational inputs, including ground- and space-based remote sensing observations covering a wide range of wavelengths from X-rays to EUV to optical to radio and in situ space-based observations of solar wind and magnetic field conditions, as well as a wide range of particle energies (keV to tens of MeV) for mainly electrons and protons. In some cases, promising model approaches have been demonstrated, but no real-time data are available to implement the model as a forecasting capability. 

With respect to forecasting coverage, the models, taken as an ensemble, provide a wide variety of outputs that hold value for forecasters aiming to predict the near-term space radiation environment. Model outputs include All Clear, probability, peak flux, fluence, and time profile predictions. Physics-based models also typically provide additional particle distribution information that, while not generally relevant to real time forecasting, holds value for furthering the understanding of the physics of particle acceleration and transport. Most of the models reviewed in the paper, however, are not transitioned into in an operational setting or require long run times and large computational resources that prevent them from being used in a forecasting context.

Currently, empirical models, which relate real-time observables to SEP forecasts, are heavily relied upon by forecast centers owning to their ability to rapidly produce results in a real-time environment. Out of the 36 models discussed in \citet{Whitman22}, 12 models are running in a real-time setting serving space weather forecasters and end-users through the SEP Scoreboard\footnote{\url{https://sep.ccmc.gsfc.nasa.gov/}}, the ESA Space Weather Service Network\footnote{\url{https://swe.ssa.esa.int}}, at NOAA SWPC, or within other government and private institutions. 
\change{For example,}{Examples include} the Proton Prediction Model \citep{balch99, balch08, kahler07, kahler17} used by NOAA SWPC and the Proton Prediction System (PPS) \citep{smart89, smart92} used by the Air Force Research Laboratory. Newer empirical models such as e.g., UMASEP \citep[University of Málaga Solar particle Event Predictor,][]{nunez2020}, RELeASE \citep[The Relativistic Electron Alert System for Exploration,][]{posner07} and SEPSTER \citep[SEP predictions based on STEReo observations,][]{richardson18} are under evaluation as part of the CCMC SEP scoreboard (see \citet{Whitman22} and references within for more details). There is broad interest in the SEP modeling community to transition models to an operational setting, which will require significant, dedicated effort from both model developers and space weather services. This worthwhile goal will need support from the various institutions that hold a stake in space weather forecasting.

However, looking forward, physics-based SEP models will be required to improve our understanding of flare and CME occurrence and resulting particle acceleration and eventually to support proton forecasting in real-time. From the point of view of space weather radiation, the majority of modeling work has focused on SEPs originated from CME-driven shocks accelerating particles in the corona and interplanetary space \citep{Ara2006, Bain2016, Wij2022}. However, coupled CME-SEP modelling is a complex problem that should include the modeling of the shock properties, SEP acceleration at CME-driven shocks as well as the transport of CMEs and SEPs in the heliosphere. 

The previously published validation results summarized in \citet{Whitman22} vary widely from model to model, ranging from extensive validation of a statistically-significant sample of events to no validation at all and everything in between. In the past, validation has not been emphasized within the research community and has been carried out non-uniformly. However, recent efforts, such as the SHINE/ISWAT/ESWW SEP Model Validation Challenge   \footnote{\url{https://ccmc.gsfc.nasa.gov/challenges/sep/}}, aim to develop standards and tools for use by the SEP modeling community and the U.S.-based space weather enterprise has developed a formalized process \footnote{\url{https://www.whitehouse.gov/wp-content/uploads/2022/03/03-2022-Space-Weather-R2O2R-Framework.pdf}} for the transitioning of models from research to operations that emphasizes the role of validation in the progression from one stage to the next. Additionally, \citet{bain21} provides an important baseline for desired model performance.

In summary, a huge number of observations \change{are}{is} required to characterize the space environments where particles are accelerated and transported, only some of which are available in real-time. Besides, a good deal of more validation is required to assess the performance of these models in their ability to replicate the space environment.

\subsection{Limitations and open questions}

\begin{sidewaysfigure}
\centering
\includegraphics[width=1.2\columnwidth]{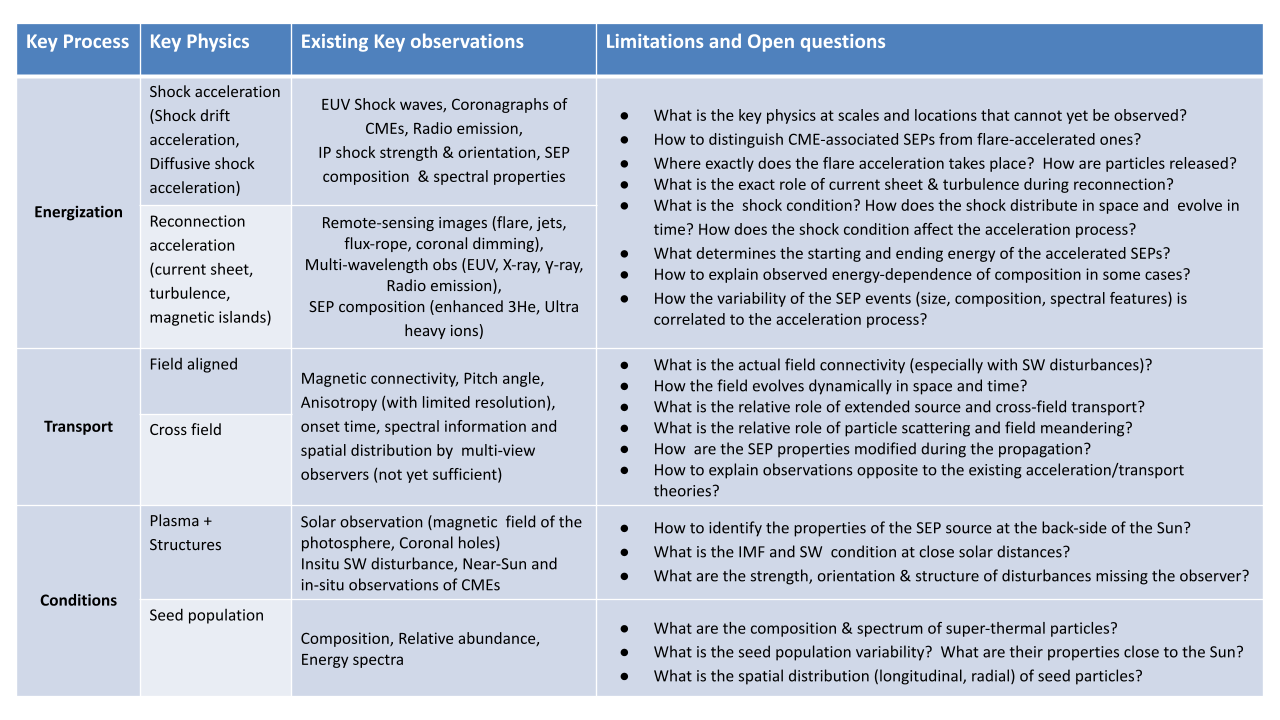}
    \caption{Summary table of the limitations and open questions related to the key physics of SEP studies}
    \label{fig:SEP_table}
\end{sidewaysfigure}

The research on SEPs still has many parts of the puzzle that remain unsolved. A recent paper by \citet{Anastasiadis2019} listed some key open issues, together with the expectations from new missions and forecasting schemes. 
Here, we discuss limitations and open questions in SEP physics by grouping them into the following three categories: energization, transport, conditions (in which the former two processes take place). Each category involves the key physics and observations as discussed in the previous sections and also summarized in the table of Fig. \ref{fig:SEP_table}. As limited by existing key observations, there are still open questions related to different processes of the SEP problem. 

On the aspect of understanding the acceleration process of SEPs, we still face the following limitations. 
\begin{itemize}
\item We still miss the key physics at scales and locations that cannot yet be observed, such as observations for particle acceleration in the low corona (seed particle populations, injection processes, turbulence, current sheet or shock properties).
\item For acceleration attributed to flare reconnection processes, we still do not know where exactly the acceleration takes place and how particles are released through the opening of magnetic field lines.
\item Due to the variability of CMEs, shocks, SEPs and their complex correlations, we cannot always distinguish the flare- or CME-associated SEPs. 
\item We cannot \add{yet} fully capture the evolution/distribution of the particle source, such as the CME-driven shock temporal evolution, and the shock's spatial structure that may give rise to different energization processes at its different parts.%, or the acceleration mechanism in twin-CME scenario.  
\item We still cannot fully explain the observed energy-dependence of SEP composition.
\item We still need to understand what the key process is for quantifying the starting and ending energy of the acceleration. 
\item We do not fully understand the big variability of the events (size, composition, spectral features) and how it is correlated to the acceleration process. 
\end{itemize}

On the aspect of understanding the transport of SEPs, we still face the following limitations. 
\begin{itemize}
\item We \change{cannot}{do not yet} accurately describe the magnetic connection between the particle sources and the observer as the actual magnetic field connectivity (especially with SW disturbances) should not always be described by nominal Parker spirals. 
\item We \change{cannot}{do not yet} understand how the magnetic field evolves dynamically in space and time (turbulence and field meandering) and how this affects SEP propagation.
\item We \change{cannot}{do not yet} fully distinguish the relative role of extended source and cross-field transport for wide-spread events.
\item We \change{cannot}{do not yet} fully characterise the way in which coronal and interplanetary transport processes modify the properties of the injected population.
\item We \change{cannot}{do not yet} clarify the roles of particle scattering and field meandering in SEP transport over the full range of SEP species and energies
\item We \change{cannot}{do not yet} explain observations which show features opposite to the two-class acceleration/transport theory, such as widely distributed $^3$He-rich events or impulsive SEPs come from very poorly connected source. 
\end{itemize}

When it comes to the conditions through which the acceleration/transport processes take place, we still have various open questions as below.
\begin{itemize}
\item There are very few observations of the solar magnetic field at the back-side of the Sun. This makes it difficult to identify the properties of the SEP source when it is located at the back-side of the Sun. 
\item Despite the recent PSP and Solar Orbiter missions, there are still limited observations of the interplanetary conditions close to the Sun, which strongly influence the early phase of the SEP transport.
\item It is difficult to quantify the strength, orientation, and structure of disturbances (ICMEs, SIRs) which did not pass the observer but have a large impact on the propagation of SEPs.
\item Regarding the seed particle population close to the Sun, we still lack observations of its composition and spectrum and their variability; we still do not know if the seed particles are constant sources (like campfires observed by Solar Orbiter) or episodic sources.
\item For the seed particle population in the interplanetary space, we do not fully understand its distribution in longitude and radial distance.
\end{itemize}

In addition, from the operational and SEP forecasting point of view, we summarize and follow Section \ref{sec:sep_forecasting_capability} to briefly list a few major limitations below.
\begin{itemize}
    \item The limitation on key observations, as discussed above, stands as a major challenge for advancing SEP models. 
    \item Most physics-based models require long run times and large computational resources that currently prevent them from being used in a forecasting context, although they are essential for advancing our understanding of the SEP physics.
    \item \add{It is difficult to use/compare models designed to reach related but different outputs. Therefore,} model validation with uniform standards is still required to assess the model ability to replicate the space environment.
    \item In many cases, real-time data are still lacking to implement the model as a forecasting tool. 
    \item SEP modeling research based on scientific interest still needs support from stake-holders within the space weather forecasting authorities to be transitioned into operations.
\end{itemize}

In addition, with the aim of better predicting the deep-space heliospheric radiation environment for planetary missions, it becomes also crucial to have multi-view real-time forecasting capabilities. An effort has been made on this aspect based on the energetic particle flux time profiles from the twin STEREO spacecraft and ACE at Earth's L1 point, in combination with SOHO outer corona images in white light and inner corona EUV images from SDO\footnote{\url{http://stereo.ssl.berkeley.edu/multistatus.php}}. 
Note that, however, STEREO-B has been out of service since October 2016. The implementation of other spacecraft at different locations for real-time monitoring would greatly benefit the purpose of $360^{\circ}$ forecast of the heliospheric radiation environment.
Recently, ESA has approved the Vigil mission which is a spacecraft deployed at the L5 point of the Sun–Earth system to enable remote sensing of the Sun and interplanetary space and in situ measurements of solar wind plasma and high energy solar particle events \citep{Vourlidas-2015-L5mission}. 
Meanwhile, Chinese scientists have proposed the Solar Ring mission, which \change{include}{includes} three spacecraft at positions somewhat shifted from the L3, L4 and L5 points to observe the Sun and carry out in situ observations \citep{wang2023ring}.
With the ongoing and future missions, we expect to see significant advances in SEP physics and forecasts in the next decade.

\section{Ground Level Enhancement}\label{sec:GLE}

\begin{figure}[ht!]
\centering
\includegraphics[trim=5 5 5 5,clip, width=0.85\textwidth]{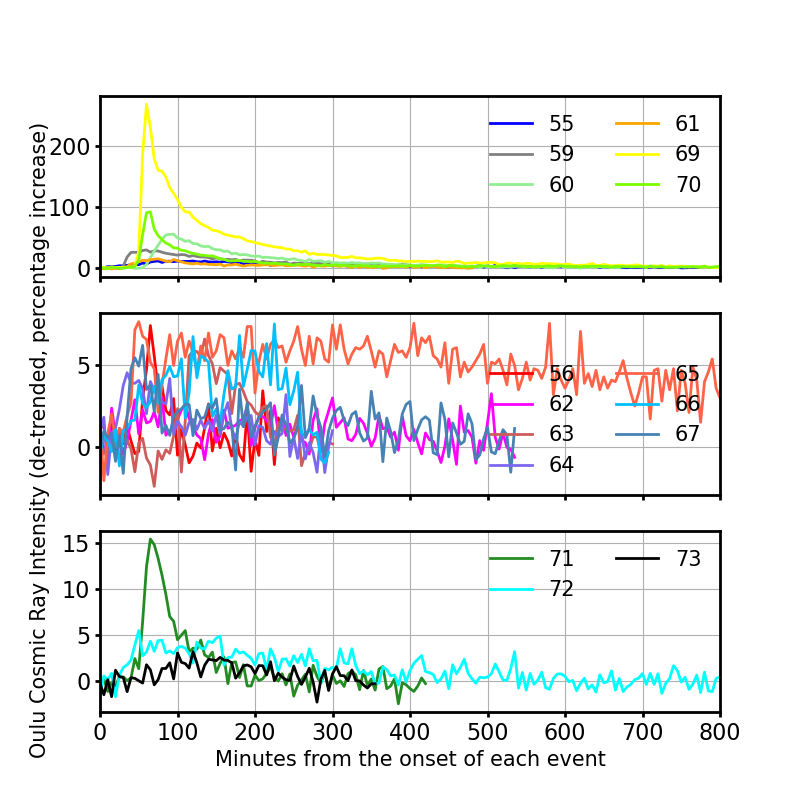}
\caption{Time profiles of the GLE events 55-73  (except for event 57, 58 and 68 \change{which had no available data}{which have caused only marginal enhancements}) as recorded by Oulu neutron-monitor station with cutoff rigidity of 0.77 GV. GLE numbers are shown in the legends. The top panel shows events during Solar Cycle (SC) 23 with peak increase above 10\%\delete{ increase}; the middle panel shows the smaller events from SC23. The bottom panel shows only 2 GLE events (GLE71 and 72) during SC24 and the one event so far (GLE73) during SC25.  Data are downloaded from the IGLED database \url{https://gle.oulu.fi} (de-trended GLE data) with 5 minute cadence.}
\label{fig:OULU_GLE}
\end{figure}

\begin{figure}[ht!]
\centering 
\includegraphics[width=\textwidth]{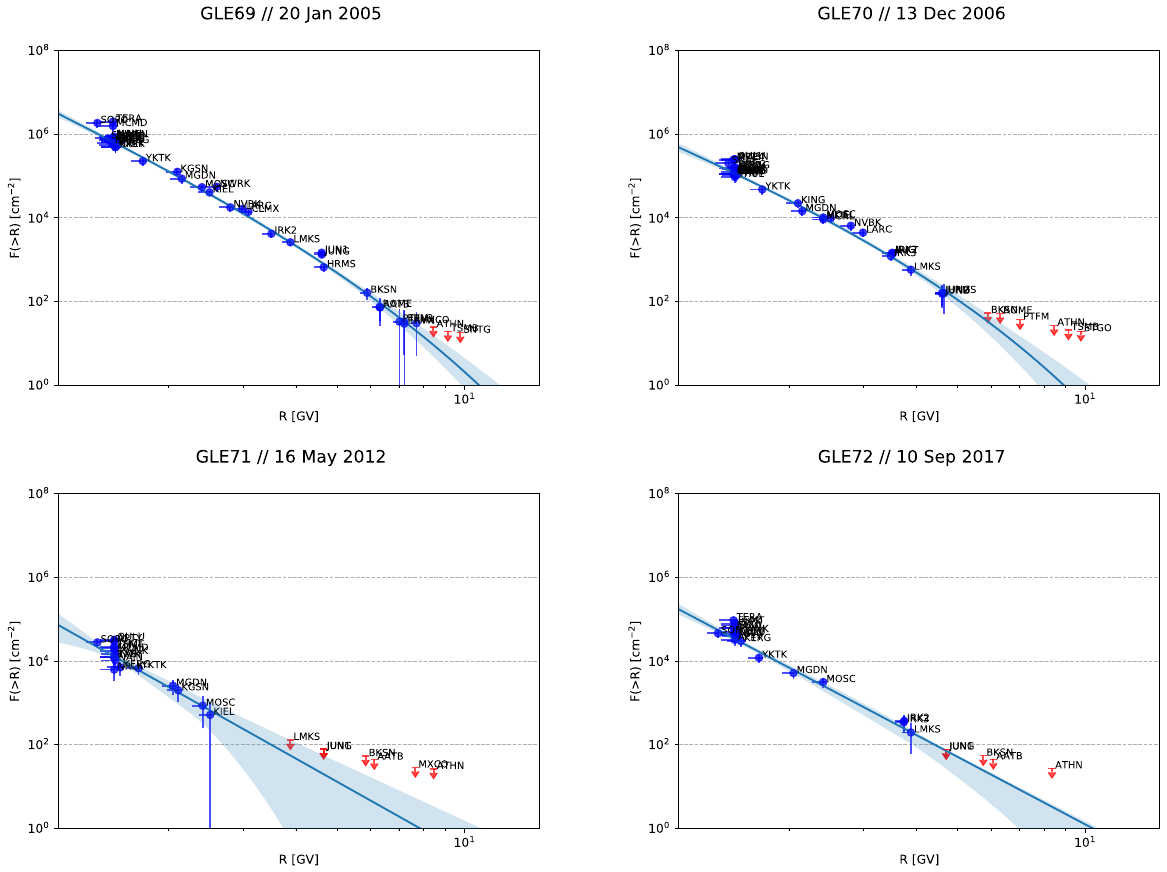}
\caption{Integral fluences reconstructed for GLE 69-72 (Credit: \citet{usoskin2020revised}, reproduced with permission from ESO). Blue points with error bars depict reconstructions of the integral fluence from individual NMs, while red arrows denote the upper limits. Error bars represent the full-range uncertainties. The thick blue curve represents the best-fit spectral approximation with 1$\sigma$ uncertainties. }
\label{fig:GLE_spec}
\end{figure}

\subsection{Introduction}\label{sec:gle_intro}
Ground-level enhancements (GLEs) occur when SEP-induced atmospheric secondary particles are registered at ground-based detectors, such as ionization chambers, muon, and neutron monitors (NMs) \citep[see, among others,][]{shea1993history, miroshnichenko2013solar}. 
GLEs are therefore related to the most energetic class of SEP events, associated with solar flares and/or coronal mass ejections (CMEs), requiring acceleration processes that produce protons with energies \delete{of at least 430 MeV upon entry in the Earth's atmosphere. This class of SEP events can}\add{high enough to provide a ground-level signature of the event. It has been studied that the threshold energy for SEP particles, upon their entry in the Earth's atmosphere, to cause GLE is about 430 MeV/n at the sea level \citep{plainaki2014ground} and about 300 MeV/n for high-altitude polar stations as South Pole \citep{mishev2021}}.
\add{After entering the the atmosphere, the primary SEP particles can} produce showers of secondary particles with sufficient intensities to exceed the GCR background \delete{providing thus a ground-level signature of the associated primaries}. Those primary SEP particles are mainly protons and, to a smaller extent, heavier ions, although some events have been also associated with the emission of solar neutrons \citep[see, for instance,][]{muraki2008detection}. 

%GLE/SEP events are rare cases of extreme space weather phenomena, favored by both the energetics of the primary solar particles and the actual geo-effectiveness of the event as a whole (e.g., the location of the particle sources on the Sun, the interplanetary magnetic field conditions in the near-Earth space, the existence (or pre-existence) of one or more ICMEs, etc.). 
Since February 1942 until October 2021, 73 GLE events have been registered. The first four GLE events were recorded by ionization chambers \add{\protect{\citep{forbush1946}}}\delete{ and their properties have been re-assessed quantitatively \protect{\citep{koldobskiy2021new}}}. 
The GLE events registered from 23 February 1956 onwards\add{, starting from GLE5,} were recorded by the worldwide network of NMs which are energy integrating detectors with cut-off rigidities depending on the actual location and altitude at which they are installed \add{[e.g.,][]\citep{miroshnichenko2018}}. 
The times of GLE events have been marked in Fig. \ref{fig:pfl} which shows their occurrence is dependent on the solar activity cycle but not solely during the solar cycle maximum years. 
Fig. \ref{fig:OULU_GLE} shows the recent GLE events of SC 23 and 24 (while the last one is from SC25), as they were registered at Oulu neutron-monitor station on the basis of 5-min data. 

NMs allow the determination, to first order, of the primary SEP spectrum during a GLE event. Most NM data are part of international databases now available to the entire scientific community, e.g., the International GLE Database \citep[IGLED][]{usoskin2020revised, vaisanen2021seven}; the Neutron Monitor Database \citep[NMDB][]{mavromichalaki2010implementation} etc. Numerous GLE-modeling efforts performed jointly with rigorous data analyses \citep[see, for instance,][and references therein]{miroshnichenko2013solar}, also at interdisciplinary level, have led to the detailed identification of the properties of the majority of the observed GLE events and to the estimation of the spectral characteristics in the circumterrestrial space of the related relativistic SEPs \citep{belov2005solar, bombardieri2008improved, plainaki2007modeling, plainaki2014ground, mishev2016analysis, mishev2018first}. 
Fig. \ref{fig:GLE_spec} shows reconstructed SEP spectra based on NMs with different cutoff rigidity \citep{usoskin2020revised}. 

In this section, we give a brief overview of the recent progress on GLE observations (and models developed in synergy with observations) and address a few issues that need to be better studied in the next years. 

\subsection{Recent Progress and current understanding}\label{sec:gle_progress}

Joint SEP/GLE studies based on data from both space borne and ground-based instruments have made major progress in the past years.
During the recent space age, GLE-observations by NMs are often complemented by various sets of space-based particle data registered during the associated SEP events \citep[see, for instance, ][]{plainaki2014ground}. High energy SEP data with some associated with GLE events have been recorded over the past years by EPHIN onboard SOHO \citep{kuhl2017solar} and by HEPAD onboard GOES \citep{vainio2017solar}. 
Although recent space borne measurements of particle fluxes above some hundreds of MeV, e.g., PAMELA \citep{adriani2015} operating from 2006 to 2016 and AMS-02 \citep{aguilar2018} in operation since 2011 have significantly broadened our view on relativistic SEP events \citep[e.g.,][]{bruno2018solar}, the value of ground-based NM observations remains essential. %This is due to a series of limitations related both to the technical characteristics of the payloads themselves and, when it is the case, to the actual low orbit of their hosting satellites. 
%Indeed, solar energetic particles are often observed only intermittently and often with limited accuracy since the satellite orbits are mostly situated within the terrestrial magnetosphere (e.g., payload onboard LEO satellites). 
At LEO altitude, Earth's magnetosphere can shield and modulate SEPs arriving at these detectors, causing the SEP flux to vary intermittently as affected by the hosting satellites' orbit.
Moreover, during such \change{extreme}{major} SEP events, saturation effects by intense particle fluxes hitting the detectors are likely to bring major uncertainties in the estimation of the event magnitudes, as already observed in the past for payload onboard GEO satellites \citep[e.g.,][]{reeves1992great, tylka1997creme96}. It should be pointed out, however, that recently the high-energy space borne data have been essentially revisited and corrected for known errors \citep[see, for instance, ][for GOES data]{raukunen2020very}.

To reveal the properties of relativistic SEPs associated to GLE events, the identification of the primary SEP spectrum is necessary and has been realized by various studies. The primary SEP spectrum reconstruction has been typically done by best-fitting NM \citep[e.g.,][]{belov2005solar, plainaki2007modeling, plainaki2009modeling, mishev2016analysis,mishev2018first, usoskin2020revised} and/or energy-dependent space borne data \citep{raukunen2018two, bruno2019spectral} to a predefined spectral shape and spatial distribution function associated to the primary SEP fluxes, containing free-parameters that characterize the ongoing physical mechanisms. 
The identification of these best-fit parameters and of their uncertainties often allows some scenarios to be eliminated regarding the primary acceleration mechanism \citep[e.g.,][]{plainaki2014ground}. 
A crucial input in these retrieval models is the NM yield-function, a topic on which important investigations have been performed in the last years \citep[e.g.,][]{clem2000neutron, mishev2020updated}. Recently, a new effective-energy analysis method to reconstruct the high-rigidity part ($\geq$1 GV) of the spectral fluence of SEPs for GLE events, based on the use of NM-data has been developed \citep{koldobskiy2018effective}. 
Moreover, combining data from low-energy ($<$ \change{430 MeV}{300 MeV/nuc}) space borne detectors located beyond the magnetosphere (e.g., at L1) with NM data it is possible to reconstruct the primary SEP spectrum and to identify the spatial characteristics of the flux (e.g., anisotropy and spatial distribution at a specific altitude within the Earth’s atmosphere) taking also into consideration the NM asymptotic directions of viewing \citep[e.g.,][]{plainaki2009neutron}. In this context the use of ground-based data is particularly important since the registration (or not) of an event at different locations on the Earth’s surface provides direct information on the minimum cut-off energy of the responsible primary particles and the dominant direction of their propagation. 

\add{In fact, before the era of ground-based detectors some very large GLE events have also left their footprint on Earth. SEPs with extremely high energies and fluxes can affect atmospheric chemistry leading to the formation of nitrates or produce cosmogenic radionuclides such as radiocarbon $^{14}$C. These signatures can be registered in the polar ice core or in the tree rings. Recent years have seen much progress in reconstructing extreme historical SEP events using such information \citep[see Section 7 of][and references therein]{cliver2022extreme} in order to understand the features and occurrence probability of these events. However, it is still unclear if different physics is involved to make such extreme events distinguished from other GLE and SEP events or they are merely large \citep{MCCRACKEN2023}. The accurate forecast of such extreme events continues to pose one of the biggest challenges for the space weather community. }

\add{In parallel to such extreme type of GLE events, there is also a growing interest to study weak and short-term events, such as sub-GLE \citep{poluianov2017sub} and  anisotropic cosmic-ray enhancement \citep[ACRE,][]{gil2018anisotropic} events. 
Sub-GLE events refer to those with a relatively weak SEP input that are not detected by sea-level neutron-monitor stations but registered by high-elevation polar regions with negligible geomagnetic and reduced atmospheric rigidity cutoffs. In comparison, a GLE event has to be registered by at least two differently located neutron monitors, including at least one neutron monitor near sea level and a corresponding enhancement in the proton flux measured by a space-borne instrument(s). 
Alternatively, the ACRE events are associated with highly anisotropic GCR flux only detected by neutron monitors with certain directions of acceptance of charged particles through Earth’s magnetosphere. It may be not related to SEPs, but caused by the local anisotropy of Forbush decreases or other disturbed heliospheric conditions.} 

\delete{GLE events start to be detected at the ground level of other planetary bodies. }With the recent progress in planetary exploration, some particle detectors have been sent to the surface of other planetary bodies, such as the Radiation Assessment Detector \citep[RAD,][]{hassler2012} on Mars and the Lunar Lander Neutron and Dosimetry instrument \citep[LND,][]{wimmer2020lunar} on the Moon. Thus our detection and knowledge of GLE events has expanded to other places other than Earth. \citet{xu2020first} reported the first SEP event detected on the Lunar surface on 2019 May 6 which had a rather low intensity. Due to the lack of an atmosphere or intrinsic magnetic field, SEPs can directly reach the lunar surface and interact with the lunar regolith to generate secondaries. Thus, the lunar surface radiation environment can be hazardous during SEP events for future lunar explorers. 
Similarly, Mars lacks an internal magnetic dynamo and its atmosphere is rather thin as compared to Earth, allowing a good portion of the SEPs to reach the surface of Mars or to cause enhanced secondary radiation on Mars. The recent two GLE events registered at Earth (GLE72 \& GLE73) were also detected at Mars \citep{guo2018September, guo2023first}. However, their properties are significantly different at two planets, for which two factors should be carefully considered: 1) different heliospheric locations of the planets mean that they may see different SEP intensities and energy spectra due to transport from the acceleration site (Section \ref{sec:SEP}); 2) different planetary environments (magnetic field, atmospheric structure and depth, regolith composition, etc.) have different influences (shielding) on the arriving SEPs. The former requires a global understanding of the SEP properties and distributions in the helisophere while the latter needs specific particle transport modeling through the planet's environment. 
Note that the space weather at other planetary bodies is a purview of the neighbour ISWAT Cluster, H4 "Space weather at planetary bodies in the Solar System" and the subject of planetary space weather will be discussed in a future separate paper.

To briefly summarize, based on the advances in recent years, we have now sufficient observation and modeling capabilities to carry out the following studies.
\begin{itemize}
\item We can assess, during SEP/GLE events, with an acceptable accuracy the spatial distribution of the radiation environment above the Earth’s atmosphere in the rigidity range approximately above 1GV.
\item We can identify during GLE events, for a predefined SEP spectrum shape, the most representative parameters characterizing the particle acceleration and transport processes (e.g., power-law spectrum index).
\item We understand, even though not exhaustively, the way atmosphere treats primary SEP fluxes during GLE events, based on back-tracing and/or forward modeling techniques.
\item We can derive with an acceptable accuracy the acceptance cones at each ground-based detector given the intensity of the magnetic field in the near-surface region.
\item \add{We start to investigate the nature of different types of GLEs, including weak sub-GLE and ACRE events as well as extreme GLE/SEP events using their cosmogenic imprints.}
\item We start detecting and understanding SEP/GLE events at other planetary bodies combined with synergistic observations near Earth and modeling approaches.
\end{itemize}

\subsection{Limitations and open questions}
\label{sec:gle_limitation}
The study of GLE events in the near future will remain important to answer a few open questions, from the pure space weather science point of view, summarized as follows: 
\begin{itemize}
\item What is the essential and necessary condition that solar eruptions need to satisfy to produce a GLE event?
\item What is the contribution of the interplanetary space structures in the SEP transport processes?
\item What are the properties of the seed population necessary for producing a GLE event? 
\item How can we predict the onset times and properties of relativistic SEPs resulting in radiation storms and GLE events? 
\item What is the distribution of the primary SEP spectrum (in near-Earth space) responsible for GLE events? 
\item \add{What different mechanisms are involved in making extreme types of SEP events?}
\end{itemize}

From an operational point of view considering the monitoring and predictability of SEP/GLE events, we can still make improvements in the following directions most of which are not only beneficial for GLE studies, but also for a wider range of space weather topics.
\begin{itemize}
\item We are in need of a denser and wider coverage of NM detectors\change{, especially in the Southern hemisphere} {\citep[see][for more detailed discussions]{mishev2020current}}. This will increase the certainty in the estimation of the modeled SEP-parameters, reducing possible code bias-effects, especially in cases of low intensity GLE events.
\item We should work towards implementing a reliable space- and ground-based network of Space Weather assets, continuously operating and maintained. In particular, \add{it is necessary to inter-calibrate ground and space measurements since the derived results do not always agree with each other \citep{miroshnichenko2018} and }there is a lack of good coverage of particle detectors in the energy range up to a few GeV. 
\item We need a fleet of spacecraft that can provide the necessary basis for testing and validating radiation models over a wide range of energies and at different locations in the inner Solar System.
\item We will benefit from a dense coverage of ground-based magnetic observatories that could provide information in real time on the actual magnetic field conditions for better interpreting and possibly predicting the direction of the major radiation hazards in space due to GLE events. 
\item \add{We need to exploit more information of historical events, even before the neutron monitor era, in order to better understand their nature and make predictions of such disastrous scenarios.}
\end{itemize}

\section{Galactic Cosmic Rays}\label{sec:GCR}

\subsection{Introduction}\label{sec:gcr_intro}
%Cosmic rays have been extensively studied at the Earth \citep{grieder2001cosmic}. Remarkably, their energy spectra have been measured with energies extending above 10$^{20}$ eV, several orders of magnitude above what the most powerful man–made particle accelerators can produce. 
%At energies above a few GeV/nuc, the energy spectra follow a power law. The spectra steepen slightly at energy near 5 $\times$ 10$^{15}$ eV, commonly referred to as the knee. At energies near 3×10$^{18}$ eV, the spectra harden again (usually called the ankle). Below this (ankle) energy the particles are most commonly interpreted as originating from supernova remnants within the galaxy \citep{biermann2001introduction} and are called galactic cosmic rays (GCRs). However, the detailed origin of these high-energy cosmic particles is still one of the most fundamental problems of modern astro-particle physics \citep{blasi2013origin}. 
%GCRs in the inner heliosphere include about 2\% electrons and 98\% atomic nuclei of which the latter are composed of about 87\% protons, 12\% helium, and $\sim$ 1\% heavier nuclei (Z$\ge$ 3) in the range of 0.1-10 GeV\citep{simpson1983}. (Abundances vary slightly depending on the phase of the solar cycle.)
GCRs originate from outside the heliosphere and can have energies far greater than SEPs. 
\change{Radiation}{Radiations} induced by GCR nuclei, especially those with high energy (e.g., protons above 100 MeV) and high charge (Z) ions \change{have been}{remain} one of the major concerns for long-term deep-space human and robotic missions \citep{Townsend1994,cucinotta2013safe,Miyake2017,Dachev2020, guo2021radiation}.

Compared to SEPs (Section \ref{sec:SEP}), the intensity and composition of GCRs are rather stable, but not constant as the charged GCR particles are affected by the varying heliospheric environment following solar activities \citep[see reviews by][]{cane1999cosmic, potgieter13, rankin2022galactic}. 
In the long term, the GCR flux was first observed to vary inversely with sunspot number \citep{forbush1954world, forbush1958cosmic} and is known as the \change{long-term}{11-year} solar modulation of GCRs. \add{Recent studies using historical cosmogenic isotope imprints have also revealed longer cycles of the modulation such as the 2400-year Hallstatt cycle, the Millennial Eddy cycle, or 210-years Suess/de Vries cyclee and so on \citep[see the review article by][]{usoskin2017history}.}
In the past decades, the decrease (or increase) in GCR intensity near solar maximum (or minimum) has been studied and quantified using data from both ground-based neutron monitors \citep[e.g.,][and Fig. \ref{fig:ACR-CR}]{usoskin2017} and high-energy particle detectors for space missions. 
As already introduced in Section \ref{sec:GLE}, NM measures the secondary neutrons generated in Earth's atmosphere by primary cosmic rays, being GCRs or occasionally SEPs with protons reaching GLE energies. 
\change{Continuous}{An example of continuous} measurement of energy-resolved particles in space is shown in the bottom panel of Fig. \ref{fig:pfl} where the SEP-subtracted background flux measured by GOES is anti-correlated with the solar-cycle sunspot number (top panel). 

\begin{figure}[ht!]
\centering
\includegraphics[width=0.9\textwidth]{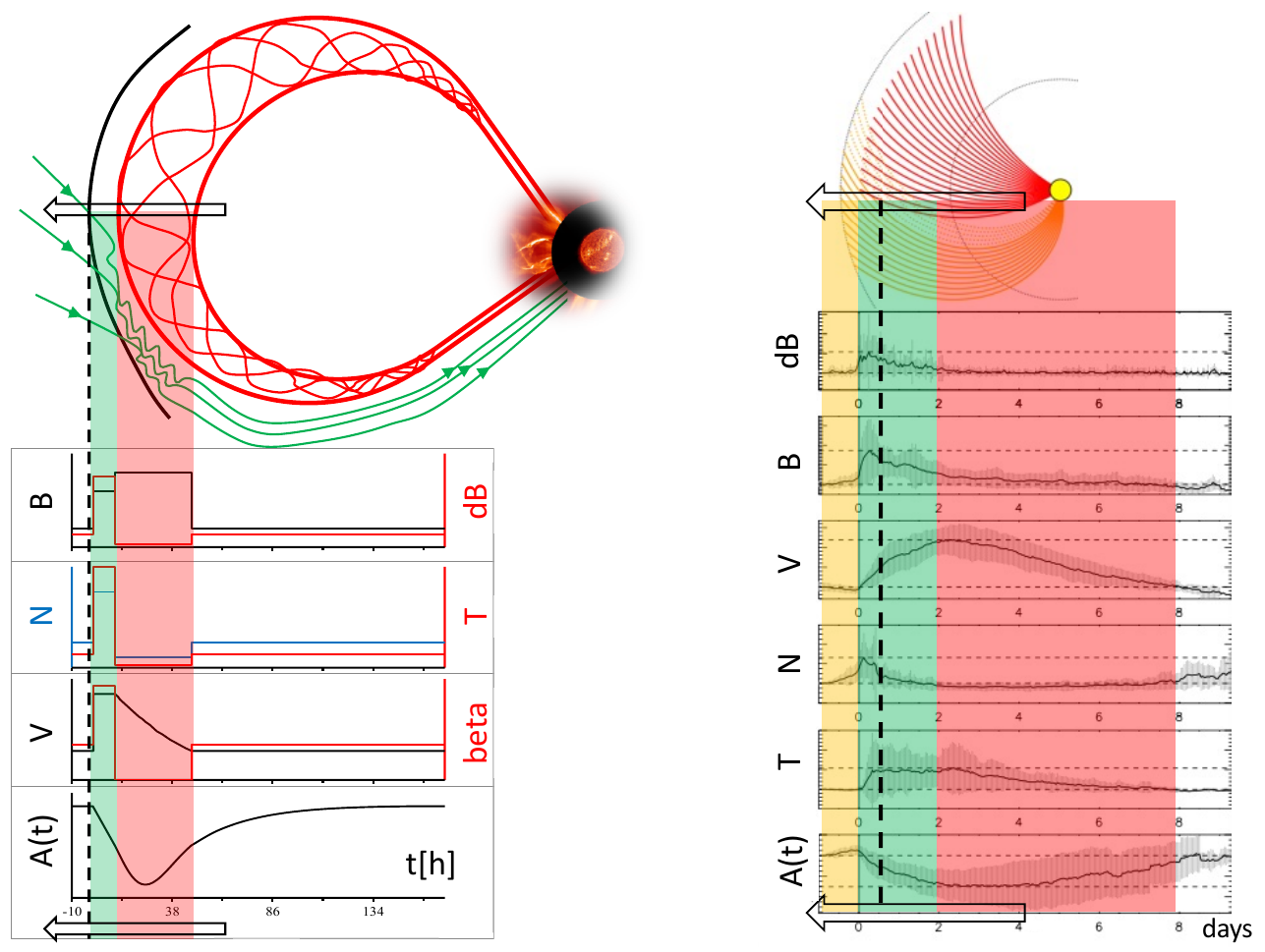}
\caption{\textit{Left:} A sketch of an ICME consisting of the magnetic ejecta/flux rope (red) and shock/sheath (black line followed by a green region) with corresponding in situ measurements observed as the spacecraft crosses through the trajectory marked by an arrow. The in situ measurements show a generic ICME profile in (top-to-bottom) magnetic field and fluctuations, density and temperature, speed and plasma beta, and cosmic ray count \citep[adapted from ][]{dumbovic20}. \textit{Right:} A sketch of a SIR with highlighted regions of slow wind (yellow), compression region (green; with stream interface marked by black dashed line), and high speed stream (red). Corresponding in situ measurements are given below observed as the spacecraft crosses through the trajectory marked by an arrow. The in situ measurements show a generic SIR profile in (top-to-bottom) magnetic field fluctuations, magnetic field strength, speed, density, temperature and cosmic ray count \citep[adapted from ][]{dumbovic22}. %The left and right bottom panels show typical Forbush decreases corresponding to ICME and SIR, respectively. 
The ICME-related generic Forbush decrease (bottom left panel) was obtained from modelling \citep[for details see][]{dumbovic20}, whereas the SIR-related \delete{recurrent} Forbush decrease generic profile (bottom right panel) was obtained using superposed epoch analysis \citep[for details see][]{dumbovic22}.}
\label{fig:FD}
\end{figure}

On short time-scales of several days, GCR modulation is dominated by interplanetary transients such as interplanetary coronal mass ejections (ICMEs) and stream interaction regions (SIRs). ICMEs are interplanetary counterparts of solar eruptive magnetic structures that evolve as they propagate in the interplanetary space and may pile-up plasma and heliospheric magnetic field in the so called sheath region as well as drive a shock \citep[see e.g. overview by][]{kilpua17}. During their passage a depression can be observed in the GCR count, called \add{a} Forbush decrease, \delete{as} first discovered by \citet{forbush37} and \citet{Hess-1937}. 
SIRs are interplanetary transients formed by the interaction of the high speed streams originating from coronal holes and the slow solar wind \citep[see e.g. overview by][]{richardson18}. Similarly as ICMEs, they can cause short-term depressions in the GCR flux. They may recur in several solar rotations, in which case they are called corotating interaction regions (CIRs) and thus their corresponding short-term Forbush decreases in the GCR flux are recurrent \add{following the 27-day solar rotation rate} \cite[][and references therein]{richardson04}. A sketch of ICME/SIR and the corresponding (recurrent) Forbush decrease is given in Fig.~\ref{fig:FD}.
For a detailed overview of ICMEs, SIRs and other solar wind disturbance in the heliosphere, readers are referred to Temmer et al. 2023 (in this special issue). 

In this section, we give a brief overview of the recent progress on GCR modulation and the physics of GCR transport\delete{which affects this modulation effect} and address the limitations and open questions of this topic. 

\subsection{Recent Progress and current understanding}\label{sec:gcr_progress}

\begin{figure}[ht!]
\centering
\includegraphics[trim=0 210 0 0, clip, width=\textwidth]{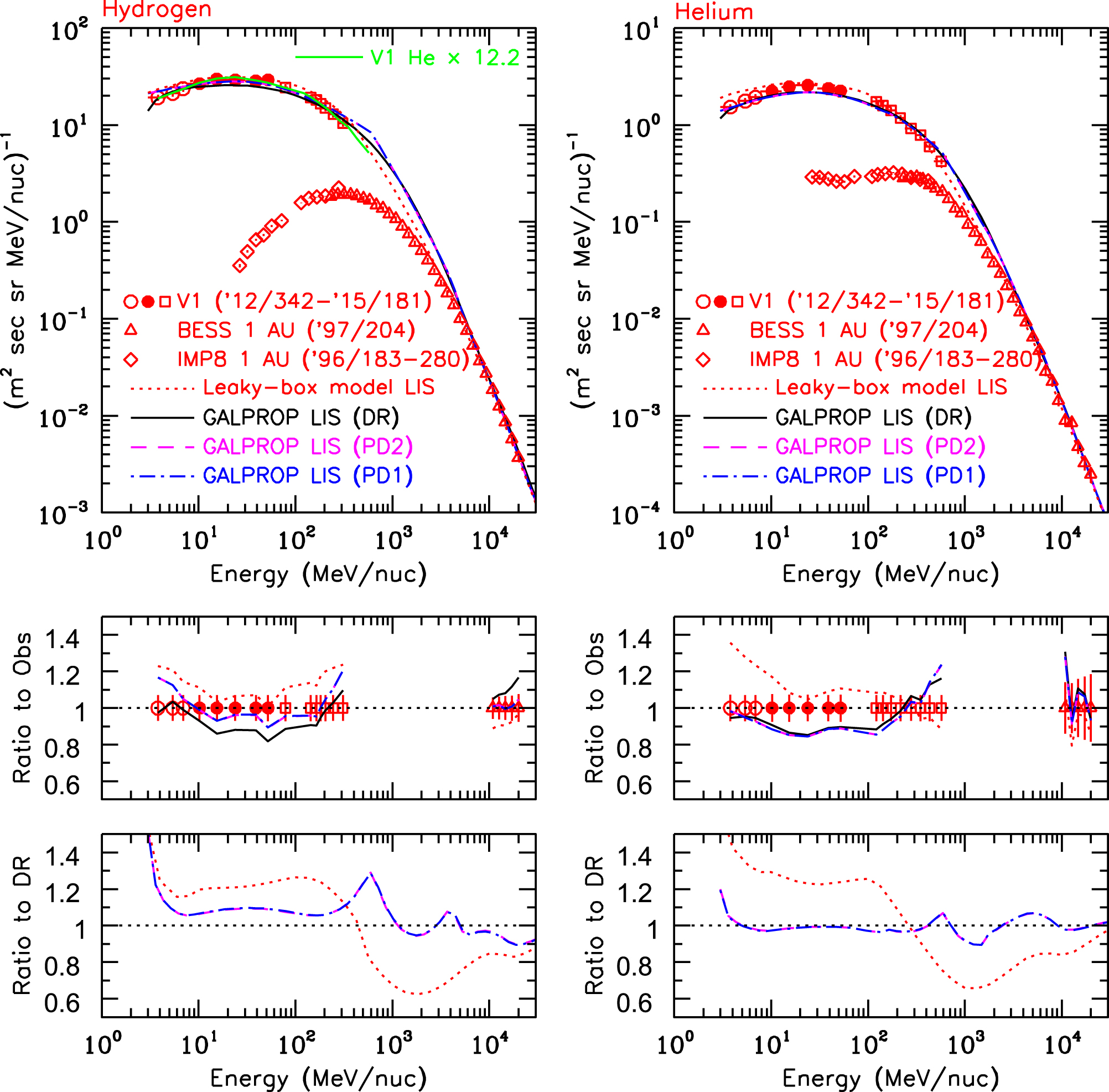}
\caption{Differential energy spectra of LIS H (left) and He (right) from V1 for the period 2012/342–2015/181, and solar-modulated spectra at 1 AU from a BESS balloon flight in 1997 and from IMP8 in the latter part of 1996. Several modeled GCR spectra are also plotted as a comparison. (Credit: \citet{Cummings2016}, reproduced with permission from AAS).}
\label{fig:LIS}
\end{figure}

In the heliosphere, GCRs propagate against the expanding solar wind and interact with embedded magnetic irregularities. 
Thus their intensities in the inner heliosphere differ from their Local Interstellar Spectra (LIS) beyond the heliopause at about 120 AU \citep{Krimigis2013}. Although LIS are the essential ingredient of the solar modulation model, our knowledge for LIS is very limited because of the difficulty of direct measurement. 
Voyager 1 crossed the heliopause on 2012 August 25 at a radial distance 121.6~AU from the Sun, and observed the low energy GCRs from the LIS \citep{Krimigis2013,Stone2013,Gloeckler2014}. 
Six years later, on 2018 November 5, Voyager 2 also entered the interstellar space and measured the LIS \citep{stone2019}.
The LIS measured by Voyager 1 shows that nuclei spectra flatten and roll over below a few hundred MeV/nuc with respect to the approximate power law dependence at higher energies, and the electrons (including positrons) intensity exceeds that of protons below $\sim$ 50 MeV \citep{Cummings2016}.
The two Voyager probes provided invaluable information on the unmodulated GCR energy spectrum as shown in Fig. \ref{fig:LIS}.
However, their energy range is very limited, from a few MeV/nuc to a few hundred MeV/nuc, so that the complete LIS are still primarily based on model predictions. %and can not determined by current directly measurements.

%GCRs undergo different processes as a function of heliocentric distance, latitude, energy, and time from the moment they gain entry to our heliosphere bubble until they reach us.

During the propagation in the heliosphere, GCRs are subjected to four major modulation processes: convection and adiabatic energy losses caused by the expanding solar wind, diffusion due to the random motion on the irregularities of the heliospheric magnetic field, and drift motions resulted from gradients and curvatures in the HMF as well as the abrupt change of the magnetic field direction above and below the heliospheric current sheet (HCS) \citep{potgieter13, moraal13}. 
The modulation effect is significant for particles with energy below $\sim$ 10 GeV with an energy dependence \citep{Gieseler2017} and varies with time and position. 
The solar-cycle variation of GCRs has been long observed at Earth by ground-based NMs and/or high-energy particle detectors onboard space missions \citep[e.g.,][]{heber06, Picozza2007, kounine2012ams,usoskin2017, Fu2021}. Recently, this modulation effect has also been observed and studied at other planets/locations in the heliosphere, such as at Mars \citep{guo2021radiation}, Saturn \citep{Roussos2020} as well as ROSETTA en route to comet 67P/Churyumov–Gerasimenko \citep{Honig2019} and New Horizons (NH) en route to Pluto \citep{wang2022variation}. 

\begin{figure}[ht!]
\centering
\includegraphics[width=0.9\textwidth]{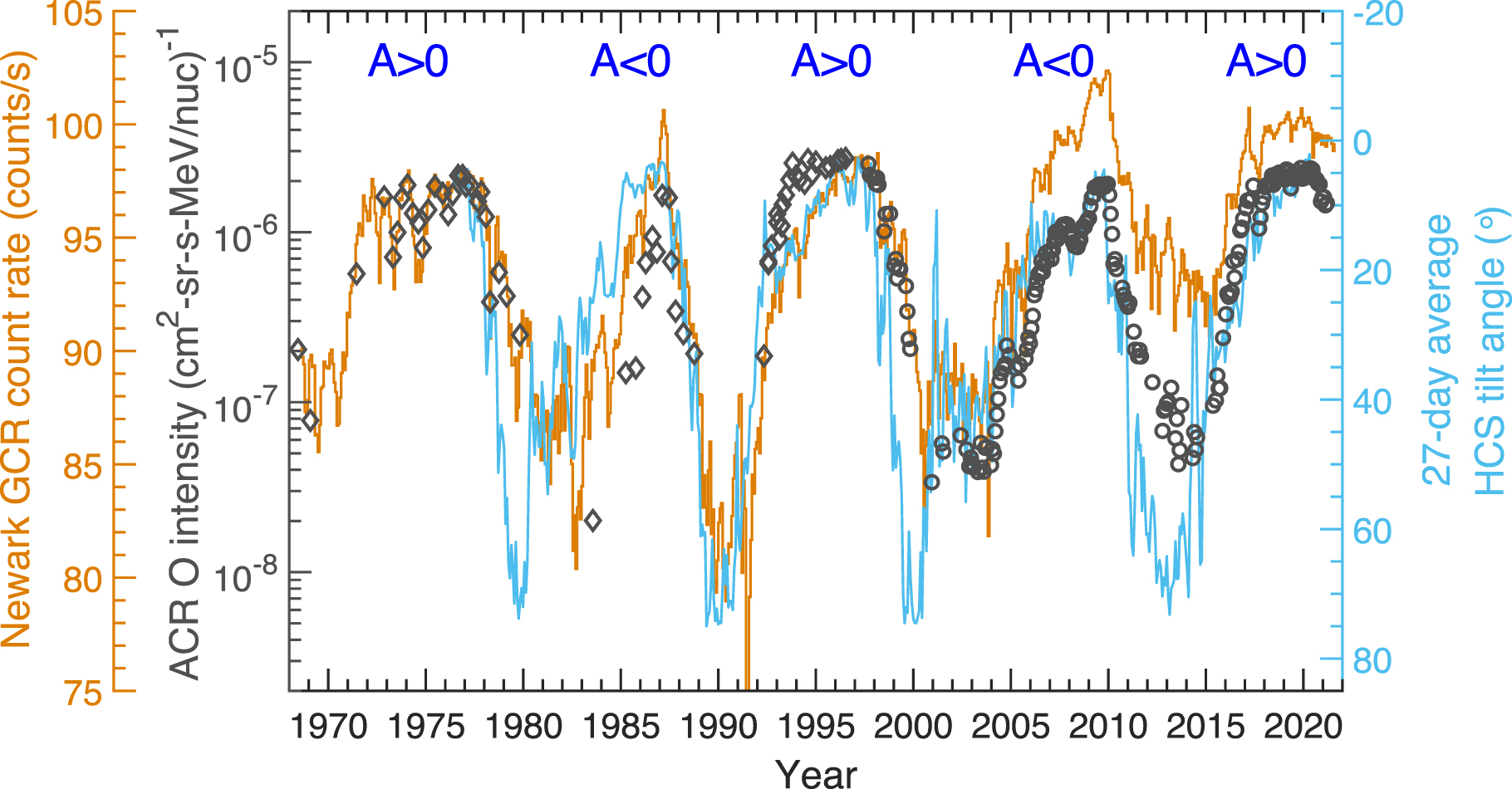}
\caption{Time profile of the count rates of secondary particles generated by GCRs as measured by the Newark neutron monitor (orange curve; first left axis), ACR oxygen intensity (black points; second left axis), and the HCS tilt angle (cyan curve; right axis), adapted from \citet{Fu2021}.}
\label{fig:ACR-CR}
\end{figure}

Since GCRs drift along opposite trajectories during opposite heliospheric magnetic field polarity cycles, their intensities not only reflect the 11-year solar activity cycle but also the 22-year solar (heliospheric) magnetic polarity cycle.  Numerous studies have explored the anti-correlation of the GCR count rate and various solar and heliospheric parameters, such as the Sunspot Number (SSN), \add{solar radio flux}, the strength and turbulence level of heliosperic magnetic field, the HCS tilt angle, the open solar magnetic flux and so on \citep[e.g.,][and Fig. \ref{fig:ACR-CR}]{heber06, usoskin_etal_1998, Cliver_Ling_2001, alanko_etal_2007, potgieter13, wang2022variation}, and empirical functions describing the GCR dependence on different solar cycle parameters have been proposed \citep[e.g.,][]{Dorman_2001, usoskin_etal_2011, wang2022variation}.
\add{Recently, the GCR anisotropy and its dependence on the 22-year magnetic cycle have also observed and are attributed to the magnetic field turbulence, solar wind convection, particle diffusion and drift processes and the combined effects with the existing GCR gradients \citep[see e.g.,][ and references therein]{Wozniak2023}.}

The transport of GCRs in the heliosphere is described by the transport equation \citep[TPE,][also referred to as the Parker transport equation]{parker65} including all major modulation processes.
%The CR intensity is a function of time, the three spatial coordinate and energy (alternatively momentum or rigidity).
In the most general case, TPE is a 3D \add{rigidity-dependent,} time-dependent and space-dependent partial differential equation with a second order diffusion tensor, which can be numerically solved \citep[see e.g. review by][]{strauss17}. 
However, solving the full TPE is highly complex, computationally expensive, and might suffer from the numerical stability problem.
Therefore, a much simpler, \add{1D} approximate solution, called the force-field approximation method, is widely used. It can be derived from TPE assuming 1D steady state, spherical symmetry\delete{, diffusion in radial direction} and neglecting adiabatic cooling and drifts \citep[for more description and discussion see e.g.][and references therein]{gleeson68,caballero-lopez04,moraal13}.
The force-field approximation is popular because it is easy to use with an analytical method describing the modulation level with a single parameter \delete{called modulation potential}.
In recent years, the stochastic differential equation (SDE) method is widely applied in solving the 2D or 3D Parker's equations due to the unconditional numerical stability and progress in computing resources \citep{Kappl2016,Potgieter2017, Boschini2018a}.
The Parker transport equation is transformed into a set of SDEs and then the solution is sampled via Monte Carlo simulations.
Regardless of whether one seeks a solution of a full TPE or a simpler but approximate solution, one needs to apply boundary conditions, i.e., information on the LIS of GCR is necessary.

Recently, the precise measurements of the energy and time dependence of GCR intensities throughout an entire solar cycle by PAMELA \citep[on board of Resurs DK1 satellite since 2006][]{bonvicini2001pamela} and AMS-02 \citep[installed on the International Space station on 2011 May][]{Bindi2017} shed light on the LIS and improve the understanding of the details of solar modulations.
These measurements cover the long solar minimum between cycle 23 and 24, the solar maximum and solar polar reversal, and the recent solar minimum in cycle 24/25.
A number of complex numerical models has been built to investigate the monthly measurements of GCR spectra by these experiments.  
The classic force-field approximation method is not sufficient to interpret these precise data, thus the modification of the force-field approximation \add{or more advanced models} are needed \citep{corti16,Cholis2016,Shen2021}.
With the developed solar modulation models and measurements from Voyagers, PAMELA and AMS-02, several new LIS \change{spectra}{models} are also constructed  \citep{Potgieter2014,corti16,Herbst2017,Gieseler2017,Zhu2018,Boschini2018b,Bisschoff2019,Wang2019,Boschini2020,Wang2022}

Studies of GCR modulation help us to conclude the following: \citep[e.g.,][]{Tomassetti2017,Felice2017,Corti2019,Wang2019,ngobeni2020,Aslam2020,Song2021,Fiandrini2021, Roussos2020,Wang2022, wang2022variation}
\begin{itemize}
\item The GCR modulation effect is rigidity dependent (stronger for particles with smaller rigidity), location dependent (generally stronger modulation at distances closer to the Sun) and time dependent (spanning from a few hours as modulated by ICMEs to more than a solar cycle \add{and even over millennia}).
\item The existence of a time lag of variation of GCR intensity relative to the solar activity, although this lag is not constant, but rather rigidity dependent, \add{polarity-dependent,} location dependent and time dependent. 
\item The radial gradient of the GCR flux within about 10 AU is mostly between 2\% and 4\%/AU, which is also rigidity dependent, location dependent and time dependent. 
\item The \delete{diffusion coefficient and the} rigidity dependence of \change{mean free paths}{diffusion coefficient} \change{vary}{varies} with time and location.
\item During the \change{solar maximum}{high solar activity epochs}, the drift effects are suppressed. 
%\item \delete{The mean free path is the same for different nuclei.} 
\end{itemize}

%On time-scales of several days/weeks, GCR flux can be modulated by interplanetary disturbances  and observed as Forbush decreases whose shape varies depending on its interplanetary cause, as well as the relative trajectory of the spacecraft through the interplanetary transient. Both ICME-caused Forbush decreases and CIR-caused recurrent ones have been extensively studied.

%The magnitude of the Forbush decrease however, not only depends on the interplanetary transient, but also where in the heliosphere it was observed and on the instrument that observed it \citep[see e.g. reviews by][]{cane00,richardson04,belov09}. Therefore, to compare observation and modelling of Forbush decreases, one must take into account the physical properties of the interplanetary transient, its evolutionary stage as well as the response of the detector \citep[see e.g.][]{dumbovic20}. 
%With new modelling efforts \citep[e.g.][]{dumbovic18b,petukhova19b,benella20,guo20}, GCRs can be used as signatures of interplanetary transients, especially where solar wind plasma and magnetic field measurements are not available. 
%It may be especially interesting to analyse the evolution of the interplanetary transient by observing the corresponding Forbush decrease at different locations in the heliosphere \citep[e.g.][]{witasse17,winslow18, guo2018September, forstner18}, however in order to do it accurately, one needs to consider different response of different detectors \citep[e.g.][]{forstner20}.

As noted earlier, on time scales of several days/weeks GCR flux is modulated by interplanetary disturbances and observed as so called Forbush decreases (Fig. \ref{fig:FD}). The variety of Forbush decreases \change{detection}{detected}, from spacecraft to ground-based observations both at Earth and other planets, such as Mars, \change{have}{has} provided further insights on the characterisation of Forbush decrease events at different locations in the heliosphere, as well as during different times in the solar cycle.
Their properties are as follows:

\begin{itemize}
\item The time evolution of the Forbush decrease is characterised by a decrease followed by a recovery period. However, the intensity of the decrease (as in, the maximum  reduction of the GCR count rate), the speed of the decrease and that of the recovery all vary depending on its interplanetary causes \citep[see e.g. review by][]{cane00, belov2008}.

\item The relative trajectory of the detector through the interplanetary transient can also affect the shape of the detected Forbush decrease.

\item The Forbush decrease magnitude depends on the interplanetary transient, the location in the heliosphere where it was observed, and the instrument that observed it \citep[e.g.][]{witasse17,winslow18, guo2018measurements, forstner18, dumbovic20, forstner20}.

\item \add{The energy spectrum of GCRs can change during Forbush decreases as being directly observed \citep[by e.g., PAMELA,][]{usoskin2015force} and the changes may depend on the state of the turbulence of the interplanetary magnetic field \citep{ALANIA2012}.}

\item With long-term spacecraft detection of transient structures such as SIRs and CMEs \add{by WIND, ACE, etc.}, as well as GCRs on the ground \add{by e.g., neutron monitors} \add{ or in space by e.g., AMS-02}, it is now possible to analyse statistically large samples of Forbush decrease events in relation with interplanetary transients. These reveal how the compression region ahead of these structures (the sheath), as well as the magnetic structure inside them (the magnetic ejecta/cloud), \change{but}{and} also the speed at which these events travel, can all play a role in driving Forbush decrease of different intensities \citep[e.g.][]{badruddinkumar2016, masias2016superposed, janvier21}. 
%However, the shock is not in itself a driver of the Forbush decrease.

\item In general, the effect can be described using the TPE approach \citep[e.g.][]{dumbovic18b,petukhova19b,benella20,vrsnak22}, though different models might use slightly different assumptions \add{and simplifications, such as the simplified force-field approximation can also be used to describe some Forbush decreases \citep{usoskin2015force}.} 
\end{itemize}

\subsection{Limitations and open questions}
\label{sec:gcr_limitation}

Although it has been well observed for about 6 solar cycles that the GCR flux can be correlated with various solar and heliospheric parameters, some observations still cannot be fully explained \citep[e.g.,][]{Ross_Chaplin_2019}. 
For instance, it has been found that the GCR intensity lags behind the sunspot number variation and this lag time during odd cycles is often longer than that during sequential even cycles. Many early studies explained this time lag as driven by the outward solar wind convection and the inward GCR transport whereby the drift direction of particles reverse during positive and negative solar (heliospheric) magnetic field polarity cycles \citep{vanAllen2000, Dorman_2001, usoskin_etal_2011, Cliver_Ling_2001, Thomas_etal_2014}. Some others suggest that this lag is primarily due to the late opening of the solar magnetic field with respect to SSN which already shows an odd-even cyclic pattern \citep{wang2022variation}.
Besides, the energy-dependence of the time lag is also poorly quantified due to the lack of continuous and accurate energy-dependent GCR data. Some recent work suggests a shorter delay for higher energy GCRs \citep{Shen2020}, an effect which may be attributed to the energy-dependent transport of GCRs in the heliosphere \citep{Moloto_Engelbrecht_2020}. But the detailed quantification is still needed based on more observations for a precise physical description of the energy-dependent transport process. 
Another important index that can be contrasted against different transport model predictions is the radial intensity gradient of GCRs. 
However, it is rather difficult to quantify due to the rareness of continuous and simultaneous measurements at different heliospheric locations and also the diverse datasets by instruments with different energy responses. Many studies are subject to various uncertainties and the further assessment at different time periods and locations is essential \citep{Roussos2020}. 

\add{Another complexity to consider is the role of the different ICME substructures (shock, sheath and magnetic ejecta) in modulating GCRs. The sheath region ahead of the ejecta, when observed further away from the Sun (e.g. with spacecraft at 1 au), combines both the shock compression and solar wind compression due to the ``piston'' effect of the moving ejecta. 
As such, sheaths are a really interesting sub-body of ICMEs that show properties different from both the solar wind and the ejecta \citep{Kilpua2021}. The shock itself is a thin barrier between the ambient solar wind and the sheath downstream. The effect of the shock only would be reported as a sudden jump downward for the FD when crossing the discontinuity, unlike the typical smooth continuous decrease (with varying slopes) often reported. Furthermore, \cite{janvier21} reported via a statistical analysis that magnetic ejecta without a sheath can also drive FDs. Then, the contribution of each substructure in driving a FD, to what intensity, and what effects when combined still needs to be fully understood. }

The limitations regarding GCR transport and modulation modelling are well summarized in reviews by \citet{vainio09},\citet{potgieter13} and \citet{Engelbrecht2022}. 
For comprehensive modelling reliable numerical schemes are needed with appropriate LIS as boundary conditions and properly estimated transport parameters. The latter still poses a huge challenge, as the diffusion tensor cannot be directly measured or easily estimated, especially given our limited understanding of the turbulence in the solar wind, the main ``ingredient'' of diffusion \citep[e.g.][]{Zank1998,matthaeus11}.
Several theories have been developed to describe the diffusion coefficient \citep[e.g.][]{Bieber1994,Teufel2003,Qin2007,Wiengarten2016,Zhao2017,Zhao2018,Shalchi2020}.
However, due to its complexity, most models still adopt simpler empirical formulas, e.g. rigidity-dependent linear or smoothly broken power law for the mean free path, and a simple spatial dependence related to heliospheric magnetic field distribution.
The empirical formulas often have some free parameters and cause \change{large free degrees}{large diversities in the underlying process} when we investigate the solar modulation effect. 
For instance, in the previously mentioned models, the monthly PAMELA and AMS-02 measurements could be reproduced with different sets of free parameters and LIS. 
Furthermore, the degeneracy among normalization of diffusion coefficient, the rigidity index of the \change{mean free path}{diffusion coefficient}, and the level of drift effect makes it hard to find the best-fit parameters.
The Markov chain Monte Carlo (MCMC) method could be somewhat helpful but can not induce distinct improvement or clear up the confusion \cite{Song2021}.
The monthly flux of protons, antiprotons, electrons and positrons, could improve the clarification on the degeneracy.
More efforts for computing the transport of turbulence and utilizing the latest diffusion theory are needed to build reliable diffusion coefficient and results in more realistic LIS and improvement for studying solar modulation processes \citep[e.g.][]{Zank2012, Engelbrecht2013,Zank2017,Zhao2017,Zhao2018,Moloto2018,Oughton2021,Engelbrecht2021,Wang2022turbulence,Adhikari2023,kleimann2023}.

Moreover, the transport of GCRs in the heliosphere is governed by the properties of heliosphere, such as the magnetic field structure, solar wind speed, HCS structure, and the heliospheric boundary. A realistic description to the dynamic heliospheric environment is essential but poses big challenges as listed below. 
\begin{itemize}
\item During the solar minimum, the solar magnetic field can be approximated as a tilted dipole filed. But during the solar maximum and the polarity reversal periods, the solar and heliospheric magnetic fields may not follow a regular form but \change{with}{have} chaotic and dynamic structures \citep[e.g.,][]{Bindi2017}. Thus the classic (or modified) Parker magnetic field structures \citep{jokipii1989,Fisk1996} are not appropriate to compute the drift velocity.
\item The solar wind speed as a main input parameter in the modulation model exhibits a distinctively latitudinal dependence in the solar minimum and is relatively uniform in the solar maximum, but such pattern is not clear during other solar activity levels \citep{McComas2002,Zurbuchen2007}. 
Most continuous solar wind measurements so far are constrained at the equatorial plane near 1 AU that could not provide a full view of solar wind speed. \add{Only the indirect method based on observations via interplanetary scintillations and backscattered Ly$\alpha$ mapping of the interplanetary H can provide the latitudinal variation of the solar wind flow \citep{Bzowski2003,sokol2013}.}
\item The large scale heliospheric magnetic structure determines the open solar flux, heliospheric magnetic field properties, HCS structures and tilt angles. However, the information of these parameters is highly model-dependent (such as based on force-free field extrapolation) and poses large uncertainties during the extreme solar activities \citep{potgieter13}.
\item It is generally accepted that the terminal shock and heliosphere structure are significantly asymmetric in terms of a nose-tail direction with a blunt shape \citep{Zank1999,Jokipii2004,Opher2009,Pogorelov2009}. In the solar modulation model, for simplicity, the terminal shock and heliopause are often modeled as a spherical structure.
\end{itemize}
The further combination of the MHD simulation and the \change{GCRs transport}{transport of GCRs} may give a more accurate description of the heliosphere condition and will refine solar modulation models \citep{Florinski2003,Ball2005,Luo2013}. 
It is also important to accumulate data of CR variations at multiple locations in the heliosphere (varying in heliospheric distance and latitude) which will improve the understanding of solar modulation \citep{Simone2011,Vos2016,Honig2019,Modzelewska2019,Knutsen2021, Roussos2020, wang2022variation}.

Regarding the short-term effect caused by CIRs and ICMEs, we still have many limitations in our current understanding.

First, there is a lack of extensive comparison between models and observations taking into account the measured particle energy range and we have limited opportunities comparing data at different heliospheric positions \change{taking into account}{limited by} different \add{energy} response of different detectors. 
For future studies, \change{implementing}{considering} energy response of the detector is of great importance and might help to resolve some long-standing issues.
%\begin{itemize}

Second, the dual nature of ICME-related Forbush decrease \change{is}{has been} known for decades: the existence of so-called two-step Forbush decreases was confirmed by many studies and it is generally accepted that their two-step nature is related to the \add{sub-structures (}shock, sheath and magnetic ejecta regions) of ICMEs \citep[for overview see e.g. ][]{cane00}. However, not all ICMEs which have both regions will produce observable two-step depressions, which in turn made some authors to question their causes \citet{jordan11}. Furthermore, it has not been fully resolved whether these two steps are a cumulative or separate effects or what determines their relative contribution to the total depression. \citet{richardson11} and more recently \citet{janvier21} found that near Earth, on average, both ejecta and sheath effects contribute approximately the same to the total depression. \citet{cane94} and later \citet{blanco13} found indications that ejecta effect weakens with time, whereas \citet{forstner20} found that sheath effect becomes more prominent in time. However, a conclusive answer on the nature of the two-step Forbush decreases remains elusive due to the differences induced by different detector responses which hamper comparisons between different instruments. Furthermore, it remains difficult to study the effect of the ejecta alone due to the hampered detection. Indeed, ICMEs without a shock and a sheath have on average smaller speeds and smaller magnetic field intensities, making them difficult to detect in the surrounding solar wind.

Besides, our limited ability to compare the Forbush decrease effect as detected by different instruments is the main reason for yet another long-standing issue: \change{how}{what} do early/late Forbush decreases look like, i.e. how do different ‘evolutionary stages’ of a Forbush decrease look like. It is suspected that the Forbush decrease should reflect the evolutionary stage of its complimentary interplanetary disturbance, however, this is yet to be shown conclusively and can only be achieved through systematic multi-spacecraft studies.

Finally, with better knowledge and understanding of Forbush decreases and especially their evolutionary stage, which presumably reflects that of their interplanetary cause – Forbush decreases can be utilized as means for ICME/SIR analysis in conditions where measurements other than cosmic ray counts are not available. %(e.g. in the pre-satellite era or on planetary surfaces). 
For instance \citet{lefevre16} and \citet{vennerstrom16} used Forbush decreases as indication of ICMEs in the pre-satellite era at Earth; \citet{moestl15} and \citet{forstner18} used Forbush decreases as indication of ICME arrival at Mars; \citet{winslow18} and \citet{witasse17} tracked ICMEs across the different locations in the heliosphere using Forbush decreases. 
With better understanding, Forbush decreases might provide more information on the ICME/SIR properties than simply arrival time. Moreover, since they can be detected by relatively cheap, small and simple detectors that can be put aboard practically any spacecraft, Forbush decreases can be easily used for space weather monitoring and diagnostic purposes, i.e., to probe and track interplanetary transients across the heliosphere.
%\end{itemize}
%set spellspellang=en_us
\section{Anomalous Cosmic Rays}\label{sec:ACR}

\subsection{Introduction}\label{sec:acr_intro}
As a result of the solar modulation, the low energy GCR intensities should decrease with decreasing particle energy. However, the analysis of the \change{GCR}{cosmic ray} spectra in the early 1970s revealed an anomalous enhancement at low energies \citep{Garcia-Munoz1973}. This enhanced component is called anomalous cosmic rays (ACRs). 

Low energy ACRs are primarily single-ionized which distinguishes them from SEPs (Sect \ref{sec:SEP}) and GCRs (Sect \ref{sec:GCR}). However, ACRs with higher charge states are present and become dominant with total energies above $\sim$ 350 MeV \citep{Klecker1995, Cummings2007}.
High energy ACRs take more time to accelerate, thus there is more ionization via charge stripping.
The abundance of multi-charged elements constraints the acceleration timescale that must be of the order of one year \citep{Mewaldt1996}.

The most abundant ACR elements are H, He, N, O, Ne and Ar\delete{. They }\add{which} have significant neutral abundance in the local interstellar medium.
\citet{Fisk1974} thus first suggested the source of ACRs is pickup ions.
Pickup ions originate from the interstellar neutral atoms, which penetrate into the heliosphere and become ionized \add{mainly} by solar radiation via photoionization and/or by solar wind ions via charge exchange. 
Then these particles are picked up by the solar wind magnetic field and convect into the outer heliosphere.
Since the pickup ion energy is $\sim$ 1 keV/nuc, they must be accelerated by about four orders of magnitude to become ACRs for which a very efficient particle acceleration mechanism is needed.

\subsection{Recent Progress and current understanding}\label{sec:acr_progress}
Before Voyagers crossed the termination shock (Section \ref{sec:GCR}), it was believed that the ACRs are accelerated through diffusive shock acceleration \add{mechanism} at the termination shock \citep{Pesses1981}. 
But Voyagers did not observe the expected power-law energy spectrum, but instead observed that in the 3-30 MeV energy range\change{,}{ and} the intensity did not increase at the shock \citep{Stone2005,Stone2008}.
For particles at \change{energy}{energies} below $\sim$ 1 MeV, called termination shock particles (TSPs), Voyagers observed a significant flux enhancement at the time of the shock crossing \citep{Decker2005,Decker2008}.  
As the Voyagers \add{further} moved into the heliosheath, the intensity of particles with energy above a few MeV continued to rise and eventually have an approximate power-law spectrum, while TSPs were observed to be \add{nearly} uniform in the heliosheath \citep{Cummings2008}.

\begin{figure}[htp]
    \centering
    \includegraphics[width=0.6\textwidth]{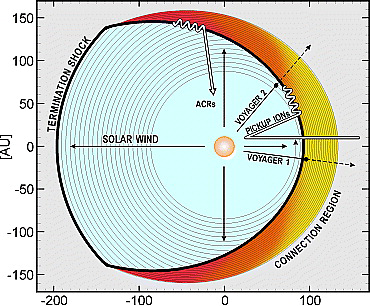}
    \caption{Schematic diagram of the blunt-shape heliosphere termination shock (Thick black line), the heliospherical magnetic field (gray line), the trajectory of pickup ions and ACRs (long arrow), and the ecliptic projection of the orbits of Voyagers. (Credit: \citet{McComas2006}, reproduced with permission from AGU).} 
    \label{fig:TS}
\end{figure}

\citet{McComas2006} first noted that the blunt-shaped structure of termination shock has a significant effect on the acceleration of particles and put forward a simple geometric interpretation to the Voyagers observations (see Figure \ref{fig:TS}). 
The intersection points between the field line and the termination shock move along the shock from the nose toward the flanks as the field line is pulled out by the solar wind into the heliosheath.
As the intersection points moving toward the flanks, pickup ions have longer time to accelerated and become ACRs \citep{Schwadron2008}.
The peak fluxes of ACRs are in the flanks, remote from the Voyagers, and these particles are transported within the heliosheath. Thus Voyagers observed the peak of ACRs in the heliosheath rather than at the shock nose.
Alternatively, TSPs are accelerated over a short time with a power law distribution and remain near the point of injection. These low energy particles are expected to peak at the nose of termination shock where Voyagers explored.

This blunt-shock concept is confirmed later by numerical computations performed by \citet{kota2008,Schwadron2008}.  
A 2D hybrid simulation suggested that accelerated TSPs at the termination shock are uniformly distributed along the termination shock, and are likely quite uniform throughout the entire heliosheath \citep{Giacalone2021}.
\citet{Cummings2019} analyzed the ACR anisotropies and found that a diffusive streaming of ACRs comes from the flanks of the heliosphere, further supporting the geometric explanation shown in Figure \ref{fig:TS}.

\change{Meanwhile}{However}, it has also been suggested that the shock interacting with large-scale plasma turbulence may result in a similar effect as the blunt-shock if the shock surface becomes rippled and the scale of the ripples is larger than the characteristic scale associated with the particle diffusion \citep{Li2006,kota2010,Guo2010}
The meandering magnetic field lines caused by the turbulent plasma upstream of the shock could also produce similar effects \citep{kota2008}.

Since in the heliosphere ACRs transport in the same plasma as GCR, they also experience the solar modulation effect and exhibit the 11-year and 22-year cycle.
\delete{A single ionized ACR has larger rigidity than GCR with the same momentum, and the propagation process is sensitive to the diffusion coefficient which is proportioned to the power law of rigidity.}
%\delete{Hence, compared to GCRs, ACRs can be a more sensitive probe for studies of the solar modulation and the interplanetary medium \citep{Panasyuk2005}.}
Figure \ref{fig:ACR-CR}, adopted from \citep{Fu2021}, shows the time profile of the ACR and GCR intensity (represented by the neutron count rate, Sect. \ref{sec:GCR}) and the Heliospheric Current Sheet (HCS) tilt angle. As shown, the ACR evolution is better correlated with the HCS tilt angle which is a popular proxy for solar activity and is closer related to the solar modulation than the sunspot number.

\subsection{Limitations and open questions}
\label{sec:acr_limitation}
\change{we}{We} still have many open questions in our current understanding of ACRs, constrained both by limited observations and lack of theories which can fully explain the observations. 

The acceleration mechanism of ACRs is still under debate. 
\citet{Fisk2009} proposed that ACRs are accelerated though diffusive compression acceleration near the heliopause and subsequently diffuse back into heliosheath.
It was also suggested magnetic reconnection is a source for ACRs acceleration \citep{Lazarian2009,Drake2010,Zhao2019}.
We refer to review paper \citep{Giacalone2012,Giacalone2022} for the detailed discussion and open questions on the acceleration mechanism of ACRs. 

Concerning the transport process, although both ACR and GCR particles are modulated by the heliospheric magnetic field, the difference in the ACR and GCR variation with solar cycle is not yet well understood.
During the last two solar minimum periods (2008-2009 with $A<0$ and in 2019-2020 with $A>0$), cosmic ray intensities were much higher than in earlier cycles due to weaker solar modulation \citep{Mewaldt2010}. However, the ACR intensities do not show the same tendency and are comparable or below their level in previous solar minimum as shown in Figure \ref{fig:ACR-CR} \citep{Fu2021}.

Progress on the cosmic ray LIS and transport in the heliosphere could help to investigate the origination and transport of the ACRs. All the improvement mentioned for GCR (Sect. \ref{sec:GCR}) are suitable for applying to ACR studies, vice versa. More efforts are still needed in the following aspects. 
\begin{itemize}
\item We should use time-dependent models including the acceleration at the acceleration site and transport in the heliosphere to compute the ACR flux and compare to observations from different spacecraft at different locations and time.
\item We should adopt the same set of heliospheric parameters to reproduce the observations for ACR and GCR energy spectra and to infer the unmodulated ACR spectra and transport parameters.
\item We should use observations to investigate ACR gradients which may improve the understanding of the transport processes.
\end{itemize}

\section{Future focuses and Recommendations}
\label{sec:Recommendations}

\begin{figure}[ht!]
    \centering
    \includegraphics[trim=50 0 300 0, clip, width=1.2\textwidth]{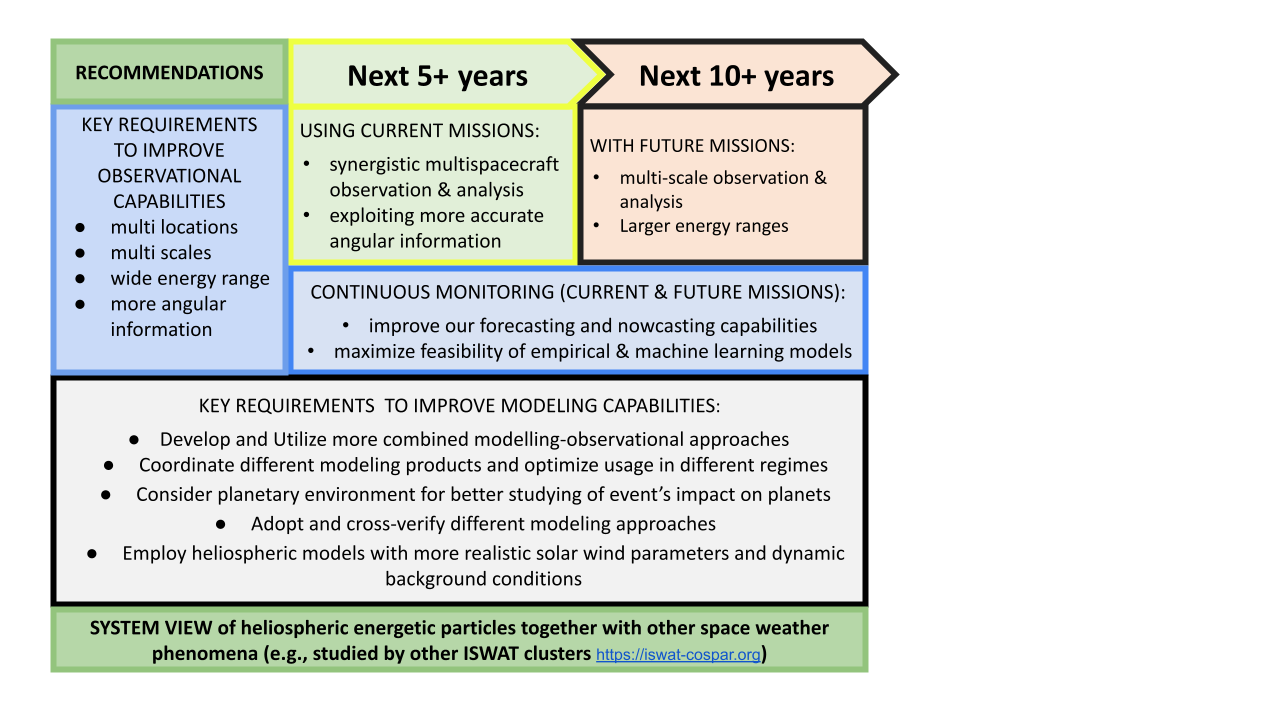}
    \caption{Recommendation for improving observational and modeling capabilities in the next 5 to 10 years}
    \label{fig:recommendation}
\end{figure}

The previous sections have reviewed the current status of our understanding of energetic particles (SEPs, GCRs, ACRs) in the heliosphere. We also pointed out our knowledge gaps in observing, modeling, physical understanding, and forecasting capabilities of the heliospheric radiation environment. In this section, we offer suggestions for narrowing these gaps and moving the field forward in the next 5 to 10 years, as illustrated in Figure \ref{fig:recommendation}. 

\subsection{Requirements of Observations}

As a first step, it is most pragmatic to further exploit existing radiation datasets from different regions in the circum-terrestrial space and in the heliosphere, e.g., the near-geospace (LEO and GEO satellites), the terrestrial NMs, the vicinity of other planets (e.g., BepiColombo, MAVEN, MSL, Juno) and near-Sun regions (Parker Solar Probe, Solar Orbiter). We stress the following aspects. 
\begin{itemize}
\item Well-organized community effort is important. For instance, the EU supported Solar EneRgetic ParticlE aNalysis plaTform for the INner hEliosphere (SERPENTINE, \url{https://serpentine-h2020.eu}) project is currently developing various open catalogs of transient phenomena related with SEP events and a series of new data products for the ongoing particle measurements from Solar Orbiter, Parker Solar Probe and BepiColombo missions. 
\item An interdisciplinary effort to combine data of different types (such as space-borne and ground-based, radiation dose and particle flux) should be also encouraged. 
\item Adopting a planetary space weather approach based on joint-science investigations for more than one planetary bodies in the heliosphere, could be of significant help for better understanding the radiation environment properties at different locations in the solar system \citep[see, for instance,][and references therein]{plainaki2020current}. 
\end{itemize}

Planning for future missions, more observations are still needed in the following aspects. 

First, at multiple well-separated locations (in longitude, latitude and radial distance) we need to maintain or increase the following observations. 
\begin{itemize}
\item Most multipoint observations of SEP fluences so far are based on 2-3 observers at ~1 AU solar distance. Recent joint observations retrieved at near-Earth and a few other locations with available spacecraft (STEREO-A, Solar Orbiter, PSP etc.) such as by \citet{Kollhoff2021} only makes it clear that we still have many questions in interpreting wide-spread SEP events and multipoint observations can greatly advance our understanding of the acceleration and transport processes.

\item Similarly, multi-spacecraft observations of ACRs and GCRs over large heliospheric scales for a continuous time period are too scarce to understand their sources and transport processes or to explain their features such as the modulation phase lagging behind the sunspot number or the energy-dependent radial gradient. Much of our knowledge on ACR and GCR source populations and transport is only based on theories and transport models as described in Sections \ref{sec:GCR} and \ref{sec:ACR}. 

\item Short-term GCR depressions directly depend on the local interplanetary conditions which are greatly influenced by transients such as SIRs and ICMEs. It is therefore recommended that multi-spacecraft measurements of both recurrent and non-recurrent Forbush decreases are employed on a regular basis to track and analyze interplanetary transients throughout the heliosphere.
\end{itemize}

Second, we also need to have more observations at various different scales (both global and local) explained as below. 
\begin{itemize}
\item SEP sources span a huge range of spatial scales: For the most energetic events, the maximum acceleration may happen within a few solar radii of the Sun, however, accelerating shocks can extend over more than 180$^{\circ}$ in heliolongitude, and persist to the outer regions of the heliosphere. Variations of the shock parameters over mesoscales (and how that affects particle acceleration) is nearly completely unknown.

\item We also need observations more densely populated at smaller scales for understanding important dynamics of the SEP acceleration (e.g., between L1 and Earth’s bow shock and at scales resolving the perpendicular shock front where acceleration takes pace).

\item Following the injection/release of SEPs, GCRs, or ACRs from their source regions, the mean free paths of particles can be of the order of one AU or much smaller. The details of their transport, in particular whether it is relatively ``scatter-free'' or dominated by diffusion and/or drifts, should clearly depend on the circumstances -including the background conditions into which the particles propagate. Multipoint measurements of pitch angle distributions, together with contextual information on the local background and the larger scale settings of the local measurements (e.g., from in-situ plasma and field, and solar imaging observations) can lead to improved concepts and models for the global (re)distributions of heliospheric particles from both internal and external sources.
\end{itemize}

Third, heliospheric particles should be monitored at a larger and continuous energy range (from superthermal to nonthermal and relativistic energies). 
\begin{itemize}
\item While superthermal particles are considered to play a key role in providing the seed population for SEP acceleration, particles in this energy range (tens of keV to a few MeV) are not routinely measured, and their composition, spectra, and spatial distributions are still poorly understood. There have been a few potentially useful instruments flown (e.g., ULEIS on ACE, SIT on STEREO), but this remains a population where a gap needs to be filled more routinely by including it in heliospheric particle detector suites.

\item At energies above hundreds of MeV, direct observations are rare and yet they are critical in terms of radiation impacts on aviation, space hardware, and humans in space. Although there are a couple of instruments (AMS02, PAMELA) at Earth, the periodic influences of Earth’s magnetosphere often prevent their usage for onset time analysis. For higher energies, low-cost ground-based neutron monitor(NM) measurements have been very useful for deriving particle spectra above the local atmospheric and magnetic cutoff energy (Section \ref{sec:GLE}). However, SEP spectra derived from NMs depend on calculated atmospheric response models which may carry unknown uncertainties. Moreover, the direct validation of these spectra is often difficult due to the lack of space observations in this high-energy range (at least above 430 MeV which is the atmospheric cutoff energy). Future Lunar outposts and Earth-orbiting space stations provide potential platforms for more regular monitoring at this energy range.

\item For studying ACRs and GCRs which have higher energy ranges than SEPs, the availability of energy-resolved observations is even less. Fluxes integrated over a wide energy range are often used and compared across different instruments. This can be misleading, e.g., when studying solar modulation of GCRs as a transient disturbance while a solar eruption passes different observers. For existing measurements, it is necessary that the energy response of different detectors are taken into account, so that the evolutionary effects can be disentangled from the detector response. More energy-resolved measurements in the future (starting at 100s of MeV/n) can better reveal the physics of the high energy particle transport or modulation process.
\end{itemize}

Besides, more accurate angular distribution information is important. 
\begin{itemize}
\item Most SEP observations provide little information on anisotropies, which can provide considerable insight on the physics. For example, counter-streaming SEP fluxes can indicate ``reservoirs'' produced by local magnetic trapping, while the pitch angle distribution indicates the role of scattering during transport to the detection site. %However, studies so far are based on at best 4 directional measurements (north, south, sunward and antisunward) which carry large uncertainties in deriving the anisotropy information. 
Modern detection techniques allowing for accurate measurements of this property can help better constrain SEP event sources and inform modeling efforts.

\item  Similarly, suprathermal particles, ACR and GCR anisotropies contain insights into their sources and transport. For example, newly created heliospheric pickup ions will retain vestiges of their initial ring-beam distribution function before they are scattered by ambient magnetic fluctuations or self-excited waves become unstable to ion cyclotron wave generation. The subsequent acceleration of these pickup ion seed populations toward their becoming ACRs can be diagnosed if the angular distribution is known. In the case of ACRs and GCRs, information about the larger heliosphere shape and state, e.g., over the solar cycle, is contained in the decadal timescale, global directional modulation of these populations. 
The Interstellar mapping and acceleration probe \citep[IMAP,][]{mccomas2018interstellar} is dedicated to increasing knowledge in these areas, but sustained measurements will hopefully be inspired given the importance/necessity of long term information.
\end{itemize}

Last but not least, for better forecasting their space weather impact of energetic particles in the heliosphere, more observations are undoubtedly needed in the above aspects (1-4:more locations, multiple scales, more energy coverage and angular information) as better understanding of the physics is essential for improving the application capabilities. 
Moreover, continuous monitoring with real time data availability from multiple locations within the inner heliosphere and at 1 AU would improve SEP forecasting and nowcasting capabilities, including for locations away from Earth such as Mars, which are increasingly important for robotic and human space exploration activities. Continuous measurements at regions where radiation information is critical for aviation and satellite operations but not yet sufficient (such as between LEO and geosynchronous orbit) and by a range of space weather monitoring instruments (when feasibility is not an issue) so that we can better nowcast the radiation effects of SEP events (due to their unpredictability) for space industries.

To summarize, we still require more observations at more locations (such as at well-separated longitudes and at distances both close to the Sun and beyond 1 AU), at both small and global scales, measuring energies extending from superthermal to relativistic energy ranges, and with improved angular resolution and coverage. Since particle detectors generally have the advantage of relative low mass, low power and low cost, we recommend carrying one on all future interplanetary missions including  Earth-monitoring spacecraft, solar observatories and even planetary missions if possible. 
For instance, the Bepi-Colombo mission to explore the planet Mercury carries the SIXS instrument which is capable of broadband measurements of X-ray, proton, and electron spectra on its way to Mercury and in orbit \citep{huovelin2020sixs}.

\subsection{Requirements of Modeling efforts}

In parallel to the need of more observations as discussed above, a more aggressive effort is still needed to bring together the modeling efforts and the existing observations with the following emphases. 
\begin{itemize}
\item Tie to observations at both ends. Observations should be used both as inputs, for constraining model parameters, and finally for testing outputs. This also echoes our previous recommendation that more observations are needed, both from particle energy and directional distribution perspective and spatial sampling perspective. 

\item Coordinate different modeling products and optimize usage in different regimes. Modeling efforts consist of several types that produce different results and have different applications. For example, some empirical SEP models produce a peak flux or event-integrated flux, while others aim to produce a complete event flux versus time profile. Similarly, some provide a single point observer view of the event, while others provide a heliosphere-wide view. Identifying the particular value(s) and challenges of each of these is important for the purposes of coordinating and prioritizing ongoing and new efforts.

\item Consider accurate and timely modeling for magnetospheric shielding of particles. Modeling efforts on how magnetic shielding affects SEP intensity and transport inside the Earth magnetosphere (and in the near-Earth region) are needed to both effectively interpret measurements such as those from the GOES geosynchronous orbit missions, as well as to predict SEP hazards for Earth-orbiting space stations under different geomagnetic activity conditions. The computational tools now exist for making realistic simulations of SEPs in Earth’s environment, but it is still difficult to put them into practice on a regular basis. In particular, the realistic modeling of this shielding requires realistic modeling of the magnetospheric response to heliospheric transients (see efforts in ISWAT-G1 cluster ``Geomagnetic environment" \url{https://www.iswat-cospar.org/g1}), which is still being developed and tested for real-time applications. Similarly, we may also need to consider modelling of particles reaching other planets with different magnetospheric conditions, such as Mercury or Mars.

\item Modelling of both SEP and GCR/ACR transport should adopt more realistic heliospheric models (\delete{see }see efforts in ISWAT-H1 cluster ``Heliospheric magnetic field and solar wind") where both large scale solar wind and IMF, as well as small-scale turbulence should be considered. Current models often rely on various ad-hoc and adjustable parameters whose values can not be directly verified but are chosen to fit the modeling results with observations. Instead, a modelling-observational approach is needed to constrain the non-observable parameters such as the diffusion tensor. %The challenges of using more realistic heliospheric models in forecasting also need to be addressed.
%Consider more realistic heliospheric conditions constraining the particle transport, including the corona, solar wind and the transients produced by CMEs and stream interactions, by using state-of-art heliospheric models (see see efforts in ISWAT-H2 cluster ``CME structure, evolution and propagation through heliosphere" \url{https://www.iswat-cospar.org/h2} and Temmer et al. 2023 in this special issue). 
For GCR/ACR models, it is also important to appropriately constrain the boundary conditions, therefore it would be prudent to further improve our knowledge of LIS through ongoing and future observations. 

\item Consider the varying background fluxes affected by interplanetary transients. I.e., ICME/SIR induced Forbush decreases of the background radiation should be incorporated within the SEP/GLE modeling to more accurately describe the temporal evolution of the radiation before, during and after the SEP events. FD modeling can benefit from space-based data in the circum-terrestrial environment joint with different ground-based NM observations as well as solar wind and IMF observations.  

\item Adopt and combine different modeling approaches. Physics-based models, being very useful for post-event studies, are often computationally expensive with limited ability to produce a prediction in real time. Forecasting for operations also needs empirical models and machine learning techniques. These models are based on existing data to identify the dependence of SEP properties on other parameters and can give rapid forecasts and are easily incorporated into forecasting and operations. To improve the forecasting capability of such models, very large data sets are needed and will directly benefit from more observations as discussed earlier; such models should also be studied and tested together with physics-based models. 
\end{itemize}

To summarize, it is critical to improve the scientific understanding of SEP events and GCR/ACR transport and use this understanding to develop and improve radiation modeling capabilities to support operations. 
In particular, \citet{Whitman22} have summarized all SEP models currently developed in the scientific community, including a description of model approach, inputs and outputs, free parameters, and any published validations or comparisons with data. They have concluded that ``The SEP modeling community has developed a rich and diverse set of SEP models that exhibit a wide array of capabilities but currently have significant limitations, in particular arising from the gaps in real-time observations. If supported with the necessary observations, and with further developments, for example in computational capabilities and the application of artificial intelligence and machine learning, the field is poised for continued growth with a great potential to contribute significantly both to space weather operations and advances in the understanding of the physics of SEP events.''

\subsection{System view of the problem in synergy with other heliospheric studies}

We need a coordinated effort to combine the results of individual Space Weather efforts in the field of radiation enabling research activities at large. Along this path, much has been already discussed throughout this section and we briefly summarize them below. 

SEPs should be studied as part of a system where the magnetic features on the Sun can evolve to drive solar eruptions such as flares, CMEs and shocks which may accelerate energetic particles that can escape into interplanetary space while being conditioned by the solar wind during their transport and can also interact with the planetary environment when arriving at a planetary body (i.e., topics studied by S: Space weather origins at the Sun,  H: Heliosphere variability teams and also G: Coupled geospace system within the ISWAT framework \url{https://iswat-cospar.org}; see related papers in this special issue).% and can be observed with different properties at different locations.

Similarly, ACRs and GCRs should also be studied as part of a system where the global magnetic field of the heliosphere (from inner-heliosphere to the outer-heliosphere and to the heliospheric boundary) is treated in a realistic and dynamic manner whereby the small-scale turbulence in the solar wind is appropriately addressed. Meanwhile, the meso-scale solar transient disturbances and short-term variations are considered as modulation factors which may have an accumulative effect on the global transport. Finally, their interaction with the planetary environment and radiation impact on long-term interplanetary missions are to be explored based on an interdisciplinary approach. %  and resolved by concurrent observations at different locations. 

% % ENA
% A future instrument can be specifically designed to observe solar ENAs at or within 1 AU, with a challengingly high sensitivity, large geometric factor and fine angular resolution (Cohen et al. 2021), especially at energies below ~1 MeV as predicted by Wang et al. (2014). Further modeling of ENA production and propagation would also include various spatial and temporal variations of CME and solar wind structures.

\section{Conclusion}
Space radiation is a significant concern for the safety of robotic and human exploration both in the near-Earth environment and towards deep space and other planets such as Mars. 
Studying energetic particle radiation in the heliosphere is essential for improving the space weather forecasting capability and also for better understanding the physics of particle energization and transport processes. 

This paper has reviewed the scientific aspects of the major sources of energetic particle radiation in the heliosphere: SEPs, GLE Events\delete{ (an extreme type of SEP events)}, GCRs and ACRs (from Sections 2 to 5). 
We have reviewed the advances in the past decades concerning the current scientific understanding and predictive capabilities. 
We also pointed out our knowledge gaps in observing, modeling, physical understanding, and forecasting capabilities of the heliospheric radiation environment. Finally we have offered considerations related to the planning of future space observations in Section \ref{sec:Recommendations}.
COSPAR’s ISWAT cluster ``H3: Radiation Environment in the Heliosphere'' (\url{https://www.iswat-cospar.org/h3}) will continue providing a scientific platform for researchers across the world to work together with the common goal to understand, characterize and predict the energetic particle radiation in the heliosphere and to mitigate radiation risks associated with areospace activities, satellite industry and human space explorations. 

\section{Acknowledgments}
We thank the three anonymous referees for their time and effort in reviewing this article.
We acknowledge the COSPAR ISWAT community and their effort in organizing this special activity, in particular the dedicated endeavour of Masha Kuznetsova and Mario Bisi.  
JG is supported by the Strategic Priority Program of the Chinese Academy of Sciences (Grant XDB41000000) and the National Natural Science Foundation of China (Grants 42074222, 42188101, 42130204).
BBW was supported in part by an NSF EPSCoR RII-Track-1 Cooperative Agreement OIA-2148653.
CMSC is supported by NASA grants NNN06AA01C, 80NSSC18K1446, 80NSSC18K0223 and grant JHU.APL173063.
SD acknowledges support from NERC (grant NE/V002864/1) and STFC (grant  ST/V000934/1).
MD acknowledges support by the Croatian Science Foundation under the project IP-2020-02-9893 (ICOHOSS).
LW is supported in part by NSFC under contracts 42225404, 42127803, 42150105. 
HMB carried out this work while supported by the Cooperative Agreement award NA17OAR4320101. The ISEP project transitioning SEP models to real time operations and the development of the SEP Scoreboard is supported by the Advanced Exploration Systems Division under the Human Exploration and Operations Mission Directorate of NASA and performed in support of the Human Health and Performance Contract for NASA (NNJ15HK11B). Part of the research described in this paper was carried out at the Jet Propulsion Laboratory, California Institute of Technology, under a contract with NASA (80NM0018D0004).
%% The Appendices part is started with the command \appendix;
%% appendix sections are then done as normal sections
%\appendix

%% If you have bibdatabase file and want bibtex to generate the
%% bibitems, please use
%%
\bibliographystyle{elsarticle-harv} 
\bibliography{Main_ref}

\begin{thebibliography}{304}
\expandafter\ifx\csname natexlab\endcsname\relax\def\natexlab#1{#1}\fi
\providecommand{\url}[1]{\texttt{#1}}
\providecommand{\href}[2]{#2}
\providecommand{\path}[1]{#1}
\providecommand{\DOIprefix}{doi:}
\providecommand{\ArXivprefix}{arXiv:}
\providecommand{\URLprefix}{URL: }
\providecommand{\Pubmedprefix}{pmid:}
\providecommand{\doi}[1]{\href{http://dx.doi.org/#1}{\path{#1}}}
\providecommand{\Pubmed}[1]{\href{pmid:#1}{\path{#1}}}
\providecommand{\bibinfo}[2]{#2}
\ifx\xfnm\relax \def\xfnm[#1]{\unskip,\space#1}\fi
%Type = Article
\bibitem[{Acu{\~n}a et~al.(1995)Acu{\~n}a, Ogilvie, Baker, Curtis, Fairfield
  and Mish}]{acuna1995wind}
\bibinfo{author}{Acu{\~n}a, M.}, \bibinfo{author}{Ogilvie, K.},
  \bibinfo{author}{Baker, D.}, \bibinfo{author}{Curtis, S.},
  \bibinfo{author}{Fairfield, D.}, \bibinfo{author}{Mish, W.},
  \bibinfo{year}{1995}.
\newblock \bibinfo{title}{The global geospace science program and its
  investigations}.
\newblock \bibinfo{journal}{Space Sci. Rev.} \bibinfo{volume}{71},
  \bibinfo{pages}{5--21}.
\newblock \DOIprefix\doi{10.1007/BF00751323}.
%Type = Article
\bibitem[{{Adhikari} et~al.(2023){Adhikari}, {Zank}, {Wang}, {Zhao}, {Telloni},
  {Pitna}, {Opher}, {Shrestha}, {McComas} and {Nykyri}}]{Adhikari2023}
\bibinfo{author}{{Adhikari}, L.}, \bibinfo{author}{{Zank}, G.P.},
  \bibinfo{author}{{Wang}, B.}, \bibinfo{author}{{Zhao}, L.},
  \bibinfo{author}{{Telloni}, D.}, \bibinfo{author}{{Pitna}, A.},
  \bibinfo{author}{{Opher}, M.}, \bibinfo{author}{{Shrestha}, B.},
  \bibinfo{author}{{McComas}, D.J.}, \bibinfo{author}{{Nykyri}, K.},
  \bibinfo{year}{2023}.
\newblock \bibinfo{title}{{Theory and Transport of Nearly Incompressible
  Magnetohydrodynamic Turbulence: High Plasma Beta Regime}}.
\newblock \bibinfo{journal}{Astrophys. J.} \bibinfo{volume}{953},
  \bibinfo{pages}{44}.
\newblock \DOIprefix\doi{10.3847/1538-4357/acde57}.
%Type = Article
\bibitem[{Adriani et~al.(2015)Adriani, Barbarino, Bazilevskaya, Bellotti,
  Boezio, Bogomolov, Bongi, Bonvicini, Bottai, Bravar et~al.}]{adriani2015}
\bibinfo{author}{Adriani, O.}, \bibinfo{author}{Barbarino, G.},
  \bibinfo{author}{Bazilevskaya, G.}, \bibinfo{author}{Bellotti, R.},
  \bibinfo{author}{Boezio, M.}, \bibinfo{author}{Bogomolov, E.},
  \bibinfo{author}{Bongi, M.}, \bibinfo{author}{Bonvicini, V.},
  \bibinfo{author}{Bottai, S.}, \bibinfo{author}{Bravar, U.}, et~al.,
  \bibinfo{year}{2015}.
\newblock \bibinfo{title}{{PAMELA}’s measurements of magnetospheric effects
  on high-energy solar particles}.
\newblock \bibinfo{journal}{Astrophys. J. Lett.} \bibinfo{volume}{801},
  \bibinfo{pages}{L3}.
\newblock \DOIprefix\doi{10.1088/2041-8205/801/1/L3}.
%Type = Article
\bibitem[{Aguilar et~al.(2018)Aguilar, Cavasonza, Alpat, Ambrosi, Arruda,
  Attig, Aupetit, Azzarello, Bachlechner, Barao et~al.}]{aguilar2018}
\bibinfo{author}{Aguilar, M.}, \bibinfo{author}{Cavasonza, L.A.},
  \bibinfo{author}{Alpat, B.}, \bibinfo{author}{Ambrosi, G.},
  \bibinfo{author}{Arruda, L.}, \bibinfo{author}{Attig, N.},
  \bibinfo{author}{Aupetit, S.}, \bibinfo{author}{Azzarello, P.},
  \bibinfo{author}{Bachlechner, A.}, \bibinfo{author}{Barao, F.}, et~al.,
  \bibinfo{year}{2018}.
\newblock \bibinfo{title}{Observation of fine time structures in the cosmic
  proton and helium fluxes with the alpha magnetic spectrometer on the
  international space station}.
\newblock \bibinfo{journal}{Phys. Rev. Lett.} \bibinfo{volume}{121},
  \bibinfo{pages}{051101}.
\newblock \DOIprefix\doi{10.1103/PhysRevLett.121.051101}.
%Type = Article
\bibitem[{Alania and Wawrzynczak(2012)}]{ALANIA2012}
\bibinfo{author}{Alania, M.V.}, \bibinfo{author}{Wawrzynczak, A.},
  \bibinfo{year}{2012}.
\newblock \bibinfo{title}{Energy dependence of the rigidity spectrum of forbush
  decrease of galactic cosmic ray intensity}.
\newblock \bibinfo{journal}{Adv. Space Res.} \bibinfo{volume}{50},
  \bibinfo{pages}{725--730}.
\newblock \DOIprefix\doi{https://doi.org/10.1016/j.asr.2011.09.027}.
  \bibinfo{note}{solar Variability, Cosmic Rays and Climate}.
%Type = Article
\bibitem[{Alanko-Huotari et~al.(2007)Alanko-Huotari, Usoskin, Mursula and
  Kovaltsov}]{alanko_etal_2007}
\bibinfo{author}{Alanko-Huotari, K.}, \bibinfo{author}{Usoskin, I.},
  \bibinfo{author}{Mursula, K.}, \bibinfo{author}{Kovaltsov, G.},
  \bibinfo{year}{2007}.
\newblock \bibinfo{title}{Cyclic variations of the heliospheric tilt angle and
  cosmic ray modulation}.
\newblock \bibinfo{journal}{Adv. in Space Res.} \bibinfo{volume}{7},
  \bibinfo{pages}{1064--1069}.
\newblock \DOIprefix\doi{10.1016/j.asr.2007.02.007}.
%Type = Article
\bibitem[{Anastasiadis et~al.(2019)Anastasiadis, Lario, Papaioannou,
  Kouloumvakos and Vourlidas}]{Anastasiadis2019}
\bibinfo{author}{Anastasiadis, A.}, \bibinfo{author}{Lario, D.},
  \bibinfo{author}{Papaioannou, A.}, \bibinfo{author}{Kouloumvakos, A.},
  \bibinfo{author}{Vourlidas, A.}, \bibinfo{year}{2019}.
\newblock \bibinfo{title}{Solar energetic particles in the inner heliosphere:
  status and open questions}.
\newblock \bibinfo{journal}{Philosophical Transactions of the Royal Society A:
  Mathematical, Physical and Engineering Sciences} \bibinfo{volume}{377},
  \bibinfo{pages}{20180100}.
\newblock \DOIprefix\doi{10.1098/rsta.2018.0100}.
%Type = Article
\bibitem[{{Aran} et~al.(2006){Aran}, {Sanahuja} and {Lario}}]{Ara2006}
\bibinfo{author}{{Aran}, A.}, \bibinfo{author}{{Sanahuja}, B.},
  \bibinfo{author}{{Lario}, D.}, \bibinfo{year}{2006}.
\newblock \bibinfo{title}{{SOLPENCO: A solar particle engineering code}}.
\newblock \bibinfo{journal}{Adv. Space Res.} \bibinfo{volume}{37},
  \bibinfo{pages}{1240--1246}.
\newblock \DOIprefix\doi{10.1016/j.asr.2005.09.019}.
%Type = Article
\bibitem[{Arnoldy et~al.(1968)Arnoldy, Kane and
  Winckler}]{arnoldy1968energetic}
\bibinfo{author}{Arnoldy, R.}, \bibinfo{author}{Kane, S.},
  \bibinfo{author}{Winckler, J.}, \bibinfo{year}{1968}.
\newblock \bibinfo{title}{Energetic solar flare x-rays observed by satellite
  and their correlation with solar radio and energetic particle emission}.
\newblock \bibinfo{journal}{Astrophys. J.} \bibinfo{volume}{151},
  \bibinfo{pages}{711}.
\newblock \DOIprefix\doi{10.1086/149470}.
%Type = Article
\bibitem[{{Aschwanden}(2012)}]{Asc2012}
\bibinfo{author}{{Aschwanden}, M.J.}, \bibinfo{year}{2012}.
\newblock \bibinfo{title}{{GeV Particle Acceleration in Solar Flares and Ground
  Level Enhancement (GLE) Events}}.
\newblock \bibinfo{journal}{Space Sci. Rev.} \bibinfo{volume}{171},
  \bibinfo{pages}{3--21}.
\newblock \DOIprefix\doi{10.1007/s11214-011-9865-x}.
%Type = Article
\bibitem[{Aslam et~al.(2021)Aslam, Bisschoff, Ngobeni, Potgieter, Munini,
  Boezio and Mikhailov}]{Aslam2020}
\bibinfo{author}{Aslam, O.P.M.}, \bibinfo{author}{Bisschoff, D.},
  \bibinfo{author}{Ngobeni, M.D.}, \bibinfo{author}{Potgieter, M.S.},
  \bibinfo{author}{Munini, R.}, \bibinfo{author}{Boezio, M.},
  \bibinfo{author}{Mikhailov, V.V.}, \bibinfo{year}{2021}.
\newblock \bibinfo{title}{{Time and Charge-sign Dependence of the Heliospheric
  Modulation of Cosmic Rays}}.
\newblock \bibinfo{journal}{Astrophys. J.} \bibinfo{volume}{909},
  \bibinfo{pages}{215}.
\newblock \DOIprefix\doi{10.3847/1538-4357/abdd35}.
%Type = Article
\bibitem[{{Badruddin}(2016)}]{badruddinkumar2016}
\bibinfo{author}{{Badruddin}, Kumar, A.}, \bibinfo{year}{2016}.
\newblock \bibinfo{title}{{Study of the Cosmic-Ray Modulation During the
  Passage of ICMEs and CIRs}}.
\newblock \bibinfo{journal}{Sol. Phys.} \bibinfo{volume}{291},
  \bibinfo{pages}{559--580}.
\newblock \DOIprefix\doi{10.1007/s11207-015-0843-4}.
%Type = Article
\bibitem[{Bain et~al.(2023a)Bain, Copeland, Onsager and
  Steenburgh}]{bain23noaa}
\bibinfo{author}{Bain, H.M.}, \bibinfo{author}{Copeland, K.},
  \bibinfo{author}{Onsager, T.G.}, \bibinfo{author}{Steenburgh, R.A.},
  \bibinfo{year}{2023}a.
\newblock \bibinfo{title}{Noaa space weather prediction center radiation
  advisories for the international civil aviation organization}.
\newblock \bibinfo{journal}{Space Weather} \bibinfo{volume}{21},
  \bibinfo{pages}{e2022SW003346}.
\newblock \DOIprefix\doi{https://doi.org/10.1029/2022SW003346}.
  \bibinfo{note}{e2022SW003346 2022SW003346}.
%Type = Article
\bibitem[{{Bain} et~al.(2016){Bain}, {Mays}, {Luhmann}, {Li}, {Jian} and
  {Odstrcil}}]{Bain2016}
\bibinfo{author}{{Bain}, H.M.}, \bibinfo{author}{{Mays}, M.L.},
  \bibinfo{author}{{Luhmann}, J.G.}, \bibinfo{author}{{Li}, Y.},
  \bibinfo{author}{{Jian}, L.K.}, \bibinfo{author}{{Odstrcil}, D.},
  \bibinfo{year}{2016}.
\newblock \bibinfo{title}{{Shock Connectivity in the August 2010 and July 2012
  Solar Energetic Particle Events Inferred from Observations and ENLIL
  Modeling}}.
\newblock \bibinfo{journal}{Astrophys. J.} \bibinfo{volume}{825},
  \bibinfo{pages}{1}.
\newblock \DOIprefix\doi{10.3847/0004-637X/825/1/1}.
%Type = Article
\bibitem[{Bain et~al.(2023b)Bain, Onsager, Mertens, Copeland, Benton, Clem,
  Mangeard, Green, Guild, Tobiska, Shelton-Mur, Zheng, Halford, Carlson and
  Pulkkinen}]{bain23improved}
\bibinfo{author}{Bain, H.M.}, \bibinfo{author}{Onsager, T.G.},
  \bibinfo{author}{Mertens, C.J.}, \bibinfo{author}{Copeland, K.},
  \bibinfo{author}{Benton, E.R.}, \bibinfo{author}{Clem, J.},
  \bibinfo{author}{Mangeard, P.S.}, \bibinfo{author}{Green, J.C.},
  \bibinfo{author}{Guild, T.B.}, \bibinfo{author}{Tobiska, W.K.},
  \bibinfo{author}{Shelton-Mur, K.}, \bibinfo{author}{Zheng, Y.},
  \bibinfo{author}{Halford, A.J.}, \bibinfo{author}{Carlson, S.},
  \bibinfo{author}{Pulkkinen, A.}, \bibinfo{year}{2023}b.
\newblock \bibinfo{title}{Improved space weather observations and modeling for
  aviation radiation}.
\newblock \bibinfo{journal}{Front. in Astron. \& Space Sci.}
  \bibinfo{volume}{10}.
\newblock \DOIprefix\doi{10.3389/fspas.2023.1149014}.
%Type = Article
\bibitem[{Bain et~al.(2021)Bain, Steenburgh, Onsager and Stitely}]{bain21}
\bibinfo{author}{Bain, H.M.}, \bibinfo{author}{Steenburgh, R.A.},
  \bibinfo{author}{Onsager, T.G.}, \bibinfo{author}{Stitely, E.M.},
  \bibinfo{year}{2021}.
\newblock \bibinfo{title}{A summary of national oceanic and atmospheric
  administration space weather prediction center proton event forecast
  performance and skill}.
\newblock \bibinfo{journal}{Space Weather} \bibinfo{volume}{19},
  \bibinfo{pages}{e2020SW002670}.
\newblock \DOIprefix\doi{10.1029/2020SW002670}.
%Type = Article
\bibitem[{Balch(1999)}]{balch99}
\bibinfo{author}{Balch, C.C.}, \bibinfo{year}{1999}.
\newblock \bibinfo{title}{{SEC proton prediction model: verification and
  analysis}}.
\newblock \bibinfo{journal}{Radiation Measurements} \bibinfo{volume}{30},
  \bibinfo{pages}{231 -- 250}.
\newblock \DOIprefix\doi{10.1016/S1350-4487(99)00052-9}.
%Type = Article
\bibitem[{Balch(2008)}]{balch08}
\bibinfo{author}{Balch, C.C.}, \bibinfo{year}{2008}.
\newblock \bibinfo{title}{Updated verification of the space weather prediction
  center's solar energetic particle prediction model}.
\newblock \bibinfo{journal}{Space Weather} \bibinfo{volume}{6}.
\newblock \DOIprefix\doi{10.1029/2007SW000337}.
%Type = Article
\bibitem[{{Ball} et~al.(2005){Ball}, {Zhang}, {Rassoul} and {Linde}}]{Ball2005}
\bibinfo{author}{{Ball}, B.}, \bibinfo{author}{{Zhang}, M.},
  \bibinfo{author}{{Rassoul}, H.}, \bibinfo{author}{{Linde}, T.},
  \bibinfo{year}{2005}.
\newblock \bibinfo{title}{{Galactic Cosmic-Ray Modulation Using a Solar Minimum
  MHD Heliosphere: A Stochastic Particle Approach}}.
\newblock \bibinfo{journal}{Astrophys. J.} \bibinfo{volume}{634},
  \bibinfo{pages}{1116--1125}.
\newblock \DOIprefix\doi{10.1086/496965}.
%Type = Article
\bibitem[{Belov(2008)}]{belov2008}
\bibinfo{author}{Belov, A.}, \bibinfo{year}{2008}.
\newblock \bibinfo{title}{Forbush effects and their connection with solar,
  interplanetary and geomagnetic phenomena}.
\newblock \bibinfo{journal}{Proceedings of the International Astronomical
  Union} \bibinfo{volume}{4}, \bibinfo{pages}{439--450}.
\newblock \DOIprefix\doi{https://doi.org/10.1017/S1743921309029676}.
%Type = Article
\bibitem[{Belov et~al.(2005)Belov, Eroshenko, Mavromichalaki, Plainaki and
  Yanke}]{belov2005solar}
\bibinfo{author}{Belov, A.}, \bibinfo{author}{Eroshenko, E.},
  \bibinfo{author}{Mavromichalaki, H.}, \bibinfo{author}{Plainaki, C.},
  \bibinfo{author}{Yanke, V.}, \bibinfo{year}{2005}.
\newblock \bibinfo{title}{Solar cosmic rays during the extremely high ground
  level enhancement on 23 february 1956}.
\newblock \bibinfo{journal}{Ann. Geophys.} \bibinfo{volume}{23},
  \bibinfo{pages}{2281--2291}.
\newblock \DOIprefix\doi{10.5194/angeo-23-2281-2005}.
%Type = Article
\bibitem[{{Benella} et~al.(2020){Benella}, {Laurenza}, {Vainio}, {Grimani},
  {Consolini}, {Hu} and {Afanasiev}}]{benella20}
\bibinfo{author}{{Benella}, S.}, \bibinfo{author}{{Laurenza}, M.},
  \bibinfo{author}{{Vainio}, R.}, \bibinfo{author}{{Grimani}, C.},
  \bibinfo{author}{{Consolini}, G.}, \bibinfo{author}{{Hu}, Q.},
  \bibinfo{author}{{Afanasiev}, A.}, \bibinfo{year}{2020}.
\newblock \bibinfo{title}{{A New Method to Model Magnetic Cloud-driven Forbush
  Decreases: The 2016 August 2 Event}}.
\newblock \bibinfo{journal}{Astrophys. J.} \bibinfo{volume}{901},
  \bibinfo{pages}{21}.
\newblock \DOIprefix\doi{10.3847/1538-4357/abac59}.
%Type = Article
\bibitem[{{Bieber} et~al.(1994){Bieber}, {Matthaeus}, {Smith}, {Wanner},
  {Kallenrode} and {Wibberenz}}]{Bieber1994}
\bibinfo{author}{{Bieber}, J.W.}, \bibinfo{author}{{Matthaeus}, W.H.},
  \bibinfo{author}{{Smith}, C.W.}, \bibinfo{author}{{Wanner}, W.},
  \bibinfo{author}{{Kallenrode}, M.B.}, \bibinfo{author}{{Wibberenz}, G.},
  \bibinfo{year}{1994}.
\newblock \bibinfo{title}{{Proton and Electron Mean Free Paths: The Palmer
  Consensus Revisited}}.
\newblock \bibinfo{journal}{Astrophys. J.} \bibinfo{volume}{420},
  \bibinfo{pages}{294}.
\newblock \DOIprefix\doi{10.1086/173559}.
%Type = Article
\bibitem[{{Bindi} et~al.(2017){Bindi}, {Corti}, {Consolandi}, {Hoffman} and
  {Whitman}}]{Bindi2017}
\bibinfo{author}{{Bindi}, V.}, \bibinfo{author}{{Corti}, C.},
  \bibinfo{author}{{Consolandi}, C.}, \bibinfo{author}{{Hoffman}, J.},
  \bibinfo{author}{{Whitman}, K.}, \bibinfo{year}{2017}.
\newblock \bibinfo{title}{{Overview of galactic cosmic ray solar modulation in
  the AMS-02 era}}.
\newblock \bibinfo{journal}{Adv. Space Res.} \bibinfo{volume}{60},
  \bibinfo{pages}{865--878}.
\newblock \DOIprefix\doi{10.1016/j.asr.2017.05.025}.
%Type = Article
\bibitem[{{Bisschoff} et~al.(2019){Bisschoff}, {Potgieter} and
  {Aslam}}]{Bisschoff2019}
\bibinfo{author}{{Bisschoff}, D.}, \bibinfo{author}{{Potgieter}, M.S.},
  \bibinfo{author}{{Aslam}, O.P.M.}, \bibinfo{year}{2019}.
\newblock \bibinfo{title}{{New Very Local Interstellar Spectra for Electrons,
  Positrons, Protons, and Light Cosmic Ray Nuclei}}.
\newblock \bibinfo{journal}{Astrophys. J.} \bibinfo{volume}{878},
  \bibinfo{pages}{59}.
\newblock \DOIprefix\doi{10.3847/1538-4357/ab1e4a}.
%Type = Article
\bibitem[{{Blanco} et~al.(2013){Blanco}, {Hidalgo}, {G{\'o}mez-Herrero},
  {Rodr{\'\i}guez-Pacheco}, {Heber}, {Wimmer-Schweingruber} and
  {Mart{\'\i}n}}]{blanco13}
\bibinfo{author}{{Blanco}, J.J.}, \bibinfo{author}{{Hidalgo}, M.A.},
  \bibinfo{author}{{G{\'o}mez-Herrero}, R.},
  \bibinfo{author}{{Rodr{\'\i}guez-Pacheco}, J.}, \bibinfo{author}{{Heber},
  B.}, \bibinfo{author}{{Wimmer-Schweingruber}, R.F.},
  \bibinfo{author}{{Mart{\'\i}n}, C.}, \bibinfo{year}{2013}.
\newblock \bibinfo{title}{{Energetic-particle-flux decreases related to
  magnetic cloud passages as observed by the Helios 1 and 2 spacecraft}}.
\newblock \bibinfo{journal}{Astron. \& Astrophys.} \bibinfo{volume}{556},
  \bibinfo{pages}{A146}.
\newblock \DOIprefix\doi{10.1051/0004-6361/201321739}.
%Type = Article
\bibitem[{Blasi(2013)}]{blasi2013origin}
\bibinfo{author}{Blasi, P.}, \bibinfo{year}{2013}.
\newblock \bibinfo{title}{The origin of galactic cosmic rays}.
\newblock \bibinfo{journal}{Astron. \& Astrophys. Rev.} \bibinfo{volume}{21},
  \bibinfo{pages}{1--73}.
\newblock \DOIprefix\doi{10.1007/s00159-013-0070-7}.
%Type = Article
\bibitem[{Bombardieri et~al.(2008)Bombardieri, Duldig, Humble and
  Michael}]{bombardieri2008improved}
\bibinfo{author}{Bombardieri, D.}, \bibinfo{author}{Duldig, M.},
  \bibinfo{author}{Humble, J.}, \bibinfo{author}{Michael, K.},
  \bibinfo{year}{2008}.
\newblock \bibinfo{title}{An improved model for relativistic solar proton
  acceleration applied to the 2005 january 20 and earlier events}.
\newblock \bibinfo{journal}{Astrophys. J.} \bibinfo{volume}{682},
  \bibinfo{pages}{1315}.
\newblock \DOIprefix\doi{10.1086/589494}.
%Type = Article
\bibitem[{Bonvicini et~al.(2001)Bonvicini, Barbiellini, Boezio, Mocchiutti,
  Schiavon, Scian, Vacchi, Zampa, Zampa, Bergstr{\"o}m
  et~al.}]{bonvicini2001pamela}
\bibinfo{author}{Bonvicini, V.}, \bibinfo{author}{Barbiellini, G.},
  \bibinfo{author}{Boezio, M.}, \bibinfo{author}{Mocchiutti, E.},
  \bibinfo{author}{Schiavon, P.}, \bibinfo{author}{Scian, G.},
  \bibinfo{author}{Vacchi, A.}, \bibinfo{author}{Zampa, G.},
  \bibinfo{author}{Zampa, N.}, \bibinfo{author}{Bergstr{\"o}m, D.}, et~al.,
  \bibinfo{year}{2001}.
\newblock \bibinfo{title}{The {PAMELA} experiment in space}.
\newblock \bibinfo{journal}{Nuclear Instruments and Methods in Physics Research
  Section A: Accelerators, Spectrometers, Detectors and Associated Equipment}
  \bibinfo{volume}{461}, \bibinfo{pages}{262--268}.
\newblock \DOIprefix\doi{10.1016/S0168-9002(00)01221-3}.
%Type = Article
\bibitem[{{Boschini} et~al.(2018a){Boschini}, {Della Torre}, {Gervasi},
  {Grandi}, {J{\'o}hannesson}, {La Vacca}, {Masi}, {Moskalenko}, {Pensotti},
  {Porter}, {Quadrani}, {Rancoita}, {Rozza} and {Tacconi}}]{Boschini2018b}
\bibinfo{author}{{Boschini}, M.J.}, \bibinfo{author}{{Della Torre}, S.},
  \bibinfo{author}{{Gervasi}, M.}, \bibinfo{author}{{Grandi}, D.},
  \bibinfo{author}{{J{\'o}hannesson}, G.}, \bibinfo{author}{{La Vacca}, G.},
  \bibinfo{author}{{Masi}, N.}, \bibinfo{author}{{Moskalenko}, I.V.},
  \bibinfo{author}{{Pensotti}, S.}, \bibinfo{author}{{Porter}, T.A.},
  \bibinfo{author}{{Quadrani}, L.}, \bibinfo{author}{{Rancoita}, P.G.},
  \bibinfo{author}{{Rozza}, D.}, \bibinfo{author}{{Tacconi}, M.},
  \bibinfo{year}{2018}a.
\newblock \bibinfo{title}{{Deciphering the Local Interstellar Spectra of
  Primary Cosmic-Ray Species with HELMOD}}.
\newblock \bibinfo{journal}{Astrophys. J.} \bibinfo{volume}{858},
  \bibinfo{pages}{61}.
\newblock \DOIprefix\doi{10.3847/1538-4357/aabc54}.
%Type = Article
\bibitem[{{Boschini} et~al.(2020){Boschini}, {Della Torre}, {Gervasi},
  {Grandi}, {J{\'o}hannesson}, {La Vacca}, {Masi}, {Moskalenko}, {Pensotti},
  {Porter}, {Quadrani}, {Rancoita}, {Rozza} and {Tacconi}}]{Boschini2020}
\bibinfo{author}{{Boschini}, M.J.}, \bibinfo{author}{{Della Torre}, S.},
  \bibinfo{author}{{Gervasi}, M.}, \bibinfo{author}{{Grandi}, D.},
  \bibinfo{author}{{J{\'o}hannesson}, G.}, \bibinfo{author}{{La Vacca}, G.},
  \bibinfo{author}{{Masi}, N.}, \bibinfo{author}{{Moskalenko}, I.V.},
  \bibinfo{author}{{Pensotti}, S.}, \bibinfo{author}{{Porter}, T.A.},
  \bibinfo{author}{{Quadrani}, L.}, \bibinfo{author}{{Rancoita}, P.G.},
  \bibinfo{author}{{Rozza}, D.}, \bibinfo{author}{{Tacconi}, M.},
  \bibinfo{year}{2020}.
\newblock \bibinfo{title}{{Inference of the Local Interstellar Spectra of
  Cosmic-Ray Nuclei Z {\ensuremath{\leq}} 28 with the GALPROP-HELMOD
  Framework}}.
\newblock \bibinfo{journal}{Astrophys. J. Suppl. Ser.} \bibinfo{volume}{250},
  \bibinfo{pages}{27}.
\newblock \DOIprefix\doi{10.3847/1538-4365/aba901}.
%Type = Article
\bibitem[{{Boschini} et~al.(2018b){Boschini}, {Della Torre}, {Gervasi}, {La
  Vacca} and {Rancoita}}]{Boschini2018a}
\bibinfo{author}{{Boschini}, M.J.}, \bibinfo{author}{{Della Torre}, S.},
  \bibinfo{author}{{Gervasi}, M.}, \bibinfo{author}{{La Vacca}, G.},
  \bibinfo{author}{{Rancoita}, P.G.}, \bibinfo{year}{2018}b.
\newblock \bibinfo{title}{{Propagation of cosmic rays in heliosphere: The
  HELMOD model}}.
\newblock \bibinfo{journal}{Adv. Space Res.} \bibinfo{volume}{62},
  \bibinfo{pages}{2859--2879}.
\newblock \DOIprefix\doi{10.1016/j.asr.2017.04.017}.
%Type = Article
\bibitem[{Bruno et~al.(2018)Bruno, Bazilevskaya, Boezio, Christian, de~Nolfo,
  Martucci, Merge, Mikhailov, Munini, Richardson et~al.}]{bruno2018solar}
\bibinfo{author}{Bruno, A.}, \bibinfo{author}{Bazilevskaya, G.},
  \bibinfo{author}{Boezio, M.}, \bibinfo{author}{Christian, E.R.},
  \bibinfo{author}{de~Nolfo, G.}, \bibinfo{author}{Martucci, M.},
  \bibinfo{author}{Merge, M.}, \bibinfo{author}{Mikhailov, V.},
  \bibinfo{author}{Munini, R.}, \bibinfo{author}{Richardson, I.}, et~al.,
  \bibinfo{year}{2018}.
\newblock \bibinfo{title}{Solar energetic particle events observed by the
  {PAMELA} mission}.
\newblock \bibinfo{journal}{Astrophys. J.} \bibinfo{volume}{862},
  \bibinfo{pages}{97}.
\newblock \DOIprefix\doi{10.3847/1538-4357/aacc26}.
%Type = Article
\bibitem[{Bruno et~al.(2019)Bruno, Christian, De~Nolfo, Richardson and
  Ryan}]{bruno2019spectral}
\bibinfo{author}{Bruno, A.}, \bibinfo{author}{Christian, E.},
  \bibinfo{author}{De~Nolfo, G.}, \bibinfo{author}{Richardson, I.G.},
  \bibinfo{author}{Ryan, J.}, \bibinfo{year}{2019}.
\newblock \bibinfo{title}{Spectral analysis of the september 2017 solar
  energetic particle events}.
\newblock \bibinfo{journal}{Space Weather} \bibinfo{volume}{17},
  \bibinfo{pages}{419--437}.
\newblock \DOIprefix\doi{10.1029/2018SW002085}.
%Type = Article
\bibitem[{{Bzowski} et~al.(2003){Bzowski}, {M{\"a}kinen}, {Kyr{\"o}l{\"a}},
  {Summanen} and {Qu{\'e}merais}}]{Bzowski2003}
\bibinfo{author}{{Bzowski}, M.}, \bibinfo{author}{{M{\"a}kinen}, T.},
  \bibinfo{author}{{Kyr{\"o}l{\"a}}, E.}, \bibinfo{author}{{Summanen}, T.},
  \bibinfo{author}{{Qu{\'e}merais}, E.}, \bibinfo{year}{2003}.
\newblock \bibinfo{title}{{Latitudinal structure and north-south asymmetry of
  the solar wind from Lyman-alpha remote sensing by SWAN}}.
\newblock \bibinfo{journal}{Astron. \& Astrophys.} \bibinfo{volume}{408},
  \bibinfo{pages}{1165--1177}.
\newblock \DOIprefix\doi{10.1051/0004-6361:20031022}.
%Type = Article
\bibitem[{{Caballero-Lopez} and {Moraal}(2004)}]{caballero-lopez04}
\bibinfo{author}{{Caballero-Lopez}, R.A.}, \bibinfo{author}{{Moraal}, H.},
  \bibinfo{year}{2004}.
\newblock \bibinfo{title}{{Limitations of the force field equation to describe
  cosmic ray modulation}}.
\newblock \bibinfo{journal}{J. Geophys. Res.: Space Phys.}
  \bibinfo{volume}{109}, \bibinfo{pages}{A01101}.
\newblock \DOIprefix\doi{10.1029/2003JA010098}.
%Type = Article
\bibitem[{Cane et~al.(2006)Cane, Mewaldt, Cohen and
  Von~Rosenvinge}]{Cane2006role}
\bibinfo{author}{Cane, H.}, \bibinfo{author}{Mewaldt, R.},
  \bibinfo{author}{Cohen, C.}, \bibinfo{author}{Von~Rosenvinge, T.},
  \bibinfo{year}{2006}.
\newblock \bibinfo{title}{Role of flares and shocks in determining solar
  energetic particle abundances}.
\newblock \bibinfo{journal}{J. Geophys. Res.: Space Phys.}
  \bibinfo{volume}{111}.
\newblock \DOIprefix\doi{10.1029/2005JA011071}.
%Type = Article
\bibitem[{Cane et~al.(1999)Cane, Wibberenz, Richardson and
  Von~Rosenvinge}]{cane1999cosmic}
\bibinfo{author}{Cane, H.}, \bibinfo{author}{Wibberenz, G.},
  \bibinfo{author}{Richardson, I.}, \bibinfo{author}{Von~Rosenvinge, T.},
  \bibinfo{year}{1999}.
\newblock \bibinfo{title}{Cosmic ray modulation and the solar magnetic field}.
\newblock \bibinfo{journal}{Geophys. Res. Lett.} \bibinfo{volume}{26},
  \bibinfo{pages}{565--568}.
\newblock \DOIprefix\doi{10.1029/1999GL900032}.
%Type = Article
\bibitem[{{Cane}(2000)}]{cane00}
\bibinfo{author}{{Cane}, H.V.}, \bibinfo{year}{2000}.
\newblock \bibinfo{title}{{Coronal Mass Ejections and Forbush Decreases}}.
\newblock \bibinfo{journal}{Space Sci. Rev.} \bibinfo{volume}{93},
  \bibinfo{pages}{55--77}.
\newblock \DOIprefix\doi{10.1023/A:1026532125747}.
%Type = Article
\bibitem[{{Cane} et~al.(1994){Cane}, {Richardson}, {von Rosenvinge} and
  {Wibberenz}}]{cane94}
\bibinfo{author}{{Cane}, H.V.}, \bibinfo{author}{{Richardson}, I.G.},
  \bibinfo{author}{{von Rosenvinge}, T.T.}, \bibinfo{author}{{Wibberenz}, G.},
  \bibinfo{year}{1994}.
\newblock \bibinfo{title}{{Cosmic ray decreases and shock structure: A
  multispacecraft study}}.
\newblock \bibinfo{journal}{J. Geophys. Res.: Space Phys.}
  \bibinfo{volume}{99}, \bibinfo{pages}{21429--21442}.
\newblock \DOIprefix\doi{10.1029/94JA01529}.
%Type = Article
\bibitem[{Cane et~al.(2003)Cane, von Rosenvinge, Cohen and Mewaldt}]{Cane2003}
\bibinfo{author}{Cane, H.V.}, \bibinfo{author}{von Rosenvinge, T.T.},
  \bibinfo{author}{Cohen, C.M.S.}, \bibinfo{author}{Mewaldt, R.A.},
  \bibinfo{year}{2003}.
\newblock \bibinfo{title}{Two components in major solar particle events}.
\newblock \bibinfo{journal}{Geophys. Res. Lett.} \bibinfo{volume}{30}.
\newblock \DOIprefix\doi{10.1029/2002GL016580}.
%Type = Article
\bibitem[{{Cholis} et~al.(2016){Cholis}, {Hooper} and {Linden}}]{Cholis2016}
\bibinfo{author}{{Cholis}, I.}, \bibinfo{author}{{Hooper}, D.},
  \bibinfo{author}{{Linden}, T.}, \bibinfo{year}{2016}.
\newblock \bibinfo{title}{{A predictive analytic model for the solar modulation
  of cosmic rays}}.
\newblock \bibinfo{journal}{Phys. Rev. D} \bibinfo{volume}{93},
  \bibinfo{pages}{043016}.
\newblock \DOIprefix\doi{10.1103/PhysRevD.93.043016}.
%Type = Article
\bibitem[{Clem and Dorman(2000)}]{clem2000neutron}
\bibinfo{author}{Clem, J.M.}, \bibinfo{author}{Dorman, L.I.},
  \bibinfo{year}{2000}.
\newblock \bibinfo{title}{Neutron monitor response functions}.
\newblock \bibinfo{journal}{Space Sci. Rev.} \bibinfo{volume}{93},
  \bibinfo{pages}{335--359}.
\newblock \DOIprefix\doi{10.1023/A:1026508915269}.
%Type = Article
\bibitem[{Cliver(2016)}]{cliver2016flare}
\bibinfo{author}{Cliver, E.}, \bibinfo{year}{2016}.
\newblock \bibinfo{title}{Flare versus shock acceleration of high-energy
  protons in solar energetic particle events}.
\newblock \bibinfo{journal}{Astrophys. J.} \bibinfo{volume}{832},
  \bibinfo{pages}{128}.
\newblock \DOIprefix\doi{10.3847/0004-637X/832/2/128}.
%Type = Article
\bibitem[{Cliver et~al.(1989)Cliver, Forrest, Cane, Reames, McGuire,
  Von~Rosenvinge, Kane and MacDowall}]{cliver1989solar}
\bibinfo{author}{Cliver, E.}, \bibinfo{author}{Forrest, D.},
  \bibinfo{author}{Cane, H.}, \bibinfo{author}{Reames, D.},
  \bibinfo{author}{McGuire, R.}, \bibinfo{author}{Von~Rosenvinge, T.},
  \bibinfo{author}{Kane, S.}, \bibinfo{author}{MacDowall, R.},
  \bibinfo{year}{1989}.
\newblock \bibinfo{title}{Solar flare nuclear gamma-rays and interplanetary
  proton events}.
\newblock \bibinfo{journal}{Astrophys. J.} \bibinfo{volume}{343},
  \bibinfo{pages}{953--970}.
\newblock \DOIprefix\doi{10.1086/167765}.
%Type = Article
\bibitem[{Cliver and Ling(2001)}]{Cliver_Ling_2001}
\bibinfo{author}{Cliver, E.W.}, \bibinfo{author}{Ling, A.G.},
  \bibinfo{year}{2001}.
\newblock \bibinfo{title}{22 year patterns in the relationship of sunspot
  number and tilt angle to cosmic-ray intensity}.
\newblock \bibinfo{journal}{Astrophys. J. Lett.} \bibinfo{volume}{551},
  \bibinfo{pages}{L189--L192}.
\newblock \DOIprefix\doi{10.1086/320022}.
%Type = Article
\bibitem[{Cliver et~al.(2022)Cliver, Schrijver, Shibata and
  Usoskin}]{cliver2022extreme}
\bibinfo{author}{Cliver, E.W.}, \bibinfo{author}{Schrijver, C.J.},
  \bibinfo{author}{Shibata, K.}, \bibinfo{author}{Usoskin, I.G.},
  \bibinfo{year}{2022}.
\newblock \bibinfo{title}{Extreme solar events}.
\newblock \bibinfo{journal}{Living Rev. Sol. Phys.} \bibinfo{volume}{19},
  \bibinfo{pages}{2}.
\newblock \DOIprefix\doi{10.1007/s41116-022-00033-8}.
%Type = Article
\bibitem[{Cohen et~al.(1999)Cohen, Mewaldt, Leske, Cummings, Stone, Wiedenbeck,
  Christian and Von~Rosenvinge}]{cohen1999new}
\bibinfo{author}{Cohen, C.}, \bibinfo{author}{Mewaldt, R.},
  \bibinfo{author}{Leske, R.}, \bibinfo{author}{Cummings, A.},
  \bibinfo{author}{Stone, E.}, \bibinfo{author}{Wiedenbeck, M.},
  \bibinfo{author}{Christian, E.}, \bibinfo{author}{Von~Rosenvinge, T.},
  \bibinfo{year}{1999}.
\newblock \bibinfo{title}{New observations of heavy-ion-rich solar particle
  events from {ACE}}.
\newblock \bibinfo{journal}{Geophys. Res. Lett.} \bibinfo{volume}{26},
  \bibinfo{pages}{2697--2700}.
\newblock \DOIprefix\doi{10.1029/1999GL900560}.
%Type = Inproceedings
\bibitem[{{Cohen} et~al.(2021){Cohen}, {Li}, {Mason}, {Shih} and
  {Wang}}]{cohen2021SEP}
\bibinfo{author}{{Cohen}, C.M.S.}, \bibinfo{author}{{Li}, G.},
  \bibinfo{author}{{Mason}, G.M.}, \bibinfo{author}{{Shih}, A.Y.},
  \bibinfo{author}{{Wang}, L.}, \bibinfo{year}{2021}.
\newblock \bibinfo{title}{{Solar Energetic Particles}}, in:
  \bibinfo{editor}{{Raouafi}, N.E.}, \bibinfo{editor}{{Vourlidas}, A.} (Eds.),
  \bibinfo{booktitle}{Solar Physics and Solar Wind}, p. \bibinfo{pages}{133}.
\newblock \DOIprefix\doi{10.1002/9781119815600.ch4}.
%Type = Article
\bibitem[{Copeland et~al.(2008)Copeland, Sauer, Duke and
  Friedberg}]{copeland08}
\bibinfo{author}{Copeland, K.}, \bibinfo{author}{Sauer, H.H.},
  \bibinfo{author}{Duke, F.E.}, \bibinfo{author}{Friedberg, W.},
  \bibinfo{year}{2008}.
\newblock \bibinfo{title}{Cosmic radiation exposure of aircraft occupants on
  simulated high-latitude flights during solar proton events from 1 january
  1986 through 1 january 2008}.
\newblock \bibinfo{journal}{Adv. Space Res.} \bibinfo{volume}{42},
  \bibinfo{pages}{1008 -- 1029}.
\newblock \DOIprefix\doi{10.1016/j.asr.2008.03.001}.
%Type = Article
\bibitem[{{Corti} et~al.(2016){Corti}, {Bindi}, {Consolandi} and
  {Whitman}}]{corti16}
\bibinfo{author}{{Corti}, C.}, \bibinfo{author}{{Bindi}, V.},
  \bibinfo{author}{{Consolandi}, C.}, \bibinfo{author}{{Whitman}, K.},
  \bibinfo{year}{2016}.
\newblock \bibinfo{title}{{Solar Modulation of the Local Interstellar Spectrum
  with Voyager 1, AMS-02, {PAMELA}, and BESS}}.
\newblock \bibinfo{journal}{Astrophys. J.} \bibinfo{volume}{829},
  \bibinfo{pages}{8}.
\newblock \DOIprefix\doi{10.3847/0004-637X/829/1/8}.
%Type = Article
\bibitem[{{Corti} et~al.(2019){Corti}, {Potgieter}, {Bindi}, {Consolandi},
  {Light}, {Palermo} and {Popkow}}]{Corti2019}
\bibinfo{author}{{Corti}, C.}, \bibinfo{author}{{Potgieter}, M.S.},
  \bibinfo{author}{{Bindi}, V.}, \bibinfo{author}{{Consolandi}, C.},
  \bibinfo{author}{{Light}, C.}, \bibinfo{author}{{Palermo}, M.},
  \bibinfo{author}{{Popkow}, A.}, \bibinfo{year}{2019}.
\newblock \bibinfo{title}{{Numerical Modeling of Galactic Cosmic-Ray Proton and
  Helium Observed by AMS-02 during the Solar Maximum of Solar Cycle 24}}.
\newblock \bibinfo{journal}{Astrophys. J.} \bibinfo{volume}{871},
  \bibinfo{pages}{253}.
\newblock \DOIprefix\doi{10.3847/1538-4357/aafac4}.
%Type = Misc
\bibitem[{Corti et~al.(2023)Corti, Whitman, Desai, Rankin, Strauss, Nitta,
  Turner and Chen}]{Corti2022}
\bibinfo{author}{Corti, C.}, \bibinfo{author}{Whitman, K.},
  \bibinfo{author}{Desai, R.}, \bibinfo{author}{Rankin, J.},
  \bibinfo{author}{Strauss, D.T.}, \bibinfo{author}{Nitta, N.},
  \bibinfo{author}{Turner, D.}, \bibinfo{author}{Chen, T.Y.},
  \bibinfo{year}{2023}.
\newblock \bibinfo{title}{Galactic cosmic rays and solar energetic particles in
  cis-lunar space: Need for contextual energetic particle measurements at
  {Earth} and supporting distributed observations}.
\newblock \DOIprefix\doi{10.48550/arXiv.2209.03635}. \bibinfo{note}{{White
  paper submitted to Decadal Survey for Solar and Space Physics (Heliophysics)
  2024-2033}}.
%Type = Article
\bibitem[{Crosby et~al.(2015)Crosby, Heynderickx, Jiggens, Aran, Sanahuja,
  Truscott, Lei, Jacobs, Poedts, Gabriel et~al.}]{crosby2015sepem}
\bibinfo{author}{Crosby, N.}, \bibinfo{author}{Heynderickx, D.},
  \bibinfo{author}{Jiggens, P.}, \bibinfo{author}{Aran, A.},
  \bibinfo{author}{Sanahuja, B.}, \bibinfo{author}{Truscott, P.},
  \bibinfo{author}{Lei, F.}, \bibinfo{author}{Jacobs, C.},
  \bibinfo{author}{Poedts, S.}, \bibinfo{author}{Gabriel, S.}, et~al.,
  \bibinfo{year}{2015}.
\newblock \bibinfo{title}{{SEPEM}: A tool for statistical modeling the solar
  energetic particle environment}.
\newblock \bibinfo{journal}{Space Weather} \bibinfo{volume}{13},
  \bibinfo{pages}{406--426}.
\newblock \DOIprefix\doi{10.1002/2013SW001008}.
%Type = Article
\bibitem[{Cucinotta and Durante(2006)}]{cucinotta2006}
\bibinfo{author}{Cucinotta, F.A.}, \bibinfo{author}{Durante, M.},
  \bibinfo{year}{2006}.
\newblock \bibinfo{title}{Cancer risk from exposure to galactic cosmic rays:
  implications for space exploration by human beings}.
\newblock \bibinfo{journal}{The Lancet Oncology} \bibinfo{volume}{7},
  \bibinfo{pages}{431 -- 435}.
\newblock \DOIprefix\doi{10.1016/S1470-2045(06)70695-7}.
%Type = Article
\bibitem[{Cucinotta et~al.(2013)Cucinotta, Kim, Chappell and
  Huff}]{cucinotta2013safe}
\bibinfo{author}{Cucinotta, F.A.}, \bibinfo{author}{Kim, M.H.Y.},
  \bibinfo{author}{Chappell, L.J.}, \bibinfo{author}{Huff, J.L.},
  \bibinfo{year}{2013}.
\newblock \bibinfo{title}{How safe is safe enough? radiation risk for a human
  mission to {Mars}}.
\newblock \bibinfo{journal}{PLoS One} \bibinfo{volume}{8},
  \bibinfo{pages}{e74988}.
\newblock \DOIprefix\doi{10.1371/journal.pone.0074988}.
%Type = Article
\bibitem[{Cucinotta et~al.(2017)Cucinotta, To and Cacao}]{CUCINOTTA20171}
\bibinfo{author}{Cucinotta, F.A.}, \bibinfo{author}{To, K.},
  \bibinfo{author}{Cacao, E.}, \bibinfo{year}{2017}.
\newblock \bibinfo{title}{Predictions of space radiation fatality risk for
  exploration missions}.
\newblock \bibinfo{journal}{Life Sci. \& Space Res.} \bibinfo{volume}{13},
  \bibinfo{pages}{1--11}.
\newblock \DOIprefix\doi{10.1016/j.lssr.2017.01.005}.
%Type = Inproceedings
\bibitem[{{Cummings} et~al.(2019){Cummings}, {Stone}, {Heikkila}, {Lal} and
  {Richardson}}]{Cummings2019}
\bibinfo{author}{{Cummings}, A.}, \bibinfo{author}{{Stone}, E.},
  \bibinfo{author}{{Heikkila}, B.C.}, \bibinfo{author}{{Lal}, N.},
  \bibinfo{author}{{Richardson}, J.}, \bibinfo{year}{2019}.
\newblock \bibinfo{title}{{Voyager 2 Observations of the Anisotropy of
  Anomalous Cosmic Rays in the Heliosheath}}, in: \bibinfo{booktitle}{36th
  International Cosmic Ray Conference (ICRC2019)}, p. \bibinfo{pages}{1071}.
%Type = Article
\bibitem[{{Cummings} and {Stone}(2007)}]{Cummings2007}
\bibinfo{author}{{Cummings}, A.C.}, \bibinfo{author}{{Stone}, E.C.},
  \bibinfo{year}{2007}.
\newblock \bibinfo{title}{{Composition of Anomalous Cosmic Rays}}.
\newblock \bibinfo{journal}{Space Sci. Rev.} \bibinfo{volume}{130},
  \bibinfo{pages}{389--399}.
\newblock \DOIprefix\doi{10.1007/s11214-007-9161-y}.
%Type = Article
\bibitem[{{Cummings} et~al.(2016){Cummings}, {Stone}, {Heikkila}, {Lal},
  {Webber}, {J{\'o}hannesson}, {Moskalenko}, {Orlando} and
  {Porter}}]{Cummings2016}
\bibinfo{author}{{Cummings}, A.C.}, \bibinfo{author}{{Stone}, E.C.},
  \bibinfo{author}{{Heikkila}, B.C.}, \bibinfo{author}{{Lal}, N.},
  \bibinfo{author}{{Webber}, W.R.}, \bibinfo{author}{{J{\'o}hannesson}, G.},
  \bibinfo{author}{{Moskalenko}, I.V.}, \bibinfo{author}{{Orlando}, E.},
  \bibinfo{author}{{Porter}, T.A.}, \bibinfo{year}{2016}.
\newblock \bibinfo{title}{{Galactic Cosmic Rays in the Local Interstellar
  Medium: Voyager 1 Observations and Model Results}}.
\newblock \bibinfo{journal}{Astrophys. J.} \bibinfo{volume}{831},
  \bibinfo{pages}{18}.
\newblock \DOIprefix\doi{10.3847/0004-637X/831/1/18}.
%Type = Inproceedings
\bibitem[{{Cummings} et~al.(2008){Cummings}, {Stone}, {McDonald}, {Heikkila},
  {Lal} and {Webber}}]{Cummings2008}
\bibinfo{author}{{Cummings}, A.C.}, \bibinfo{author}{{Stone}, E.C.},
  \bibinfo{author}{{McDonald}, F.B.}, \bibinfo{author}{{Heikkila}, B.C.},
  \bibinfo{author}{{Lal}, N.}, \bibinfo{author}{{Webber}, W.R.},
  \bibinfo{year}{2008}.
\newblock \bibinfo{title}{{Anomalous Cosmic Rays in the Heliosheath}}, in:
  \bibinfo{editor}{{Li}, G.}, \bibinfo{editor}{{Hu}, Q.},
  \bibinfo{editor}{{Verkhoglyadova}, O.}, \bibinfo{editor}{{Zank}, G.P.},
  \bibinfo{editor}{{Lin}, R.P.}, \bibinfo{editor}{{Luhmann}, J.} (Eds.),
  \bibinfo{booktitle}{Particle Acceleration and Transport in the Heliosphere
  and Beyond: 7th Annual International AstroPhysics Conference}, pp.
  \bibinfo{pages}{343--348}.
\newblock \DOIprefix\doi{10.1063/1.2982469}.
%Type = Article
\bibitem[{{Dachev} et~al.(2020){Dachev}, {Tomov}, {Matviichuk}, {Dimitrov},
  {Semkova}, {Koleva}, {Jordanova}, {Bankov}, {Shurshakov} and
  {Benghin}}]{Dachev2020}
\bibinfo{author}{{Dachev}, T.P.}, \bibinfo{author}{{Tomov}, B.T.},
  \bibinfo{author}{{Matviichuk}, Y.N.}, \bibinfo{author}{{Dimitrov}, P.G.},
  \bibinfo{author}{{Semkova}, J.V.}, \bibinfo{author}{{Koleva}, R.T.},
  \bibinfo{author}{{Jordanova}, M.M.}, \bibinfo{author}{{Bankov}, N.G.},
  \bibinfo{author}{{Shurshakov}, V.A.}, \bibinfo{author}{{Benghin}, V.V.},
  \bibinfo{year}{2020}.
\newblock \bibinfo{title}{{Solar modulation of the GCR flux and dose rate,
  observed in space between 1991 and 2019}}.
\newblock \bibinfo{journal}{Life Sci. \& Space Res.} \bibinfo{volume}{26},
  \bibinfo{pages}{114--124}.
\newblock \DOIprefix\doi{10.1016/j.lssr.2020.06.002}.
%Type = Article
\bibitem[{{Dalla} et~al.(2003){Dalla}, {Balogh}, {Krucker}, {Posner},
  {M{\"u}ller-Mellin}, {Anglin}, {Hofer}, {Marsden}, {Sanderson}, {Tranquille},
  {Heber}, {Zhang} and {McKibben}}]{Dal2003a}
\bibinfo{author}{{Dalla}, S.}, \bibinfo{author}{{Balogh}, A.},
  \bibinfo{author}{{Krucker}, S.}, \bibinfo{author}{{Posner}, A.},
  \bibinfo{author}{{M{\"u}ller-Mellin}, R.}, \bibinfo{author}{{Anglin}, J.D.},
  \bibinfo{author}{{Hofer}, M.Y.}, \bibinfo{author}{{Marsden}, R.G.},
  \bibinfo{author}{{Sanderson}, T.R.}, \bibinfo{author}{{Tranquille}, C.},
  \bibinfo{author}{{Heber}, B.}, \bibinfo{author}{{Zhang}, M.},
  \bibinfo{author}{{McKibben}, R.B.}, \bibinfo{year}{2003}.
\newblock \bibinfo{title}{{Properties of high heliolatitude solar energetic
  particle events and constraints on models of acceleration and propagation}}.
\newblock \bibinfo{journal}{Geophys. Res. Lett.} \bibinfo{volume}{30},
  \bibinfo{pages}{8035}.
\newblock \DOIprefix\doi{10.1029/2003GL017139}.
%Type = Article
\bibitem[{{Dalla} et~al.(2020){Dalla}, {de Nolfo}, {Bruno}, {Giacalone},
  {Laitinen}, {Thomas}, {Battarbee} and {Marsh}}]{Dal2020}
\bibinfo{author}{{Dalla}, S.}, \bibinfo{author}{{de Nolfo}, G.A.},
  \bibinfo{author}{{Bruno}, A.}, \bibinfo{author}{{Giacalone}, J.},
  \bibinfo{author}{{Laitinen}, T.}, \bibinfo{author}{{Thomas}, S.},
  \bibinfo{author}{{Battarbee}, M.}, \bibinfo{author}{{Marsh}, M.S.},
  \bibinfo{year}{2020}.
\newblock \bibinfo{title}{{3D propagation of relativistic solar protons through
  interplanetary space}}.
\newblock \bibinfo{journal}{Astron. \& Astrophys.} \bibinfo{volume}{639},
  \bibinfo{pages}{A105}.
\newblock \DOIprefix\doi{10.1051/0004-6361/201937338}.
%Type = Article
\bibitem[{{Dalla} et~al.(2013){Dalla}, {Marsh}, {Kelly} and
  {Laitinen}}]{Dal2013}
\bibinfo{author}{{Dalla}, S.}, \bibinfo{author}{{Marsh}, M.S.},
  \bibinfo{author}{{Kelly}, J.}, \bibinfo{author}{{Laitinen}, T.},
  \bibinfo{year}{2013}.
\newblock \bibinfo{title}{{Solar energetic particle drifts in the Parker
  spiral}}.
\newblock \bibinfo{journal}{J. Geophys. Res.: Space Phys.}
  \bibinfo{volume}{118}, \bibinfo{pages}{5979--5985}.
\newblock \DOIprefix\doi{10.1002/jgra.50589}.
%Type = Article
\bibitem[{{De Simone} et~al.(2011){De Simone}, {Di Felice}, {Gieseler},
  {Boezio}, {Casolino}, {Picozza}, {Heber} and {{PAMELA}
  Collaboration}}]{Simone2011}
\bibinfo{author}{{De Simone}, N.}, \bibinfo{author}{{Di Felice}, V.},
  \bibinfo{author}{{Gieseler}, J.}, \bibinfo{author}{{Boezio}, M.},
  \bibinfo{author}{{Casolino}, M.}, \bibinfo{author}{{Picozza}, P.},
  \bibinfo{author}{{Heber}, B.}, \bibinfo{author}{{{PAMELA} Collaboration}},
  \bibinfo{year}{2011}.
\newblock \bibinfo{title}{{Latitudinal and radial gradients of galactic cosmic
  ray protons in the inner heliosphere - {PAMELA} and Ulysses observations}}.
\newblock \bibinfo{journal}{Astrophy. Space Sci. Transactions}
  \bibinfo{volume}{7}, \bibinfo{pages}{425--434}.
\newblock \DOIprefix\doi{10.5194/astra-7-425-2011}.
%Type = Article
\bibitem[{{Decker} et~al.(2005){Decker}, {Krimigis}, {Roelof}, {Hill},
  {Armstrong}, {Gloeckler}, {Hamilton} and {Lanzerotti}}]{Decker2005}
\bibinfo{author}{{Decker}, R.B.}, \bibinfo{author}{{Krimigis}, S.M.},
  \bibinfo{author}{{Roelof}, E.C.}, \bibinfo{author}{{Hill}, M.E.},
  \bibinfo{author}{{Armstrong}, T.P.}, \bibinfo{author}{{Gloeckler}, G.},
  \bibinfo{author}{{Hamilton}, D.C.}, \bibinfo{author}{{Lanzerotti}, L.J.},
  \bibinfo{year}{2005}.
\newblock \bibinfo{title}{{Voyager 1 in the Foreshock, Termination Shock, and
  Heliosheath}}.
\newblock \bibinfo{journal}{Science} \bibinfo{volume}{309},
  \bibinfo{pages}{2020--2024}.
\newblock \DOIprefix\doi{10.1126/science.1117569}.
%Type = Article
\bibitem[{{Decker} et~al.(2008){Decker}, {Krimigis}, {Roelof}, {Hill},
  {Armstrong}, {Gloeckler}, {Hamilton} and {Lanzerotti}}]{Decker2008}
\bibinfo{author}{{Decker}, R.B.}, \bibinfo{author}{{Krimigis}, S.M.},
  \bibinfo{author}{{Roelof}, E.C.}, \bibinfo{author}{{Hill}, M.E.},
  \bibinfo{author}{{Armstrong}, T.P.}, \bibinfo{author}{{Gloeckler}, G.},
  \bibinfo{author}{{Hamilton}, D.C.}, \bibinfo{author}{{Lanzerotti}, L.J.},
  \bibinfo{year}{2008}.
\newblock \bibinfo{title}{{Mediation of the solar wind termination shock by
  non-thermal ions}}.
\newblock \bibinfo{journal}{Nature} \bibinfo{volume}{454},
  \bibinfo{pages}{67--70}.
\newblock \DOIprefix\doi{10.1038/nature07030}.
%Type = Article
\bibitem[{{Desai} and {Giacalone}(2016)}]{desai2016large}
\bibinfo{author}{{Desai}, M.}, \bibinfo{author}{{Giacalone}, J.},
  \bibinfo{year}{2016}.
\newblock \bibinfo{title}{{Large gradual solar energetic particle events}}.
\newblock \bibinfo{journal}{Living Rev. Sol. Phys.} \bibinfo{volume}{13},
  \bibinfo{pages}{3}.
\newblock \DOIprefix\doi{10.1007/s41116-016-0002-5}.
%Type = Article
\bibitem[{{Di Felice} et~al.(2017){Di Felice}, {Munini}, {Vos} and
  {Potgieter}}]{Felice2017}
\bibinfo{author}{{Di Felice}, V.}, \bibinfo{author}{{Munini}, R.},
  \bibinfo{author}{{Vos}, E.E.}, \bibinfo{author}{{Potgieter}, M.S.},
  \bibinfo{year}{2017}.
\newblock \bibinfo{title}{{New Evidence for Charge-sign-dependent Modulation
  During the Solar Minimum of 2006 to 2009}}.
\newblock \bibinfo{journal}{Astrophys. J.} \bibinfo{volume}{834},
  \bibinfo{pages}{89}.
\newblock \DOIprefix\doi{10.3847/1538-4357/834/1/89}.
%Type = Article
\bibitem[{{Dierckxsens} et~al.(2015){Dierckxsens}, {Tziotziou}, {Dalla},
  {Patsou}, {Marsh}, {Crosby}, {Malandraki} and {Tsiropoula}}]{Dier2015}
\bibinfo{author}{{Dierckxsens}, M.}, \bibinfo{author}{{Tziotziou}, K.},
  \bibinfo{author}{{Dalla}, S.}, \bibinfo{author}{{Patsou}, I.},
  \bibinfo{author}{{Marsh}, M.S.}, \bibinfo{author}{{Crosby}, N.B.},
  \bibinfo{author}{{Malandraki}, O.}, \bibinfo{author}{{Tsiropoula}, G.},
  \bibinfo{year}{2015}.
\newblock \bibinfo{title}{{Relationship between Solar Energetic Particles and
  Properties of Flares and CMEs: Statistical Analysis of Solar Cycle 23
  Events}}.
\newblock \bibinfo{journal}{Sol. Phys.} \bibinfo{volume}{290},
  \bibinfo{pages}{841--874}.
\newblock \DOIprefix\doi{10.1007/s11207-014-0641-4}.
%Type = Article
\bibitem[{Dobynde et~al.(2023)Dobynde, Harikumaran, Guo, Wheeler, Galea and
  Buticchi}]{dobynde2023aircraft}
\bibinfo{author}{Dobynde, M.}, \bibinfo{author}{Harikumaran, J.},
  \bibinfo{author}{Guo, J.}, \bibinfo{author}{Wheeler, P.},
  \bibinfo{author}{Galea, M.}, \bibinfo{author}{Buticchi, G.},
  \bibinfo{year}{2023}.
\newblock \bibinfo{title}{Cosmic radiation reliability analysis for aircraft
  power electronics}.
\newblock \bibinfo{journal}{IEEE Transactions on Transportation
  Electrification} ,
  \bibinfo{pages}{1--1}\DOIprefix\doi{10.1109/TTE.2023.3278319}.
%Type = Article
\bibitem[{Dobynde et~al.(2021)Dobynde, Shprits, Drozdov, Hoffman and
  Li}]{dobynde2021beating}
\bibinfo{author}{Dobynde, M.}, \bibinfo{author}{Shprits, Y.},
  \bibinfo{author}{Drozdov, A.}, \bibinfo{author}{Hoffman, J.},
  \bibinfo{author}{Li, J.}, \bibinfo{year}{2021}.
\newblock \bibinfo{title}{Beating 1 sievert: Optimal radiation shielding of
  astronauts on a mission to mars}.
\newblock \bibinfo{journal}{Space Weather} \bibinfo{volume}{19},
  \bibinfo{pages}{e2021SW002749}.
\newblock \DOIprefix\doi{10.1029/2021SW002749}.
%Type = Article
\bibitem[{Domingo et~al.(1995)Domingo, Fleck and Poland}]{domingo1995soho}
\bibinfo{author}{Domingo, V.}, \bibinfo{author}{Fleck, B.},
  \bibinfo{author}{Poland, A.}, \bibinfo{year}{1995}.
\newblock \bibinfo{title}{{SOHO}: the solar and heliospheric observatory}.
\newblock \bibinfo{journal}{Space Sci. Rev.} \bibinfo{volume}{72},
  \bibinfo{pages}{81--84}.
\newblock \DOIprefix\doi{10.1007/BF00768758}.
%Type = Article
\bibitem[{Dorman(2001)}]{Dorman_2001}
\bibinfo{author}{Dorman, L.I.}, \bibinfo{year}{2001}.
\newblock \bibinfo{title}{Cosmic ray long-term variation: even-odd cycle
  effect, role of drifts, and the onset of cycle 23}.
\newblock \bibinfo{journal}{Adv. Space Res.} \bibinfo{volume}{27},
  \bibinfo{pages}{601--606}.
\newblock \DOIprefix\doi{10.1016/S0273-1177(01)00088-6}.
%Type = Article
\bibitem[{{Drake} et~al.(2010){Drake}, {Opher}, {Swisdak} and
  {Chamoun}}]{Drake2010}
\bibinfo{author}{{Drake}, J.F.}, \bibinfo{author}{{Opher}, M.},
  \bibinfo{author}{{Swisdak}, M.}, \bibinfo{author}{{Chamoun}, J.N.},
  \bibinfo{year}{2010}.
\newblock \bibinfo{title}{{A Magnetic Reconnection Mechanism for the Generation
  of Anomalous Cosmic Rays}}.
\newblock \bibinfo{journal}{Astrophys. J.} \bibinfo{volume}{709},
  \bibinfo{pages}{963--974}.
\newblock \DOIprefix\doi{10.1088/0004-637X/709/2/963}.
%Type = Article
\bibitem[{Dresing et~al.(2014)Dresing, G{\'o}mez-Herrero, Heber, Klassen,
  Malandraki, Dr{\"o}ge and Kartavykh}]{dresing2014statistical}
\bibinfo{author}{Dresing, N.}, \bibinfo{author}{G{\'o}mez-Herrero, R.},
  \bibinfo{author}{Heber, B.}, \bibinfo{author}{Klassen, A.},
  \bibinfo{author}{Malandraki, O.}, \bibinfo{author}{Dr{\"o}ge, W.},
  \bibinfo{author}{Kartavykh, Y.}, \bibinfo{year}{2014}.
\newblock \bibinfo{title}{Statistical survey of widely spread out solar
  electron events observed with stereo and ace with special attention to
  anisotropies}.
\newblock \bibinfo{journal}{Astron. \& Astrophys.} \bibinfo{volume}{567},
  \bibinfo{pages}{A27}.
\newblock \DOIprefix\doi{10.1051/0004-6361/201423789}.
%Type = Article
\bibitem[{{Dumbovi{\'c}} et~al.(2018){Dumbovi{\'c}}, {Heber}, {Vr{\v s}nak},
  {Temmer} and {Kirin}}]{dumbovic18b}
\bibinfo{author}{{Dumbovi{\'c}}, M.}, \bibinfo{author}{{Heber}, B.},
  \bibinfo{author}{{Vr{\v s}nak}, B.}, \bibinfo{author}{{Temmer}, M.},
  \bibinfo{author}{{Kirin}, A.}, \bibinfo{year}{2018}.
\newblock \bibinfo{title}{{An Analytical Diffusion-Expansion Model for Forbush
  Decreases Caused by Flux Ropes}}.
\newblock \bibinfo{journal}{Astrophys. J.} \bibinfo{volume}{860},
  \bibinfo{pages}{71}.
\newblock \DOIprefix\doi{10.3847/1538-4357/aac2de}.
%Type = Article
\bibitem[{{Dumbovi{\'c}} et~al.(2020){Dumbovi{\'c}}, {Vr{\v{s}}nak}, {Guo},
  {Heber}, {Dissauer}, {Carcaboso}, {Temmer}, {Veronig}, {Podladchikova},
  {M{\"o}stl}, {Amerstorfer} and {Kirin}}]{dumbovic20}
\bibinfo{author}{{Dumbovi{\'c}}, M.}, \bibinfo{author}{{Vr{\v{s}}nak}, B.},
  \bibinfo{author}{{Guo}, J.}, \bibinfo{author}{{Heber}, B.},
  \bibinfo{author}{{Dissauer}, K.}, \bibinfo{author}{{Carcaboso}, F.},
  \bibinfo{author}{{Temmer}, M.}, \bibinfo{author}{{Veronig}, A.},
  \bibinfo{author}{{Podladchikova}, T.}, \bibinfo{author}{{M{\"o}stl}, C.},
  \bibinfo{author}{{Amerstorfer}, T.}, \bibinfo{author}{{Kirin}, A.},
  \bibinfo{year}{2020}.
\newblock \bibinfo{title}{{Evolution of Coronal Mass Ejections and the
  Corresponding Forbush Decreases: Modeling vs. Multi-Spacecraft
  Observations}}.
\newblock \bibinfo{journal}{Sol. Phys.} \bibinfo{volume}{295},
  \bibinfo{pages}{104}.
\newblock \DOIprefix\doi{10.1007/s11207-020-01671-7}.
%Type = Article
\bibitem[{{Dumbovi{\'c}} et~al.(2022){Dumbovi{\'c}}, {Vr{\v{s}}nak}, {Temmer},
  {Heber} and {K{\"u}hl}}]{dumbovic22}
\bibinfo{author}{{Dumbovi{\'c}}, M.}, \bibinfo{author}{{Vr{\v{s}}nak}, B.},
  \bibinfo{author}{{Temmer}, M.}, \bibinfo{author}{{Heber}, B.},
  \bibinfo{author}{{K{\"u}hl}, P.}, \bibinfo{year}{2022}.
\newblock \bibinfo{title}{{Generic profile of a long-lived corotating
  interaction region and associated recurrent Forbush decrease}}.
\newblock \bibinfo{journal}{Astron. \& Astrophys.} \bibinfo{volume}{658},
  \bibinfo{pages}{A187}.
\newblock \DOIprefix\doi{10.1051/0004-6361/202140861}.
%Type = Article
\bibitem[{Ellison and Ramaty(1985)}]{ellison1985shock}
\bibinfo{author}{Ellison, D.C.}, \bibinfo{author}{Ramaty, R.},
  \bibinfo{year}{1985}.
\newblock \bibinfo{title}{Shock acceleration of electrons and ions in solar
  flares}.
\newblock \bibinfo{journal}{Astrophys. J.} \bibinfo{volume}{298},
  \bibinfo{pages}{400--408}.
\newblock \DOIprefix\doi{10.1086/163623}.
%Type = Article
\bibitem[{{Engelbrecht} and {Burger}(2013)}]{Engelbrecht2013}
\bibinfo{author}{{Engelbrecht}, N.E.}, \bibinfo{author}{{Burger}, R.A.},
  \bibinfo{year}{2013}.
\newblock \bibinfo{title}{{An Ab Initio Model for Cosmic-ray Modulation}}.
\newblock \bibinfo{journal}{Astrophys. J.} \bibinfo{volume}{772},
  \bibinfo{pages}{46}.
\newblock \DOIprefix\doi{10.1088/0004-637X/772/1/46}.
%Type = Article
\bibitem[{{Engelbrecht} et~al.(2022){Engelbrecht}, {Effenberger}, {Florinski},
  {Potgieter}, {Ruffolo}, {Chhiber}, {Usmanov}, {Rankin} and
  {Els}}]{Engelbrecht2022}
\bibinfo{author}{{Engelbrecht}, N.E.}, \bibinfo{author}{{Effenberger}, F.},
  \bibinfo{author}{{Florinski}, V.}, \bibinfo{author}{{Potgieter}, M.S.},
  \bibinfo{author}{{Ruffolo}, D.}, \bibinfo{author}{{Chhiber}, R.},
  \bibinfo{author}{{Usmanov}, A.V.}, \bibinfo{author}{{Rankin}, J.S.},
  \bibinfo{author}{{Els}, P.L.}, \bibinfo{year}{2022}.
\newblock \bibinfo{title}{{Theory of Cosmic Ray Transport in the Heliosphere}}.
\newblock \bibinfo{journal}{Space Sci. Rev.} \bibinfo{volume}{218},
  \bibinfo{pages}{33}.
\newblock \DOIprefix\doi{10.1007/s11214-022-00896-1}.
%Type = Article
\bibitem[{{Engelbrecht} and {Moloto}(2021)}]{Engelbrecht2021}
\bibinfo{author}{{Engelbrecht}, N.E.}, \bibinfo{author}{{Moloto}, K.D.},
  \bibinfo{year}{2021}.
\newblock \bibinfo{title}{{An Ab Initio Approach to Antiproton Modulation in
  the Inner Heliosphere}}.
\newblock \bibinfo{journal}{Astrophys. J.} \bibinfo{volume}{908},
  \bibinfo{pages}{167}.
\newblock \DOIprefix\doi{10.3847/1538-4357/abd3a5}.
%Type = Article
\bibitem[{{Fiandrini} et~al.(2021){Fiandrini}, {Tomassetti}, {Bertucci},
  {Donnini}, {Graziani}, {Khiali} and {Reina Conde}}]{Fiandrini2021}
\bibinfo{author}{{Fiandrini}, E.}, \bibinfo{author}{{Tomassetti}, N.},
  \bibinfo{author}{{Bertucci}, B.}, \bibinfo{author}{{Donnini}, F.},
  \bibinfo{author}{{Graziani}, M.}, \bibinfo{author}{{Khiali}, B.},
  \bibinfo{author}{{Reina Conde}, A.}, \bibinfo{year}{2021}.
\newblock \bibinfo{title}{{Numerical modeling of cosmic rays in the
  heliosphere: Analysis of proton data from AMS-02 and {PAMELA}}}.
\newblock \bibinfo{journal}{Phys. Rev. D} \bibinfo{volume}{104},
  \bibinfo{pages}{023012}.
\newblock \DOIprefix\doi{10.1103/PhysRevD.104.023012}.
%Type = Article
\bibitem[{{Fisk}(1996)}]{Fisk1996}
\bibinfo{author}{{Fisk}, L.A.}, \bibinfo{year}{1996}.
\newblock \bibinfo{title}{{Motion of the footpoints of heliospheric magnetic
  field lines at the Sun: Implications for recurrent energetic particle events
  at high heliographic latitudes}}.
\newblock \bibinfo{journal}{J. Geophys. Res.: Space Phys.}
  \bibinfo{volume}{101}, \bibinfo{pages}{15547--15554}.
\newblock \DOIprefix\doi{10.1029/96JA01005}.
%Type = Article
\bibitem[{{Fisk} and {Gloeckler}(2009)}]{Fisk2009}
\bibinfo{author}{{Fisk}, L.A.}, \bibinfo{author}{{Gloeckler}, G.},
  \bibinfo{year}{2009}.
\newblock \bibinfo{title}{{The acceleration of Anomalous Cosmic Rays by
  stochastic acceleration in the heliosheath}}.
\newblock \bibinfo{journal}{Adv. Space Res.} \bibinfo{volume}{43},
  \bibinfo{pages}{1471--1478}.
\newblock \DOIprefix\doi{10.1016/j.asr.2009.02.010}.
%Type = Article
\bibitem[{{Fisk} et~al.(1974){Fisk}, {Kozlovsky} and {Ramaty}}]{Fisk1974}
\bibinfo{author}{{Fisk}, L.A.}, \bibinfo{author}{{Kozlovsky}, B.},
  \bibinfo{author}{{Ramaty}, R.}, \bibinfo{year}{1974}.
\newblock \bibinfo{title}{{An Interpretation of the Observed Oxygen and
  Nitrogen Enhancements in Low-Energy Cosmic Rays}}.
\newblock \bibinfo{journal}{Astrophys. J. Lett.} \bibinfo{volume}{190},
  \bibinfo{pages}{L35}.
\newblock \DOIprefix\doi{10.1086/181498}.
%Type = Article
\bibitem[{{Florinski} et~al.(2003){Florinski}, {Zank} and
  {Pogorelov}}]{Florinski2003}
\bibinfo{author}{{Florinski}, V.}, \bibinfo{author}{{Zank}, G.P.},
  \bibinfo{author}{{Pogorelov}, N.V.}, \bibinfo{year}{2003}.
\newblock \bibinfo{title}{{Galactic cosmic ray transport in the global
  heliosphere}}.
\newblock \bibinfo{journal}{J. Geophys. Res.: Space Phys.}
  \bibinfo{volume}{108}, \bibinfo{pages}{1228}.
\newblock \DOIprefix\doi{10.1029/2002JA009695}.
%Type = Article
\bibitem[{Forbush(1937)}]{forbush37}
\bibinfo{author}{Forbush, S.E.}, \bibinfo{year}{1937}.
\newblock \bibinfo{title}{On the effects in cosmic-ray intensity observed
  during the recent magnetic storm}.
\newblock \bibinfo{journal}{Phys. Rev.} \bibinfo{volume}{51},
  \bibinfo{pages}{1108--1109}.
\newblock \DOIprefix\doi{10.1103/PhysRev.51.1108.3}.
%Type = Article
\bibitem[{{Forbush}(1946)}]{forbush1946}
\bibinfo{author}{{Forbush}, S.E.}, \bibinfo{year}{1946}.
\newblock \bibinfo{title}{{Three Unusual Cosmic-Ray Increases Possibly Due to
  Charged Particles from the Sun}}.
\newblock \bibinfo{journal}{Phys. Rev.} \bibinfo{volume}{70},
  \bibinfo{pages}{771--772}.
\newblock \DOIprefix\doi{10.1103/PhysRev.70.771}.
%Type = Article
\bibitem[{Forbush(1954)}]{forbush1954world}
\bibinfo{author}{Forbush, S.E.}, \bibinfo{year}{1954}.
\newblock \bibinfo{title}{World-wide cosmic ray variations, 1937--1952}.
\newblock \bibinfo{journal}{J. Geophys. Res.} \bibinfo{volume}{59},
  \bibinfo{pages}{525--542}.
\newblock \DOIprefix\doi{10.1029/JZ059i004p00525}.
%Type = Article
\bibitem[{Forbush(1958)}]{forbush1958cosmic}
\bibinfo{author}{Forbush, S.E.}, \bibinfo{year}{1958}.
\newblock \bibinfo{title}{Cosmic-ray intensity variations during two solar
  cycles}.
\newblock \bibinfo{journal}{J. Geophys. Res.} \bibinfo{volume}{63},
  \bibinfo{pages}{651--669}.
\newblock \DOIprefix\doi{10.1029/JZ063i004p00651}.
%Type = Article
\bibitem[{Freiherr~von Forstner et~al.(2018)Freiherr~von Forstner, Guo,
  Wimmer‐Schweingruber, Hassler, Temmer, Dumbović, Jian, Appel, Čalogović,
  Ehresmann, Heber, Lohf, Posner, Steigies, Vršnak and Zeitlin}]{forstner18}
\bibinfo{author}{Freiherr~von Forstner, J.L.}, \bibinfo{author}{Guo, J.},
  \bibinfo{author}{Wimmer‐Schweingruber, R.F.}, \bibinfo{author}{Hassler,
  D.M.}, \bibinfo{author}{Temmer, M.}, \bibinfo{author}{Dumbović, M.},
  \bibinfo{author}{Jian, L.K.}, \bibinfo{author}{Appel, J.K.},
  \bibinfo{author}{Čalogović, J.}, \bibinfo{author}{Ehresmann, B.},
  \bibinfo{author}{Heber, B.}, \bibinfo{author}{Lohf, H.},
  \bibinfo{author}{Posner, A.}, \bibinfo{author}{Steigies, C.T.},
  \bibinfo{author}{Vršnak, B.}, \bibinfo{author}{Zeitlin, C.J.},
  \bibinfo{year}{2018}.
\newblock \bibinfo{title}{Using forbush decreases to derive the transit time of
  icmes propagating from 1 au to {Mars}}.
\newblock \bibinfo{journal}{J. Geophys. Res.: Space Phys.}
  \bibinfo{volume}{123}, \bibinfo{pages}{39--56}.
\newblock \DOIprefix\doi{10.1002/2017JA024700}.
%Type = Article
\bibitem[{{Fox} et~al.(2016){Fox}, {Velli}, {Bale}, {Decker}, {Driesman},
  {Howard}, {Kasper}, {Kinnison}, {Kusterer}, {Lario}, {Lockwood}, {McComas},
  {Raouafi} and {Szabo}}]{fox2016psp}
\bibinfo{author}{{Fox}, N.J.}, \bibinfo{author}{{Velli}, M.C.},
  \bibinfo{author}{{Bale}, S.D.}, \bibinfo{author}{{Decker}, R.},
  \bibinfo{author}{{Driesman}, A.}, \bibinfo{author}{{Howard}, R.A.},
  \bibinfo{author}{{Kasper}, J.C.}, \bibinfo{author}{{Kinnison}, J.},
  \bibinfo{author}{{Kusterer}, M.}, \bibinfo{author}{{Lario}, D.},
  \bibinfo{author}{{Lockwood}, M.K.}, \bibinfo{author}{{McComas}, D.J.},
  \bibinfo{author}{{Raouafi}, N.E.}, \bibinfo{author}{{Szabo}, A.},
  \bibinfo{year}{2016}.
\newblock \bibinfo{title}{{The Solar Probe Plus Mission: Humanity's First Visit
  to Our Star}}.
\newblock \bibinfo{journal}{Space Sci. Rev.} \bibinfo{volume}{204},
  \bibinfo{pages}{7--48}.
\newblock \DOIprefix\doi{10.1007/s11214-015-0211-6}.
%Type = Article
\bibitem[{{Freiherr von Forstner} et~al.(2020){Freiherr von Forstner}, {Guo},
  {Wimmer-Schweingruber}, {Dumbovi{\'c}}, {Janvier}, {D{\'e}moulin}, {Veronig},
  {Temmer}, {Papaioannou}, {Dasso}, {Hassler} and {Zeitlin}}]{forstner20}
\bibinfo{author}{{Freiherr von Forstner}, J.L.}, \bibinfo{author}{{Guo}, J.},
  \bibinfo{author}{{Wimmer-Schweingruber}, R.F.},
  \bibinfo{author}{{Dumbovi{\'c}}, M.}, \bibinfo{author}{{Janvier}, M.},
  \bibinfo{author}{{D{\'e}moulin}, P.}, \bibinfo{author}{{Veronig}, A.},
  \bibinfo{author}{{Temmer}, M.}, \bibinfo{author}{{Papaioannou}, A.},
  \bibinfo{author}{{Dasso}, S.}, \bibinfo{author}{{Hassler}, D.M.},
  \bibinfo{author}{{Zeitlin}, C.J.}, \bibinfo{year}{2020}.
\newblock \bibinfo{title}{{Comparing the Properties of ICME-Induced Forbush
  Decreases at Earth and Mars}}.
\newblock \bibinfo{journal}{J. Geophys. Res.: Space Phys.}
  \bibinfo{volume}{125}, \bibinfo{pages}{e27662}.
\newblock \DOIprefix\doi{10.1029/2019JA027662}.
%Type = Article
\bibitem[{Fu et~al.(2022)Fu, Ding, Zhang, Zhang, Li, Li, Tang, Zhang, Xu, Wang
  et~al.}]{fu2022first}
\bibinfo{author}{Fu, S.}, \bibinfo{author}{Ding, Z.}, \bibinfo{author}{Zhang,
  Y.}, \bibinfo{author}{Zhang, X.}, \bibinfo{author}{Li, C.},
  \bibinfo{author}{Li, G.}, \bibinfo{author}{Tang, S.}, \bibinfo{author}{Zhang,
  H.}, \bibinfo{author}{Xu, Y.}, \bibinfo{author}{Wang, Y.}, et~al.,
  \bibinfo{year}{2022}.
\newblock \bibinfo{title}{First report of a solar energetic particle event
  observed by china’s tianwen-1 mission in transit to mars}.
\newblock \bibinfo{journal}{Astrophys. J. Lett.} \bibinfo{volume}{934},
  \bibinfo{pages}{L15}.
%Type = Article
\bibitem[{{Fu} et~al.(2021){Fu}, {Zhao}, {Zhang}, {Luo} and {Li}}]{Fu2021}
\bibinfo{author}{{Fu}, S.}, \bibinfo{author}{{Zhao}, L.},
  \bibinfo{author}{{Zhang}, X.}, \bibinfo{author}{{Luo}, P.},
  \bibinfo{author}{{Li}, Y.}, \bibinfo{year}{2021}.
\newblock \bibinfo{title}{{Comparison of Anomalous and Galactic Cosmic-Ray
  Oxygen at 1 au during 1997-2020}}.
\newblock \bibinfo{journal}{Astrophys. J. Lett.} \bibinfo{volume}{920},
  \bibinfo{pages}{L12}.
\newblock \DOIprefix\doi{10.3847/2041-8213/ac29b9}.
%Type = Article
\bibitem[{{Garcia-Munoz} et~al.(1973){Garcia-Munoz}, {Mason} and
  {Simpson}}]{Garcia-Munoz1973}
\bibinfo{author}{{Garcia-Munoz}, M.}, \bibinfo{author}{{Mason}, G.M.},
  \bibinfo{author}{{Simpson}, J.A.}, \bibinfo{year}{1973}.
\newblock \bibinfo{title}{{A New Test for Solar Modulation Theory: the 1972
  May-July Low-Energy Galactic Cosmic-Ray Proton and Helium Spectra}}.
\newblock \bibinfo{journal}{Astrophys. J. Lett.} \bibinfo{volume}{182},
  \bibinfo{pages}{L81}.
\newblock \DOIprefix\doi{10.1086/181224}.
%Type = Article
\bibitem[{{Giacalone} et~al.(2012){Giacalone}, {Drake} and
  {Jokipii}}]{Giacalone2012}
\bibinfo{author}{{Giacalone}, J.}, \bibinfo{author}{{Drake}, J.F.},
  \bibinfo{author}{{Jokipii}, J.R.}, \bibinfo{year}{2012}.
\newblock \bibinfo{title}{{The Acceleration Mechanism of Anomalous Cosmic
  Rays}}.
\newblock \bibinfo{journal}{Space Sci. Rev.} \bibinfo{volume}{173},
  \bibinfo{pages}{283--307}.
\newblock \DOIprefix\doi{10.1007/s11214-012-9915-z}.
%Type = Article
\bibitem[{{Giacalone} et~al.(2022){Giacalone}, {Fahr}, {Fichtner}, {Florinski},
  {Heber}, {Hill}, {K{\'o}ta}, {Leske}, {Potgieter} and
  {Rankin}}]{Giacalone2022}
\bibinfo{author}{{Giacalone}, J.}, \bibinfo{author}{{Fahr}, H.},
  \bibinfo{author}{{Fichtner}, H.}, \bibinfo{author}{{Florinski}, V.},
  \bibinfo{author}{{Heber}, B.}, \bibinfo{author}{{Hill}, M.E.},
  \bibinfo{author}{{K{\'o}ta}, J.}, \bibinfo{author}{{Leske}, R.A.},
  \bibinfo{author}{{Potgieter}, M.S.}, \bibinfo{author}{{Rankin}, J.S.},
  \bibinfo{year}{2022}.
\newblock \bibinfo{title}{{Anomalous Cosmic Rays and Heliospheric Energetic
  Particles}}.
\newblock \bibinfo{journal}{Space Sci. Rev.} \bibinfo{volume}{218},
  \bibinfo{pages}{22}.
\newblock \DOIprefix\doi{10.1007/s11214-022-00890-7}.
%Type = Article
\bibitem[{{Giacalone} et~al.(2021){Giacalone}, {Nakanotani}, {Zank},
  {K{\`o}ta}, {Opher} and {Richardson}}]{Giacalone2021}
\bibinfo{author}{{Giacalone}, J.}, \bibinfo{author}{{Nakanotani}, M.},
  \bibinfo{author}{{Zank}, G.P.}, \bibinfo{author}{{K{\`o}ta}, J.},
  \bibinfo{author}{{Opher}, M.}, \bibinfo{author}{{Richardson}, J.D.},
  \bibinfo{year}{2021}.
\newblock \bibinfo{title}{{Hybrid Simulations of Interstellar Pickup Protons
  Accelerated at the Solar-wind Termination Shock at Multiple Locations}}.
\newblock \bibinfo{journal}{Astrophys. J.} \bibinfo{volume}{911},
  \bibinfo{pages}{27}.
\newblock \DOIprefix\doi{10.3847/1538-4357/abe93a}.
%Type = Article
\bibitem[{{Gieseler} et~al.(2017){Gieseler}, {Heber} and
  {Herbst}}]{Gieseler2017}
\bibinfo{author}{{Gieseler}, J.}, \bibinfo{author}{{Heber}, B.},
  \bibinfo{author}{{Herbst}, K.}, \bibinfo{year}{2017}.
\newblock \bibinfo{title}{{An Empirical Modification of the Force Field
  Approach to Describe the Modulation of Galactic Cosmic Rays Close to Earth in
  a Broad Range of Rigidities}}.
\newblock \bibinfo{journal}{J. Geophys. Res.: Space Phys.}
  \bibinfo{volume}{122}, \bibinfo{pages}{10,964--10,979}.
\newblock \DOIprefix\doi{10.1002/2017JA024763}.
%Type = Article
\bibitem[{Gil et~al.(2018)Gil, Kovaltsov, Mikhailov, Mishev, Poluianov and
  Usoskin}]{gil2018anisotropic}
\bibinfo{author}{Gil, A.}, \bibinfo{author}{Kovaltsov, G.A.},
  \bibinfo{author}{Mikhailov, V.V.}, \bibinfo{author}{Mishev, A.},
  \bibinfo{author}{Poluianov, S.}, \bibinfo{author}{Usoskin, I.G.},
  \bibinfo{year}{2018}.
\newblock \bibinfo{title}{An anisotropic cosmic-ray enhancement event on
  07-june-2015: A possible origin}.
\newblock \bibinfo{journal}{Sol. Phys.} \bibinfo{volume}{293},
  \bibinfo{pages}{154}.
\newblock \DOIprefix\doi{10.1007/s11207-018-1375-5}.
%Type = Article
\bibitem[{{Gleeson} and {Axford}(1968)}]{gleeson68}
\bibinfo{author}{{Gleeson}, L.J.}, \bibinfo{author}{{Axford}, W.I.},
  \bibinfo{year}{1968}.
\newblock \bibinfo{title}{{Solar Modulation of Galactic Cosmic Rays}}.
\newblock \bibinfo{journal}{Astrophys. J.} \bibinfo{volume}{154},
  \bibinfo{pages}{1011}.
\newblock \DOIprefix\doi{10.1086/149822}.
%Type = Article
\bibitem[{{Gloeckler} and {Fisk}(2014)}]{Gloeckler2014}
\bibinfo{author}{{Gloeckler}, G.}, \bibinfo{author}{{Fisk}, L.A.},
  \bibinfo{year}{2014}.
\newblock \bibinfo{title}{{A test for whether or not Voyager 1 has crossed the
  heliopause}}.
\newblock \bibinfo{journal}{Geophys. Res. Lett.} \bibinfo{volume}{41},
  \bibinfo{pages}{5325--5330}.
\newblock \DOIprefix\doi{10.1002/2014GL060781}.
%Type = Article
\bibitem[{G{\'o}mez-Herrero et~al.(2015)G{\'o}mez-Herrero, Dresing, Klassen,
  Heber, Lario, Agueda, Malandraki, Blanco, Rodr{\'\i}guez-Pacheco and
  Banjac}]{gomez2015circumsolar}
\bibinfo{author}{G{\'o}mez-Herrero, R.}, \bibinfo{author}{Dresing, N.},
  \bibinfo{author}{Klassen, A.}, \bibinfo{author}{Heber, B.},
  \bibinfo{author}{Lario, D.}, \bibinfo{author}{Agueda, N.},
  \bibinfo{author}{Malandraki, O.E.}, \bibinfo{author}{Blanco, J.},
  \bibinfo{author}{Rodr{\'\i}guez-Pacheco, J.}, \bibinfo{author}{Banjac, S.},
  \bibinfo{year}{2015}.
\newblock \bibinfo{title}{Circumsolar energetic particle distribution on 2011
  november 3}.
\newblock \bibinfo{journal}{Astrophys. J.} \bibinfo{volume}{799},
  \bibinfo{pages}{55}.
\newblock \DOIprefix\doi{10.1088/0004-637X/799/1/55}.
%Type = Article
\bibitem[{{Guo} et~al.(2010){Guo}, {Jokipii} and {Kota}}]{Guo2010}
\bibinfo{author}{{Guo}, F.}, \bibinfo{author}{{Jokipii}, J.R.},
  \bibinfo{author}{{Kota}, J.}, \bibinfo{year}{2010}.
\newblock \bibinfo{title}{{Particle Acceleration by Collisionless Shocks
  Containing Large-scale Magnetic-field Variations}}.
\newblock \bibinfo{journal}{Astrophys. J.} \bibinfo{volume}{725},
  \bibinfo{pages}{128--133}.
\newblock \DOIprefix\doi{10.1088/0004-637X/725/1/128}.
%Type = Article
\bibitem[{Guo et~al.(2018a)Guo, Dumbovi{\'c}, Wimmer-Schweingruber, Temmer,
  Lohf, Wang, Veronig, Hassler, Mays, Zeitlin et~al.}]{guo2018September}
\bibinfo{author}{Guo, J.}, \bibinfo{author}{Dumbovi{\'c}, M.},
  \bibinfo{author}{Wimmer-Schweingruber, R.F.}, \bibinfo{author}{Temmer, M.},
  \bibinfo{author}{Lohf, H.}, \bibinfo{author}{Wang, Y.},
  \bibinfo{author}{Veronig, A.}, \bibinfo{author}{Hassler, D.M.},
  \bibinfo{author}{Mays, L.M.}, \bibinfo{author}{Zeitlin, C.}, et~al.,
  \bibinfo{year}{2018}a.
\newblock \bibinfo{title}{Modeling the evolution and propagation of 10
  september 2017 cmes and seps arriving at {Mars} constrained by remote sensing
  and in situ measurement}.
\newblock \bibinfo{journal}{Space Weather} \bibinfo{volume}{16},
  \bibinfo{pages}{1156–1169}.
\newblock \DOIprefix\doi{10.1029/2018SW001973}.
%Type = Article
\bibitem[{Guo et~al.(2023)Guo, Li, Zhang, Dobynde, Wang, Xu, Berger, Semkova,
  Wimmer-Schweingruber, Hassler et~al.}]{guo2023first}
\bibinfo{author}{Guo, J.}, \bibinfo{author}{Li, X.}, \bibinfo{author}{Zhang,
  J.}, \bibinfo{author}{Dobynde, M.I.}, \bibinfo{author}{Wang, Y.},
  \bibinfo{author}{Xu, Z.}, \bibinfo{author}{Berger, T.},
  \bibinfo{author}{Semkova, J.}, \bibinfo{author}{Wimmer-Schweingruber, R.F.},
  \bibinfo{author}{Hassler, D.M.}, et~al., \bibinfo{year}{2023}.
\newblock \bibinfo{title}{The first ground level enhancement seen on three
  planetary surfaces: Earth, moon, and mars}.
\newblock \bibinfo{journal}{Geophysical Research Letters} \bibinfo{volume}{50},
  \bibinfo{pages}{e2023GL103069}.
\newblock \DOIprefix\doi{10.1029/2023GL103069}.
%Type = Article
\bibitem[{Guo et~al.(2018b)Guo, Lillis, Wimmer-Schweingruber, Zeitlin,
  Simonson, Rahmati, Posner, Papaioannou, Lundt, Lee
  et~al.}]{guo2018measurements}
\bibinfo{author}{Guo, J.}, \bibinfo{author}{Lillis, R.},
  \bibinfo{author}{Wimmer-Schweingruber, R.F.}, \bibinfo{author}{Zeitlin, C.},
  \bibinfo{author}{Simonson, P.}, \bibinfo{author}{Rahmati, A.},
  \bibinfo{author}{Posner, A.}, \bibinfo{author}{Papaioannou, A.},
  \bibinfo{author}{Lundt, N.}, \bibinfo{author}{Lee, C.O.}, et~al.,
  \bibinfo{year}{2018}b.
\newblock \bibinfo{title}{Measurements of forbush decreases at {Mars}: both by
  {MSL} on ground and by maven in orbit}.
\newblock \bibinfo{journal}{Astron. \& Astrophys.} \bibinfo{volume}{611},
  \bibinfo{pages}{A79}.
\newblock \DOIprefix\doi{10.1051/0004-6361/201732087}.
%Type = Article
\bibitem[{{Guo} et~al.(2021){Guo}, {Zeitlin}, {Wimmer-Schweingruber},
  {Hassler}, {Ehresmann}, {Rafkin}, {Freiherr von Forstner}, {Khaksarighiri},
  {Liu} and {Wang}}]{guo2021radiation}
\bibinfo{author}{{Guo}, J.}, \bibinfo{author}{{Zeitlin}, C.},
  \bibinfo{author}{{Wimmer-Schweingruber}, R.F.}, \bibinfo{author}{{Hassler},
  D.M.}, \bibinfo{author}{{Ehresmann}, B.}, \bibinfo{author}{{Rafkin}, S.},
  \bibinfo{author}{{Freiherr von Forstner}, J.L.},
  \bibinfo{author}{{Khaksarighiri}, S.}, \bibinfo{author}{{Liu}, W.},
  \bibinfo{author}{{Wang}, Y.}, \bibinfo{year}{2021}.
\newblock \bibinfo{title}{{Radiation environment for future human exploration
  on the surface of Mars: the current understanding based on MSL/RAD dose
  measurements}}.
\newblock \bibinfo{journal}{Astron. \& Astrophys. Rev.} \bibinfo{volume}{29},
  \bibinfo{pages}{8}.
\newblock \DOIprefix\doi{10.1007/s00159-021-00136-5}.
%Type = Article
\bibitem[{Hands et~al.(2022)Hands, Lei, Davis, Clewer, Dyer and
  Ryden}]{hands2022new}
\bibinfo{author}{Hands, A.}, \bibinfo{author}{Lei, F.}, \bibinfo{author}{Davis,
  C.}, \bibinfo{author}{Clewer, B.}, \bibinfo{author}{Dyer, C.},
  \bibinfo{author}{Ryden, K.}, \bibinfo{year}{2022}.
\newblock \bibinfo{title}{A new model for nowcasting the aviation radiation
  environment with comparisons to in situ measurements during gles}.
\newblock \bibinfo{journal}{Space Weather} \bibinfo{volume}{20},
  \bibinfo{pages}{e2022SW003155}.
\newblock \DOIprefix\doi{10.1029/2022SW003155}.
%Type = Article
\bibitem[{Hassler et~al.(2012)Hassler, Zeitlin, Wimmer-Schweingruber,
  B{\"o}ttcher, Martin, Andrews, B{\"o}hm, Brinza, Bullock, Burmeister
  et~al.}]{hassler2012}
\bibinfo{author}{Hassler, D.M.}, \bibinfo{author}{Zeitlin, C.},
  \bibinfo{author}{Wimmer-Schweingruber, R.F.}, \bibinfo{author}{B{\"o}ttcher,
  S.I.}, \bibinfo{author}{Martin, C.}, \bibinfo{author}{Andrews, J.},
  \bibinfo{author}{B{\"o}hm, E.}, \bibinfo{author}{Brinza, D.},
  \bibinfo{author}{Bullock, M.}, \bibinfo{author}{Burmeister, S.}, et~al.,
  \bibinfo{year}{2012}.
\newblock \bibinfo{title}{The {Radiation Assessment Detector} ({RAD})
  investigation}.
\newblock \bibinfo{journal}{Space Sci. Rev.} \bibinfo{volume}{170},
  \bibinfo{pages}{503--558}.
\newblock \DOIprefix\doi{10.1007/s11214-012-9913-1}.
%Type = Article
\bibitem[{{Heber} et~al.(2006){Heber}, {Fichtner} and {Scherer}}]{heber06}
\bibinfo{author}{{Heber}, B.}, \bibinfo{author}{{Fichtner}, H.},
  \bibinfo{author}{{Scherer}, K.}, \bibinfo{year}{2006}.
\newblock \bibinfo{title}{{Solar and Heliospheric Modulation of Galactic Cosmic
  Rays}}.
\newblock \bibinfo{journal}{Space Sci. Rev.} \bibinfo{volume}{125},
  \bibinfo{pages}{81--93}.
\newblock \DOIprefix\doi{10.1007/s11214-006-9048-3}.
%Type = Article
\bibitem[{{Herbst} et~al.(2017){Herbst}, {Muscheler} and {Heber}}]{Herbst2017}
\bibinfo{author}{{Herbst}, K.}, \bibinfo{author}{{Muscheler}, R.},
  \bibinfo{author}{{Heber}, B.}, \bibinfo{year}{2017}.
\newblock \bibinfo{title}{{The new local interstellar spectra and their
  influence on the production rates of the cosmogenic radionuclides $^{10}$Be
  and $^{14}$C}}.
\newblock \bibinfo{journal}{J. Geophys. Res.: Space Phys.}
  \bibinfo{volume}{122}, \bibinfo{pages}{23--34}.
\newblock \DOIprefix\doi{10.1002/2016JA023207}.
%Type = Article
\bibitem[{{Hess} and {Demmelmair}(1937)}]{Hess-1937}
\bibinfo{author}{{Hess}, V.F.}, \bibinfo{author}{{Demmelmair}, A.},
  \bibinfo{year}{1937}.
\newblock \bibinfo{title}{{World-wide Effect in Cosmic Ray Intensity, as
  Observed during a Recent Magnetic Storm}}.
\newblock \bibinfo{journal}{Nature} \bibinfo{volume}{140},
  \bibinfo{pages}{316--317}.
\newblock \DOIprefix\doi{10.1038/140316a0}.
%Type = Article
\bibitem[{{Honig} et~al.(2019){Honig}, {Witasse}, {Evans}, {Nieminen},
  {Kuulkers}, {Taylor}, {Heber}, {Guo} and {S{\'a}nchez-Cano}}]{Honig2019}
\bibinfo{author}{{Honig}, T.}, \bibinfo{author}{{Witasse}, O.G.},
  \bibinfo{author}{{Evans}, H.}, \bibinfo{author}{{Nieminen}, P.},
  \bibinfo{author}{{Kuulkers}, E.}, \bibinfo{author}{{Taylor}, M.G.G.T.},
  \bibinfo{author}{{Heber}, B.}, \bibinfo{author}{{Guo}, J.},
  \bibinfo{author}{{S{\'a}nchez-Cano}, B.}, \bibinfo{year}{2019}.
\newblock \bibinfo{title}{{Multi-point galactic cosmic ray measurements between
  1 and 4.5 AU over a full solar cycle}}.
\newblock \bibinfo{journal}{Ann. Geophys.} \bibinfo{volume}{37},
  \bibinfo{pages}{903--918}.
\newblock \DOIprefix\doi{10.5194/angeo-37-903-2019}.
%Type = Article
\bibitem[{{Huovelin} et~al.(2020){Huovelin}, {Vainio}, {Kilpua}, {Lehtolainen},
  {Korpela}, {Esko}, {Muinonen}, {Bunce}, {Martindale}, {Grande}, {Andersson},
  {Nenonen}, {Lehti}, {Schmidt}, {Genzer}, {Vihavainen}, {Saari}, {Peltonen},
  {Valtonen}, {Talvioja}, {Portin}, {Narendranath}, {Jarvinen}, {Okada},
  {Milillo}, {Laurenza}, {Heino} and {Oleynik}}]{huovelin2020sixs}
\bibinfo{author}{{Huovelin}, J.}, \bibinfo{author}{{Vainio}, R.},
  \bibinfo{author}{{Kilpua}, E.}, \bibinfo{author}{{Lehtolainen}, A.},
  \bibinfo{author}{{Korpela}, S.}, \bibinfo{author}{{Esko}, E.},
  \bibinfo{author}{{Muinonen}, K.}, \bibinfo{author}{{Bunce}, E.},
  \bibinfo{author}{{Martindale}, A.}, \bibinfo{author}{{Grande}, M.},
  \bibinfo{author}{{Andersson}, H.}, \bibinfo{author}{{Nenonen}, S.},
  \bibinfo{author}{{Lehti}, J.}, \bibinfo{author}{{Schmidt}, W.},
  \bibinfo{author}{{Genzer}, M.}, \bibinfo{author}{{Vihavainen}, T.},
  \bibinfo{author}{{Saari}, J.}, \bibinfo{author}{{Peltonen}, J.},
  \bibinfo{author}{{Valtonen}, E.}, \bibinfo{author}{{Talvioja}, M.},
  \bibinfo{author}{{Portin}, P.}, \bibinfo{author}{{Narendranath}, S.},
  \bibinfo{author}{{Jarvinen}, R.}, \bibinfo{author}{{Okada}, T.},
  \bibinfo{author}{{Milillo}, A.}, \bibinfo{author}{{Laurenza}, M.},
  \bibinfo{author}{{Heino}, E.}, \bibinfo{author}{{Oleynik}, P.},
  \bibinfo{year}{2020}.
\newblock \bibinfo{title}{{Solar Intensity X-Ray and Particle Spectrometer
  SIXS: Instrument Design and First Results}}.
\newblock \bibinfo{journal}{Space Sci. Rev.} \bibinfo{volume}{216},
  \bibinfo{pages}{94}.
\newblock \DOIprefix\doi{10.1007/s11214-020-00717-3}.
%Type = Article
\bibitem[{Iancu et~al.(2018)Iancu, Boutros, Olsen, Davis, Stewart, Eiwaz,
  Marzulla, Belknap, Fallgren, Edmondson et~al.}]{iancu2018space}
\bibinfo{author}{Iancu, O.D.}, \bibinfo{author}{Boutros, S.W.},
  \bibinfo{author}{Olsen, R.H.}, \bibinfo{author}{Davis, M.J.},
  \bibinfo{author}{Stewart, B.}, \bibinfo{author}{Eiwaz, M.},
  \bibinfo{author}{Marzulla, T.}, \bibinfo{author}{Belknap, J.},
  \bibinfo{author}{Fallgren, C.M.}, \bibinfo{author}{Edmondson, E.F.}, et~al.,
  \bibinfo{year}{2018}.
\newblock \bibinfo{title}{Space radiation alters genotype--phenotype
  correlations in fear learning and memory tests}.
\newblock \bibinfo{journal}{Front. in Genetics} \bibinfo{volume}{9},
  \bibinfo{pages}{404}.
\newblock \DOIprefix\doi{10.3389/fgene.2018.00404}.
%Type = Article
\bibitem[{{Janvier} et~al.(2021){Janvier}, {D{\'e}moulin}, {Guo}, {Dasso},
  {Regnault}, {Topsi-Moutesidou}, {Gutierrez} and {Perri}}]{janvier21}
\bibinfo{author}{{Janvier}, M.}, \bibinfo{author}{{D{\'e}moulin}, P.},
  \bibinfo{author}{{Guo}, J.}, \bibinfo{author}{{Dasso}, S.},
  \bibinfo{author}{{Regnault}, F.}, \bibinfo{author}{{Topsi-Moutesidou}, S.},
  \bibinfo{author}{{Gutierrez}, C.}, \bibinfo{author}{{Perri}, B.},
  \bibinfo{year}{2021}.
\newblock \bibinfo{title}{{The Two-step Forbush Decrease: A Tale of Two
  Substructures Modulating Galactic Cosmic Rays within Coronal Mass
  Ejections}}.
\newblock \bibinfo{journal}{Astrophys. J.} \bibinfo{volume}{922},
  \bibinfo{pages}{216}.
\newblock \DOIprefix\doi{10.3847/1538-4357/ac2b9b}.
%Type = Article
\bibitem[{Jokipii and Kota(1989)}]{jokipii1989}
\bibinfo{author}{Jokipii, J.}, \bibinfo{author}{Kota, J.},
  \bibinfo{year}{1989}.
\newblock \bibinfo{title}{The polar heliospheric magnetic field}.
\newblock \bibinfo{journal}{Geophys. Res. Lett.} \bibinfo{volume}{16},
  \bibinfo{pages}{1--4}.
\newblock \DOIprefix\doi{10.1029/GL016i001p00001}.
%Type = Article
\bibitem[{{Jokipii} et~al.(2004){Jokipii}, {Giacalone} and
  {K{\'o}ta}}]{Jokipii2004}
\bibinfo{author}{{Jokipii}, J.R.}, \bibinfo{author}{{Giacalone}, J.},
  \bibinfo{author}{{K{\'o}ta}, J.}, \bibinfo{year}{2004}.
\newblock \bibinfo{title}{{Transverse Streaming Anisotropies of Charged
  Particles Accelerated at the Solar Wind Termination Shock}}.
\newblock \bibinfo{journal}{Astrophys. J. Lett.} \bibinfo{volume}{611},
  \bibinfo{pages}{L141--L144}.
\newblock \DOIprefix\doi{10.1086/423993}.
%Type = Article
\bibitem[{{Jordan} et~al.(2011){Jordan}, {Spence}, {Blake} and
  {Shaul}}]{jordan11}
\bibinfo{author}{{Jordan}, A.P.}, \bibinfo{author}{{Spence}, H.E.},
  \bibinfo{author}{{Blake}, J.B.}, \bibinfo{author}{{Shaul}, D.N.A.},
  \bibinfo{year}{2011}.
\newblock \bibinfo{title}{{Revisiting two-step Forbush decreases}}.
\newblock \bibinfo{journal}{J. Geophys. Res.: Space Phys.}
  \bibinfo{volume}{116}, \bibinfo{pages}{A11103}.
\newblock \DOIprefix\doi{10.1029/2011JA016791}.
%Type = Article
\bibitem[{Kahler(1994)}]{kahler1994injection}
\bibinfo{author}{Kahler, S.}, \bibinfo{year}{1994}.
\newblock \bibinfo{title}{Injection profiles of solar energetic particles as
  functions of coronal mass ejection heights}.
\newblock \bibinfo{journal}{Astrophys. J.} \bibinfo{volume}{428},
  \bibinfo{pages}{837--842}.
\newblock \DOIprefix\doi{10.1086/174292}.
%Type = Article
\bibitem[{{Kahler} et~al.(2007){Kahler}, {Cliver} and {Ling}}]{kahler07}
\bibinfo{author}{{Kahler}, S.W.}, \bibinfo{author}{{Cliver}, E.W.},
  \bibinfo{author}{{Ling}, A.G.}, \bibinfo{year}{2007}.
\newblock \bibinfo{title}{{Validating the proton prediction system (PPS)}}.
\newblock \bibinfo{journal}{J. Atmos. Solar-Terres. Phys.}
  \bibinfo{volume}{69}, \bibinfo{pages}{43--49}.
\newblock \DOIprefix\doi{10.1016/j.jastp.2006.06.009}.
%Type = Article
\bibitem[{{Kahler} et~al.(2017){Kahler}, {White} and {Ling}}]{kahler17}
\bibinfo{author}{{Kahler}, S.W.}, \bibinfo{author}{{White}, S.M.},
  \bibinfo{author}{{Ling}, A.G.}, \bibinfo{year}{2017}.
\newblock \bibinfo{title}{{Forecasting E $>$ 50-MeV proton events with the
  proton prediction system (PPS)}}.
\newblock \bibinfo{journal}{J. Space Weather \& Space Clim.}
  \bibinfo{volume}{7}, \bibinfo{pages}{A27}.
\newblock \DOIprefix\doi{10.1051/swsc/2017025}.
%Type = Article
\bibitem[{{Kappl}(2016)}]{Kappl2016}
\bibinfo{author}{{Kappl}, R.}, \bibinfo{year}{2016}.
\newblock \bibinfo{title}{{SOLARPROP: Charge-sign dependent solar modulation
  for everyone}}.
\newblock \bibinfo{journal}{Computer Physics Communications}
  \bibinfo{volume}{207}, \bibinfo{pages}{386--399}.
\newblock \DOIprefix\doi{10.1016/j.cpc.2016.05.025}.
%Type = Article
\bibitem[{Kennedy(2014)}]{Kennedy2014}
\bibinfo{author}{Kennedy, A.R.}, \bibinfo{year}{2014}.
\newblock \bibinfo{title}{Biological effects of space radiation and development
  of effective countermeasures}.
\newblock \bibinfo{journal}{Life Sci. \& Space Res.} \bibinfo{volume}{1},
  \bibinfo{pages}{10--43}.
\newblock \DOIprefix\doi{10.1016/j.lssr.2014.02.004}.
%Type = Article
\bibitem[{{Kilpua} et~al.(2017){Kilpua}, {Koskinen} and {Pulkkinen}}]{kilpua17}
\bibinfo{author}{{Kilpua}, E.}, \bibinfo{author}{{Koskinen}, H.E.J.},
  \bibinfo{author}{{Pulkkinen}, T.I.}, \bibinfo{year}{2017}.
\newblock \bibinfo{title}{{Coronal mass ejections and their sheath regions in
  interplanetary space}}.
\newblock \bibinfo{journal}{Living Rev. Sol. Phys.} \bibinfo{volume}{14},
  \bibinfo{pages}{5}.
\newblock \DOIprefix\doi{10.1007/s41116-017-0009-6}.
%Type = Article
\bibitem[{{Kilpua} et~al.(2021){Kilpua}, {Good}, {Ala-Lahti}, {Osmane},
  {Fontaine}, {Hadid}, {Janvier} and {Yordanova}}]{Kilpua2021}
\bibinfo{author}{{Kilpua}, E.K.J.}, \bibinfo{author}{{Good}, S.W.},
  \bibinfo{author}{{Ala-Lahti}, M.}, \bibinfo{author}{{Osmane}, A.},
  \bibinfo{author}{{Fontaine}, D.}, \bibinfo{author}{{Hadid}, L.},
  \bibinfo{author}{{Janvier}, M.}, \bibinfo{author}{{Yordanova}, E.},
  \bibinfo{year}{2021}.
\newblock \bibinfo{title}{{Statistical analysis of magnetic field fluctuations
  in CME-driven sheath regions}}.
\newblock \bibinfo{journal}{Front. in Astron. \& Space Sci.}
  \bibinfo{volume}{7}, \bibinfo{pages}{109}.
\newblock \DOIprefix\doi{10.3389/fspas.2020.610278}.
%Type = Article
\bibitem[{{Klecker} et~al.(1995){Klecker}, {McNab}, {Blake}, {Hamilton},
  {Hovestadt}, {Kaestle}, {Looper}, {Mason}, {Mazur} and
  {Scholer}}]{Klecker1995}
\bibinfo{author}{{Klecker}, B.}, \bibinfo{author}{{McNab}, M.C.},
  \bibinfo{author}{{Blake}, J.B.}, \bibinfo{author}{{Hamilton}, D.C.},
  \bibinfo{author}{{Hovestadt}, D.}, \bibinfo{author}{{Kaestle}, H.},
  \bibinfo{author}{{Looper}, M.D.}, \bibinfo{author}{{Mason}, G.M.},
  \bibinfo{author}{{Mazur}, J.E.}, \bibinfo{author}{{Scholer}, M.},
  \bibinfo{year}{1995}.
\newblock \bibinfo{title}{{Charge State of Anomalous Cosmic-Ray Nitrogen,
  Oxygen, and Neon: SAMPEX Observations}}.
\newblock \bibinfo{journal}{Astrophys. J. Lett.} \bibinfo{volume}{442},
  \bibinfo{pages}{L69}.
\newblock \DOIprefix\doi{10.1086/187818}.
%Type = Article
\bibitem[{Kleimann et~al.(2023)Kleimann, Oughton, Fichtner and
  Scherer}]{kleimann2023}
\bibinfo{author}{Kleimann, J.}, \bibinfo{author}{Oughton, S.},
  \bibinfo{author}{Fichtner, H.}, \bibinfo{author}{Scherer, K.},
  \bibinfo{year}{2023}.
\newblock \bibinfo{title}{A three-dimensional model for the evolution of
  magnetohydrodynamic turbulence in the outer heliosphere}.
\newblock \bibinfo{journal}{Astrophys. J.} \bibinfo{volume}{953},
  \bibinfo{pages}{133}.
\newblock \DOIprefix\doi{10.3847/1538-4357/acd84e}.
%Type = Article
\bibitem[{{Klein} and {Dalla}(2017)}]{klein2017acceleration}
\bibinfo{author}{{Klein}, K.L.}, \bibinfo{author}{{Dalla}, S.},
  \bibinfo{year}{2017}.
\newblock \bibinfo{title}{{Acceleration and Propagation of Solar Energetic
  Particles}}.
\newblock \bibinfo{journal}{Space Sci. Rev.} \bibinfo{volume}{212},
  \bibinfo{pages}{1107--1136}.
\newblock \DOIprefix\doi{10.1007/s11214-017-0382-4}.
%Type = Article
\bibitem[{Klein et~al.(2022)Klein, Musset, Vilmer, Briand, Krucker, Battaglia,
  Dresing, Palmroos and Gary}]{klein2022relativistic}
\bibinfo{author}{Klein, K.L.}, \bibinfo{author}{Musset, S.},
  \bibinfo{author}{Vilmer, N.}, \bibinfo{author}{Briand, C.},
  \bibinfo{author}{Krucker, S.}, \bibinfo{author}{Battaglia, A.F.},
  \bibinfo{author}{Dresing, N.}, \bibinfo{author}{Palmroos, C.},
  \bibinfo{author}{Gary, D.E.}, \bibinfo{year}{2022}.
\newblock \bibinfo{title}{The relativistic solar particle event on 28 october
  2021: Evidence of particle acceleration within and escape from the solar
  corona}.
\newblock \bibinfo{journal}{Astron. \& Astrophys.} \bibinfo{volume}{663},
  \bibinfo{pages}{A173}.
\newblock \DOIprefix\doi{10.1051/0004-6361/202243903}.
%Type = Article
\bibitem[{{Knutsen} et~al.(2021){Knutsen}, {Witasse}, {Sanchez-Cano}, {Lester},
  {Wimmer-Schweingruber}, {Denis}, {Godfrey} and {Johnstone}}]{Knutsen2021}
\bibinfo{author}{{Knutsen}, E.W.}, \bibinfo{author}{{Witasse}, O.},
  \bibinfo{author}{{Sanchez-Cano}, B.}, \bibinfo{author}{{Lester}, M.},
  \bibinfo{author}{{Wimmer-Schweingruber}, R.F.}, \bibinfo{author}{{Denis},
  M.}, \bibinfo{author}{{Godfrey}, J.}, \bibinfo{author}{{Johnstone}, A.},
  \bibinfo{year}{2021}.
\newblock \bibinfo{title}{{Galactic cosmic ray modulation at Mars and beyond
  measured with EDACs on Mars Express and Rosetta}}.
\newblock \bibinfo{journal}{Astron. \& Astrophys.} \bibinfo{volume}{650},
  \bibinfo{pages}{A165}.
\newblock \DOIprefix\doi{10.1051/0004-6361/202140767}.
%Type = Article
\bibitem[{Kohl et~al.(2006)Kohl, Noci, Cranmer and
  Raymond}]{kohl2006ultraviolet}
\bibinfo{author}{Kohl, J.L.}, \bibinfo{author}{Noci, G.},
  \bibinfo{author}{Cranmer, S.R.}, \bibinfo{author}{Raymond, J.C.},
  \bibinfo{year}{2006}.
\newblock \bibinfo{title}{Ultraviolet spectroscopy of the extended solar
  corona}.
\newblock \bibinfo{journal}{Astron. \& Astrophys. Rev.} \bibinfo{volume}{13},
  \bibinfo{pages}{31--157}.
\newblock \DOIprefix\doi{10.1007/s00159-005-0026-7}.
%Type = Article
\bibitem[{Koldobskiy et~al.(2021)Koldobskiy, Raukunen, Vainio, Kovaltsov and
  Usoskin}]{koldobskiy2021new}
\bibinfo{author}{Koldobskiy, S.}, \bibinfo{author}{Raukunen, O.},
  \bibinfo{author}{Vainio, R.}, \bibinfo{author}{Kovaltsov, G.A.},
  \bibinfo{author}{Usoskin, I.}, \bibinfo{year}{2021}.
\newblock \bibinfo{title}{New reconstruction of event-integrated spectra
  (spectral fluences) for major solar energetic particle events}.
\newblock \bibinfo{journal}{Astron. \& Astrophys.} \bibinfo{volume}{647},
  \bibinfo{pages}{A132}.
\newblock \DOIprefix\doi{10.1051/0004-6361/202040058}.
%Type = Article
\bibitem[{Koldobskiy et~al.(2018)Koldobskiy, Kovaltsov and
  Usoskin}]{koldobskiy2018effective}
\bibinfo{author}{Koldobskiy, S.A.}, \bibinfo{author}{Kovaltsov, G.A.},
  \bibinfo{author}{Usoskin, I.G.}, \bibinfo{year}{2018}.
\newblock \bibinfo{title}{Effective rigidity of a polar neutron monitor for
  recording ground-level enhancements}.
\newblock \bibinfo{journal}{Sol. Phys.} \bibinfo{volume}{293},
  \bibinfo{pages}{1--9}.
\newblock \DOIprefix\doi{10.1007/s11207-018-1326-1}.
%Type = Article
\bibitem[{{Kollhoff} et~al.(2021){Kollhoff}, {Kouloumvakos}, {Lario},
  {Dresing}, {G{\'o}mez-Herrero}, {Rodr{\'\i}guez-Garc{\'\i}a}, {Malandraki},
  {Richardson}, {Posner} and {Klein}}]{Kollhoff2021}
\bibinfo{author}{{Kollhoff}, A.}, \bibinfo{author}{{Kouloumvakos}, A.},
  \bibinfo{author}{{Lario}, D.}, \bibinfo{author}{{Dresing}, N.},
  \bibinfo{author}{{G{\'o}mez-Herrero}, R.},
  \bibinfo{author}{{Rodr{\'\i}guez-Garc{\'\i}a}, L.},
  \bibinfo{author}{{Malandraki}, O.E.}, \bibinfo{author}{{Richardson}, I.G.},
  \bibinfo{author}{{Posner}, A.}, \bibinfo{author}{{Klein}, K.L.e.a.},
  \bibinfo{year}{2021}.
\newblock \bibinfo{title}{{The first widespread solar energetic particle event
  observed by Solar Orbiter on 2020 November 29}}.
\newblock \bibinfo{journal}{Astron. \& Astrophys.} \bibinfo{volume}{656},
  \bibinfo{pages}{A20}.
\newblock \DOIprefix\doi{10.1051/0004-6361/202140937}.
%Type = Article
\bibitem[{{K{\'o}ta}(2010)}]{kota2010}
\bibinfo{author}{{K{\'o}ta}, J.}, \bibinfo{year}{2010}.
\newblock \bibinfo{title}{{Particle Acceleration at Near-perpendicular Shocks:
  The Role of Field-line Topology}}.
\newblock \bibinfo{journal}{Astrophys. J.} \bibinfo{volume}{723},
  \bibinfo{pages}{393--397}.
\newblock \DOIprefix\doi{10.1088/0004-637X/723/1/393}.
%Type = Inproceedings
\bibitem[{{K{\'o}ta} and {Jokipii}(2008)}]{kota2008}
\bibinfo{author}{{K{\'o}ta}, J.}, \bibinfo{author}{{Jokipii}, J.R.},
  \bibinfo{year}{2008}.
\newblock \bibinfo{title}{{Anomalous Cosmic Rays in the Heliosheath: Simulation
  with a Blunt Termination Shock}}, in: \bibinfo{editor}{{Li}, G.},
  \bibinfo{editor}{{Hu}, Q.}, \bibinfo{editor}{{Verkhoglyadova}, O.},
  \bibinfo{editor}{{Zank}, G.P.}, \bibinfo{editor}{{Lin}, R.P.},
  \bibinfo{editor}{{Luhmann}, J.} (Eds.), \bibinfo{booktitle}{Particle
  Acceleration and Transport in the Heliosphere and Beyond: 7th Annual
  International AstroPhysics Conference}, pp. \bibinfo{pages}{397--403}.
\newblock \DOIprefix\doi{10.1063/1.2982477}.
%Type = Article
\bibitem[{Kouloumvakos et~al.(2015)Kouloumvakos, Nindos, Valtonen,
  Alissandrakis, Malandraki, Tsitsipis, Kontogeorgos, Moussas and
  Hillaris}]{kouloumvakos2015properties}
\bibinfo{author}{Kouloumvakos, A.}, \bibinfo{author}{Nindos, A.},
  \bibinfo{author}{Valtonen, E.}, \bibinfo{author}{Alissandrakis, C.},
  \bibinfo{author}{Malandraki, O.}, \bibinfo{author}{Tsitsipis, P.},
  \bibinfo{author}{Kontogeorgos, A.}, \bibinfo{author}{Moussas, X.},
  \bibinfo{author}{Hillaris, A.}, \bibinfo{year}{2015}.
\newblock \bibinfo{title}{Properties of solar energetic particle events
  inferred from their associated radio emission}.
\newblock \bibinfo{journal}{Astron. \& Astrophys.} \bibinfo{volume}{580}.
\newblock \DOIprefix\doi{10.1051/0004-6361/201424397}.
%Type = Article
\bibitem[{{Kounine}(2012)}]{kounine2012ams}
\bibinfo{author}{{Kounine}, A.}, \bibinfo{year}{2012}.
\newblock \bibinfo{title}{{The Alpha Magnetic Spectrometer on the International
  Space Station}}.
\newblock \bibinfo{journal}{International J. of Modern Phys. E}
  \bibinfo{volume}{21}, \bibinfo{pages}{1230005}.
\newblock \DOIprefix\doi{10.1142/S0218301312300056}.
%Type = Incollection
\bibitem[{Kress et~al.(2020)Kress, Rodriguez and Onsager}]{kress20}
\bibinfo{author}{Kress, B.T.}, \bibinfo{author}{Rodriguez, J.V.},
  \bibinfo{author}{Onsager, T.G.}, \bibinfo{year}{2020}.
\newblock \bibinfo{title}{{Chapter 20 - The GOES-R Space Environment In Situ
  Suite (SEISS): Measurement of Energetic Particles in Geospace}}, in:
  \bibinfo{editor}{Goodman, S.J.}, \bibinfo{editor}{Schmit, T.J.},
  \bibinfo{editor}{Daniels, J.}, \bibinfo{editor}{Redmon, R.J.} (Eds.),
  \bibinfo{booktitle}{{The GOES-R Series}}. \bibinfo{publisher}{Elsevier}, pp.
  \bibinfo{pages}{243 -- 250}.
\newblock \DOIprefix\doi{10.1016/B978-0-12-814327-8.00020-2}.
%Type = Article
\bibitem[{{Krimigis} et~al.(2013){Krimigis}, {Decker}, {Roelof}, {Hill},
  {Armstrong}, {Gloeckler}, {Hamilton} and {Lanzerotti}}]{Krimigis2013}
\bibinfo{author}{{Krimigis}, S.M.}, \bibinfo{author}{{Decker}, R.B.},
  \bibinfo{author}{{Roelof}, E.C.}, \bibinfo{author}{{Hill}, M.E.},
  \bibinfo{author}{{Armstrong}, T.P.}, \bibinfo{author}{{Gloeckler}, G.},
  \bibinfo{author}{{Hamilton}, D.C.}, \bibinfo{author}{{Lanzerotti}, L.J.},
  \bibinfo{year}{2013}.
\newblock \bibinfo{title}{{Search for the Exit: Voyager 1 at
  Heliosphere{\textquoteright}s Border with the Galaxy}}.
\newblock \bibinfo{journal}{Science} \bibinfo{volume}{341},
  \bibinfo{pages}{144--147}.
\newblock \DOIprefix\doi{10.1126/science.1235721}.
%Type = Article
\bibitem[{Krucker et~al.(1999)Krucker, Larson, Lin and
  Thompson}]{krucker1999origin}
\bibinfo{author}{Krucker, S.}, \bibinfo{author}{Larson, D.E.},
  \bibinfo{author}{Lin, R.P.}, \bibinfo{author}{Thompson, B.J.},
  \bibinfo{year}{1999}.
\newblock \bibinfo{title}{On the origin of impulsive electron events observed
  at 1 au}.
\newblock \bibinfo{journal}{Astrophys. J.} \bibinfo{volume}{519},
  \bibinfo{pages}{864}.
\newblock \DOIprefix\doi{10.1086/307415}.
%Type = Article
\bibitem[{K{\"u}hl et~al.(2017)K{\"u}hl, Dresing, Heber and
  Klassen}]{kuhl2017solar}
\bibinfo{author}{K{\"u}hl, P.}, \bibinfo{author}{Dresing, N.},
  \bibinfo{author}{Heber, B.}, \bibinfo{author}{Klassen, A.},
  \bibinfo{year}{2017}.
\newblock \bibinfo{title}{Solar energetic particle events with protons above
  500 mev between 1995 and 2015 measured with {SOHO/EPHIN}}.
\newblock \bibinfo{journal}{Sol. Phys.} \bibinfo{volume}{292},
  \bibinfo{pages}{1--13}.
\newblock \DOIprefix\doi{10.1007/s11207-016-1033-8}.
%Type = Article
\bibitem[{Laitinen et~al.(2015)Laitinen, Huttunen-Heikinmaa, Valtonen and
  Dalla}]{laitinen2015correcting}
\bibinfo{author}{Laitinen, T.}, \bibinfo{author}{Huttunen-Heikinmaa, K.},
  \bibinfo{author}{Valtonen, E.}, \bibinfo{author}{Dalla, S.},
  \bibinfo{year}{2015}.
\newblock \bibinfo{title}{Correcting for interplanetary scattering in velocity
  dispersion analysis of solar energetic particles}.
\newblock \bibinfo{journal}{Astrophys. J.} \bibinfo{volume}{806},
  \bibinfo{pages}{114}.
\newblock \DOIprefix\doi{10.1088/0004-637X/806/1/114}.
%Type = Article
\bibitem[{{Laitinen} et~al.(2016){Laitinen}, {Kopp}, {Effenberger}, {Dalla} and
  {Marsh}}]{Lai2016}
\bibinfo{author}{{Laitinen}, T.}, \bibinfo{author}{{Kopp}, A.},
  \bibinfo{author}{{Effenberger}, F.}, \bibinfo{author}{{Dalla}, S.},
  \bibinfo{author}{{Marsh}, M.S.}, \bibinfo{year}{2016}.
\newblock \bibinfo{title}{{Solar energetic particle access to distant
  longitudes through turbulent field-line meandering}}.
\newblock \bibinfo{journal}{Astron. \& Astrophys.} \bibinfo{volume}{591},
  \bibinfo{pages}{A18}.
\newblock \DOIprefix\doi{10.1051/0004-6361/201527801}.
%Type = Article
\bibitem[{Lario et~al.(2013)Lario, Aran, G{\'o}mez-Herrero, Dresing, Heber, Ho,
  Decker and Roelof}]{lario2013longitudinal}
\bibinfo{author}{Lario, D.}, \bibinfo{author}{Aran, A.},
  \bibinfo{author}{G{\'o}mez-Herrero, R.}, \bibinfo{author}{Dresing, N.},
  \bibinfo{author}{Heber, B.}, \bibinfo{author}{Ho, G.},
  \bibinfo{author}{Decker, R.}, \bibinfo{author}{Roelof, E.},
  \bibinfo{year}{2013}.
\newblock \bibinfo{title}{Longitudinal and radial dependence of solar energetic
  particle peak intensities: Stereo, ace, soho, goes, and messenger
  observations}.
\newblock \bibinfo{journal}{Astrophys. J.} \bibinfo{volume}{767},
  \bibinfo{pages}{41}.
\newblock \DOIprefix\doi{10.1088/0004-637X/767/1/41}.
%Type = Article
\bibitem[{Lario et~al.(2006)Lario, Kallenrode, Decker, Roelof, Krimigis, Aran
  and Sanahuja}]{lario2006radial}
\bibinfo{author}{Lario, D.}, \bibinfo{author}{Kallenrode, M.B.},
  \bibinfo{author}{Decker, R.}, \bibinfo{author}{Roelof, E.},
  \bibinfo{author}{Krimigis, S.}, \bibinfo{author}{Aran, A.},
  \bibinfo{author}{Sanahuja, B.}, \bibinfo{year}{2006}.
\newblock \bibinfo{title}{Radial and longitudinal dependence of solar 4-13 mev
  and 27-37 mev proton peak intensities and fluences: Helios and imp 8
  observations}.
\newblock \bibinfo{journal}{Astrophys. J.} \bibinfo{volume}{653},
  \bibinfo{pages}{1531}.
\newblock \DOIprefix\doi{10.1086/508982}.
%Type = Article
\bibitem[{Lario et~al.(2016)Lario, Kwon, Vourlidas, Raouafi, Haggerty, Ho,
  Anderson, Papaioannou, G{\'o}mez-Herrero, Dresing
  et~al.}]{lario2016longitudinal}
\bibinfo{author}{Lario, D.}, \bibinfo{author}{Kwon, R.Y.},
  \bibinfo{author}{Vourlidas, A.}, \bibinfo{author}{Raouafi, N.},
  \bibinfo{author}{Haggerty, D.}, \bibinfo{author}{Ho, G.},
  \bibinfo{author}{Anderson, B.}, \bibinfo{author}{Papaioannou, A.},
  \bibinfo{author}{G{\'o}mez-Herrero, R.}, \bibinfo{author}{Dresing, N.},
  et~al., \bibinfo{year}{2016}.
\newblock \bibinfo{title}{Longitudinal properties of a widespread solar
  energetic particle event on 2014 february 25: evolution of the associated cme
  shock}.
\newblock \bibinfo{journal}{Astrophys. J.} \bibinfo{volume}{819},
  \bibinfo{pages}{72}.
\newblock \DOIprefix\doi{10.3847/0004-637X/819/1/72}.
%Type = Article
\bibitem[{{Lazarian} and {Opher}(2009)}]{Lazarian2009}
\bibinfo{author}{{Lazarian}, A.}, \bibinfo{author}{{Opher}, M.},
  \bibinfo{year}{2009}.
\newblock \bibinfo{title}{{A Model of Acceleration of Anomalous Cosmic Rays by
  Reconnection in the Heliosheath}}.
\newblock \bibinfo{journal}{Astrophys. J.} \bibinfo{volume}{703},
  \bibinfo{pages}{8--21}.
\newblock \DOIprefix\doi{10.1088/0004-637X/703/1/8}.
%Type = Article
\bibitem[{{Lef{\`e}vre} et~al.(2016){Lef{\`e}vre}, {Vennerstr{\o}m},
  {Dumbovi{\'c}}, {Vr{\v{s}}nak}, {Sudar}, {Arlt}, {Clette} and
  {Crosby}}]{lefevre16}
\bibinfo{author}{{Lef{\`e}vre}, L.}, \bibinfo{author}{{Vennerstr{\o}m}, S.},
  \bibinfo{author}{{Dumbovi{\'c}}, M.}, \bibinfo{author}{{Vr{\v{s}}nak}, B.},
  \bibinfo{author}{{Sudar}, D.}, \bibinfo{author}{{Arlt}, R.},
  \bibinfo{author}{{Clette}, F.}, \bibinfo{author}{{Crosby}, N.},
  \bibinfo{year}{2016}.
\newblock \bibinfo{title}{{Detailed Analysis of Solar Data Related to
  Historical Extreme Geomagnetic Storms: 1868 - 2010}}.
\newblock \bibinfo{journal}{Sol. Phys.} \bibinfo{volume}{291},
  \bibinfo{pages}{1483--1531}.
\newblock \DOIprefix\doi{10.1007/s11207-016-0892-3}.
%Type = Article
\bibitem[{Li and Lee(2015)}]{li2015scatter}
\bibinfo{author}{Li, G.}, \bibinfo{author}{Lee, M.A.}, \bibinfo{year}{2015}.
\newblock \bibinfo{title}{Scatter-dominated interplanetary transport of solar
  energetic particles in large gradual events and the formation of double
  power-law differential fluence spectra of ground-level events during solar
  cycle 23}.
\newblock \bibinfo{journal}{Astrophys. J.} \bibinfo{volume}{810},
  \bibinfo{pages}{82}.
\newblock \DOIprefix\doi{10.1088/0004-637X/810/1/82}.
%Type = Inproceedings
\bibitem[{{Li} and {Zank}(2006)}]{Li2006}
\bibinfo{author}{{Li}, G.}, \bibinfo{author}{{Zank}, G.P.},
  \bibinfo{year}{2006}.
\newblock \bibinfo{title}{{Particle acceleration at a rippling termination
  shock}}, in: \bibinfo{editor}{{Heerikhuisen}, J.},
  \bibinfo{editor}{{Florinski}, V.}, \bibinfo{editor}{{Zank}, G.P.},
  \bibinfo{editor}{{Pogorelov}, N.V.} (Eds.), \bibinfo{booktitle}{Physics of
  the Inner Heliosheath}, pp. \bibinfo{pages}{183--189}.
\newblock \DOIprefix\doi{10.1063/1.2359325}.
%Type = Article
\bibitem[{Lin et~al.(1995)Lin, Anderson, Ashford, Carlson, Curtis, Ergun,
  Larson, McFadden, McCarthy, Parks et~al.}]{lin1995wind}
\bibinfo{author}{Lin, R.}, \bibinfo{author}{Anderson, K.},
  \bibinfo{author}{Ashford, S.}, \bibinfo{author}{Carlson, C.},
  \bibinfo{author}{Curtis, D.}, \bibinfo{author}{Ergun, R.},
  \bibinfo{author}{Larson, D.}, \bibinfo{author}{McFadden, J.},
  \bibinfo{author}{McCarthy, M.}, \bibinfo{author}{Parks, G.}, et~al.,
  \bibinfo{year}{1995}.
\newblock \bibinfo{title}{A three-dimensional plasma and energetic particle
  investigation for the wind spacecraft}.
\newblock \bibinfo{journal}{Space Sci. Rev.} \bibinfo{volume}{71},
  \bibinfo{pages}{125--153}.
\newblock \DOIprefix\doi{10.1007/BF00751328}.
%Type = Article
\bibitem[{Lin et~al.(2006)Lin, Szabo, Antiochos, Bale, Davila
  et~al.}]{lin2006solar}
\bibinfo{author}{Lin, R.}, \bibinfo{author}{Szabo, A.},
  \bibinfo{author}{Antiochos, S.}, \bibinfo{author}{Bale, S.},
  \bibinfo{author}{Davila, J.}, et~al., \bibinfo{year}{2006}.
\newblock \bibinfo{title}{{Solar Sentinels}: Report of the science and
  technology definition team}.
\newblock \bibinfo{journal}{NASA/TM} \bibinfo{volume}{2006214137}.
\newblock \URLprefix
  \url{https://ntrs.nasa.gov/api/citations/20090024212/downloads/20090024212.pdf}.
%Type = Article
\bibitem[{{Liu} et~al.(2006){Liu}, {Petrosian} and {Mason}}]{liu2006}
\bibinfo{author}{{Liu}, S.}, \bibinfo{author}{{Petrosian}, V.},
  \bibinfo{author}{{Mason}, G.M.}, \bibinfo{year}{2006}.
\newblock \bibinfo{title}{{Stochastic Acceleration of $^{3}$He and $^{4}$He in
  Solar Flares by Parallel-propagating Plasma Waves: General Results}}.
\newblock \bibinfo{journal}{Astrophys. J.} \bibinfo{volume}{636},
  \bibinfo{pages}{462--474}.
\newblock \DOIprefix\doi{10.1086/497883}.
%Type = Article
\bibitem[{{Luo} et~al.(2013){Luo}, {Zhang}, {Rassoul}, {Pogorelov} and
  {Heerikhuisen}}]{Luo2013}
\bibinfo{author}{{Luo}, X.}, \bibinfo{author}{{Zhang}, M.},
  \bibinfo{author}{{Rassoul}, H.K.}, \bibinfo{author}{{Pogorelov}, N.V.},
  \bibinfo{author}{{Heerikhuisen}, J.}, \bibinfo{year}{2013}.
\newblock \bibinfo{title}{{Galactic Cosmic-Ray Modulation in a Realistic Global
  Magnetohydrodynamic Heliosphere}}.
\newblock \bibinfo{journal}{Astrophys. J.} \bibinfo{volume}{764},
  \bibinfo{pages}{85}.
\newblock \DOIprefix\doi{10.1088/0004-637X/764/1/85}.
%Type = Article
\bibitem[{Maliniemi et~al.(2022)Maliniemi, Arsenovic, Sepp\"al\"a and
  Nesse~Tyss{\o}y}]{Maliniemi2022}
\bibinfo{author}{Maliniemi, V.}, \bibinfo{author}{Arsenovic, P.},
  \bibinfo{author}{Sepp\"al\"a, A.}, \bibinfo{author}{Nesse~Tyss{\o}y, H.},
  \bibinfo{year}{2022}.
\newblock \bibinfo{title}{The influence of energetic particle precipitation on
  antarctic stratospheric chlorine and ozone over the 20th century}.
\newblock \bibinfo{journal}{Atmospheric Chemistry and Physics}
  \bibinfo{volume}{22}, \bibinfo{pages}{8137--8149}.
\newblock \DOIprefix\doi{10.5194/acp-22-8137-2022}.
%Type = Article
\bibitem[{{Marsh} et~al.(2013){Marsh}, {Dalla}, {Kelly} and
  {Laitinen}}]{Mar2013}
\bibinfo{author}{{Marsh}, M.S.}, \bibinfo{author}{{Dalla}, S.},
  \bibinfo{author}{{Kelly}, J.}, \bibinfo{author}{{Laitinen}, T.},
  \bibinfo{year}{2013}.
\newblock \bibinfo{title}{{Drift-induced Perpendicular Transport of Solar
  Energetic Particles}}.
\newblock \bibinfo{journal}{Astrophys. J.} \bibinfo{volume}{774},
  \bibinfo{pages}{4}.
\newblock \DOIprefix\doi{10.1088/0004-637X/774/1/4}.
%Type = Article
\bibitem[{Mas{\'\i}as-Meza et~al.(2016)Mas{\'\i}as-Meza, Dasso, D{\'e}moulin,
  Rodriguez and Janvier}]{masias2016superposed}
\bibinfo{author}{Mas{\'\i}as-Meza, J.}, \bibinfo{author}{Dasso, S.},
  \bibinfo{author}{D{\'e}moulin, P.}, \bibinfo{author}{Rodriguez, L.},
  \bibinfo{author}{Janvier, M.}, \bibinfo{year}{2016}.
\newblock \bibinfo{title}{Superposed epoch study of icme sub-structures near
  {Earth} and their effects on galactic cosmic rays}.
\newblock \bibinfo{journal}{Astron. \& Astrophys.} \bibinfo{volume}{592},
  \bibinfo{pages}{A118}.
\newblock \DOIprefix\doi{10.1051/0004-6361/201628571}.
%Type = Article
\bibitem[{Mason et~al.(1999)Mason, Cohen, Cummings, Dwyer, Gold, Krimigis,
  Leske, Mazur, Mewaldt, M{\"o}bius et~al.}]{mason1999particle}
\bibinfo{author}{Mason, G.}, \bibinfo{author}{Cohen, C.},
  \bibinfo{author}{Cummings, A.}, \bibinfo{author}{Dwyer, J.},
  \bibinfo{author}{Gold, R.}, \bibinfo{author}{Krimigis, S.},
  \bibinfo{author}{Leske, R.}, \bibinfo{author}{Mazur, J.},
  \bibinfo{author}{Mewaldt, R.}, \bibinfo{author}{M{\"o}bius, E.}, et~al.,
  \bibinfo{year}{1999}.
\newblock \bibinfo{title}{Particle acceleration and sources in the november
  1997 solar energetic particle events}.
\newblock \bibinfo{journal}{Geophys. Res. Lett.} \bibinfo{volume}{26},
  \bibinfo{pages}{141--144}.
\newblock \DOIprefix\doi{10.1029/1998GL900235}.
%Type = Article
\bibitem[{Mason et~al.(2012)Mason, Li, Cohen, Desai, Haggerty, Leske, Mewaldt
  and Zank}]{mason2012}
\bibinfo{author}{Mason, G.}, \bibinfo{author}{Li, G.}, \bibinfo{author}{Cohen,
  C.}, \bibinfo{author}{Desai, M.}, \bibinfo{author}{Haggerty, D.},
  \bibinfo{author}{Leske, R.}, \bibinfo{author}{Mewaldt, R.},
  \bibinfo{author}{Zank, G.}, \bibinfo{year}{2012}.
\newblock \bibinfo{title}{Interplanetary propagation of solar energetic
  particle heavy ions observed at 1 au and the role of energy scaling}.
\newblock \bibinfo{journal}{Astrophys. J.} \bibinfo{volume}{761},
  \bibinfo{pages}{104}.
\newblock \DOIprefix\doi{10.1088/0004-637X/761/2/104}.
%Type = Article
\bibitem[{{Mason}(2007)}]{mason2007}
\bibinfo{author}{{Mason}, G.M.}, \bibinfo{year}{2007}.
\newblock \bibinfo{title}{{$^{3}$He-Rich Solar Energetic Particle Events}}.
\newblock \bibinfo{journal}{Space Sci. Rev.} \bibinfo{volume}{130},
  \bibinfo{pages}{231--242}.
\newblock \DOIprefix\doi{10.1007/s11214-007-9156-8}.
%Type = Article
\bibitem[{{Mason} and {Gloeckler}(2012)}]{mason2012ssr}
\bibinfo{author}{{Mason}, G.M.}, \bibinfo{author}{{Gloeckler}, G.},
  \bibinfo{year}{2012}.
\newblock \bibinfo{title}{{Power Law Distributions of Suprathermal Ions in the
  Quiet Solar Wind}}.
\newblock \bibinfo{journal}{Space Sci. Rev.} \bibinfo{volume}{172},
  \bibinfo{pages}{241--251}.
\newblock \DOIprefix\doi{10.1007/s11214-010-9741-0}.
%Type = Article
\bibitem[{{Masson} et~al.(2009){Masson}, {Klein}, {B{\"u}tikofer},
  {Fl{\"u}ckiger}, {Kurt}, {Yushkov} and {Krucker}}]{Mas2009}
\bibinfo{author}{{Masson}, S.}, \bibinfo{author}{{Klein}, K.L.},
  \bibinfo{author}{{B{\"u}tikofer}, R.}, \bibinfo{author}{{Fl{\"u}ckiger}, E.},
  \bibinfo{author}{{Kurt}, V.}, \bibinfo{author}{{Yushkov}, B.},
  \bibinfo{author}{{Krucker}, S.}, \bibinfo{year}{2009}.
\newblock \bibinfo{title}{{Acceleration of Relativistic Protons During the 20
  January 2005 Flare and CME}}.
\newblock \bibinfo{journal}{Sol. Phys.} \bibinfo{volume}{257},
  \bibinfo{pages}{305--322}.
\newblock \DOIprefix\doi{10.1007/s11207-009-9377-y}.
%Type = Article
\bibitem[{{Matthaeus} and {Velli}(2011)}]{matthaeus11}
\bibinfo{author}{{Matthaeus}, W.H.}, \bibinfo{author}{{Velli}, M.},
  \bibinfo{year}{2011}.
\newblock \bibinfo{title}{{Who Needs Turbulence?. A Review of Turbulence
  Effects in the Heliosphere and on the Fundamental Process of Reconnection}}.
\newblock \bibinfo{journal}{Space Sci. Rev.} \bibinfo{volume}{160},
  \bibinfo{pages}{145--168}.
\newblock \DOIprefix\doi{10.1007/s11214-011-9793-9}.
%Type = Article
\bibitem[{Mavromichalaki et~al.(2010)Mavromichalaki, Souvatzoglou, Sarlanis,
  Mariatos, Papaioannou, Belov, Eroshenko, Yanke
  et~al.}]{mavromichalaki2010implementation}
\bibinfo{author}{Mavromichalaki, H.}, \bibinfo{author}{Souvatzoglou, G.},
  \bibinfo{author}{Sarlanis, C.}, \bibinfo{author}{Mariatos, G.},
  \bibinfo{author}{Papaioannou, A.}, \bibinfo{author}{Belov, A.},
  \bibinfo{author}{Eroshenko, E.}, \bibinfo{author}{Yanke, V.}, et~al.,
  \bibinfo{year}{2010}.
\newblock \bibinfo{title}{Implementation of the ground level enhancement alert
  software at nmdb database}.
\newblock \bibinfo{journal}{New Astronomy} \bibinfo{volume}{15},
  \bibinfo{pages}{744--748}.
\newblock \DOIprefix\doi{10.1016/j.newast.2010.05.009}.
%Type = Article
\bibitem[{McComas et~al.(2016)McComas, Alexander, Angold, Bale, Beebe,
  Birdwell, Boyle, Burgum, Burnham, Christian et~al.}]{mccomas2016integrated}
\bibinfo{author}{McComas, D.}, \bibinfo{author}{Alexander, N.},
  \bibinfo{author}{Angold, N.}, \bibinfo{author}{Bale, S.},
  \bibinfo{author}{Beebe, C.}, \bibinfo{author}{Birdwell, B.},
  \bibinfo{author}{Boyle, M.}, \bibinfo{author}{Burgum, J.},
  \bibinfo{author}{Burnham, J.}, \bibinfo{author}{Christian, E.}, et~al.,
  \bibinfo{year}{2016}.
\newblock \bibinfo{title}{{Integrated Science Investigation of the Sun (ISIS):
  Design of the energetic particle investigation}}.
\newblock \bibinfo{journal}{Space Sci. Rev.} \bibinfo{volume}{204},
  \bibinfo{pages}{187--256}.
\newblock \DOIprefix\doi{10.1007/s11214-014-0059-1}.
%Type = Article
\bibitem[{McComas et~al.(2018)McComas, Christian, Schwadron, Fox, Westlake,
  Allegrini, Baker, Biesecker, Bzowski, Clark et~al.}]{mccomas2018interstellar}
\bibinfo{author}{McComas, D.}, \bibinfo{author}{Christian, E.R.},
  \bibinfo{author}{Schwadron, N.A.}, \bibinfo{author}{Fox, N.},
  \bibinfo{author}{Westlake, J.}, \bibinfo{author}{Allegrini, F.},
  \bibinfo{author}{Baker, D.}, \bibinfo{author}{Biesecker, D.},
  \bibinfo{author}{Bzowski, M.}, \bibinfo{author}{Clark, G.}, et~al.,
  \bibinfo{year}{2018}.
\newblock \bibinfo{title}{{Interstellar mapping and acceleration probe (IMAP):
  A new NASA mission}}.
\newblock \bibinfo{journal}{Space Sci. Rev.} \bibinfo{volume}{214},
  \bibinfo{pages}{1--54}.
\newblock \DOIprefix\doi{10.1007/s11214-018-0550-1}.
%Type = Article
\bibitem[{{McComas} et~al.(2002){McComas}, {Elliott}, {Gosling}, {Reisenfeld},
  {Skoug}, {Goldstein}, {Neugebauer} and {Balogh}}]{McComas2002}
\bibinfo{author}{{McComas}, D.J.}, \bibinfo{author}{{Elliott}, H.A.},
  \bibinfo{author}{{Gosling}, J.T.}, \bibinfo{author}{{Reisenfeld}, D.B.},
  \bibinfo{author}{{Skoug}, R.M.}, \bibinfo{author}{{Goldstein}, B.E.},
  \bibinfo{author}{{Neugebauer}, M.}, \bibinfo{author}{{Balogh}, A.},
  \bibinfo{year}{2002}.
\newblock \bibinfo{title}{{Ulysses' second fast-latitude scan: Complexity near
  solar maximum and the reformation of polar coronal holes}}.
\newblock \bibinfo{journal}{Geophys. Res. Lett.} \bibinfo{volume}{29},
  \bibinfo{pages}{1290}.
\newblock \DOIprefix\doi{10.1029/2001GL014164}.
%Type = Article
\bibitem[{{McComas} and {Schwadron}(2006)}]{McComas2006}
\bibinfo{author}{{McComas}, D.J.}, \bibinfo{author}{{Schwadron}, N.A.},
  \bibinfo{year}{2006}.
\newblock \bibinfo{title}{{An explanation of the Voyager paradox: Particle
  acceleration at a blunt termination shock}}.
\newblock \bibinfo{journal}{Geophys. Res. Lett.} \bibinfo{volume}{33},
  \bibinfo{pages}{L04102}.
\newblock \DOIprefix\doi{10.1029/2005GL025437}.
%Type = Article
\bibitem[{McCracken et~al.(2023)McCracken, Shea and Smart}]{MCCRACKEN2023}
\bibinfo{author}{McCracken, K.}, \bibinfo{author}{Shea, M.},
  \bibinfo{author}{Smart, D.}, \bibinfo{year}{2023}.
\newblock \bibinfo{title}{A high time-resolution analysis of the ground-level
  enhancement (gle) of 23 february 1956 in terms of the cshkp standard flare
  model}.
\newblock \bibinfo{journal}{Adv. Space Res.}
  \DOIprefix\doi{https://doi.org/10.1016/j.asr.2023.06.049}.
%Type = Article
\bibitem[{Meier et~al.(2020)Meier, Copeland, Kl{\"o}ble, Matthi{\"a},
  Plettenberg, Schennetten, Wirtz and Hellweg}]{meier2020radiation}
\bibinfo{author}{Meier, M.M.}, \bibinfo{author}{Copeland, K.},
  \bibinfo{author}{Kl{\"o}ble, K.E.}, \bibinfo{author}{Matthi{\"a}, D.},
  \bibinfo{author}{Plettenberg, M.C.}, \bibinfo{author}{Schennetten, K.},
  \bibinfo{author}{Wirtz, M.}, \bibinfo{author}{Hellweg, C.E.},
  \bibinfo{year}{2020}.
\newblock \bibinfo{title}{Radiation in the atmosphere—a hazard to aviation
  safety?}
\newblock \bibinfo{journal}{Atmosphere} \bibinfo{volume}{11},
  \bibinfo{pages}{1358}.
\newblock \DOIprefix\doi{10.3390/atmos11121358}.
%Type = Article
\bibitem[{Mertens and Slaba(2019)}]{mertens2019}
\bibinfo{author}{Mertens, C.J.}, \bibinfo{author}{Slaba, T.C.},
  \bibinfo{year}{2019}.
\newblock \bibinfo{title}{Characterization of solar energetic particle
  radiation dose to astronaut crew on deep-space exploration missions}.
\newblock \bibinfo{journal}{Space Weather} \bibinfo{volume}{17},
  \bibinfo{pages}{1650--1658}.
\newblock \DOIprefix\doi{10.1029/2019SW002363}.
%Type = Article
\bibitem[{{Mewaldt} et~al.(2010){Mewaldt}, {Davis}, {Lave}, {Leske}, {Stone},
  {Wiedenbeck}, {Binns}, {Christian}, {Cummings}, {de Nolfo}, {Israel},
  {Labrador} and {von Rosenvinge}}]{Mewaldt2010}
\bibinfo{author}{{Mewaldt}, R.A.}, \bibinfo{author}{{Davis}, A.J.},
  \bibinfo{author}{{Lave}, K.A.}, \bibinfo{author}{{Leske}, R.A.},
  \bibinfo{author}{{Stone}, E.C.}, \bibinfo{author}{{Wiedenbeck}, M.E.},
  \bibinfo{author}{{Binns}, W.R.}, \bibinfo{author}{{Christian}, E.R.},
  \bibinfo{author}{{Cummings}, A.C.}, \bibinfo{author}{{de Nolfo}, G.A.},
  \bibinfo{author}{{Israel}, M.H.}, \bibinfo{author}{{Labrador}, A.W.},
  \bibinfo{author}{{von Rosenvinge}, T.T.}, \bibinfo{year}{2010}.
\newblock \bibinfo{title}{{Record-setting Cosmic-ray Intensities in 2009 and
  2010}}.
\newblock \bibinfo{journal}{Astrophys. J. Lett.} \bibinfo{volume}{723},
  \bibinfo{pages}{L1--L6}.
\newblock \DOIprefix\doi{10.1088/2041-8205/723/1/L1}.
%Type = Article
\bibitem[{{Mewaldt} et~al.(1996){Mewaldt}, {Selesnick}, {Cummings}, {Stone} and
  {von Rosenvinge}}]{Mewaldt1996}
\bibinfo{author}{{Mewaldt}, R.A.}, \bibinfo{author}{{Selesnick}, R.S.},
  \bibinfo{author}{{Cummings}, J.R.}, \bibinfo{author}{{Stone}, E.C.},
  \bibinfo{author}{{von Rosenvinge}, T.T.}, \bibinfo{year}{1996}.
\newblock \bibinfo{title}{{Evidence for Multiply Charged Anomalous Cosmic
  Rays}}.
\newblock \bibinfo{journal}{Astrophys. J. Lett.} \bibinfo{volume}{466},
  \bibinfo{pages}{L43}.
\newblock \DOIprefix\doi{10.1086/310169}.
%Type = Article
\bibitem[{Milillo et~al.(2020)Milillo, Fujimoto, Murakami, Benkhoff, Zender,
  Aizawa, D{\'o}sa, Griton, Heyner, Ho et~al.}]{milillo2020}
\bibinfo{author}{Milillo, A.}, \bibinfo{author}{Fujimoto, M.},
  \bibinfo{author}{Murakami, G.}, \bibinfo{author}{Benkhoff, J.},
  \bibinfo{author}{Zender, J.}, \bibinfo{author}{Aizawa, S.},
  \bibinfo{author}{D{\'o}sa, M.}, \bibinfo{author}{Griton, L.},
  \bibinfo{author}{Heyner, D.}, \bibinfo{author}{Ho, G.}, et~al.,
  \bibinfo{year}{2020}.
\newblock \bibinfo{title}{Investigating {Mercury’s} environment with the
  two-spacecraft {BepiColombo} mission}.
\newblock \bibinfo{journal}{Space Sci. Rev.} \bibinfo{volume}{216},
  \bibinfo{pages}{1--78}.
\newblock \DOIprefix\doi{10.1007/s11214-020-00712-8}.
%Type = Article
\bibitem[{Miroshnichenko et~al.(2013)Miroshnichenko, Vashenyuk and
  P{\'e}rez-Peraza}]{miroshnichenko2013solar}
\bibinfo{author}{Miroshnichenko, L.}, \bibinfo{author}{Vashenyuk, E.},
  \bibinfo{author}{P{\'e}rez-Peraza, J.}, \bibinfo{year}{2013}.
\newblock \bibinfo{title}{Solar cosmic rays: 70 years of ground-based
  observations}.
\newblock \bibinfo{journal}{Geomagnetism and Aeronomy} \bibinfo{volume}{53},
  \bibinfo{pages}{541--560}.
\newblock \DOIprefix\doi{10.1134/S0016793213050125}.
%Type = Article
\bibitem[{{Miroshnichenko, LI}(2018)}]{miroshnichenko2018}
\bibinfo{author}{{Miroshnichenko, LI}}, \bibinfo{year}{2018}.
\newblock \bibinfo{title}{Retrospective analysis of {GLEs} and estimates of
  radiation risks}.
\newblock \bibinfo{journal}{J. Space Weather \& Space Clim.}
  \bibinfo{volume}{8}, \bibinfo{pages}{A52}.
\newblock \DOIprefix\doi{10.1051/swsc/2018042}.
%Type = Article
\bibitem[{Mishev and Usoskin(2016)}]{mishev2016analysis}
\bibinfo{author}{Mishev, A.}, \bibinfo{author}{Usoskin, I.},
  \bibinfo{year}{2016}.
\newblock \bibinfo{title}{Analysis of the ground-level enhancements on 14 july
  2000 and 13 december 2006 using neutron monitor data}.
\newblock \bibinfo{journal}{Sol. Phys.} \bibinfo{volume}{291},
  \bibinfo{pages}{1225--1239}.
\newblock \DOIprefix\doi{10.1007/s11207-016-0877-2}.
%Type = Article
\bibitem[{Mishev and Usoskin(2020)}]{mishev2020current}
\bibinfo{author}{Mishev, A.}, \bibinfo{author}{Usoskin, I.},
  \bibinfo{year}{2020}.
\newblock \bibinfo{title}{Current status and possible extension of the global
  neutron monitor network}.
\newblock \bibinfo{journal}{Journal of Space Weather and Space Climate}
  \bibinfo{volume}{10}, \bibinfo{pages}{17}.
\newblock \DOIprefix\doi{10.1051/swsc/2020020}.
%Type = Article
\bibitem[{Mishev et~al.(2018)Mishev, Usoskin, Raukunen, Paassilta, Valtonen,
  Kocharov and Vainio}]{mishev2018first}
\bibinfo{author}{Mishev, A.}, \bibinfo{author}{Usoskin, I.},
  \bibinfo{author}{Raukunen, O.}, \bibinfo{author}{Paassilta, M.},
  \bibinfo{author}{Valtonen, E.}, \bibinfo{author}{Kocharov, L.},
  \bibinfo{author}{Vainio, R.}, \bibinfo{year}{2018}.
\newblock \bibinfo{title}{First analysis of ground-level enhancement {(GLE)} 72
  on 10 september 2017: Spectral and anisotropy characteristics}.
\newblock \bibinfo{journal}{Sol. Phys.} \bibinfo{volume}{293},
  \bibinfo{pages}{1--15}.
\newblock \DOIprefix\doi{10.1007/s11207-018-1354-x}.
%Type = Article
\bibitem[{Mishev et~al.(2020)Mishev, Koldobskiy, Kovaltsov, Gil and
  Usoskin}]{mishev2020updated}
\bibinfo{author}{Mishev, A.L.}, \bibinfo{author}{Koldobskiy, S.A.},
  \bibinfo{author}{Kovaltsov, G.A.}, \bibinfo{author}{Gil, A.},
  \bibinfo{author}{Usoskin, I.G.}, \bibinfo{year}{2020}.
\newblock \bibinfo{title}{Updated neutron-monitor yield function: Bridging
  between in situ and ground-based cosmic ray measurements}.
\newblock \bibinfo{journal}{J. Geophys. Res.: Space Phys.}
  \bibinfo{volume}{125}, \bibinfo{pages}{e2019JA027433}.
\newblock \DOIprefix\doi{10.1029/2019JA027433}.
%Type = Article
\bibitem[{Mishev and Poluianov(2021)}]{mishev2021}
\bibinfo{author}{Mishev, A.L.}, \bibinfo{author}{Poluianov, S.},
  \bibinfo{year}{2021}.
\newblock \bibinfo{title}{About the altitude profile of the atmospheric cut-off
  of cosmic rays : new revised assessment}.
\newblock \bibinfo{journal}{Sol. Phys.} \bibinfo{volume}{296}.
\newblock \DOIprefix\doi{10.1007/s11207-021-01875-5}.
%Type = Article
\bibitem[{{Miyake} et~al.(2017){Miyake}, {Kataoka} and {Sato}}]{Miyake2017}
\bibinfo{author}{{Miyake}, S.}, \bibinfo{author}{{Kataoka}, R.},
  \bibinfo{author}{{Sato}, T.}, \bibinfo{year}{2017}.
\newblock \bibinfo{title}{{Cosmic ray modulation and radiation dose of aircrews
  during the solar cycle 24/25}}.
\newblock \bibinfo{journal}{Space Weather} \bibinfo{volume}{15},
  \bibinfo{pages}{589--605}.
\newblock \DOIprefix\doi{10.1002/2016SW001588}.
%Type = Article
\bibitem[{{Modzelewska} et~al.(2019){Modzelewska}, {Iskra}, {Wozniak},
  {Siluszyk} and {Alania}}]{Modzelewska2019}
\bibinfo{author}{{Modzelewska}, R.}, \bibinfo{author}{{Iskra}, K.},
  \bibinfo{author}{{Wozniak}, W.}, \bibinfo{author}{{Siluszyk}, M.},
  \bibinfo{author}{{Alania}, M.V.}, \bibinfo{year}{2019}.
\newblock \bibinfo{title}{{Features of the Galactic Cosmic Ray Anisotropy in
  Solar Cycle 24 and Solar Minima 23/24 and 24/25}}.
\newblock \bibinfo{journal}{Sol. Phys.} \bibinfo{volume}{294},
  \bibinfo{pages}{148}.
\newblock \DOIprefix\doi{10.1007/s11207-019-1540-5}.
%Type = Article
\bibitem[{Moloto and Engelbrecht(2020)}]{Moloto_Engelbrecht_2020}
\bibinfo{author}{Moloto, K.}, \bibinfo{author}{Engelbrecht, N.E.},
  \bibinfo{year}{2020}.
\newblock \bibinfo{title}{A fully time-dependent {Ab} initio cosmic-ray
  modulation model applied to historical cosmic-ray modulation}.
\newblock \bibinfo{journal}{Astrophys. J.} \bibinfo{volume}{894},
  \bibinfo{pages}{121}.
\newblock \DOIprefix\doi{10.3847/1538-4357/ab87a2}.
%Type = Article
\bibitem[{{Moloto} et~al.(2018){Moloto}, {Engelbrecht} and
  {Burger}}]{Moloto2018}
\bibinfo{author}{{Moloto}, K.D.}, \bibinfo{author}{{Engelbrecht}, N.E.},
  \bibinfo{author}{{Burger}, R.A.}, \bibinfo{year}{2018}.
\newblock \bibinfo{title}{{A Simplified Ab Initio Cosmic-ray Modulation Model
  with Simulated Time Dependence and Predictive Capability}}.
\newblock \bibinfo{journal}{Astrophys. J.} \bibinfo{volume}{859},
  \bibinfo{pages}{107}.
\newblock \DOIprefix\doi{10.3847/1538-4357/aac174}.
%Type = Article
\bibitem[{{Moraal}(2013)}]{moraal13}
\bibinfo{author}{{Moraal}, H.}, \bibinfo{year}{2013}.
\newblock \bibinfo{title}{{Cosmic-Ray Modulation Equations}}.
\newblock \bibinfo{journal}{Space Sci. Rev.} \bibinfo{volume}{176},
  \bibinfo{pages}{299--319}.
\newblock \DOIprefix\doi{10.1007/s11214-011-9819-3}.
%Type = Article
\bibitem[{{M{\"o}stl} et~al.(2015){M{\"o}stl}, {Rollett}, {Frahm}, {Liu},
  {Long}, {Colaninno}, {Reiss}, {Temmer}, {Farrugia}, {Posner}, {Dumbovi{\'c}},
  {Janvier}, {D{\'e}moulin}, {Boakes}, {Devos}, {Kraaikamp}, {Mays} and
  {Vr{\v{s}}nak}}]{moestl15}
\bibinfo{author}{{M{\"o}stl}, C.}, \bibinfo{author}{{Rollett}, T.},
  \bibinfo{author}{{Frahm}, R.A.}, \bibinfo{author}{{Liu}, Y.D.},
  \bibinfo{author}{{Long}, D.M.}, \bibinfo{author}{{Colaninno}, R.C.},
  \bibinfo{author}{{Reiss}, M.A.}, \bibinfo{author}{{Temmer}, M.},
  \bibinfo{author}{{Farrugia}, C.J.}, \bibinfo{author}{{Posner}, A.},
  \bibinfo{author}{{Dumbovi{\'c}}, M.}, \bibinfo{author}{{Janvier}, M.},
  \bibinfo{author}{{D{\'e}moulin}, P.}, \bibinfo{author}{{Boakes}, P.},
  \bibinfo{author}{{Devos}, A.}, \bibinfo{author}{{Kraaikamp}, E.},
  \bibinfo{author}{{Mays}, M.L.}, \bibinfo{author}{{Vr{\v{s}}nak}, B.},
  \bibinfo{year}{2015}.
\newblock \bibinfo{title}{{Strong coronal channelling and interplanetary
  evolution of a solar storm up to Earth and Mars}}.
\newblock \bibinfo{journal}{Nature Commun.} \bibinfo{volume}{6},
  \bibinfo{pages}{7135}.
\newblock \DOIprefix\doi{10.1038/ncomms8135}.
%Type = Article
\bibitem[{M{\"u}ller et~al.(2020)M{\"u}ller, Cyr, Zouganelis, Gilbert, Marsden,
  Nieves-Chinchilla, Antonucci, Auch{\`e}re, Berghmans, Horbury
  et~al.}]{muller2020solar}
\bibinfo{author}{M{\"u}ller, D.}, \bibinfo{author}{Cyr, O.S.},
  \bibinfo{author}{Zouganelis, I.}, \bibinfo{author}{Gilbert, H.R.},
  \bibinfo{author}{Marsden, R.}, \bibinfo{author}{Nieves-Chinchilla, T.},
  \bibinfo{author}{Antonucci, E.}, \bibinfo{author}{Auch{\`e}re, F.},
  \bibinfo{author}{Berghmans, D.}, \bibinfo{author}{Horbury, T.}, et~al.,
  \bibinfo{year}{2020}.
\newblock \bibinfo{title}{The solar orbiter mission-science overview}.
\newblock \bibinfo{journal}{Astron. \& Astrophys.} \bibinfo{volume}{642},
  \bibinfo{pages}{A1}.
\newblock \DOIprefix\doi{10.1051/0004-6361/202038467}.
%Type = Article
\bibitem[{Muraki et~al.(2008)Muraki, Matsubara, Masuda, Sakakibara, Sako,
  Watanabe, B{\"u}tikofer, Fl{\"u}ckiger, Chilingarian, Hovsepyan
  et~al.}]{muraki2008detection}
\bibinfo{author}{Muraki, Y.}, \bibinfo{author}{Matsubara, Y.},
  \bibinfo{author}{Masuda, S.}, \bibinfo{author}{Sakakibara, S.},
  \bibinfo{author}{Sako, T.}, \bibinfo{author}{Watanabe, K.},
  \bibinfo{author}{B{\"u}tikofer, R.}, \bibinfo{author}{Fl{\"u}ckiger, E.},
  \bibinfo{author}{Chilingarian, A.}, \bibinfo{author}{Hovsepyan, G.}, et~al.,
  \bibinfo{year}{2008}.
\newblock \bibinfo{title}{Detection of high-energy solar neutrons and protons
  by ground level detectors on {April} 15, 2001}.
\newblock \bibinfo{journal}{Astroparticle Physics} \bibinfo{volume}{29},
  \bibinfo{pages}{229--242}.
\newblock \DOIprefix\doi{10.1016/j.astropartphys.2007.12.007}.
%Type = Book
\bibitem[{{National Research Council,}(2006)}]{NRC2006}
\bibinfo{author}{{National Research Council,}}, \bibinfo{year}{2006}.
\newblock \bibinfo{title}{Space Radiation Hazards and the Vision for Space
  Exploration: Report of a Workshop}.
\newblock \bibinfo{publisher}{The National Academies Press},
  \bibinfo{address}{Washington, DC}.
\newblock \URLprefix
  \url{https://nap.nationalacademies.org/catalog/11760/space-radiation-hazards-and-the-vision-for-space-exploration-report},
  \DOIprefix\doi{10.17226/11760}.
%Type = Article
\bibitem[{Ngobeni et~al.(2020)Ngobeni, Aslam, Bisschoff, Potgieter, Ndiitwani,
  Boezio, Marcelli, Munini, Mikhailov and Koldobskiy}]{ngobeni2020}
\bibinfo{author}{Ngobeni, M.}, \bibinfo{author}{Aslam, O.},
  \bibinfo{author}{Bisschoff, D.}, \bibinfo{author}{Potgieter, M.},
  \bibinfo{author}{Ndiitwani, D.}, \bibinfo{author}{Boezio, M.},
  \bibinfo{author}{Marcelli, N.}, \bibinfo{author}{Munini, R.},
  \bibinfo{author}{Mikhailov, V.}, \bibinfo{author}{Koldobskiy, S.},
  \bibinfo{year}{2020}.
\newblock \bibinfo{title}{The {3D} numerical modeling of the solar modulation
  of galactic protons and helium nuclei related to observations by {PAMELA}
  between 2006 and 2009}.
\newblock \bibinfo{journal}{Astrophys. Space Sci.} \bibinfo{volume}{365},
  \bibinfo{pages}{1--18}.
\newblock \DOIprefix\doi{10.1007/s10509-020-03896-1}.
%Type = Article
\bibitem[{{N{\'u}{\~n}ez}(2022)}]{nunez2020}
\bibinfo{author}{{N{\'u}{\~n}ez}, M.}, \bibinfo{year}{2022}.
\newblock \bibinfo{title}{{Evaluation of the UMASEP-10 Version 2 Tool for
  Predicting All $>$10 MeV SEP Events of Solar Cycles 22, 23 and 24}}.
\newblock \bibinfo{journal}{Universe} \bibinfo{volume}{8}, \bibinfo{pages}{35}.
\newblock \DOIprefix\doi{10.3390/universe8010035}.
%Type = Article
\bibitem[{{Opher} et~al.(2009){Opher}, {Bibi}, {Toth}, {Richardson},
  {Izmodenov} and {Gombosi}}]{Opher2009}
\bibinfo{author}{{Opher}, M.}, \bibinfo{author}{{Bibi}, F.A.},
  \bibinfo{author}{{Toth}, G.}, \bibinfo{author}{{Richardson}, J.D.},
  \bibinfo{author}{{Izmodenov}, V.V.}, \bibinfo{author}{{Gombosi}, T.I.},
  \bibinfo{year}{2009}.
\newblock \bibinfo{title}{{A strong, highly-tilted interstellar magnetic field
  near the Solar System}}.
\newblock \bibinfo{journal}{Nature} \bibinfo{volume}{462},
  \bibinfo{pages}{1036--1038}.
\newblock \DOIprefix\doi{10.1038/nature08567}.
%Type = Article
\bibitem[{{Oughton} and {Engelbrecht}(2021)}]{Oughton2021}
\bibinfo{author}{{Oughton}, S.}, \bibinfo{author}{{Engelbrecht}, N.E.},
  \bibinfo{year}{2021}.
\newblock \bibinfo{title}{{Solar wind turbulence: Connections with energetic
  particles}}.
\newblock \bibinfo{journal}{New Astron.} \bibinfo{volume}{83},
  \bibinfo{pages}{101507}.
\newblock \DOIprefix\doi{10.1016/j.newast.2020.101507}.
%Type = Article
\bibitem[{{Parker}(1965)}]{parker65}
\bibinfo{author}{{Parker}, E.N.}, \bibinfo{year}{1965}.
\newblock \bibinfo{title}{{The passage of energetic charged particles through
  interplanetary space}}.
\newblock \bibinfo{journal}{Planet. Space Sci.} \bibinfo{volume}{13},
  \bibinfo{pages}{9--+}.
\newblock \DOIprefix\doi{10.1016/0032-0633(65)90131-5}.
%Type = Article
\bibitem[{{Pesses} et~al.(1981){Pesses}, {Jokipii} and {Eichler}}]{Pesses1981}
\bibinfo{author}{{Pesses}, M.E.}, \bibinfo{author}{{Jokipii}, J.R.},
  \bibinfo{author}{{Eichler}, D.}, \bibinfo{year}{1981}.
\newblock \bibinfo{title}{{Cosmic ray drift, shock wave acceleration, and the
  anomalous component of cosmic rays}}.
\newblock \bibinfo{journal}{Astrophys. J. Lett.} \bibinfo{volume}{246},
  \bibinfo{pages}{L85--L88}.
\newblock \DOIprefix\doi{10.1086/183559}.
%Type = Article
\bibitem[{{Petukhova} et~al.(2019){Petukhova}, {Petukhov} and
  {Petukhov}}]{petukhova19b}
\bibinfo{author}{{Petukhova}, A.S.}, \bibinfo{author}{{Petukhov}, I.S.},
  \bibinfo{author}{{Petukhov}, S.I.}, \bibinfo{year}{2019}.
\newblock \bibinfo{title}{{Theory of the Formation of Forbush Decrease in a
  Magnetic Cloud: Dependence of Forbush Decrease Characteristics on Magnetic
  Cloud Parameters}}.
\newblock \bibinfo{journal}{Astrophys. J.} \bibinfo{volume}{880},
  \bibinfo{pages}{17}.
\newblock \DOIprefix\doi{10.3847/1538-4357/ab2889}.
%Type = Article
\bibitem[{{Picozza} et~al.(2007){Picozza}, {Galper}, {Castellini}, {Adriani},
  {Altamura}, {Ambriola}, {Barbarino}, {Basili}, {Bazilevskaja}, {Bencardino},
  {Boezio}, {Bogomolov}, {Bonechi}, {Bongi}, {Bongiorno}, {Bonvicini},
  {Cafagna}, {Campana}, {Carlson}, {Casolino}, {de Marzo}, {de Pascale}, {de
  Rosa}, {Fedele}, {Hofverberg}, {Koldashov}, {Krutkov}, {Kvashnin}, {Lund},
  {Lundquist}, {Maksumov}, {Malvezzi}, {Marcelli}, {Menn}, {Mikhailov},
  {Minori}, {Misin}, {Mocchiutti}, {Morselli}, {Nikonov}, {Orsi}, {Osteria},
  {Papini}, {Pearce}, {Ricci}, {Ricciarini}, {Runtso}, {Russo}, {Simon},
  {Sparvoli}, {Spillantini}, {Stozhkov}, {Taddei}, {Vacchi}, {Vannuccini},
  {Voronov}, {Yurkin}, {Zampa}, {Zampa} and {Zverev}}]{Picozza2007}
\bibinfo{author}{{Picozza}, P.}, \bibinfo{author}{{Galper}, A.M.},
  \bibinfo{author}{{Castellini}, G.}, \bibinfo{author}{{Adriani}, O.},
  \bibinfo{author}{{Altamura}, F.}, \bibinfo{author}{{Ambriola}, M.},
  \bibinfo{author}{{Barbarino}, G.C.}, \bibinfo{author}{{Basili}, A.},
  \bibinfo{author}{{Bazilevskaja}, G.A.}, \bibinfo{author}{{Bencardino}, R.},
  \bibinfo{author}{{Boezio}, M.}, \bibinfo{author}{{Bogomolov}, E.A.},
  \bibinfo{author}{{Bonechi}, L.}, \bibinfo{author}{{Bongi}, M.},
  \bibinfo{author}{{Bongiorno}, L.}, \bibinfo{author}{{Bonvicini}, V.},
  \bibinfo{author}{{Cafagna}, F.}, \bibinfo{author}{{Campana}, D.},
  \bibinfo{author}{{Carlson}, P.}, \bibinfo{author}{{Casolino}, M.},
  \bibinfo{author}{{de Marzo}, C.}, \bibinfo{author}{{de Pascale}, M.P.},
  \bibinfo{author}{{de Rosa}, G.}, \bibinfo{author}{{Fedele}, D.},
  \bibinfo{author}{{Hofverberg}, P.}, \bibinfo{author}{{Koldashov}, S.V.},
  \bibinfo{author}{{Krutkov}, S.Y.}, \bibinfo{author}{{Kvashnin}, A.N.},
  \bibinfo{author}{{Lund}, J.}, \bibinfo{author}{{Lundquist}, J.},
  \bibinfo{author}{{Maksumov}, O.}, \bibinfo{author}{{Malvezzi}, V.},
  \bibinfo{author}{{Marcelli}, L.}, \bibinfo{author}{{Menn}, W.},
  \bibinfo{author}{{Mikhailov}, V.V.}, \bibinfo{author}{{Minori}, M.},
  \bibinfo{author}{{Misin}, S.}, \bibinfo{author}{{Mocchiutti}, E.},
  \bibinfo{author}{{Morselli}, A.}, \bibinfo{author}{{Nikonov}, N.N.},
  \bibinfo{author}{{Orsi}, S.}, \bibinfo{author}{{Osteria}, G.},
  \bibinfo{author}{{Papini}, P.}, \bibinfo{author}{{Pearce}, M.},
  \bibinfo{author}{{Ricci}, M.}, \bibinfo{author}{{Ricciarini}, S.B.},
  \bibinfo{author}{{Runtso}, M.F.}, \bibinfo{author}{{Russo}, S.},
  \bibinfo{author}{{Simon}, M.}, \bibinfo{author}{{Sparvoli}, R.},
  \bibinfo{author}{{Spillantini}, P.}, \bibinfo{author}{{Stozhkov}, Y.I.},
  \bibinfo{author}{{Taddei}, E.}, \bibinfo{author}{{Vacchi}, A.},
  \bibinfo{author}{{Vannuccini}, E.}, \bibinfo{author}{{Voronov}, S.A.},
  \bibinfo{author}{{Yurkin}, Y.T.}, \bibinfo{author}{{Zampa}, G.},
  \bibinfo{author}{{Zampa}, N.}, \bibinfo{author}{{Zverev}, V.G.},
  \bibinfo{year}{2007}.
\newblock \bibinfo{title}{{{PAMELA} A payload for antimatter matter exploration
  and light-nuclei astrophysics}}.
\newblock \bibinfo{journal}{Astroparticle Phys.} \bibinfo{volume}{27},
  \bibinfo{pages}{296--315}.
\newblock \DOIprefix\doi{10.1016/j.astropartphys.2006.12.002}.
%Type = Article
\bibitem[{Plainaki et~al.(2020)Plainaki, Antonucci, Bemporad, Berrilli,
  Bertucci, Castronuovo, De~Michelis, Giardino, Iuppa, Laurenza
  et~al.}]{plainaki2020current}
\bibinfo{author}{Plainaki, C.}, \bibinfo{author}{Antonucci, M.},
  \bibinfo{author}{Bemporad, A.}, \bibinfo{author}{Berrilli, F.},
  \bibinfo{author}{Bertucci, B.}, \bibinfo{author}{Castronuovo, M.},
  \bibinfo{author}{De~Michelis, P.}, \bibinfo{author}{Giardino, M.},
  \bibinfo{author}{Iuppa, R.}, \bibinfo{author}{Laurenza, M.}, et~al.,
  \bibinfo{year}{2020}.
\newblock \bibinfo{title}{Current state and perspectives of space weather
  science in italy}.
\newblock \bibinfo{journal}{J. Space Weather \& Space Clim.}
  \bibinfo{volume}{10}, \bibinfo{pages}{6}.
\newblock \DOIprefix\doi{10.1051/swsc/2020003}.
%Type = Article
\bibitem[{Plainaki et~al.(2007)Plainaki, Belov, Eroshenko, Mavromichalaki and
  Yanke}]{plainaki2007modeling}
\bibinfo{author}{Plainaki, C.}, \bibinfo{author}{Belov, A.},
  \bibinfo{author}{Eroshenko, E.}, \bibinfo{author}{Mavromichalaki, H.},
  \bibinfo{author}{Yanke, V.}, \bibinfo{year}{2007}.
\newblock \bibinfo{title}{Modeling ground level enhancements: Event of 20
  january 2005}.
\newblock \bibinfo{journal}{J. Geophys. Res.: Space Phys.}
  \bibinfo{volume}{112}.
\newblock \DOIprefix\doi{10.1029/2006JA011926}.
%Type = Article
\bibitem[{Plainaki et~al.(2009a)Plainaki, Mavromichalaki, Belov, Eroshenko and
  Yanke}]{plainaki2009modeling}
\bibinfo{author}{Plainaki, C.}, \bibinfo{author}{Mavromichalaki, H.},
  \bibinfo{author}{Belov, A.}, \bibinfo{author}{Eroshenko, E.},
  \bibinfo{author}{Yanke, V.}, \bibinfo{year}{2009}a.
\newblock \bibinfo{title}{Modeling the solar cosmic ray event of 13 december
  2006 using ground level neutron monitor data}.
\newblock \bibinfo{journal}{Adv. Space Res.} \bibinfo{volume}{43},
  \bibinfo{pages}{474--479}.
\newblock \DOIprefix\doi{10.1016/j.asr.2008.07.011}.
%Type = Article
\bibitem[{Plainaki et~al.(2009b)Plainaki, Mavromichalaki, Belov, Eroshenko and
  Yanke}]{plainaki2009neutron}
\bibinfo{author}{Plainaki, C.}, \bibinfo{author}{Mavromichalaki, H.},
  \bibinfo{author}{Belov, A.}, \bibinfo{author}{Eroshenko, E.},
  \bibinfo{author}{Yanke, V.}, \bibinfo{year}{2009}b.
\newblock \bibinfo{title}{Neutron monitor asymptotic directions of viewing
  during the event of 13 december 2006}.
\newblock \bibinfo{journal}{Adv. Space Res.} \bibinfo{volume}{43},
  \bibinfo{pages}{518--522}.
\newblock \DOIprefix\doi{10.1016/j.asr.2008.09.007}.
%Type = Article
\bibitem[{Plainaki et~al.(2014)Plainaki, Mavromichalaki, Laurenza, Gerontidou,
  Kanellakopoulos and Storini}]{plainaki2014ground}
\bibinfo{author}{Plainaki, C.}, \bibinfo{author}{Mavromichalaki, H.},
  \bibinfo{author}{Laurenza, M.}, \bibinfo{author}{Gerontidou, M.},
  \bibinfo{author}{Kanellakopoulos, A.}, \bibinfo{author}{Storini, M.},
  \bibinfo{year}{2014}.
\newblock \bibinfo{title}{The ground-level enhancement of 2012 may 17:
  derivation of solar proton event properties through the application of the
  nmbangle ppola model}.
\newblock \bibinfo{journal}{Astrophys. J.} \bibinfo{volume}{785},
  \bibinfo{pages}{160}.
\newblock \DOIprefix\doi{10.1088/0004-637X/785/2/160}.
%Type = Article
\bibitem[{{Pogorelov} et~al.(2009){Pogorelov}, {Heerikhuisen}, {Zank},
  {Mitchell} and {Cairns}}]{Pogorelov2009}
\bibinfo{author}{{Pogorelov}, N.V.}, \bibinfo{author}{{Heerikhuisen}, J.},
  \bibinfo{author}{{Zank}, G.P.}, \bibinfo{author}{{Mitchell}, J.J.},
  \bibinfo{author}{{Cairns}, I.H.}, \bibinfo{year}{2009}.
\newblock \bibinfo{title}{{Heliospheric asymmetries due to the action of the
  interstellar magnetic field}}.
\newblock \bibinfo{journal}{Adv. Space Res.} \bibinfo{volume}{44},
  \bibinfo{pages}{1337--1344}.
\newblock \DOIprefix\doi{10.1016/j.asr.2009.07.019}.
%Type = Article
\bibitem[{Poluianov et~al.(2017)Poluianov, Usoskin, Mishev, Shea and
  Smart}]{poluianov2017sub}
\bibinfo{author}{Poluianov, S.}, \bibinfo{author}{Usoskin, I.},
  \bibinfo{author}{Mishev, A.}, \bibinfo{author}{Shea, M.},
  \bibinfo{author}{Smart, D.}, \bibinfo{year}{2017}.
\newblock \bibinfo{title}{Gle and sub-gle redefinition in the light of
  high-altitude polar neutron monitors}.
\newblock \bibinfo{journal}{Sol. Phys.} \bibinfo{volume}{292},
  \bibinfo{pages}{176}.
\newblock \DOIprefix\doi{10.1007/s11207-017-1202-4}.
%Type = Article
\bibitem[{{Posner}(2007)}]{posner07}
\bibinfo{author}{{Posner}, A.}, \bibinfo{year}{2007}.
\newblock \bibinfo{title}{Up to 1-hour forecasting of radiation hazards from
  solar energetic ion events with relativistic electrons}.
\newblock \bibinfo{journal}{Space Weather} \bibinfo{volume}{5},
  \bibinfo{pages}{05001}.
\newblock \DOIprefix\doi{10.1029/2006SW000268}.
%Type = Article
\bibitem[{{Potgieter}(2014)}]{Potgieter2014}
\bibinfo{author}{{Potgieter}, M.}, \bibinfo{year}{2014}.
\newblock \bibinfo{title}{{Very Local Interstellar Spectra for Galactic
  Electrons, Protons and Helium}}.
\newblock \bibinfo{journal}{Brazilian J. of Phys.} \bibinfo{volume}{44},
  \bibinfo{pages}{581--588}.
\newblock \DOIprefix\doi{10.1007/s13538-014-0238-2}.
%Type = Article
\bibitem[{{Potgieter}(2013)}]{potgieter13}
\bibinfo{author}{{Potgieter}, M.S.}, \bibinfo{year}{2013}.
\newblock \bibinfo{title}{{Solar Modulation of Cosmic Rays}}.
\newblock \bibinfo{journal}{Living Rev. Sol. Phys.} \bibinfo{volume}{10},
  \bibinfo{pages}{3}.
\newblock \DOIprefix\doi{10.12942/lrsp-2013-3}.
%Type = Article
\bibitem[{{Potgieter}(2017)}]{Potgieter2017}
\bibinfo{author}{{Potgieter}, M.S.}, \bibinfo{year}{2017}.
\newblock \bibinfo{title}{{The global modulation of cosmic rays during a quiet
  heliosphere: A modeling perspective}}.
\newblock \bibinfo{journal}{Adv. Space Res.} \bibinfo{volume}{60},
  \bibinfo{pages}{848--864}.
\newblock \DOIprefix\doi{10.1016/j.asr.2016.09.003}.
%Type = Article
\bibitem[{{Qin}(2007)}]{Qin2007}
\bibinfo{author}{{Qin}, G.}, \bibinfo{year}{2007}.
\newblock \bibinfo{title}{{Nonlinear Parallel Diffusion of Charged Particles:
  Extension to the Nonlinear Guiding Center Theory}}.
\newblock \bibinfo{journal}{Astrophys. J.} \bibinfo{volume}{656},
  \bibinfo{pages}{217--221}.
\newblock \DOIprefix\doi{10.1086/510510}.
%Type = Article
\bibitem[{Rankin et~al.(2022)Rankin, Bindi, Bykov, Cummings, Della~Torre,
  Florinski, Heber, Potgieter, Stone and Zhang}]{rankin2022galactic}
\bibinfo{author}{Rankin, J.S.}, \bibinfo{author}{Bindi, V.},
  \bibinfo{author}{Bykov, A.M.}, \bibinfo{author}{Cummings, A.C.},
  \bibinfo{author}{Della~Torre, S.}, \bibinfo{author}{Florinski, V.},
  \bibinfo{author}{Heber, B.}, \bibinfo{author}{Potgieter, M.S.},
  \bibinfo{author}{Stone, E.C.}, \bibinfo{author}{Zhang, M.},
  \bibinfo{year}{2022}.
\newblock \bibinfo{title}{Galactic cosmic rays throughout the heliosphere and
  in the very local interstellar medium}.
\newblock \bibinfo{journal}{Space Sci. Rev.} \bibinfo{volume}{218},
  \bibinfo{pages}{42}.
\newblock \DOIprefix\doi{10.1007/s11214-022-00912-4}.
%Type = Article
\bibitem[{Raukunen et~al.(2020)Raukunen, Paassilta, Vainio, Rodriguez, Eronen,
  Crosby, Dierckxsens, Jiggens, Heynderickx and Sandberg}]{raukunen2020very}
\bibinfo{author}{Raukunen, O.}, \bibinfo{author}{Paassilta, M.},
  \bibinfo{author}{Vainio, R.}, \bibinfo{author}{Rodriguez, J.V.},
  \bibinfo{author}{Eronen, T.}, \bibinfo{author}{Crosby, N.},
  \bibinfo{author}{Dierckxsens, M.}, \bibinfo{author}{Jiggens, P.},
  \bibinfo{author}{Heynderickx, D.}, \bibinfo{author}{Sandberg, I.},
  \bibinfo{year}{2020}.
\newblock \bibinfo{title}{Very high energy proton peak flux model}.
\newblock \bibinfo{journal}{J. Space Weather \& Space Clim.}
  \bibinfo{volume}{10}, \bibinfo{pages}{24}.
\newblock \DOIprefix\doi{10.1051/swsc/2020024}.
%Type = Article
\bibitem[{Raukunen et~al.(2018)Raukunen, Vainio, Tylka, Dietrich, Jiggens,
  Heynderickx, Dierckxsens, Crosby, Ganse and Siipola}]{raukunen2018two}
\bibinfo{author}{Raukunen, O.}, \bibinfo{author}{Vainio, R.},
  \bibinfo{author}{Tylka, A.J.}, \bibinfo{author}{Dietrich, W.F.},
  \bibinfo{author}{Jiggens, P.}, \bibinfo{author}{Heynderickx, D.},
  \bibinfo{author}{Dierckxsens, M.}, \bibinfo{author}{Crosby, N.},
  \bibinfo{author}{Ganse, U.}, \bibinfo{author}{Siipola, R.},
  \bibinfo{year}{2018}.
\newblock \bibinfo{title}{Two solar proton fluence models based on ground level
  enhancement observations}.
\newblock \bibinfo{journal}{J. Space Weather \& Space Clim.}
  \bibinfo{volume}{8}, \bibinfo{pages}{A04}.
\newblock \DOIprefix\doi{10.1051/swsc/2017031}.
%Type = Article
\bibitem[{Reames(1999)}]{reames1999particle}
\bibinfo{author}{Reames, D.V.}, \bibinfo{year}{1999}.
\newblock \bibinfo{title}{Particle acceleration at the sun and in the
  heliosphere}.
\newblock \bibinfo{journal}{Space Sci. Rev.} \bibinfo{volume}{90},
  \bibinfo{pages}{413--491}.
\newblock \DOIprefix\doi{10.1023/A:1005105831781}.
%Type = Article
\bibitem[{Reames(2013)}]{reames2013two}
\bibinfo{author}{Reames, D.V.}, \bibinfo{year}{2013}.
\newblock \bibinfo{title}{The two sources of solar energetic particles}.
\newblock \bibinfo{journal}{Space Sci. Rev.} \bibinfo{volume}{175},
  \bibinfo{pages}{53--92}.
\newblock \DOIprefix\doi{10.1007/s11214-013-9958-9}.
%Type = Article
\bibitem[{Reames(2015)}]{reames2015sources}
\bibinfo{author}{Reames, D.V.}, \bibinfo{year}{2015}.
\newblock \bibinfo{title}{What are the sources of solar energetic particles?
  element abundances and source plasma temperatures}.
\newblock \bibinfo{journal}{Space Science Reviews} \bibinfo{volume}{194},
  \bibinfo{pages}{303--327}.
\newblock \DOIprefix\doi{10.1007/s11214-015-0210-7}.
%Type = Article
\bibitem[{Reeves et~al.(1992)Reeves, Cayton, Gary and Belian}]{reeves1992great}
\bibinfo{author}{Reeves, G.}, \bibinfo{author}{Cayton, T.},
  \bibinfo{author}{Gary, S.}, \bibinfo{author}{Belian, R.},
  \bibinfo{year}{1992}.
\newblock \bibinfo{title}{The great solar energetic particle events of 1989
  observed from geosynchronous orbit}.
\newblock \bibinfo{journal}{J. Geophys. Res.: Space Phys.}
  \bibinfo{volume}{97}, \bibinfo{pages}{6219--6226}.
\newblock \DOIprefix\doi{10.1029/91JA03102}.
%Type = Article
\bibitem[{Richardson et~al.(2014)Richardson, von Rosenvinge, Cane, Christian,
  Cohen, Labrador, Leske, Mewaldt, Wiedenbeck and Stone}]{richardson2014}
\bibinfo{author}{Richardson, I.}, \bibinfo{author}{von Rosenvinge, T.},
  \bibinfo{author}{Cane, H.}, \bibinfo{author}{Christian, E.},
  \bibinfo{author}{Cohen, C.}, \bibinfo{author}{Labrador, A.},
  \bibinfo{author}{Leske, R.}, \bibinfo{author}{Mewaldt, R.},
  \bibinfo{author}{Wiedenbeck, M.}, \bibinfo{author}{Stone, E.},
  \bibinfo{year}{2014}.
\newblock \bibinfo{title}{> 25 mev proton events observed by the high energy
  telescopes on the stereo a and b spacecraft and/or at {Earth} during the
  first~ seven years of the stereo mission}.
\newblock \bibinfo{journal}{Sol. Phys.} \bibinfo{volume}{289},
  \bibinfo{pages}{3059--3107}.
\newblock \DOIprefix\doi{10.1007/978-1-4939-2038-9_27}.
%Type = Article
\bibitem[{{Richardson}(2004)}]{richardson04}
\bibinfo{author}{{Richardson}, I.G.}, \bibinfo{year}{2004}.
\newblock \bibinfo{title}{{Energetic Particles and Corotating Interaction
  Regions in the Solar Wind}}.
\newblock \bibinfo{journal}{Space Sci. Rev.} \bibinfo{volume}{111},
  \bibinfo{pages}{267--376}.
\newblock \DOIprefix\doi{10.1023/B:SPAC.0000032689.52830.3e}.
%Type = Article
\bibitem[{{Richardson}(2018)}]{richardson18}
\bibinfo{author}{{Richardson}, I.G.}, \bibinfo{year}{2018}.
\newblock \bibinfo{title}{{Solar wind stream interaction regions throughout the
  heliosphere}}.
\newblock \bibinfo{journal}{Living Rev. Sol. Phys.} \bibinfo{volume}{15},
  \bibinfo{pages}{1}.
\newblock \DOIprefix\doi{10.1007/s41116-017-0011-z}.
%Type = Article
\bibitem[{{Richardson} and {Cane}(2011)}]{richardson11}
\bibinfo{author}{{Richardson}, I.G.}, \bibinfo{author}{{Cane}, H.V.},
  \bibinfo{year}{2011}.
\newblock \bibinfo{title}{{Galactic Cosmic Ray Intensity Response to
  Interplanetary Coronal Mass Ejections/Magnetic Clouds in 1995 - 2009}}.
\newblock \bibinfo{journal}{Sol. Phys.} \bibinfo{volume}{270},
  \bibinfo{pages}{609--627}.
\newblock \DOIprefix\doi{10.1007/s11207-011-9774-x}.
%Type = Article
\bibitem[{Rodriguez et~al.(2014)Rodriguez, Krosschell and Green}]{Rodriguez14}
\bibinfo{author}{Rodriguez, J.V.}, \bibinfo{author}{Krosschell, J.C.},
  \bibinfo{author}{Green, J.C.}, \bibinfo{year}{2014}.
\newblock \bibinfo{title}{{Intercalibration of GOES 8–15 solar proton
  detectors}}.
\newblock \bibinfo{journal}{Space Weather} \bibinfo{volume}{12},
  \bibinfo{pages}{92--109}.
\newblock \DOIprefix\doi{10.1002/2013SW000996}.
%Type = Article
\bibitem[{Rodr{\'\i}guez-Pacheco et~al.(2020)Rodr{\'\i}guez-Pacheco,
  Wimmer-Schweingruber, Mason, Ho, S{\'a}nchez-Prieto, Prieto, Mart{\'\i}n,
  Seifert, Andrews, Kulkarni et~al.}]{rodriguez2020energetic}
\bibinfo{author}{Rodr{\'\i}guez-Pacheco, J.},
  \bibinfo{author}{Wimmer-Schweingruber, R.}, \bibinfo{author}{Mason, G.},
  \bibinfo{author}{Ho, G.}, \bibinfo{author}{S{\'a}nchez-Prieto, S.},
  \bibinfo{author}{Prieto, M.}, \bibinfo{author}{Mart{\'\i}n, C.},
  \bibinfo{author}{Seifert, H.}, \bibinfo{author}{Andrews, G.},
  \bibinfo{author}{Kulkarni, S.}, et~al., \bibinfo{year}{2020}.
\newblock \bibinfo{title}{The energetic particle detector-energetic particle
  instrument suite for the solar orbiter mission}.
\newblock \bibinfo{journal}{Astron. \& Astrophys.} \bibinfo{volume}{642},
  \bibinfo{pages}{A7}.
\newblock \DOIprefix\doi{10.1051/0004-6361/201935287}.
%Type = Article
\bibitem[{Ross and Chaplin(2019)}]{Ross_Chaplin_2019}
\bibinfo{author}{Ross, E.}, \bibinfo{author}{Chaplin, W.J.},
  \bibinfo{year}{2019}.
\newblock \bibinfo{title}{The behaviour of galactic cosmic-ray intensity during
  solar activity cycle 24}.
\newblock \bibinfo{journal}{Sol. Phys.} \bibinfo{volume}{294},
  \bibinfo{pages}{8}.
\newblock \DOIprefix\doi{10.1007/s11207-019-1397-7}.
%Type = Article
\bibitem[{Rouillard et~al.(2012)Rouillard, Sheeley, Tylka, Vourlidas, Ng,
  Rakowski, Cohen, Mewaldt, Mason, Reames, Savani, StCyr and
  Szabo}]{Rouillard2012}
\bibinfo{author}{Rouillard, A.P.}, \bibinfo{author}{Sheeley, N.R.},
  \bibinfo{author}{Tylka, A.}, \bibinfo{author}{Vourlidas, A.},
  \bibinfo{author}{Ng, C.K.}, \bibinfo{author}{Rakowski, C.},
  \bibinfo{author}{Cohen, C.M.S.}, \bibinfo{author}{Mewaldt, R.A.},
  \bibinfo{author}{Mason, G.M.}, \bibinfo{author}{Reames, D.},
  \bibinfo{author}{Savani, N.P.}, \bibinfo{author}{StCyr, O.C.},
  \bibinfo{author}{Szabo, A.}, \bibinfo{year}{2012}.
\newblock \bibinfo{title}{{THE} {LONGITUDINAL} {PROPERTIES} {OF} a {SOLAR}
  {ENERGETIC} {PARTICLE} {EVENT} {INVESTIGATED} {USING} {MODERN} {SOLAR}
  {IMAGING}}.
\newblock \bibinfo{journal}{Astrophys. J.} \bibinfo{volume}{752},
  \bibinfo{pages}{44}.
\newblock \DOIprefix\doi{10.1088/0004-637x/752/1/44}.
%Type = Article
\bibitem[{Roussos et~al.(2020)Roussos, Dialynas, Krupp, Kollmann, Paranicas,
  Roelof, Yuan, Mitchell and Krimigis}]{Roussos2020}
\bibinfo{author}{Roussos, E.}, \bibinfo{author}{Dialynas, K.},
  \bibinfo{author}{Krupp, N.}, \bibinfo{author}{Kollmann, P.},
  \bibinfo{author}{Paranicas, C.}, \bibinfo{author}{Roelof, E.C.},
  \bibinfo{author}{Yuan, C.}, \bibinfo{author}{Mitchell, D.G.},
  \bibinfo{author}{Krimigis, S.M.}, \bibinfo{year}{2020}.
\newblock \bibinfo{title}{Long- and short-term variability of galactic
  cosmic-ray radial intensity gradients between 1 and 9.5 au: Observations by
  cassini, {BESS}, {BESS}-polar, {PAMELA}, and {AMS}-02}.
\newblock \bibinfo{journal}{Astrophys. J.} \bibinfo{volume}{904},
  \bibinfo{pages}{165}.
\newblock \DOIprefix\doi{10.3847/1538-4357/abc346}.
%Type = Article
\bibitem[{Sandberg et~al.(2014)Sandberg, Jiggens, Heynderickx and
  Daglis}]{sandberg2014cross}
\bibinfo{author}{Sandberg, I.}, \bibinfo{author}{Jiggens, P.},
  \bibinfo{author}{Heynderickx, D.}, \bibinfo{author}{Daglis, I.},
  \bibinfo{year}{2014}.
\newblock \bibinfo{title}{Cross calibration of {NOAA GOES} solar proton
  detectors using corrected {NASA IMP-8/GME} data}.
\newblock \bibinfo{journal}{Geophys. Res. Lett.} \bibinfo{volume}{41},
  \bibinfo{pages}{4435--4441}.
\newblock \DOIprefix\doi{10.1002/2014GL060469}.
%Type = Inproceedings
\bibitem[{{Sauer}(1989)}]{sauer89}
\bibinfo{author}{{Sauer}, H.H.}, \bibinfo{year}{1989}.
\newblock \bibinfo{title}{{SEL monitoring of the Earth's energetic particle
  radiation environment}}, in: \bibinfo{editor}{{Rester}, A.~C., J.},
  \bibinfo{editor}{{Trombka}, J.I.} (Eds.), \bibinfo{booktitle}{High-Energy
  Radiation Background in Space}, pp. \bibinfo{pages}{216--221}.
\newblock \DOIprefix\doi{10.1063/1.38171}.
%Type = Article
\bibitem[{{Schwadron} et~al.(2008){Schwadron}, {Lee} and
  {McComas}}]{Schwadron2008}
\bibinfo{author}{{Schwadron}, N.A.}, \bibinfo{author}{{Lee}, M.A.},
  \bibinfo{author}{{McComas}, D.J.}, \bibinfo{year}{2008}.
\newblock \bibinfo{title}{{Diffusive Acceleration at the Blunt Termination
  Shock}}.
\newblock \bibinfo{journal}{Astrophys. J.} \bibinfo{volume}{675},
  \bibinfo{pages}{1584--1600}.
\newblock \DOIprefix\doi{10.1086/527026}.
%Type = Article
\bibitem[{{Shalchi}(2020)}]{Shalchi2020}
\bibinfo{author}{{Shalchi}, A.}, \bibinfo{year}{2020}.
\newblock \bibinfo{title}{{Perpendicular Transport of Energetic Particles in
  Magnetic Turbulence}}.
\newblock \bibinfo{journal}{Space Sci. Rev.} \bibinfo{volume}{216},
  \bibinfo{pages}{23}.
\newblock \DOIprefix\doi{10.1007/s11214-020-0644-4}.
%Type = Inproceedings
\bibitem[{Shea and Smart(1993)}]{shea1993history}
\bibinfo{author}{Shea, M.}, \bibinfo{author}{Smart, D.}, \bibinfo{year}{1993}.
\newblock \bibinfo{title}{History of energetic solar protons for the past three
  solar cycles including cycle 22 update}, in: \bibinfo{booktitle}{Biological
  Effects and Physics of Solar and Galactic Cosmic Radiation},
  \bibinfo{organization}{Springer}. pp. \bibinfo{pages}{37--71}.
%Type = Article
\bibitem[{Shen et~al.(2020)Shen, Qin, Zuo, Wei and Xu}]{Shen2020}
\bibinfo{author}{Shen, Z.}, \bibinfo{author}{Qin, G.}, \bibinfo{author}{Zuo,
  P.}, \bibinfo{author}{Wei, F.}, \bibinfo{author}{Xu, X.},
  \bibinfo{year}{2020}.
\newblock \bibinfo{title}{A study of variations of galactic cosmic-ray
  intensity based on a hybrid data-processing method}.
\newblock \bibinfo{journal}{Astrophys. J.} \bibinfo{volume}{900},
  \bibinfo{pages}{143}.
\newblock \DOIprefix\doi{10.3847/1538-4357/abac60}.
%Type = Article
\bibitem[{{Shen} et~al.(2021){Shen}, {Yang}, {Zuo}, {Qin}, {Wei}, {Xu} and
  {Xie}}]{Shen2021}
\bibinfo{author}{{Shen}, Z.}, \bibinfo{author}{{Yang}, H.},
  \bibinfo{author}{{Zuo}, P.}, \bibinfo{author}{{Qin}, G.},
  \bibinfo{author}{{Wei}, F.}, \bibinfo{author}{{Xu}, X.},
  \bibinfo{author}{{Xie}, Y.}, \bibinfo{year}{2021}.
\newblock \bibinfo{title}{{Solar Modulation of Galactic Cosmic-Ray Protons
  Based on a Modified Force-field Approach}}.
\newblock \bibinfo{journal}{Astrophys. J.} \bibinfo{volume}{921},
  \bibinfo{pages}{109}.
\newblock \DOIprefix\doi{10.3847/1538-4357/ac1fe8}.
%Type = Article
\bibitem[{Sihver et~al.(2015)Sihver, Ploc, Puchalska, Ambro{\v{z}}ov{\'a},
  Kuban{\v{c}}{\'a}k, Kyselov{\'a} and Shurshakov}]{sihver2015radiation}
\bibinfo{author}{Sihver, L.}, \bibinfo{author}{Ploc, O.},
  \bibinfo{author}{Puchalska, M.}, \bibinfo{author}{Ambro{\v{z}}ov{\'a}, I.},
  \bibinfo{author}{Kuban{\v{c}}{\'a}k, J.}, \bibinfo{author}{Kyselov{\'a}, D.},
  \bibinfo{author}{Shurshakov, V.}, \bibinfo{year}{2015}.
\newblock \bibinfo{title}{Radiation environment at aviation altitudes and in
  space}.
\newblock \bibinfo{journal}{Radiation protection dosimetry}
  \bibinfo{volume}{164}, \bibinfo{pages}{477--483}.
\newblock \DOIprefix\doi{10.1093/rpd/ncv330}.
%Type = Article
\bibitem[{Slaba and Blattnig(2014)}]{slaba2014gcr}
\bibinfo{author}{Slaba, T.C.}, \bibinfo{author}{Blattnig, S.R.},
  \bibinfo{year}{2014}.
\newblock \bibinfo{title}{{GCR environmental models I}: Sensitivity analysis
  for gcr environments}.
\newblock \bibinfo{journal}{Space Weather} \bibinfo{volume}{12},
  \bibinfo{pages}{217--224}.
\newblock \DOIprefix\doi{10.1002/2013SW001025}.
%Type = Article
\bibitem[{Smart and Shea(1989)}]{smart89}
\bibinfo{author}{Smart, D.}, \bibinfo{author}{Shea, M.}, \bibinfo{year}{1989}.
\newblock \bibinfo{title}{Pps-87: A new event oriented solar proton prediction
  model}.
\newblock \bibinfo{journal}{Adv. Space Res.} \bibinfo{volume}{9},
  \bibinfo{pages}{281--284}.
\newblock \DOIprefix\doi{10.1016/0273-1177(89)90450-X}.
%Type = Article
\bibitem[{Smart and Shea(1992)}]{smart92}
\bibinfo{author}{Smart, D.}, \bibinfo{author}{Shea, M.}, \bibinfo{year}{1992}.
\newblock \bibinfo{title}{Modeling the time-intensity profile of solar flare
  generated particle fluxes in the inner heliosphere}.
\newblock \bibinfo{journal}{Adv. Space Res.} \bibinfo{volume}{12},
  \bibinfo{pages}{303--312}.
\newblock \DOIprefix\doi{10.1016/0273-1177(92)90120-M}.
%Type = Article
\bibitem[{{Sok{\'o}{\l}} et~al.(2013){Sok{\'o}{\l}}, {Bzowski}, {Tokumaru},
  {Fujiki} and {McComas}}]{sokol2013}
\bibinfo{author}{{Sok{\'o}{\l}}, J.M.}, \bibinfo{author}{{Bzowski}, M.},
  \bibinfo{author}{{Tokumaru}, M.}, \bibinfo{author}{{Fujiki}, K.},
  \bibinfo{author}{{McComas}, D.J.}, \bibinfo{year}{2013}.
\newblock \bibinfo{title}{{Heliolatitude and Time Variations of Solar Wind
  Structure from in situ Measurements and Interplanetary Scintillation
  Observations}}.
\newblock \bibinfo{journal}{Sol. Phys.} \bibinfo{volume}{285},
  \bibinfo{pages}{167--200}.
\newblock \DOIprefix\doi{10.1007/s11207-012-9993-9},
  \href{http://arxiv.org/abs/1112.5249}{{\tt arXiv:1112.5249}}.
%Type = Article
\bibitem[{{Song} et~al.(2021){Song}, {Luo}, {Potgieter}, {Liu} and
  {Geng}}]{Song2021}
\bibinfo{author}{{Song}, X.}, \bibinfo{author}{{Luo}, X.},
  \bibinfo{author}{{Potgieter}, M.S.}, \bibinfo{author}{{Liu}, X.},
  \bibinfo{author}{{Geng}, Z.}, \bibinfo{year}{2021}.
\newblock \bibinfo{title}{{A Numerical Study of the Solar Modulation of
  Galactic Protons and Helium from 2006 to 2017}}.
\newblock \bibinfo{journal}{Astrophys. J. Suppl. Ser.} \bibinfo{volume}{257},
  \bibinfo{pages}{48}.
\newblock \DOIprefix\doi{10.3847/1538-4365/ac281c}.
%Type = Article
\bibitem[{Stassinopoulos and Raymond(1988)}]{stassinopoulos1988space}
\bibinfo{author}{Stassinopoulos, E.}, \bibinfo{author}{Raymond, J.P.},
  \bibinfo{year}{1988}.
\newblock \bibinfo{title}{The space radiation environment for electronics}.
\newblock \bibinfo{journal}{Proceedings of the IEEE} \bibinfo{volume}{76},
  \bibinfo{pages}{1423--1442}.
\newblock \DOIprefix\doi{10.1109/5.90113}.
%Type = Article
\bibitem[{{Stone} et~al.(2019){Stone}, {Cummings}, {Heikkila} and
  {Lal}}]{stone2019}
\bibinfo{author}{{Stone}, E.C.}, \bibinfo{author}{{Cummings}, A.C.},
  \bibinfo{author}{{Heikkila}, B.C.}, \bibinfo{author}{{Lal}, N.},
  \bibinfo{year}{2019}.
\newblock \bibinfo{title}{{Cosmic ray measurements from Voyager 2 as it crossed
  into interstellar space}}.
\newblock \bibinfo{journal}{Nature Astronomy} \bibinfo{volume}{3},
  \bibinfo{pages}{1013--1018}.
\newblock \DOIprefix\doi{10.1038/s41550-019-0928-3}.
%Type = Article
\bibitem[{{Stone} et~al.(2005){Stone}, {Cummings}, {McDonald}, {Heikkila},
  {Lal} and {Webber}}]{Stone2005}
\bibinfo{author}{{Stone}, E.C.}, \bibinfo{author}{{Cummings}, A.C.},
  \bibinfo{author}{{McDonald}, F.B.}, \bibinfo{author}{{Heikkila}, B.C.},
  \bibinfo{author}{{Lal}, N.}, \bibinfo{author}{{Webber}, W.R.},
  \bibinfo{year}{2005}.
\newblock \bibinfo{title}{{Voyager 1 Explores the Termination Shock Region and
  the Heliosheath Beyond}}.
\newblock \bibinfo{journal}{Science} \bibinfo{volume}{309},
  \bibinfo{pages}{2017--2020}.
\newblock \DOIprefix\doi{10.1126/science.1117684}.
%Type = Article
\bibitem[{{Stone} et~al.(2008){Stone}, {Cummings}, {McDonald}, {Heikkila},
  {Lal} and {Webber}}]{Stone2008}
\bibinfo{author}{{Stone}, E.C.}, \bibinfo{author}{{Cummings}, A.C.},
  \bibinfo{author}{{McDonald}, F.B.}, \bibinfo{author}{{Heikkila}, B.C.},
  \bibinfo{author}{{Lal}, N.}, \bibinfo{author}{{Webber}, W.R.},
  \bibinfo{year}{2008}.
\newblock \bibinfo{title}{{An asymmetric solar wind termination shock}}.
\newblock \bibinfo{journal}{Nature} \bibinfo{volume}{454},
  \bibinfo{pages}{71--74}.
\newblock \DOIprefix\doi{10.1038/nature07022}.
%Type = Article
\bibitem[{{Stone} et~al.(2013){Stone}, {Cummings}, {McDonald}, {Heikkila},
  {Lal} and {Webber}}]{Stone2013}
\bibinfo{author}{{Stone}, E.C.}, \bibinfo{author}{{Cummings}, A.C.},
  \bibinfo{author}{{McDonald}, F.B.}, \bibinfo{author}{{Heikkila}, B.C.},
  \bibinfo{author}{{Lal}, N.}, \bibinfo{author}{{Webber}, W.R.},
  \bibinfo{year}{2013}.
\newblock \bibinfo{title}{{Voyager 1 Observes Low-Energy Galactic Cosmic Rays
  in a Region Depleted of Heliospheric Ions}}.
\newblock \bibinfo{journal}{Science} \bibinfo{volume}{341},
  \bibinfo{pages}{150--153}.
\newblock \DOIprefix\doi{10.1126/science.1236408}.
%Type = Article
\bibitem[{Stone et~al.(1998)Stone, Frandsen, Mewaldt, Christian, Margolies,
  Ormes and Snow}]{stone1998ace}
\bibinfo{author}{Stone, E.C.}, \bibinfo{author}{Frandsen, A.},
  \bibinfo{author}{Mewaldt, R.}, \bibinfo{author}{Christian, E.},
  \bibinfo{author}{Margolies, D.}, \bibinfo{author}{Ormes, J.},
  \bibinfo{author}{Snow, F.}, \bibinfo{year}{1998}.
\newblock \bibinfo{title}{The advanced composition explorer}.
\newblock \bibinfo{journal}{Space Sci. Rev.} \bibinfo{volume}{86},
  \bibinfo{pages}{1--22}.
\newblock \DOIprefix\doi{10.1023/A:1005082526237}.
%Type = Article
\bibitem[{Strauss and Fichtner(2015)}]{strauss2015aspects}
\bibinfo{author}{Strauss, R.}, \bibinfo{author}{Fichtner, H.},
  \bibinfo{year}{2015}.
\newblock \bibinfo{title}{On aspects pertaining to the perpendicular diffusion
  of solar energetic particles}.
\newblock \bibinfo{journal}{Astrophys. J.} \bibinfo{volume}{801},
  \bibinfo{pages}{29}.
\newblock \DOIprefix\doi{10.1088/0004-637X/801/1/29}.
%Type = Article
\bibitem[{Strauss et~al.(2016)Strauss, le~Roux, Engelbrecht, Ruffolo and
  Dunzlaff}]{strauss2016}
\bibinfo{author}{Strauss, R.D.}, \bibinfo{author}{le~Roux, J.A.},
  \bibinfo{author}{Engelbrecht, N.E.}, \bibinfo{author}{Ruffolo, D.},
  \bibinfo{author}{Dunzlaff, P.}, \bibinfo{year}{2016}.
\newblock \bibinfo{title}{Non-axisymmetric perpendicular diffusion of charged
  particles and their transport across tangential magnetic discontinuities}.
\newblock \bibinfo{journal}{Astrophys. J.} \bibinfo{volume}{825},
  \bibinfo{pages}{43}.
\newblock \DOIprefix\doi{10.3847/0004-637X/825/1/43}.
%Type = Article
\bibitem[{{Strauss} and {Effenberger}(2017)}]{strauss17}
\bibinfo{author}{{Strauss}, R.D.T.}, \bibinfo{author}{{Effenberger}, F.},
  \bibinfo{year}{2017}.
\newblock \bibinfo{title}{{A Hitch-hiker's Guide to Stochastic Differential
  Equations. Solution Methods for Energetic Particle Transport in Space Physics
  and Astrophysics}}.
\newblock \bibinfo{journal}{Space Sci. Rev.} \bibinfo{volume}{212},
  \bibinfo{pages}{151--192}.
\newblock \DOIprefix\doi{10.1007/s11214-017-0351-y}.
%Type = Article
\bibitem[{{Teufel} and {Schlickeiser}(2003)}]{Teufel2003}
\bibinfo{author}{{Teufel}, A.}, \bibinfo{author}{{Schlickeiser}, R.},
  \bibinfo{year}{2003}.
\newblock \bibinfo{title}{{Analytic calculation of the parallel mean free path
  of heliospheric cosmic rays. II. Dynamical magnetic slab turbulence and
  random sweeping slab turbulence with finite wave power at small
  wavenumbers}}.
\newblock \bibinfo{journal}{Astron. \& Astrophys.} \bibinfo{volume}{397},
  \bibinfo{pages}{15--25}.
\newblock \DOIprefix\doi{10.1051/0004-6361:20021471}.
%Type = Article
\bibitem[{Thomas et~al.(2014)Thomas, Owens and Lockwood}]{Thomas_etal_2014}
\bibinfo{author}{Thomas, S.R.}, \bibinfo{author}{Owens, M.J.},
  \bibinfo{author}{Lockwood, M.}, \bibinfo{year}{2014}.
\newblock \bibinfo{title}{The 22-year hale cycle in cosmic ray flux--evidence
  for direct heliospheric modulation}.
\newblock \bibinfo{journal}{Sol. Phys.} \bibinfo{volume}{289},
  \bibinfo{pages}{407--421}.
\newblock \DOIprefix\doi{10.1007/s11207-013-0341-5}.
%Type = Article
\bibitem[{{Tomassetti} et~al.(2017){Tomassetti}, {Orcinha}, {Bar{\~a}o} and
  {Bertucci}}]{Tomassetti2017}
\bibinfo{author}{{Tomassetti}, N.}, \bibinfo{author}{{Orcinha}, M.},
  \bibinfo{author}{{Bar{\~a}o}, F.}, \bibinfo{author}{{Bertucci}, B.},
  \bibinfo{year}{2017}.
\newblock \bibinfo{title}{{Evidence for a Time Lag in Solar Modulation of
  Galactic Cosmic Rays}}.
\newblock \bibinfo{journal}{Astrophys. J. Lett.} \bibinfo{volume}{849},
  \bibinfo{pages}{L32}.
\newblock \DOIprefix\doi{10.3847/2041-8213/aa9373}.
%Type = Article
\bibitem[{{Townsend} et~al.(1994){Townsend}, {Cucinotta}, {Wilson}, {Shinn} and
  {Badhwar}}]{Townsend1994}
\bibinfo{author}{{Townsend}, L.W.}, \bibinfo{author}{{Cucinotta}, C.F.},
  \bibinfo{author}{{Wilson}, J.W.}, \bibinfo{author}{{Shinn}, J.L.},
  \bibinfo{author}{{Badhwar}, G.}, \bibinfo{year}{1994}.
\newblock \bibinfo{title}{{Solar modulation and nuclear fragmentation effects
  in galactic cosmic ray transport through shielding}}.
\newblock \bibinfo{journal}{Adv. Space Res.} \bibinfo{volume}{14},
  \bibinfo{pages}{853--861}.
\newblock \DOIprefix\doi{10.1016/0273-1177(94)90550-9}.
%Type = Article
\bibitem[{Tylka et~al.(2005)Tylka, Cohen, Dietrich, Lee, Maclennan, Mewaldt, Ng
  and Reames}]{tylka2005shock}
\bibinfo{author}{Tylka, A.}, \bibinfo{author}{Cohen, C.},
  \bibinfo{author}{Dietrich, W.}, \bibinfo{author}{Lee, M.},
  \bibinfo{author}{Maclennan, C.}, \bibinfo{author}{Mewaldt, R.},
  \bibinfo{author}{Ng, C.}, \bibinfo{author}{Reames, D.}, \bibinfo{year}{2005}.
\newblock \bibinfo{title}{Shock geometry, seed populations, and the origin of
  variable elemental composition at high energies in large gradual solar
  particle events}.
\newblock \bibinfo{journal}{Astrophys. J.} \bibinfo{volume}{625},
  \bibinfo{pages}{474}.
\newblock \DOIprefix\doi{10.1086/429384}.
%Type = Article
\bibitem[{Tylka et~al.(1997)Tylka, Adams, Boberg, Brownstein, Dietrich,
  Flueckiger, Petersen, Shea, Smart and Smith}]{tylka1997creme96}
\bibinfo{author}{Tylka, A.J.}, \bibinfo{author}{Adams, J.H.},
  \bibinfo{author}{Boberg, P.R.}, \bibinfo{author}{Brownstein, B.},
  \bibinfo{author}{Dietrich, W.F.}, \bibinfo{author}{Flueckiger, E.O.},
  \bibinfo{author}{Petersen, E.L.}, \bibinfo{author}{Shea, M.A.},
  \bibinfo{author}{Smart, D.F.}, \bibinfo{author}{Smith, E.C.},
  \bibinfo{year}{1997}.
\newblock \bibinfo{title}{{CREME96}: A revision of the cosmic ray effects on
  micro-electronics code}.
\newblock \bibinfo{journal}{IEEE Trans. on Nuclear Sci.} \bibinfo{volume}{44},
  \bibinfo{pages}{2150--2160}.
\newblock \DOIprefix\doi{10.1109/23.659030}.
%Type = Article
\bibitem[{Usoskin et~al.(2011)Usoskin, Bazilevskaya and
  Kovaltsov}]{usoskin_etal_2011}
\bibinfo{author}{Usoskin, I.}, \bibinfo{author}{Bazilevskaya, G.},
  \bibinfo{author}{Kovaltsov, G.}, \bibinfo{year}{2011}.
\newblock \bibinfo{title}{Solar modulation parameter for cosmic rays since 1936
  reconstructed from ground-based neutron monitors and ionization chambers}.
\newblock \bibinfo{journal}{J. Geophys. Res.: Space Phys.}
  \bibinfo{volume}{116}, \bibinfo{pages}{A02104}.
\newblock \DOIprefix\doi{10.1029/2010JA016105}.
%Type = Article
\bibitem[{Usoskin et~al.(1998)Usoskin, Kananen, Mursula, Tanskanen and
  Kovaltsov}]{usoskin_etal_1998}
\bibinfo{author}{Usoskin, I.}, \bibinfo{author}{Kananen, H.},
  \bibinfo{author}{Mursula, K.}, \bibinfo{author}{Tanskanen, P.},
  \bibinfo{author}{Kovaltsov, G.}, \bibinfo{year}{1998}.
\newblock \bibinfo{title}{Correlative study of solar activity and cosmic ray
  intensity}.
\newblock \bibinfo{journal}{J. Geophys. Res.: Space Phys.}
  \bibinfo{volume}{103}, \bibinfo{pages}{9567--9574}.
\newblock \DOIprefix\doi{10.1029/97JA03782}.
%Type = Article
\bibitem[{Usoskin et~al.(2020)Usoskin, Koldobskiy, Kovaltsov, Gil, Usoskina,
  Willamo and Ibragimov}]{usoskin2020revised}
\bibinfo{author}{Usoskin, I.}, \bibinfo{author}{Koldobskiy, S.},
  \bibinfo{author}{Kovaltsov, G.}, \bibinfo{author}{Gil, A.},
  \bibinfo{author}{Usoskina, I.}, \bibinfo{author}{Willamo, T.},
  \bibinfo{author}{Ibragimov, A.}, \bibinfo{year}{2020}.
\newblock \bibinfo{title}{Revised {GLE} database: Fluences of solar energetic
  particles as measured by the neutron-monitor network since 1956}.
\newblock \bibinfo{journal}{Astron. \& Astrophys.} \bibinfo{volume}{640},
  \bibinfo{pages}{A17}.
\newblock \DOIprefix\doi{10.1051/0004-6361/202038272}.
%Type = Article
\bibitem[{Usoskin et~al.(2015)Usoskin, Kovaltsov, Adriani, Barbarino,
  Bazilevskaya, Bellotti, Boezio, Bogomolov, Bongi, Bonvicini
  et~al.}]{usoskin2015force}
\bibinfo{author}{Usoskin, I.}, \bibinfo{author}{Kovaltsov, G.},
  \bibinfo{author}{Adriani, O.}, \bibinfo{author}{Barbarino, G.},
  \bibinfo{author}{Bazilevskaya, G.}, \bibinfo{author}{Bellotti, R.},
  \bibinfo{author}{Boezio, M.}, \bibinfo{author}{Bogomolov, E.},
  \bibinfo{author}{Bongi, M.}, \bibinfo{author}{Bonvicini, V.}, et~al.,
  \bibinfo{year}{2015}.
\newblock \bibinfo{title}{Force-field parameterization of the galactic cosmic
  ray spectrum: Validation for forbush decreases}.
\newblock \bibinfo{journal}{Adv. Space Res.} \bibinfo{volume}{55},
  \bibinfo{pages}{2940--2945}.
%Type = Article
\bibitem[{Usoskin(2017)}]{usoskin2017history}
\bibinfo{author}{Usoskin, I.G.}, \bibinfo{year}{2017}.
\newblock \bibinfo{title}{A history of solar activity over millennia}.
\newblock \bibinfo{journal}{Living Reviews in Solar Physics}
  \bibinfo{volume}{14}, \bibinfo{pages}{3}.
\newblock \DOIprefix\doi{10.1007/s41116-017-0006-9}.
%Type = Article
\bibitem[{Usoskin et~al.(2017)Usoskin, Gil, Kovaltsov, Mishev and
  Mikhailov}]{usoskin2017}
\bibinfo{author}{Usoskin, I.G.}, \bibinfo{author}{Gil, A.},
  \bibinfo{author}{Kovaltsov, G.A.}, \bibinfo{author}{Mishev, A.L.},
  \bibinfo{author}{Mikhailov, V.V.}, \bibinfo{year}{2017}.
\newblock \bibinfo{title}{Heliospheric modulation of cosmic rays during the
  neutron monitor era: Calibration using {PAMELA} data for 2006–2010}.
\newblock \bibinfo{journal}{J. Geophys. Res.: Space Phys.}
  \bibinfo{volume}{122}, \bibinfo{pages}{3875--3887}.
\newblock \DOIprefix\doi{10.1002/2016JA023819}.
%Type = Article
\bibitem[{{Vainio} et~al.(2009){Vainio}, {Desorgher}, {Heynderickx}, {Storini},
  {Fl{\"u}ckiger}, {Horne}, {Kovaltsov}, {Kudela}, {Laurenza},
  {McKenna-Lawlor}, {Rothkaehl} and {Usoskin}}]{vainio09}
\bibinfo{author}{{Vainio}, R.}, \bibinfo{author}{{Desorgher}, L.},
  \bibinfo{author}{{Heynderickx}, D.}, \bibinfo{author}{{Storini}, M.},
  \bibinfo{author}{{Fl{\"u}ckiger}, E.}, \bibinfo{author}{{Horne}, R.B.},
  \bibinfo{author}{{Kovaltsov}, G.A.}, \bibinfo{author}{{Kudela}, K.},
  \bibinfo{author}{{Laurenza}, M.}, \bibinfo{author}{{McKenna-Lawlor}, S.},
  \bibinfo{author}{{Rothkaehl}, H.}, \bibinfo{author}{{Usoskin}, I.G.},
  \bibinfo{year}{2009}.
\newblock \bibinfo{title}{{Dynamics of the Earth's Particle Radiation
  Environment}}.
\newblock \bibinfo{journal}{Space Sci. Rev.} \bibinfo{volume}{147},
  \bibinfo{pages}{187--231}.
\newblock \DOIprefix\doi{10.1007/s11214-009-9496-7}.
%Type = Article
\bibitem[{Vainio et~al.(2017)Vainio, Raukunen, Tylka, Dietrich and
  Afanasiev}]{vainio2017solar}
\bibinfo{author}{Vainio, R.}, \bibinfo{author}{Raukunen, O.},
  \bibinfo{author}{Tylka, A.J.}, \bibinfo{author}{Dietrich, W.F.},
  \bibinfo{author}{Afanasiev, A.}, \bibinfo{year}{2017}.
\newblock \bibinfo{title}{Why is solar cycle 24 an inefficient producer of
  high-energy particle events?}
\newblock \bibinfo{journal}{Astron. \& Astrophys.} \bibinfo{volume}{604},
  \bibinfo{pages}{A47}.
\newblock \DOIprefix\doi{10.1051/0004-6361/201730547}.
%Type = Article
\bibitem[{V{\"a}is{\"a}nen et~al.(2021)V{\"a}is{\"a}nen, Usoskin and
  Mursula}]{vaisanen2021seven}
\bibinfo{author}{V{\"a}is{\"a}nen, P.}, \bibinfo{author}{Usoskin, I.},
  \bibinfo{author}{Mursula, K.}, \bibinfo{year}{2021}.
\newblock \bibinfo{title}{Seven decades of neutron monitors (1951--2019):
  Overview and evaluation of data sources}.
\newblock \bibinfo{journal}{J. Geophys. Res.: Space Phys.}
  \bibinfo{volume}{126}, \bibinfo{pages}{e2020JA028941}.
\newblock \DOIprefix\doi{10.1029/2020JA028941}.
%Type = Article
\bibitem[{Van~Allen(2000)}]{vanAllen2000}
\bibinfo{author}{Van~Allen, J.A.}, \bibinfo{year}{2000}.
\newblock \bibinfo{title}{On the modulation of galactic cosmic ray intensity
  during solar activity cycles 19, 20, 21, 22 and early 23}.
\newblock \bibinfo{journal}{Geophys. Res. Lett.} \bibinfo{volume}{27},
  \bibinfo{pages}{2453--2456}.
\newblock \DOIprefix\doi{10.1029/2000GL003792}.
%Type = Article
\bibitem[{{van den Berg} et~al.(2020){van den Berg}, {Strauss} and
  {Effenberger}}]{Ber2020}
\bibinfo{author}{{van den Berg}, J.}, \bibinfo{author}{{Strauss}, D.T.},
  \bibinfo{author}{{Effenberger}, F.}, \bibinfo{year}{2020}.
\newblock \bibinfo{title}{{A Primer on Focused Solar Energetic Particle
  Transport}}.
\newblock \bibinfo{journal}{Space Sci. Rev.} \bibinfo{volume}{216},
  \bibinfo{pages}{146}.
\newblock \DOIprefix\doi{10.1007/s11214-020-00771-x}.
%Type = Article
\bibitem[{{Vennerstrom} et~al.(2016){Vennerstrom}, {Lefevre}, {Dumbovi{\'c}},
  {Crosby}, {Malandraki}, {Patsou}, {Clette}, {Veronig}, {Vr{\v{s}}nak}, {Leer}
  and {Moretto}}]{vennerstrom16}
\bibinfo{author}{{Vennerstrom}, S.}, \bibinfo{author}{{Lefevre}, L.},
  \bibinfo{author}{{Dumbovi{\'c}}, M.}, \bibinfo{author}{{Crosby}, N.},
  \bibinfo{author}{{Malandraki}, O.}, \bibinfo{author}{{Patsou}, I.},
  \bibinfo{author}{{Clette}, F.}, \bibinfo{author}{{Veronig}, A.},
  \bibinfo{author}{{Vr{\v{s}}nak}, B.}, \bibinfo{author}{{Leer}, K.},
  \bibinfo{author}{{Moretto}, T.}, \bibinfo{year}{2016}.
\newblock \bibinfo{title}{{Extreme Geomagnetic Storms - 1868 - 2010}}.
\newblock \bibinfo{journal}{Sol. Phys.} \bibinfo{volume}{291},
  \bibinfo{pages}{1447--1481}.
\newblock \DOIprefix\doi{10.1007/s11207-016-0897-y}.
%Type = Article
\bibitem[{{Vos} and {Potgieter}(2016)}]{Vos2016}
\bibinfo{author}{{Vos}, E.E.}, \bibinfo{author}{{Potgieter}, M.S.},
  \bibinfo{year}{2016}.
\newblock \bibinfo{title}{{Global Gradients for Cosmic-Ray Protons in the
  Heliosphere During the Solar Minimum of Cycle 23/24}}.
\newblock \bibinfo{journal}{Sol. Phys.} \bibinfo{volume}{291},
  \bibinfo{pages}{2181--2195}.
\newblock \DOIprefix\doi{10.1007/s11207-016-0945-7}.
%Type = Article
\bibitem[{Vourlidas(2015)}]{Vourlidas-2015-L5mission}
\bibinfo{author}{Vourlidas, A.}, \bibinfo{year}{2015}.
\newblock \bibinfo{title}{Mission to the {Sun}-{Earth} {L5 Lagrangian} point:
  An optimal platform for space weather research}.
\newblock \bibinfo{journal}{Space Weather} \bibinfo{volume}{13},
  \bibinfo{pages}{197--201}.
\newblock \DOIprefix\doi{10.1002/2015SW001173}.
%Type = Article
\bibitem[{{Vr{\v{s}}nak} et~al.(2022){Vr{\v{s}}nak}, {Dumbovi{\'c}}, {Heber}
  and {Kirin}}]{vrsnak22}
\bibinfo{author}{{Vr{\v{s}}nak}, B.}, \bibinfo{author}{{Dumbovi{\'c}}, M.},
  \bibinfo{author}{{Heber}, B.}, \bibinfo{author}{{Kirin}, A.},
  \bibinfo{year}{2022}.
\newblock \bibinfo{title}{{Analytic modeling of recurrent Forbush decreases
  caused by corotating interaction regions}}.
\newblock \bibinfo{journal}{Astron. \& Astrophys.} \bibinfo{volume}{658},
  \bibinfo{pages}{A186}.
\newblock \DOIprefix\doi{10.1051/0004-6361/202140846}.
%Type = Article
\bibitem[{{Wang} et~al.(2019){Wang}, {Bi}, {Fang}, {Lin} and {Yin}}]{Wang2019}
\bibinfo{author}{{Wang}, B.B.}, \bibinfo{author}{{Bi}, X.J.},
  \bibinfo{author}{{Fang}, K.}, \bibinfo{author}{{Lin}, S.J.},
  \bibinfo{author}{{Yin}, P.F.}, \bibinfo{year}{2019}.
\newblock \bibinfo{title}{{Time-dependent solar modulation of cosmic rays from
  solar minimum to solar maximum}}.
\newblock \bibinfo{journal}{Phys. Rev. D} \bibinfo{volume}{100},
  \bibinfo{pages}{063006}.
\newblock \DOIprefix\doi{10.1103/PhysRevD.100.063006}.
%Type = Article
\bibitem[{{Wang} et~al.(2022a){Wang}, {Bi}, {Fang}, {Lin} and {Yin}}]{Wang2022}
\bibinfo{author}{{Wang}, B.B.}, \bibinfo{author}{{Bi}, X.J.},
  \bibinfo{author}{{Fang}, K.}, \bibinfo{author}{{Lin}, S.J.},
  \bibinfo{author}{{Yin}, P.F.}, \bibinfo{year}{2022}a.
\newblock \bibinfo{title}{{Solar modulation of cosmic proton and helium with
  AMS-02}}.
\newblock \bibinfo{journal}{Phys. Rev. D} \bibinfo{volume}{106},
  \bibinfo{pages}{063006}.
\newblock \DOIprefix\doi{10.1103/PhysRevD.106.063006}.
%Type = Article
\bibitem[{{Wang} et~al.(2022b){Wang}, {Zank}, {Adhikari} and
  {Zhao}}]{Wang2022turbulence}
\bibinfo{author}{{Wang}, B.B.}, \bibinfo{author}{{Zank}, G.P.},
  \bibinfo{author}{{Adhikari}, L.}, \bibinfo{author}{{Zhao}, L.L.},
  \bibinfo{year}{2022}b.
\newblock \bibinfo{title}{{On the Conservation of Turbulence Energy in
  Turbulence Transport Models}}.
\newblock \bibinfo{journal}{Astrophys. J.} \bibinfo{volume}{928},
  \bibinfo{pages}{176}.
\newblock \DOIprefix\doi{10.3847/1538-4357/ac596e}.
%Type = Article
\bibitem[{Wang et~al.(2012)Wang, Lin, Krucker and Mason}]{wangLY2012}
\bibinfo{author}{Wang, L.}, \bibinfo{author}{Lin, R.},
  \bibinfo{author}{Krucker, S.}, \bibinfo{author}{Mason, G.M.},
  \bibinfo{year}{2012}.
\newblock \bibinfo{title}{A statistical study of solar electron events over one
  solar cycle}.
\newblock \bibinfo{journal}{Astrophys. J.} \bibinfo{volume}{759},
  \bibinfo{pages}{69}.
\newblock \DOIprefix\doi{10.1088/0004-637X/759/1/69}.
%Type = Article
\bibitem[{{Wang} et~al.(2015){Wang}, {Yang}, {He}, {Tu}, {Pei},
  {Wimmer-Schweingruber} and {Bale}}]{wang2015}
\bibinfo{author}{{Wang}, L.}, \bibinfo{author}{{Yang}, L.},
  \bibinfo{author}{{He}, J.}, \bibinfo{author}{{Tu}, C.},
  \bibinfo{author}{{Pei}, Z.}, \bibinfo{author}{{Wimmer-Schweingruber}, R.F.},
  \bibinfo{author}{{Bale}, S.D.}, \bibinfo{year}{2015}.
\newblock \bibinfo{title}{{Solar Wind {\ensuremath{\sim}}20-200 keV Superhalo
  Electrons at Quiet Times}}.
\newblock \bibinfo{journal}{Astrophys. J. Lett.} \bibinfo{volume}{803},
  \bibinfo{pages}{L2}.
\newblock \DOIprefix\doi{10.1088/2041-8205/803/1/L2}.
%Type = Article
\bibitem[{Wang et~al.(2023)Wang, Bai, Chen, Chen, Cheng, Deng, Deng, Deng,
  Feng, Gou, Guo, Guo, Hao, He, Hou, Huang, Huang, Ji, Jiang, Jiang, Jin, Li,
  Li, Liu, Liu, Liu, Liu, Liu, Qiu, Shen, Shen, Shen, Shi, Su, Su, Su, Sun,
  Tan, Tian, Wang, Xia, Xie, Xiong, Xu, Yan, Yan, Yang, Yang, Zhang, Zhang,
  Zhang, Zhao, Zhou and Zou}]{wang2023ring}
\bibinfo{author}{Wang, Y.}, \bibinfo{author}{Bai, X.}, \bibinfo{author}{Chen,
  C.}, \bibinfo{author}{Chen, L.}, \bibinfo{author}{Cheng, X.},
  \bibinfo{author}{Deng, L.}, \bibinfo{author}{Deng, L.},
  \bibinfo{author}{Deng, Y.}, \bibinfo{author}{Feng, L.}, \bibinfo{author}{Gou,
  T.}, \bibinfo{author}{Guo, J.}, \bibinfo{author}{Guo, Y.},
  \bibinfo{author}{Hao, X.}, \bibinfo{author}{He, J.}, \bibinfo{author}{Hou,
  J.}, \bibinfo{author}{Huang, J.}, \bibinfo{author}{Huang, Z.},
  \bibinfo{author}{Ji, H.}, \bibinfo{author}{Jiang, C.},
  \bibinfo{author}{Jiang, J.}, \bibinfo{author}{Jin, C.}, \bibinfo{author}{Li,
  X.}, \bibinfo{author}{Li, Y.}, \bibinfo{author}{Liu, J.},
  \bibinfo{author}{Liu, K.}, \bibinfo{author}{Liu, L.}, \bibinfo{author}{Liu,
  R.}, \bibinfo{author}{Liu, R.}, \bibinfo{author}{Qiu, C.},
  \bibinfo{author}{Shen, C.}, \bibinfo{author}{Shen, F.},
  \bibinfo{author}{Shen, Y.}, \bibinfo{author}{Shi, X.}, \bibinfo{author}{Su,
  J.}, \bibinfo{author}{Su, Y.}, \bibinfo{author}{Su, Y.},
  \bibinfo{author}{Sun, M.}, \bibinfo{author}{Tan, B.}, \bibinfo{author}{Tian,
  H.}, \bibinfo{author}{Wang, Y.}, \bibinfo{author}{Xia, L.},
  \bibinfo{author}{Xie, J.}, \bibinfo{author}{Xiong, M.}, \bibinfo{author}{Xu,
  M.}, \bibinfo{author}{Yan, X.}, \bibinfo{author}{Yan, Y.},
  \bibinfo{author}{Yang, S.}, \bibinfo{author}{Yang, S.},
  \bibinfo{author}{Zhang, S.}, \bibinfo{author}{Zhang, Q.},
  \bibinfo{author}{Zhang, Y.}, \bibinfo{author}{Zhao, J.},
  \bibinfo{author}{Zhou, G.}, \bibinfo{author}{Zou, H.}, \bibinfo{year}{2023}.
\newblock \bibinfo{title}{Solar ring mission: Building a panorama of the sun
  and inner-heliosphere}.
\newblock \bibinfo{journal}{Adv. Space Res.} \bibinfo{volume}{71},
  \bibinfo{pages}{1146--1164}.
\newblock \DOIprefix\doi{10.1016/j.asr.2022.10.045}.
%Type = Article
\bibitem[{Wang et~al.(2022)Wang, Guo, Li, Roussos and Zhao}]{wang2022variation}
\bibinfo{author}{Wang, Y.}, \bibinfo{author}{Guo, J.}, \bibinfo{author}{Li,
  G.}, \bibinfo{author}{Roussos, E.}, \bibinfo{author}{Zhao, J.},
  \bibinfo{year}{2022}.
\newblock \bibinfo{title}{Variation in cosmic-ray intensity lags sunspot
  number: Implications of late opening of solar magnetic field}.
\newblock \bibinfo{journal}{Astrophys. J.} \bibinfo{volume}{928},
  \bibinfo{pages}{157}.
\newblock \DOIprefix\doi{10.3847/1538-4357/ac5896}.
%Type = Article
\bibitem[{{Waterfall} et~al.(2022){Waterfall}, {Dalla}, {Laitinen},
  {Hutchinson} and {Marsh}}]{Wat2022}
\bibinfo{author}{{Waterfall}, C.O.G.}, \bibinfo{author}{{Dalla}, S.},
  \bibinfo{author}{{Laitinen}, T.}, \bibinfo{author}{{Hutchinson}, A.},
  \bibinfo{author}{{Marsh}, M.}, \bibinfo{year}{2022}.
\newblock \bibinfo{title}{{Modeling the Transport of Relativistic Solar Protons
  along a Heliospheric Current Sheet during Historic GLE Events}}.
\newblock \bibinfo{journal}{Astrophys. J.} \bibinfo{volume}{934},
  \bibinfo{pages}{82}.
\newblock \DOIprefix\doi{10.3847/1538-4357/ac795d}.
%Type = Article
\bibitem[{Whitman et~al.(2022)Whitman, Egeland, Richardson, Allison
  et~al.}]{Whitman22}
\bibinfo{author}{Whitman, K.}, \bibinfo{author}{Egeland, R.},
  \bibinfo{author}{Richardson, I.G.}, \bibinfo{author}{Allison, C.}, et~al.,
  \bibinfo{year}{2022}.
\newblock \bibinfo{title}{Review of solar energetic particle models}.
\newblock \bibinfo{journal}{Adv. Space Res.}
  \DOIprefix\doi{10.1016/j.asr.2022.08.006}.
%Type = Article
\bibitem[{Wibberenz and Cane(2006)}]{wibberenz2006multi}
\bibinfo{author}{Wibberenz, G.}, \bibinfo{author}{Cane, H.},
  \bibinfo{year}{2006}.
\newblock \bibinfo{title}{Multi-spacecraft observations of solar flare
  particles in the inner heliosphere}.
\newblock \bibinfo{journal}{Astrophys. J.} \bibinfo{volume}{650},
  \bibinfo{pages}{1199}.
\newblock \DOIprefix\doi{10.1086/506598}.
%Type = Article
\bibitem[{Wiedenbeck et~al.(2012)Wiedenbeck, Mason, Cohen, Nitta,
  G{\'{o}}mez-Herrero and Haggerty}]{Wiedenbeck2012}
\bibinfo{author}{Wiedenbeck, M.E.}, \bibinfo{author}{Mason, G.M.},
  \bibinfo{author}{Cohen, C.M.S.}, \bibinfo{author}{Nitta, N.V.},
  \bibinfo{author}{G{\'{o}}mez-Herrero, R.}, \bibinfo{author}{Haggerty, D.K.},
  \bibinfo{year}{2012}.
\newblock \bibinfo{title}{Observations of solar energetic particles from
  3he-rich events over a wide range of heliographic longitude}.
\newblock \bibinfo{journal}{Astrophys. J.} \bibinfo{volume}{762},
  \bibinfo{pages}{54}.
\newblock \DOIprefix\doi{10.1088/0004-637x/762/1/54}.
%Type = Article
\bibitem[{{Wiengarten} et~al.(2016){Wiengarten}, {Oughton}, {Engelbrecht},
  {Fichtner}, {Kleimann} and {Scherer}}]{Wiengarten2016}
\bibinfo{author}{{Wiengarten}, T.}, \bibinfo{author}{{Oughton}, S.},
  \bibinfo{author}{{Engelbrecht}, N.E.}, \bibinfo{author}{{Fichtner}, H.},
  \bibinfo{author}{{Kleimann}, J.}, \bibinfo{author}{{Scherer}, K.},
  \bibinfo{year}{2016}.
\newblock \bibinfo{title}{{A Generalized Two-component Model of Solar Wind
  Turbulence and ab initio Diffusion Mean-Free Paths and Drift Lengthscales of
  Cosmic Rays}}.
\newblock \bibinfo{journal}{Astrophys. J.} \bibinfo{volume}{833},
  \bibinfo{pages}{17}.
\newblock \DOIprefix\doi{10.3847/0004-637X/833/1/17},
  \href{http://arxiv.org/abs/1609.08271}{{\tt arXiv:1609.08271}}.
%Type = Article
\bibitem[{{Wijsen} et~al.(2022){Wijsen}, {Aran}, {Scolini}, {Lario},
  {Afanasiev}, {Vainio}, {Sanahuja}, {Pomoell} and {Poedts}}]{Wij2022}
\bibinfo{author}{{Wijsen}, N.}, \bibinfo{author}{{Aran}, A.},
  \bibinfo{author}{{Scolini}, C.}, \bibinfo{author}{{Lario}, D.},
  \bibinfo{author}{{Afanasiev}, A.}, \bibinfo{author}{{Vainio}, R.},
  \bibinfo{author}{{Sanahuja}, B.}, \bibinfo{author}{{Pomoell}, J.},
  \bibinfo{author}{{Poedts}, S.}, \bibinfo{year}{2022}.
\newblock \bibinfo{title}{{Observation-based modelling of the energetic storm
  particle event of 14 July 2012}}.
\newblock \bibinfo{journal}{Astron. \& Astrophys.} \bibinfo{volume}{659},
  \bibinfo{pages}{A187}.
\newblock \DOIprefix\doi{10.1051/0004-6361/202142698}.
%Type = Article
\bibitem[{Wimmer-Schweingruber et~al.(2020)Wimmer-Schweingruber, Yu,
  B{\"o}ttcher, Zhang, Burmeister, Lohf, Guo, Xu, Schuster, Seimetz
  et~al.}]{wimmer2020lunar}
\bibinfo{author}{Wimmer-Schweingruber, R.F.}, \bibinfo{author}{Yu, J.},
  \bibinfo{author}{B{\"o}ttcher, S.I.}, \bibinfo{author}{Zhang, S.},
  \bibinfo{author}{Burmeister, S.}, \bibinfo{author}{Lohf, H.},
  \bibinfo{author}{Guo, J.}, \bibinfo{author}{Xu, Z.},
  \bibinfo{author}{Schuster, B.}, \bibinfo{author}{Seimetz, L.}, et~al.,
  \bibinfo{year}{2020}.
\newblock \bibinfo{title}{\protect{The Lunar Lander Neutron and Dosimetry (LND)
  Experiment on Chang’E 4}}.
\newblock \bibinfo{journal}{Space Sci. Rev.} \bibinfo{volume}{216},
  \bibinfo{pages}{1--40}.
\newblock \DOIprefix\doi{10.1007/s11214-020-00725-3}.
%Type = Article
\bibitem[{{Winslow} et~al.(2018){Winslow}, {Schwadron}, {Lugaz}, {Guo},
  {Joyce}, {Jordan}, {Wilson}, {Spence}, {Lawrence}, {Wimmer-Schweingruber} and
  {Mays}}]{winslow18}
\bibinfo{author}{{Winslow}, R.M.}, \bibinfo{author}{{Schwadron}, N.A.},
  \bibinfo{author}{{Lugaz}, N.}, \bibinfo{author}{{Guo}, J.},
  \bibinfo{author}{{Joyce}, C.J.}, \bibinfo{author}{{Jordan}, A.P.},
  \bibinfo{author}{{Wilson}, J.K.}, \bibinfo{author}{{Spence}, H.E.},
  \bibinfo{author}{{Lawrence}, D.J.}, \bibinfo{author}{{Wimmer-Schweingruber},
  R.F.}, \bibinfo{author}{{Mays}, M.L.}, \bibinfo{year}{2018}.
\newblock \bibinfo{title}{{Opening a Window on ICME-driven GCR Modulation in
  the Inner Solar System}}.
\newblock \bibinfo{journal}{Astrophys. J.} \bibinfo{volume}{856},
  \bibinfo{pages}{139}.
\newblock \DOIprefix\doi{10.3847/1538-4357/aab098}.
%Type = Article
\bibitem[{{Witasse} et~al.(2017){Witasse}, {S{\'a}nchez-Cano}, {Mays},
  {Kajdi{\v c}}, {Opgenoorth}, {Elliott}, {Richardson}, {Zouganelis}, {Zender},
  {Wimmer-Schweingruber}, {Turc}, {Taylor}, {Roussos}, {Rouillard}, {Richter},
  {Richardson}, {Ramstad}, {Provan}, {Posner}, {Plaut}, {Odstrcil}, {Nilsson},
  {Niemenen}, {Milan}, {Mandt}, {Lohf}, {Lester}, {Lebreton}, {Kuulkers},
  {Krupp}, {Koenders}, {James}, {Intzekara}, {Holmstrom}, {Hassler}, {Hall},
  {Guo}, {Goldstein}, {Goetz}, {Glassmeier}, {G{\'e}not}, {Evans}, {Espley},
  {Edberg}, {Dougherty}, {Cowley}, {Burch}, {Behar}, {Barabash}, {Andrews} and
  {Altobelli}}]{witasse17}
\bibinfo{author}{{Witasse}, O.}, \bibinfo{author}{{S{\'a}nchez-Cano}, B.},
  \bibinfo{author}{{Mays}, M.L.}, \bibinfo{author}{{Kajdi{\v c}}, P.},
  \bibinfo{author}{{Opgenoorth}, H.}, \bibinfo{author}{{Elliott}, H.A.},
  \bibinfo{author}{{Richardson}, I.G.}, \bibinfo{author}{{Zouganelis}, I.},
  \bibinfo{author}{{Zender}, J.}, \bibinfo{author}{{Wimmer-Schweingruber},
  R.F.}, \bibinfo{author}{{Turc}, L.}, \bibinfo{author}{{Taylor}, M.G.G.T.},
  \bibinfo{author}{{Roussos}, E.}, \bibinfo{author}{{Rouillard}, A.},
  \bibinfo{author}{{Richter}, I.}, \bibinfo{author}{{Richardson}, J.D.},
  \bibinfo{author}{{Ramstad}, R.}, \bibinfo{author}{{Provan}, G.},
  \bibinfo{author}{{Posner}, A.}, \bibinfo{author}{{Plaut}, J.J.},
  \bibinfo{author}{{Odstrcil}, D.}, \bibinfo{author}{{Nilsson}, H.},
  \bibinfo{author}{{Niemenen}, P.}, \bibinfo{author}{{Milan}, S.E.},
  \bibinfo{author}{{Mandt}, K.}, \bibinfo{author}{{Lohf}, H.},
  \bibinfo{author}{{Lester}, M.}, \bibinfo{author}{{Lebreton}, J.P.},
  \bibinfo{author}{{Kuulkers}, E.}, \bibinfo{author}{{Krupp}, N.},
  \bibinfo{author}{{Koenders}, C.}, \bibinfo{author}{{James}, M.K.},
  \bibinfo{author}{{Intzekara}, D.}, \bibinfo{author}{{Holmstrom}, M.},
  \bibinfo{author}{{Hassler}, D.M.}, \bibinfo{author}{{Hall}, B.E.S.},
  \bibinfo{author}{{Guo}, J.}, \bibinfo{author}{{Goldstein}, R.},
  \bibinfo{author}{{Goetz}, C.}, \bibinfo{author}{{Glassmeier}, K.H.},
  \bibinfo{author}{{G{\'e}not}, V.}, \bibinfo{author}{{Evans}, H.},
  \bibinfo{author}{{Espley}, J.}, \bibinfo{author}{{Edberg}, N.J.T.},
  \bibinfo{author}{{Dougherty}, M.}, \bibinfo{author}{{Cowley}, S.W.H.},
  \bibinfo{author}{{Burch}, J.}, \bibinfo{author}{{Behar}, E.},
  \bibinfo{author}{{Barabash}, S.}, \bibinfo{author}{{Andrews}, D.J.},
  \bibinfo{author}{{Altobelli}, N.}, \bibinfo{year}{2017}.
\newblock \bibinfo{title}{{Interplanetary coronal mass ejection observed at
  STEREO-A, Mars, comet 67P/Churyumov-Gerasimenko, Saturn, and New Horizons en
  route to Pluto: Comparison of its Forbush decreases at 1.4, 3.1, and 9.9
  AU}}.
\newblock \bibinfo{journal}{J. Geophys. Res.: Space Phys.}
  \bibinfo{volume}{122}, \bibinfo{pages}{7865--7890}.
\newblock \DOIprefix\doi{10.1002/2017JA023884}.
%Type = Article
\bibitem[{{Wozniak} et~al.(2023){Wozniak}, {Iskra}, {Modzelewska} and
  {Siluszyk}}]{Wozniak2023}
\bibinfo{author}{{Wozniak}, W.}, \bibinfo{author}{{Iskra}, K.},
  \bibinfo{author}{{Modzelewska}, R.}, \bibinfo{author}{{Siluszyk}, M.},
  \bibinfo{year}{2023}.
\newblock \bibinfo{title}{{Analysis of Galactic Cosmic Ray Anisotropy During
  the Time Period from 1996 to 2020}}.
\newblock \bibinfo{journal}{Sol. Phys.} \bibinfo{volume}{298},
  \bibinfo{pages}{28}.
\newblock \DOIprefix\doi{10.1007/s11207-023-02120-x}.
%Type = Article
\bibitem[{Xie et~al.(2019)Xie, St.~Cyr, Mäkelä and Gopalswamy}]{Xie2019}
\bibinfo{author}{Xie, H.}, \bibinfo{author}{St.~Cyr, O.C.},
  \bibinfo{author}{Mäkelä, P.}, \bibinfo{author}{Gopalswamy, N.},
  \bibinfo{year}{2019}.
\newblock \bibinfo{title}{Statistical study on multispacecraft widespread solar
  energetic particle events during solar cycle 24}.
\newblock \bibinfo{journal}{J. Geophys. Res.: Space Phys.}
  \bibinfo{volume}{124}, \bibinfo{pages}{6384--6402}.
\newblock \DOIprefix\doi{10.1029/2019JA026832}.
%Type = Article
\bibitem[{Xu et~al.(2020)Xu, Guo, Wimmer-Schweingruber, von Forstner, Wang,
  Dresing, Lohf, Zhang, Heber and Yang}]{xu2020first}
\bibinfo{author}{Xu, Z.}, \bibinfo{author}{Guo, J.},
  \bibinfo{author}{Wimmer-Schweingruber, R.F.}, \bibinfo{author}{von Forstner,
  J.L.F.}, \bibinfo{author}{Wang, Y.}, \bibinfo{author}{Dresing, N.},
  \bibinfo{author}{Lohf, H.}, \bibinfo{author}{Zhang, S.},
  \bibinfo{author}{Heber, B.}, \bibinfo{author}{Yang, M.},
  \bibinfo{year}{2020}.
\newblock \bibinfo{title}{\protect{First Solar energetic particles measured on
  the Lunar far-side}}.
\newblock \bibinfo{journal}{Astrophys. J. Lett.} \bibinfo{volume}{902},
  \bibinfo{pages}{L30}.
\newblock \DOIprefix\doi{10.3847/2041-8213/abbccc}.
%Type = Article
\bibitem[{{Zank}(1999)}]{Zank1999}
\bibinfo{author}{{Zank}, G.P.}, \bibinfo{year}{1999}.
\newblock \bibinfo{title}{{Interaction of the solar wind with the local
  interstellar medium: a theoretical perspective}}.
\newblock \bibinfo{journal}{Space Sci. Rev.} \bibinfo{volume}{89},
  \bibinfo{pages}{413--688}.
\newblock \DOIprefix\doi{10.1023/A:1005155601277}.
%Type = Article
\bibitem[{{Zank} et~al.(2017){Zank}, {Adhikari}, {Hunana}, {Shiota}, {Bruno}
  and {Telloni}}]{Zank2017}
\bibinfo{author}{{Zank}, G.P.}, \bibinfo{author}{{Adhikari}, L.},
  \bibinfo{author}{{Hunana}, P.}, \bibinfo{author}{{Shiota}, D.},
  \bibinfo{author}{{Bruno}, R.}, \bibinfo{author}{{Telloni}, D.},
  \bibinfo{year}{2017}.
\newblock \bibinfo{title}{{Theory and Transport of Nearly Incompressible
  Magnetohydrodynamic Turbulence}}.
\newblock \bibinfo{journal}{Astrophys. J.} \bibinfo{volume}{835},
  \bibinfo{pages}{147}.
\newblock \DOIprefix\doi{10.3847/1538-4357/835/2/147}.
%Type = Article
\bibitem[{{Zank} et~al.(2012){Zank}, {Dosch}, {Hunana}, {Florinski},
  {Matthaeus} and {Webb}}]{Zank2012}
\bibinfo{author}{{Zank}, G.P.}, \bibinfo{author}{{Dosch}, A.},
  \bibinfo{author}{{Hunana}, P.}, \bibinfo{author}{{Florinski}, V.},
  \bibinfo{author}{{Matthaeus}, W.H.}, \bibinfo{author}{{Webb}, G.M.},
  \bibinfo{year}{2012}.
\newblock \bibinfo{title}{{The Transport of Low-frequency Turbulence in
  Astrophysical Flows. I. Governing Equations}}.
\newblock \bibinfo{journal}{Astrophys. J.} \bibinfo{volume}{745},
  \bibinfo{pages}{35}.
\newblock \DOIprefix\doi{10.1088/0004-637X/745/1/35}.
%Type = Article
\bibitem[{{Zank} et~al.(1998){Zank}, {Matthaeus}, {Bieber} and
  {Moraal}}]{Zank1998}
\bibinfo{author}{{Zank}, G.P.}, \bibinfo{author}{{Matthaeus}, W.H.},
  \bibinfo{author}{{Bieber}, J.W.}, \bibinfo{author}{{Moraal}, H.},
  \bibinfo{year}{1998}.
\newblock \bibinfo{title}{{The radial and latitudinal dependence of the cosmic
  ray diffusion tensor in the heliosphere}}.
\newblock \bibinfo{journal}{J. Geophys. Res.: Space Phys.}
  \bibinfo{volume}{103}, \bibinfo{pages}{2085--2098}.
\newblock \DOIprefix\doi{10.1029/97JA03013}.
%Type = Article
\bibitem[{{Zhao} et~al.(2017){Zhao}, {Adhikari}, {Zank}, {Hu} and
  {Feng}}]{Zhao2017}
\bibinfo{author}{{Zhao}, L.L.}, \bibinfo{author}{{Adhikari}, L.},
  \bibinfo{author}{{Zank}, G.P.}, \bibinfo{author}{{Hu}, Q.},
  \bibinfo{author}{{Feng}, X.S.}, \bibinfo{year}{2017}.
\newblock \bibinfo{title}{{Cosmic Ray Diffusion Tensor throughout the
  Heliosphere Derived from a Nearly Incompressible Magnetohydrodynamic
  Turbulence Model}}.
\newblock \bibinfo{journal}{Astrophys. J.} \bibinfo{volume}{849},
  \bibinfo{pages}{88}.
\newblock \DOIprefix\doi{10.3847/1538-4357/aa932a}.
%Type = Article
\bibitem[{{Zhao} et~al.(2018){Zhao}, {Adhikari}, {Zank}, {Hu} and
  {Feng}}]{Zhao2018}
\bibinfo{author}{{Zhao}, L.L.}, \bibinfo{author}{{Adhikari}, L.},
  \bibinfo{author}{{Zank}, G.P.}, \bibinfo{author}{{Hu}, Q.},
  \bibinfo{author}{{Feng}, X.S.}, \bibinfo{year}{2018}.
\newblock \bibinfo{title}{{Influence of the Solar Cycle on Turbulence
  Properties and Cosmic-Ray Diffusion}}.
\newblock \bibinfo{journal}{Astrophys. J.} \bibinfo{volume}{856},
  \bibinfo{pages}{94}.
\newblock \DOIprefix\doi{10.3847/1538-4357/aab362}.
%Type = Article
\bibitem[{{Zhao} et~al.(2019){Zhao}, {Zank}, {Hu}, {Chen}, {Adhikari},
  {leRoux}, {Cummings}, {Stone} and {Burlaga}}]{Zhao2019}
\bibinfo{author}{{Zhao}, L.L.}, \bibinfo{author}{{Zank}, G.P.},
  \bibinfo{author}{{Hu}, Q.}, \bibinfo{author}{{Chen}, Y.},
  \bibinfo{author}{{Adhikari}, L.}, \bibinfo{author}{{leRoux}, J.A.},
  \bibinfo{author}{{Cummings}, A.}, \bibinfo{author}{{Stone}, E.},
  \bibinfo{author}{{Burlaga}, L.F.}, \bibinfo{year}{2019}.
\newblock \bibinfo{title}{{ACR Proton Acceleration Associated with Reconnection
  Processes beyond the Heliospheric Termination Shock}}.
\newblock \bibinfo{journal}{Astrophys. J.} \bibinfo{volume}{886},
  \bibinfo{pages}{144}.
\newblock \DOIprefix\doi{10.3847/1538-4357/ab4db4}.
%Type = Article
\bibitem[{{Zhu} et~al.(2018){Zhu}, {Yuan} and {Wei}}]{Zhu2018}
\bibinfo{author}{{Zhu}, C.R.}, \bibinfo{author}{{Yuan}, Q.},
  \bibinfo{author}{{Wei}, D.M.}, \bibinfo{year}{2018}.
\newblock \bibinfo{title}{{Studies on Cosmic-Ray Nuclei with Voyager, ACE, and
  AMS-02. I. Local Interstellar Spectra and Solar Modulation}}.
\newblock \bibinfo{journal}{Astrophys. J.} \bibinfo{volume}{863},
  \bibinfo{pages}{119}.
\newblock \DOIprefix\doi{10.3847/1538-4357/aacff9}.
%Type = Article
\bibitem[{{Zurbuchen}(2007)}]{Zurbuchen2007}
\bibinfo{author}{{Zurbuchen}, T.H.}, \bibinfo{year}{2007}.
\newblock \bibinfo{title}{{A New View of the Coupling of the Sun and the
  Heliosphere}}.
\newblock \bibinfo{journal}{Annu. Rev. Astron. Astrophys.}
  \bibinfo{volume}{45}, \bibinfo{pages}{297--338}.
\newblock \DOIprefix\doi{10.1146/annurev.astro.45.010807.154030}.

\end{thebibliography}

%% else use the following coding to input the bibitems directly in the
%% TeX file.

% \begin{thebibliography}{00}

% %% \bibitem{label}
% %% Text of bibliographic item

% \bibitem{}

% \end{thebibliography}
\end{document}